\documentclass[
twoside,
cleardoublepage=empty,
12pt,
bibliography=totoc,
headings=big,
captions=tableheading,
chapterprefix=true,
]{scrreprt}

\usepackage[
a4paper,
left=30mm,
right=30mm,
top=25mm,
bottom=30mm,
bindingoffset=7mm,
]{geometry}

\usepackage[utf8]{inputenc}
\usepackage[T1]{fontenc}

\usepackage{graphicx,url}
\usepackage{titlesec}
\usepackage{multirow}
\usepackage{longtable}
\usepackage{float}
\usepackage[british]{babel}
\usepackage{colortbl}
\usepackage{acronym}
\usepackage{calc}
\usepackage{bbding} %
\usepackage{import}
\usepackage{varwidth}
\usepackage{minibox}
\usepackage{enumerate}
\usepackage{setspace}
\usepackage{amsmath} %
\usepackage[flushleft]{threeparttable}

\usepackage{color}
\usepackage[table]{xcolor}

\usepackage[printwatermark]{xwatermark}

\usepackage[linesnumbered,ruled,algochapter]{algorithm2e}
\DeclareOldFontCommand{\rm}{\normalfont\rmfamily}{\mathrm}
\DeclareOldFontCommand{\sf}{\normalfont\sffamily}{\mathsf}
\DeclareOldFontCommand{\tt}{\normalfont\ttfamily}{\mathtt}
\DeclareOldFontCommand{\bf}{\normalfont\bfseries}{\mathbf}
\DeclareOldFontCommand{\it}{\normalfont\itshape}{\mathit}
\DeclareOldFontCommand{\sl}{\normalfont\slshape}{\@nomath\sl}
\DeclareOldFontCommand{\sc}{\normalfont\scshape}{\@nomath\sc}
\DeclareRobustCommand*\cal{\@fontswitch\relax\mathcal}
\DeclareRobustCommand*\mit{\@fontswitch\relax\mathnormal}

\usepackage{breakcites}
\usepackage[breaklinks=true]{hyperref}
\hypersetup{
  colorlinks   = true, %
  urlcolor     = black, %
  linkcolor    = black, %
  citecolor   = black %
}
\newif\iflatextortf  %
\latextortffalse  %
\usepackage{lmodern}
\addtokomafont{chapterprefix}{\raggedleft}
\addtokomafont{section}{\Large}
\addtokomafont{subsection}{\large}
\addtokomafont{subsubsection}{\normalsize }

\usepackage[nohints]{minitoc}

\newcommand{\monthname}[1]{%
\ifcase#1
\or January%
\or February%
\or March%
\or April%
\or May%
\or June%
\or July%
\or August%
\or September%
\or October%
\or November%
\or December%
\fi}

\renewcommand{\title}{Mechanisms for Resilient \\ Video Transmission }
\newcommand{\titlePT}{Mecanismos para transmiss\~{a}o \\ resiliente de v\'{i}deos}
\renewcommand{\author}{Roger Kreutz Immich}

\newcommand{\FirstAdvisor}{Prof. Dr. Marilia Pascoal Curado}
\newcommand{\SecondAdvisor}{Prof. Dr. Eduardo Cerqueira}
\newcommand{\submissiondate}{February, 2017} %

 \newenvironment{lyxlist}[1]
   {\begin{list}{}
     {\settowidth{\labelwidth}{#1}
      \setlength{\leftmargin}{\labelwidth}
      \addtolength{\leftmargin}{\labelsep}
      }}
   {\end{list}}

\usepackage{fancyhdr}
\fancypagestyle{plain}{%
	\fancyhf{}
	\fancyfoot[C]{\thepage}
	\renewcommand{\headrulewidth}{0pt}
}

\usepackage[parfill]{parskip}

\usepackage{lettrine}

\usepackage{wallpaper}

\begin{document} \sloppy
\pagestyle{empty}
\pagenumbering{roman}

\newif\ifBW
\newif\ifDRAFT
\ifDRAFT \else

\ULCornerWallPaper{1}{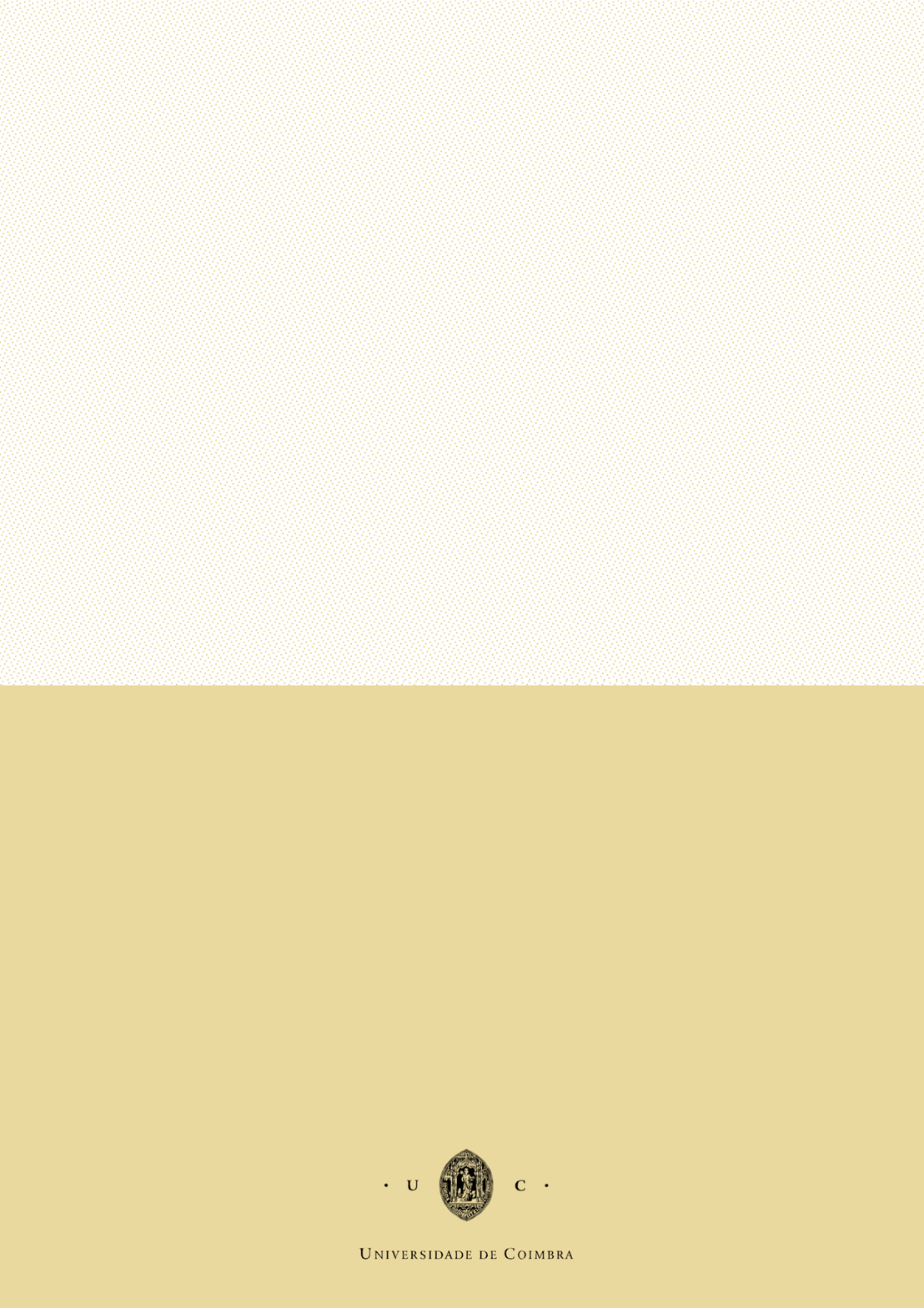}

\newgeometry{left=0cm,right=0cm,bottom=0cm,top=0cm}

\begin{figure}[!htb]
	\includegraphics[width=210mm]{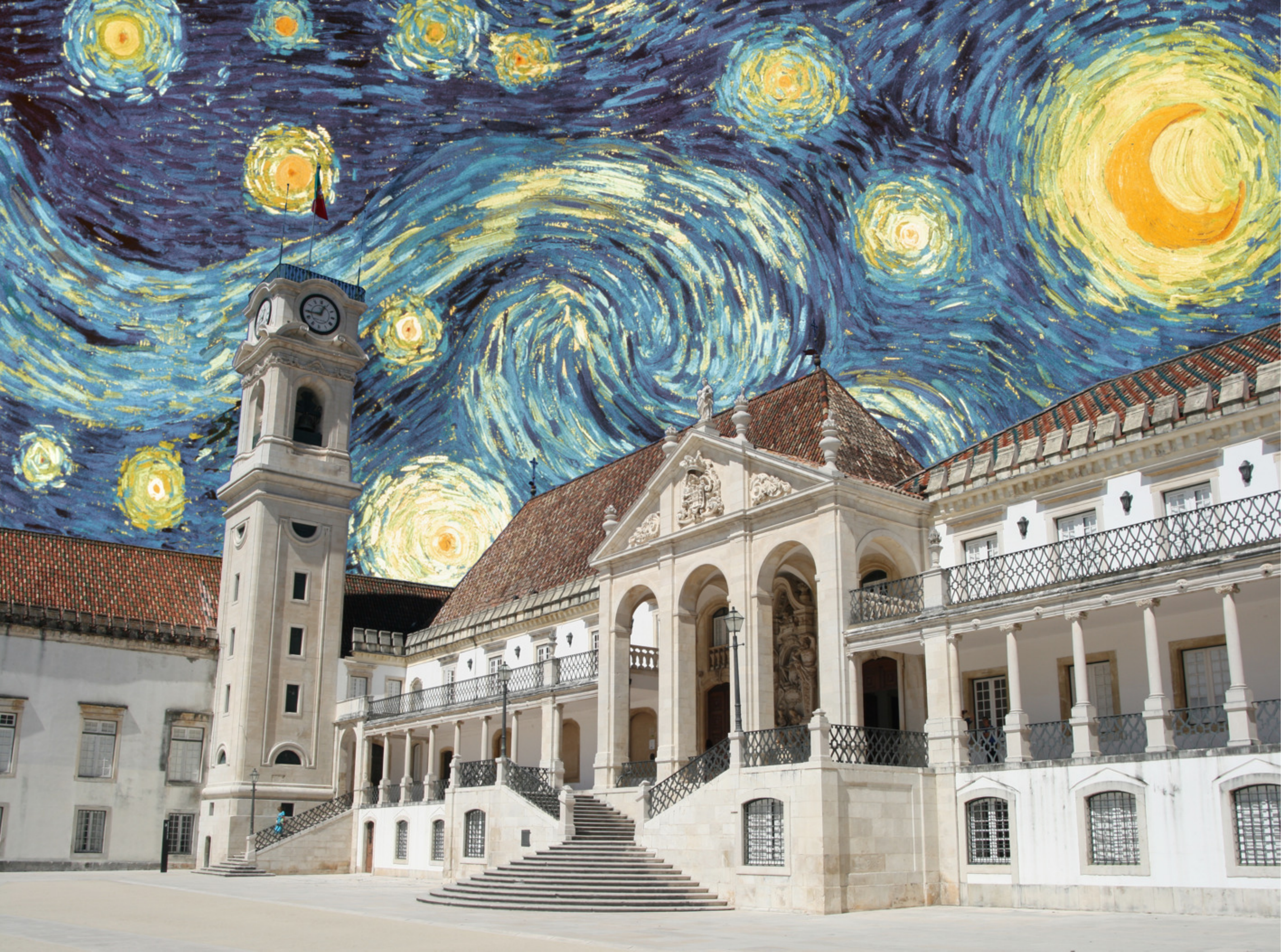}
\end{figure}

\vskip 1cm

\begin{center}
	\begin{minipage}{12cm}
	\thispagestyle{empty}

	\begin{center}

	{\fontsize{14}{20}\selectfont{} Roger Kreutz Immich}
	\vskip 0.6cm

	\begin{spacing}{2}
		{\fontsize{24}{20}\selectfont{} Mechanisms for Resilient \\ Video Transmission}
	\end{spacing}
	\end{center}

	\vskip 0.6cm

	{\fontsize{10}{12}\selectfont{} 
	Tese de Doutoramento do Programa de Doutoramento em Ci\^{e}ncias e Tecnologias da Informa\c{c}\~{a}o, orientada pela Professora Doutora Marilia Curado e pelo Professor Doutor Eduardo Cerqueira e apresentada ao Departamento de Engenharia Inform\'{a}tica da Faculdade de Ci\^{e}ncias e Tecnologia da Universidade de Coimbra
	}

	\vskip 0.8cm

	\begin{center}
	{\fontsize{10}{20}\selectfont{}Fevereiro de 2017}
	\end{center}

	\end{minipage}
\end{center}

\newpage
\ClearWallPaper
\restoregeometry
 	\cleardoublepage
\newenvironment{FrontpageEN}%
{
	\cleardoublepage
	\thispagestyle{empty}
}
{\vfill\null}
\begin{FrontpageEN}

\begin{center}
\begin{spacing}{1.3}

\begin{figure}[!htb]
	\begin{center}
		\includegraphics[width=80mm]{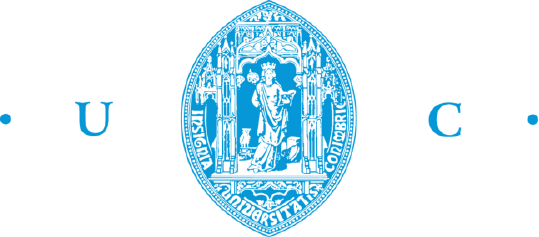}
	\end{center}
\end{figure}

\vspace{-0.5cm}

{\sc\normalsize
	Department of Informatics Engineering \\
	Faculty of Sciences and Technology \\
	University of Coimbra\\
}

\vskip 0.7cm

\begin{spacing}{2}
	{\bf\sc\huge\title} \\
\end{spacing}

\vskip 1.5cm

{\large\author} \\

\vskip 2cm

Doctoral Program in Information Science and Technology\\
PhD Thesis submitted to the University of Coimbra

\vskip 1cm

\normalsize
Advised by \FirstAdvisor \\
and by \SecondAdvisor 

\vskip 15mm

\vskip 0.8cm

\submissiondate

\end{spacing} 
\end{center}

\end{FrontpageEN}

 	\cleardoublepage

\newenvironment{FrontpagePT}%
{
	\cleardoublepage
	\thispagestyle{empty}
}
{\vfill\null}
\begin{FrontpagePT}

\begin{center}
\begin{spacing}{1.3}

\begin{figure}[!htb]
	\begin{center}
        \includegraphics[width=80mm]{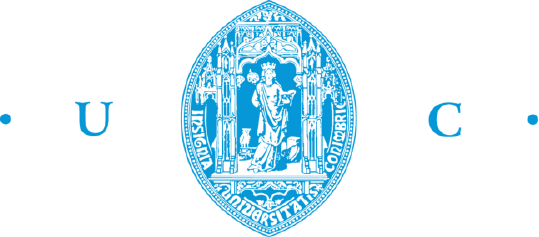}
    \end{center}
\end{figure}

\vspace{-0.5cm}

{\sc\normalsize
	Departamento de Engenharia Inform\'{a}tica \\
	Faculdade de Ci\^{e}ncias e Tecnologia \\
	Universidade de Coimbra \\
}

\vskip 0.7cm

\begin{spacing}{2}
	{\bf\sc\huge\titlePT} \\
\end{spacing}

\vskip 1.5cm

{\large\author} \\

\vskip 2cm

Programa de Doutoramento em Ci\^{e}ncias e Tecnologias da Informa\c{c}\~{a}o\\
Tese de Doutoramento apresentada \`{a} Universidade de Coimbra

\vskip 1cm

\normalsize
Orientado pela \FirstAdvisor \\
e pelo \SecondAdvisor 

\vskip 15mm

\vskip 0.8cm

\submissiondate

\end{spacing} 
\end{center}

\end{FrontpagePT} 	\cleardoublepage

\newenvironment{grants}%
{
	\cleardoublepage
	\thispagestyle{empty}
}
{\vfill\null}
\begin{grants}

\vspace*{18cm}

This work was partially supported by Portuguese Foundation for Science and Technology~(FCT) under the grant PTDC/EEATEL/105472/2008 and SFRH/BD/79094/2011, as well as by Brazilian National Counsel of Technological and Scientific Development~(CNPq) under the grant 238062/2012-0.

\vspace{0.5cm}

\begin{figure}[!hb]	
		\includegraphics[width=80mm]{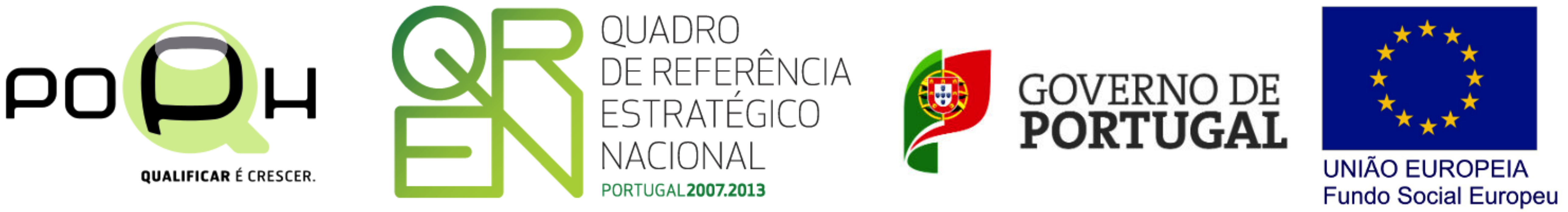} \ \ 
		\includegraphics[width=25mm]{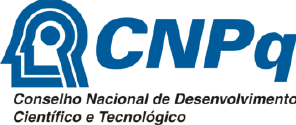}
\end{figure}

\end{grants}
 	\cleardoublepage

\begin{minipage}{14cm}
\thispagestyle{empty}

\vskip 21cm

\hfill
\href{http://en.wikipedia.org/wiki/Caffeine}{1,3,7-Trimethylxanthine}, \textit{thank you}

\end{minipage}
 	\cleardoublepage

\fi
\renewcommand*{\raggedsection}{\raggedleft}
\addtokomafont{chapter}{\fontsize{20}{20}\selectfont}

\ifDRAFT \else

\newenvironment{Acknowledgements}%
{
	\cleardoublepage
	\thispagestyle{empty}
	\phantomsection \label{Acknowledgements}
	\addcontentsline{toc}{chapter}{Acknowledgements}
	\begin{center}%
		{\bf\LARGE Acknowledgements}
	\end{center}
}
{\vfill\null}
\begin{Acknowledgements}
	~\newline
	~\newline

First of all, I would like to say thank you to everyone that was involved in one way or another in this journey.

I would like to give a special mention to both of my advisors, Marilia Curado and Eduardo Cerqueira, for the trust put in me in the first place and also for all the insights, support, and patience. Thanks for making me feel like home since the beginning.

To all my friends in Coimbra, that I will not name here because it would require so many additional pages to do it right. Having said that, one friend I have to name here. Bruno, not only this but thanks for holding the fort while I was away and for facing all the bureaucratic and lagging processes in my name. 

Thanks, mom, dad, and sis for always, always, always believing in me unconditionally. 

Eve, thanks for giving me the space that I need to finish the Thesis and also for doing everything you could to help me.%

\end{Acknowledgements}
 	\cleardoublepage

\newenvironment{Abstract}%
{
	\cleardoublepage
	\thispagestyle{empty}
	\phantomsection \label{Abstract}
	\addcontentsline{toc}{chapter}{Abstract}
	\begin{center}%
		{\bf\LARGE Abstract}
	\end{center}
}
{\vfill\null}
\begin{Abstract}
	~\newline
	~\newline

\lettrine[lines=3]{\color{gray}\bf{W}}{} ireless networks are envisaged to be one of the most important technologies to provide cost-efficient content delivery, including for video applications.
They will allow thousands of thousands of fixed and mobile users to access, produce, share, and consume video content in an ubiquitous way.
Real-time video services over these networks are becoming a part of everyday life and have been used to spread information ranging from education to entertainment content. 
However, the challenge of dealing with the fluctuating bandwidth, scarce resources, and time-varying error rate of these networks, highlights the need for error-resilient video transmission. 

In this context, the combination of Forward Error Correction~(FEC) and Unequal Error Protection~(UEP) approaches is known to provide the distribution of video applications for wireless users with Quality of Experience~(QoE) assurance. 
In order to correctly perform the UEP it is necessary to identify the most important parts in the video sequences. 

To tackle this issue, this thesis proposed a procedure to assess the video characteristics, 
such as the codec type, the frame type and size, the length and format of the group of pictures as well as the motion vectors, and their related impact on the perceived quality to end-users.
Furthermore, as the video content plays an important role on the perceived quality, this thesis also proposes a method to characterise the video's motion intensity. 
This involves conducting an exploratory data analysis in bootstrap time and then the use of several techniques in real-time to use the found results.
The purpose of the above-mentioned proposals is to give support for the main goal of this thesis, which is to propose mechanisms for resilient video transmission.

Taking everything into consideration, this thesis proposes a series of cross-layer video-aware and FEC-based mechanisms with UEP to enhance video transmission in several types of wireless networks.
A number of methods to set an adaptive amount of redundancy were used in these mechanisms, such as heuristic techniques, random neural networks, ant colony optimisation, and fuzzy logic. 

In the first one, heuristic techniques, the mechanisms rely on human experience to define the best strategy. 
In addition, the aim is not to reach a perfect solution, but a practical one with satisfactory results.
In the random neural networks methods, the neurones are trained and validated before run-time until they are able to perform an adequate numeric categorisation.
The ant colony optimisation techniques use a defined metaheuristic based on the ant's pheromones and pre-set rules to compute a precise amount of redundancy.
The last one, fuzzy logic techniques, the mechanisms depend on fuzzy rules and sets to find an adequate redundancy ratio.

The advantages and drawbacks of the proposed mechanisms were demonstrated in realistic simulations using real video sequences and actual network traces.
The assessments were conducted with well-known QoE metrics.
The results show that all the proposed mechanisms were able to outperform the competitors on both perceived video quality and network footprint.

~\newline
~\newline

\end{Abstract}
 	\cleardoublepage

\newenvironment{Resumo}%
{
	\cleardoublepage
	\thispagestyle{empty}
	\phantomsection \label{Resumo}
	\addcontentsline{toc}{chapter}{Resumo}
	\begin{center}%
		{\bf\LARGE Resumo}
	\end{center}
}
{\vfill\null}
\begin{Resumo}
	~\newline
	~\newline

\lettrine[lines=3]{\color{gray}\bf{A}}{} s redes sem fios estão entre as tecnologias mais importantes para prover a entrega de conteúdos a um custo acessível, inclusive no caso de aplicações de vídeo.
Esta tecnologia vai permitir que milhares de utilizadores móveis tenham acesso, produzam, partilhem e consumam conteúdo de vídeo de uma maneira ubíqua.
Nestas redes, os serviços de distribuição de vídeos em tempo real já se tornaram parte do dia-a-dia e são utilizados para difundir desde conteúdo educacional como também de entretenimento.
Entretanto, devido às dificuldades de gerir a flutuação da largura de banda disponível, os recursos escassos, e os erros que variam de tempo-em-tempo, revela-se a necessidade de uma transmissão de vídeo resistente a perdas.

Neste contexto, a utilização conjunta de correção antecipada de erros~(do Inglês \textit{Forward Error Correction} - FEC) e também da proteção desigual contra erros~(do Ingês \textit{Unequal Error Protection} - UEP) podem auxiliar na distribuição de serviços de vídeo para utilizadores de rede sem fios com garantia de qualidade de experiência~(do Inglês \textit{Quality of Experience} - QoE).
Procurando aplicar de forma adequada a proteção desigual de erros é necessário identificar corretamente as partes mais importantes das sequências de vídeo.

Para resolver esta situação, esta tese propôs um procedimento para avaliar as características dos vídeos, tais como o tipo de codec, o tipo e tamanho dos quadros, o comprimento do grupo de imagens e também os vetores de movimentações, bem como o impacto destas características na qualidade percebida pelos utilizadores finais.
O conteúdo do vídeo também é importante na definição da qualidade. 
Esta tese também propôs um método para melhor caracterizar a intensidade de movimentação. 
Isto envolve realizar uma analise exploratória de dados antes de iniciar o sistema, assim como a utilização de diversas técnicas para aceder aos resultados obtidos em tempo real. 
As técnicas acima foram propostas para dar suporte ao principal tema desta tese, que é o de projetar e construir mecanismos para a transmissão resiliente de vídeos.

Levando em consideração as informações acima, esta tese propôs diversos mecanismos utilizando técnicas de ``cross-layer'' e ``video-aware'' baseadas na correção antecipada dos erros e com proteção desigual da informação. 
O objetivo principal é melhorar a qualidade dos vídeos transmitidos em diversos tipos de redes sem fio.
Para chegar a este resultado, diversos mecanismos adaptativos foram utilizados, tais como técnicas heurísticas, redes neuronais aleatórias, otimização por colónias de formigas e lógica difusa.

As técnicas heurísticas referem-se a métodos que utilizam a experiência humana sobre o assunto para definir a melhor estratégia. 
É importante frisar que o principal objetivo não é encontrar a solução perfeita, mas sim uma solução prática que tenha resultados satisfatórios.
Outra técnica utilizada são as redes neuronais aleatórias, onde os neurónios são treinados e validados ante da execução do mecanismo até que estes atinjam a capacidade de realizar adequadamente a categorização numérica das informações que recebem.
No método de otimização por colónia de formigas, a quantidade de redundância é calculada utilizando uma meta-heurística que é baseada nos feromônios das formigas e em regras pré-definidas.
O último método é a lógica difusa que utiliza regras e um conjunto de dados para encontrar a quantidade mais adequada de redundância.

As vantagens e desvantagens dos mecanismos propostos foram demonstradas através de simulações realísticas com a utilização de sequências de vídeo reais e arquivos de registos de redes.
A avaliação dos mecanismos foram realizadas através de métricas conhecidas de qualidade de experiência.
Os resultados obtidos demonstram que os mecanismos propostos foram capazes de obter uma melhor performance que os competidores tanto na questão da qualidade dos vídeos como na questão da sobrecarga da rede.

~\newline
~\newline
{\bf\large Palavras-chave:}
Correção antecipada de erros~(FEC); FEC orientado ao vídeo; Qualidade de experiência~(QoE); Concepção cross-layer; Proteção desigual dos dados

\end{Resumo}
 	\cleardoublepage

\newenvironment{Foreword}%
{
	\cleardoublepage
	\thispagestyle{empty}
	\phantomsection \label{Foreword}
	\addcontentsline{toc}{chapter}{Foreword}
	\begin{center}%
	{\bf\LARGE Foreword}
	\end{center}
}
{\vfill\null}
\begin{Foreword}
~\newline
~\newline

\lettrine[lines=3]{\color{gray}\bf{T}}{} he work detailed in this thesis was accomplished at the Laboratory of Communication and Telematics~(LCT) of the Centre for Informatics and Systems of the University of Coimbra~(CISUC), within the context of the following projects:

\vspace{5mm}

\begin{description}
	
	\item[Project UBIQUIMESH] - Portuguese Foundation for Science and Technology (FCT) UBIQUIMESH project (PTDC / EEATEL / 105472 / 2008). The aim of this project was to optimise the communications in wireless mesh networks. The work performed in the project was related to QoE metrics and mechanisms to optimise the video transmission over wireless networks, which resulted in several publications.

	\item[Project iCIS] - Intelligent Computing in the Internet of Services~(iCIS) project (CENTRO-07-ST24-FEDER-002003), co-financed by QREN, in the scope of the ``Mais Centro'' Program.
	The goal of this project was to research intelligent computing for the Internet of Services and Things, focusing on information capture and management as well as data analysis and knowledge extraction for smart Internet services. 
	The activities include QoE-related issues and reliable video delivery.
	The achieved results were published in several conference papers and journals.

\end{description}

\vspace{6mm}

This work was funded by the following grants:

\vspace{6mm}

\begin{description}
	
	\item[Project grant] Portuguese Foundation for Science and Technology (FCT) UBIQUIMESH project (PTDC/EEATEL/105472/2008)
	
	\item[Doctoral grant] Portuguese Foundation for Science and Technology (FCT) (SFRH/BD/79094/2011)
	
	\item[Doctoral grant] Brazilian National Counsel of Technological and Scientific Development (CNPq) (238062/2012-0)

\end{description}

\vspace{6mm}

The outcome of the design, experiments, and assessments of several mechanisms on the course of this work resulted in the following publications:

\vspace{6mm}
\textbf{Journal papers:}
\begin{itemize}

	\item {Immich}, R. and Cerqueira, E. and Curado, M., ``\textbf{Shielding video streaming against packet losses over VANETs}'', The Journal of Mobile Communication, Computation and Information, Wireless Networks, Volume 22, Issue 8, pp 2563-2577, Springer, 2015
	
	\item {Immich}, R. and Borges, P. and Cerqueira, E. and Curado, M., ``\textbf{QoE-driven video delivery improvement using packet loss prediction}'', International Journal of Parallel, Emergent and Distributed Systems, Volume 30, Issue 6, pp 478-493, Taylor \& Francis, 2015
	
	\item Ros\'{a}rio, D. and Cerqueira, E. and Neto, A. and Riker, A. and {Immich}, R. and Curado, M., ``\textbf{A QoE handover architecture for converged heterogeneous wireless networks}'', The Journal of Mobile Communication, Computation and Information, Wireless Networks, Volume 19, Issue 8, pp 2005–2020, Springer, 2013
	
\end{itemize}

\bigbreak

\textbf{Book chapters:}
\begin{itemize}

	\item {Immich}, R. and Cerqueira, E. and Curado, M., ``\textbf{Improving video QoE in Unmanned Aerial Vehicles using an adaptive FEC mechanism}'', in Wireless Networking for Moving Objects: Models, Approaches, Techniques, Protocols, Architectures, Tools, Applications and Services, Volume 8611, pp 198-216, Springer LNCS, 2014
	
	\item {Immich}, R. and Cerqueira, E. and Curado, M., ``\textbf{Cross-layer FEC-based Mechanism for Packet Loss Resilient Video Transmission}'', in Data Traffic Monitoring and Analysis: From measurement, classification and anomaly detection to Quality of experience, Volume 7754, pp 320-336, Springer LNCS, 2013
	
\end{itemize}

\bigbreak

\textbf{Conference papers:}
\begin{itemize}

	\item {Immich}, R. and Cerqueira, E. and Curado, M., ``\textbf{Towards a QoE-driven Mechanism for Improved H.265 Video Delivery}'', in the 15th IFIP Annual Mediterranean Ad Hoc Networking Workshop (MED-HOC-NET), 2016
	
	\item {Immich}, R. and Cerqueira, E. and Curado, M., ``\textbf{Adaptive QoE-driven video transmission over Vehicular Ad-hoc Networks}'', in the IEEE Conference on Computer Communications Workshops (INFOCOM), 2015
	
	\item {Immich}, R. and Cerqueira, E. and Curado, M., ``\textbf{Towards the Enhancement of UAV Video Transmission with Motion Intensity Awareness}'', in the IEEE IFIP Wireless Days, 2014
	
	\item {Immich}, R. and Borges, P. and Cerqueira, E. and Curado, M., ``\textbf{Adaptive Motion-aware FEC-based Mechanism to Ensure Video Transmission}'', in the 19th IEEE Symposium on Computers and Communications (ISCC), 2014
	
	\item {Immich}, R. and Borges, P. and Cerqueira, E. and Curado,M., ``\textbf{Ensuring QoE in Wireless Networks with Adaptive FEC and Fuzzy Logic-based Mechanisms}'', in the IEEE International Conference on Communications (ICC), 2014
	
	\item {Immich}, R. and Borges, P. and Cerqueira, E. and Curado, M., ``\textbf{AntMind: Enhancing Error Protection for Video Streaming in Wireless Networks}'', in the 5th IEEE International Conference on Smart Communications in Network Technologies (SaCoNET), 2014
	
	\item Cerqueira, E. and Quadros, C. and Neto, A. and Riker, A. and {Immich}, R. and Curado, M. and Pescap\'{e}e, A., ``\textbf{A Quality of Experience Handover System for Heterogeneous Multimedia Wireless Networks}'', in the IEEE International Conference on Computing, Networking and Communications (ICNC), 2013
	
	\item {Immich}, R. and Cerqueira, E. and Curado, M., ``\textbf{Adaptive Video-Aware FEC-based Mechanism with Unequal Error Protection Scheme}'', in the 28th Annual ACM Symposium on Applied Computing (SAC), 2013
	
	\item Zhao, Z. and Braun, T. and Ros\'{a}rio, D. and Cerqueira, E. and {Immich}, R. and Curado, M., ``\textbf{QoE-aware FEC Mechanism for Intrusion Detection in Multi-tier Wireless Multimedia Sensor Networks}'', in the International Workshop on Wireless Multimedia Sensor Networks (WMSN), 2012
	
	\item Cerqueira, E. and Curado, M. and Neto, A. and Riker, A. and {Immich}, R. and Quadros, C., ``\textbf{A Mobile QoE Architecture for Heterogeneous Multimedia Wireless Networks}'', in the GC'12 Workshop: The 4th IEEE International Workshop on Mobility Management in the Networks of the Future World (MobiWorld), 2012
	
	\item Cerqueira, E. and Neto, A. and {Immich}, R. and Curado, M. and Riker, A. Barros, H., ``\textbf{A Parametric QoE Video Quality Estimator for Wireless Networks}'', in the GC'12 Workshop: IEEE Workshop on Quality of Experience for Multimedia Communications (QoEMC), 2012
	
\end{itemize}

\bigbreak

\textbf{Co-advisor of MSc thesis:}
\begin{itemize}
	
	\item Borges, P. and {Immich}, R. and Curado, M., ``\textbf{Mechanisms for resilient video transmission in wireless networks}", University of Coimbra, 2014
	
\end{itemize}

\bigbreak

\textbf{Special session presentations:}
\begin{itemize}
	
	\item Borges, P. and {Immich}, R. and Cerqueira, E. and Curado, M., ``\textbf{Mechanisms for resilient video transmission in wireless networks: Adaptive FEC mechanism with random neural network classification and ant colony optimization}''. 18 Semin\'{a}rio Rede Tem\'{a}tica de Comunica\c{c}\~{o}es M\'{o}veis (RTCM), 2014
	
	\item  {Immich}, R., Cerqueira, E. and Curado, M., ``\textbf{Cross-layer FEC-based Mechanism to ensure Quality of Experience in Video Transmission}''. 15 Semin\'{a}rio Rede Tem\'{a}tica de Comunica\c{c}\~{o}es M\'{o}veis (RTCM), 2012
	
\end{itemize}
\nocite{Immich2015a,Immich2015,Immich2015b,Borges2014,Immich2014a,Immich2014b,Immich2014,Immich2014c,Immich2014d,Immich2013,Immich2013a,Quadros2013,Rosario2013,Cerqueira2012,Immich2012,Quadros2012,Zhao2012,Immich2016}

\end{Foreword}
 	\cleardoublepage
	
\fi
\fancyhead{}
\fancyhead[RO,LE]{\leftmark}
\pagestyle{fancy}

\setcounter{secnumdepth}{3}
\setcounter{tocdepth}{2}
\phantomsection \label{contents}
\dominitoc
\tableofcontents
\cleardoublepage

\phantomsection \label{listoffig}
\listoffigures %
\addcontentsline{toc}{chapter}{List of Figures}
\cleardoublepage

\phantomsection \label{listofalgo}
\listofalgorithms
\addcontentsline{toc}{chapter}{List of Algorithms}
\cleardoublepage

\phantomsection \label{listoftab}
\listoftables
\addcontentsline{toc}{chapter}{List of Tables}
\cleardoublepage

\phantomsection \label{listofacro}
\addcontentsline{toc}{chapter}{Abbreviations and Acronyms}
\markboth{Abbreviations and Acronyms}{Abbreviations and Acronyms}

\chapter*{Abbreviations and Acronyms}

\begin{lyxlist}{00.00.0000}

\item [\textbf{4G}] 4th Generation Networks
\item [\textbf{5G}] 5th Generation Networks
\item [\textbf{A-STAR}] Anchor-based Street and Traffic Aware Routing
\item [\textbf{ACK}] Acknowledgement
\item [\textbf{ACR}] Absolute Category Rating
\item [\textbf{AC}] Access Category
\item [\textbf{AODV}] Ad-hoc On-demand Distance Vector
\item [\textbf{AP}] Access Points
\item [\textbf{ARQ}] Automatic Repeat reQuest
\item [\textbf{AU}] Application unit
\item [\textbf{AWARE}] Autonomous self-deploying and operation of Wireless sensor-actuator networks cooperating with AeRial objEcts
\item [\textbf{B-Frames}] Bi-directionally predictive coded frames 
\item [\textbf{BFP}] Bentley Faust Preparata
\item [\textbf{CATT}] Contention-Aware Transmission Time
\item [\textbf{CBR}] Constant Bit Rate 
\item [\textbf{CIF}] Common Intermediate Format
\item [\textbf{CSMA}] Carrier Sense Multiple Access
\item [\textbf{DCR}] Degradation Category Rating
\item [\textbf{DOLSR}] Directional Optimized Link State Routing Protocol
\item [\textbf{DSCQS}] Double-Stimulus Continuous Quality-Scale
\item [\textbf{DSR}] Dynamic Source Routing
\item [\textbf{ECC}] Error-Correcting Code
\item [\textbf{EC}] Error Correction
\item [\textbf{EDCA}] Enhanced Distributed Channel Access
\item [\textbf{ETT}] Expected Transmission Time
\item [\textbf{ETX}] Expected Transmission Count
\item [\textbf{FANETs}] Flying Ad-Hoc Networks
\item [\textbf{FEC}] Forward Error Correction
\item [\textbf{FLC}] Fuzzy Logic Controller
\item [\textbf{GGR}] Guaranteed Geocast Routing
\item [\textbf{GPMOR}] Geographic Position Mobility Oriented Routing
\item [\textbf{GPSR}] Greedy Perimeter Stateless Routing
\item [\textbf{GSR}] Geographic Source Routing
\item [\textbf{GoP}] Group of Picture 
\item [\textbf{HEVC}] High Efficiency Video Coding
\item [\textbf{HFS}] Hierarchical Fuzzy System
\item [\textbf{HVS}] Human Visual System
\item [\textbf{HWMP}] Hybrid Wireless Mesh Protocol
\item [\textbf{I-Frames}] Intra-coded frames
\item [\textbf{ITS}] Intelligent Transportation Systems
\item [\textbf{ITU-R}] ITU Radiocommunications Sector
\item [\textbf{ITU-T}] ITU Telecommunication Standardization Sector
\item [\textbf{ITU}] International Telecommunication Union
\item [\textbf{LAETT}] Load Aware Expected Transmission Time
\item [\textbf{LAN}] Local Area Network
\item [\textbf{LDPC}] Low-Density Parity-Check
\item [\textbf{LD}] Linkage Distance 
\item [\textbf{LTE}] Long-Term Evolution
\item [\textbf{LTE-Advanced}] Long-Term Evolution Advanced
\item [\textbf{MANET}] Mobile Ad-Hoc Network
\item [\textbf{MIC}] Metric of Interference and Channel-switching
\item [\textbf{MIND}] Metric for INterference and channel Diversity
\item [\textbf{MIROSE}] Minimum Interference Route Selection
\item [\textbf{MMSE}] Minimum Mean Squared Error
\item [\textbf{MOS}] Mean Opinion Score
\item [\textbf{MPCA}] Mobility Prediction Clustering Algorithm
\item [\textbf{MPEG}] Moving Picture Experts Group
\item [\textbf{MSE}] Mean Squared Error
\item [\textbf{NACK}] Negative Acknowledgement
\item [\textbf{NSDVCR}] Network State Dependent Video Compression Rate
\item [\textbf{OBU}] On Board Unit
\item [\textbf{OLSR}] Optimized Link State Routing Protocol
\item [\textbf{OSI}] Open Systems Interconnection
\item [\textbf{P-Frames}] Predictive coded frames
\item [\textbf{PSNR}] Peak Signal to Noise Ratio
\item [\textbf{QoE}] Quality of Experience
\item [\textbf{QoS}] Quality of Service
\item [\textbf{RF}] Radio Frequency
\item [\textbf{RNN}] Random Neural Networks
\item [\textbf{RSSI}] Received Signal Strength Indication
\item [\textbf{RSU}] RoadSide Unit
\item [\textbf{RS}] Reed-Solomon 
\item [\textbf{RTP}] Real-time Transport Protocol
\item [\textbf{RTSP}] Real Time Streaming Protocol 
\item [\textbf{RTT}] Round-Trip Time 
\item [\textbf{SDN}] Software-Defined Networks
\item [\textbf{SI13}] 13-pixel Spatial Information filter
\item [\textbf{SNR}] Signal-to-Noise Ratio
\item [\textbf{SSIM}] Structural Similarity
\item [\textbf{TACR}] Trust dependent Ant Colony Routing
\item [\textbf{TCP}] Transmission Control Protocol
\item [\textbf{TS}]	Transport Stream
\item [\textbf{UAV}] Unmanned Aerial Vehicle
\item [\textbf{UDP}] User Datagram Protocol
\item [\textbf{UEP}] Unequal Error Protection 
\item [\textbf{UGC}] User-Generated Content
\item [\textbf{V2D}] Vehicle-to-Device
\item [\textbf{V2G}] Vehicle-to-Grid
\item [\textbf{V2I}] Vehicle-to-Infrastructure
\item [\textbf{V2P}] Vehicle-to-Pedestrian
\item [\textbf{V2V}] Vehicle-to-Vehicle
\item [\textbf{V2X}] Vehicle-to-Everything
\item [\textbf{VANETs}] Vehicular Ad-Hoc Networks
\item [\textbf{VQEG}] Video Quality Experts Group 
\item [\textbf{VQMT}] MSU Video Quality Measurement Tool
\item [\textbf{VQM}] Video Quality Metric
\item [\textbf{VoD}] Video-on-Demand
\item [\textbf{WANETs}] Wireless Ad-Hoc Networks
\item [\textbf{WAVE}] Wireless Access for Vehicular Environments
\item [\textbf{WCETT-LB}] Weighted Cumulative Expected Transmission Time with Load Balancing
\item [\textbf{WLAN}] Wireless Local Area Network
\item [\textbf{WMNs}] Wireless Mesh Networks

\end{lyxlist}
 \cleardoublepage

\renewcommand*{\raggedsection}{\raggedright}
\addtokomafont{chapter}{\fontsize{25}{28}\selectfont\filleft}

\fancypagestyle{plain}{%
	\fancyhf{}
	\fancyfoot[C]{\textemdash~\thepage~\textemdash}
	\renewcommand{\headrulewidth}{0pt}
}
\fancyfoot{}
\fancyfoot[C]{\textemdash~\thepage~\textemdash}
\pagestyle{fancy}

\pagenumbering{arabic}

\setcounter{mtc}{8}
\chapter{Introduction}
\label{ch:Introduction}

\renewcommand*{\dictumwidth}{.45\textwidth}
\renewcommand*{\dictumauthorformat}[1]{({#1})\bigskip}
\dictum[Isaac Asimov, Book of Science and Nature Quotations]{The saddest aspect of life right now is that science gathers knowledge faster than society gathers wisdom.}
\minitoc

\lettrine[lines=3]{\color{gray}\bf{T}}{} his thesis contemplates the necessity of providing high-quality video delivery for end-users.
In order to do that, several different approaches were adopted solving a considerable number of current issues.
In this process, the contributions reached include an approach to quantify the impact of the video characteristics on its quality, a method to characterise the motion intensity of video sequences, as well as several mechanisms to improve the video quality over diverse network scenarios.
The research background and motivation of this thesis are presented and investigated followed by the discussion of the proposed objectives together with the respective contributions, as well as the thesis outline.

\section{Background and Motivation}
\label{sec:intro:motivation}

In the last few years, there has been a rapid proliferation of a wide range of real-time video services and applications. 
This growth can be partly attributed to the recent development and improvement of mobile devices, such as notebooks, tablets, and smartphones, as well as advances in wireless networks~\cite{comScore2013,Adobe2014}. 
This improvement allows an increased number of services and applications to be available to the end-users.
The adoption of video services can be used as means to provide users with both information and entertainment content. 
Furthermore, the whole Internet, and not just the mobile end-users, is experiencing a considerable growth in traffic that is in part led by these novel real-time video services.
According to Cisco, by 2020 the global Internet traffic will be 95 times higher than it was in 2005.
Additionally, the video IP traffic will represent over 82\% of the global IP traffic~\cite{Cisco2016}. 

This situation is easily explained by the large amount of new forms of information and entertainment that are being released every day by thousands of users.
This includes, but is not limited to, User-Generated Content~(UGC), news websites, social networking, and e-learning.
Moreover, a considerable number of television studios, film production companies, and recording studios are also increasing the variety of video content that is available over the Internet. 
Just to give one example, by 2020 it is predicted that every second almost a million minutes of video content will be sent over the global Internet~\cite{Cisco2016}.

It is clear that the distribution of video content over the Internet is rapidly expanding. 
End-users are increasingly consuming this type of content that might be anything from homemade videos to professionally developed programming. 
However, real-time video transmission imposes more challenges than non-real-time data transfer. 
Keeping a satisfactory perception of video quality requires the provision of sufficient bandwidth, low interference, as well as low delay and jitter throughout the video transmission.

In the light of this, several multi-hop wireless networking technologies are being increasingly employed~\cite{Meng2016}, such as Wireless Mesh Networks~(WMNs), Flying Ad-Hoc Networks~(FANETs), and Vehicular Ad-Hoc Networks~(VANETs).

The use of these wireless networks introduces additional challenges that have to be addressed to improve the video quality from the end-users perspective.
Several of these challenges can be attributed to a more dynamic and error-prone environment. 
First of all, the wireless channels are shared among the nodes, and because of this, they must collaborate to allow simultaneous video transmissions, as well as other network/Internet traffic. 

A further aggravating factor is that this method of transmission is susceptible to severe physical conditions, such as multipath fading, Radio Frequency~(RF) interference, shadowing, and background noise. 
In contrast with the wired links, packet loss does not necessarily mean network congestion, and can often be related to random physical causes, leading to time-varying communications impairments and recurring link interruption. 
Since most of the video services are real-time applications, they need a steady and continuous flow of packets, which can be affected by several of the above-mentioned factors in wireless environments. 
These challenges become even more serious in multi-hop wireless networks.

In addition, all the video transmissions in a wireless network are subjected to a shared channel medium and also exposed to a high packet drop rate, delay, and jitter during the transmission.
Additionally, the underlying routing protocol might impose some message exchange overhead due to route updates. 
Burst losses are also fairly common since packets may be sent on non-working routes before the routing protocol detects the disruption. 

To make the matter worst, the growth of the network leads to higher node density which can increase packet collisions, due to the concurrent transmission. 
Nevertheless, one of the main challenges in this type of network is how to evenly distribute the available bandwidth among the requesting nodes for real-time traffic~\cite{Liu2009}.
The network related issues and features are further discussed in Chapter~\ref{ch:networkStuff}.

In addition to the network parameters, there are other issues that might influence the video quality. 
Some arise from the video characteristics, such as the frame and codec type, bitrate, video format and size of the Group of Picture~(GoP), and even the content of the video~\cite{Yuan2006,Khan2010}. 
Apart from that, not all the packets have an equal impact on perceptual quality. 
There is a correspondence between the type of information that the packet carries and the impact it has on the user perception of video quality~\cite{Greengrass2009}.

Taking everything into consideration, is necessary a flexible and adjustable mechanism to improve the video quality.
Hence, these requirements are best met by using a Unequal Error Protection~(UEP) scheme. 
Through a UEP-based scheme, it is possible to assign different amounts of redundancy to the original data. 
This helps to better protect the most important information, allowing the transmission of videos with high quality, while introducing less redundant information, therefore decreasing the network overhead.
An elaborate review about the video characteristics and the impact on its quality can be found in Chapter~\ref{ch:videoStuff}.

Quality of Experience~(QoE) metrics are commonly used to assess the video quality level, allowing also to identify in what situations the flaws are more noticeable~\cite{Serral-Gracia2010}.
QoE metrics measure the video degradation and can be defined as how users subjectively perceive the quality of an application or service~\cite{Piamrat2009}. 
This means that the performance should be measured end-to-end and must reflect the user standpoint. 
QoE is related to Quality of Service~(QoS) metrics, but differs from it by providing a more comprehensive assessment and was designed to provide an appraisal of human expectations in regards to the video service.

QoE metrics can be classified into two general approaches, namely subjective and objective.
Subjective evaluations are based on human individuals rating the video quality.
Because of this, it is possible to capture a great variety of details that might impact on the perceived quality.
To apply these metrics, however, tend to be time-consuming, expensive, and hard to employ in real-time, as well as the results are not reproducible.
On the other hand, objective metrics try to mimic the human vision system by using mathematical models to identify and measure the video impairments.
Therefore, they do not rely on human interventions and are both verifiable and reproducible.
More details about QoE metrics can be found in Chapter~\ref{ch:videoStuff}.

In the light of the aforementioned issues and features, 
the employment of an adaptive data protection is essential to overcome the transmission challenges in wireless networks, providing both high-perceived video quality and low network overhead.
One way to achieve this level of protection is by resending packets if lost or dropped, i.e., Automatic Repeat reQuest~(ARQ) method.
Another possibility is to add redundant information to the original data set, in this case, if some information is lost it can be rebuilt, i.e. Forward Error Correction~(FEC) schemes.
Both techniques have advantages and disadvantages which are explained and discussed in Chapter~\ref{ch:rw}.

A number of different solutions have been proposed to address the aforementioned challenges. 
Several of these proposals are reviewed and summarised in Chapter~\ref{ch:rw}.  
However, as far as the literature goes, none of them was able to present a holistic mechanism which takes into consideration QoE-sensitive information, video details, as well as the network conditions. 
To this end, this thesis proposes, discusses, and assesses several mechanisms, tailored to diverse wireless network characteristics, which improve the video quality perceived by the end-users in wireless networks while reducing the network overhead.

\section{Objectives and Contributions}
\label{sec:intro:objectives}

The main goals of this thesis are to define a method which enables the characterization of the motion intensity in arbitrary video sequences and also to design, implement, and assess several FEC-based mechanisms with content-awareness to enhance the quality of video delivery from the point-of-view of end-users over diverse types of wireless networks. 
This is accompanied by a comprehensive analysis of the state of the art of video transmission optimisation mechanisms, with a particular focus on multi-hop wireless networks. 

In addition, the proposed mechanisms were evaluated by using the Network Simulator~3~(NS-3) due to the necessity of large-scale networks with a large~(and expensive) variety of hardware equipment. 
The simulator offers a practical feedback allowing the designer to investigate the correctness and efficiency of the proposed mechanisms in a controlled and reproducible environment.
Therefore, it is easier to explore the unforeseen interaction between the multiple elements involved in the experiments~\cite{Breslau2000}.
In addition, it also allows a straight and fair comparison of results in a transverse manner among several research efforts.
Despite the use of a simulator, real video sequences and traces of real network operations and characteristics were used. 
This provides closer results to real world implementations.

The specific goals of this thesis are as follows:

\begin{description}
	
	\item[Goal 1] - Define a method to provide video content characterization according to its motion intensity, as well as the video parameters;
	
	\item[Goal 2] - Propose a set of adaptive FEC-based content- and video-aware mechanisms, which aim to support video distribution to wireless users over error-prone networks. 
	These mechanisms should assure the QoE and optimise the usage of the wireless channels resources;
	
	\item[Goal 3] - Study the performance of the proposed mechanisms in various network environments. The assessment should always include how the video quality is perceived by the end-users and also the network overhead impact.
	
\end{description}

Taking into consideration the specific goals, this thesis has succeeded in producing the follow main contributions:
\begin{description}%
	
	\item[\protect{%
		\parbox[t][2\baselineskip][t]{\textwidth}{
	Contribution 1, Assessing the impact of the video characteristics on the video quality level}
	}
	]
	This contribution is related to the study of video sequences details, such as the frame type and size, the codec type, the length and format of the group of pictures, as well as the motion vectors, and their impact on the perceived video quality. 
	All methods and mechanisms proposed in this thesis, from Chapter~\ref{ch:MESH} to~\ref{ch:VANET}, benefit from the findings generated in this contribution;

	\item[Contribution 2, A method to characterise the motion intensity of videos]~\\
	The motion intensity plays an important role on the amount of redundancy needed to improve or even maintaining a high video quality over networks with high error rates.
	The proposed method encompasses several procedures, such as an exploratory data analysis using a hierarchical clustering approach to separate and group video sequences with similar motion intensities. 
	After the clustering process, several techniques can be applied to provide an easy way to access the cluster's characteristics, such as a heuristic routine~(as seen in Chapter~\ref{ch:MESH}), a random neural network~(as seen in Chapter~\ref{ch:MESH}), and fuzzy logic sets and rules~(as seen in Chapters~\ref{ch:MESH}, \ref{ch:UAV}, and~\ref{ch:VANET}).
	The characterisation method is always performed offline, where a database, sets, and rules are generated to be used in the real-time mechanisms;
		
	\item[Contribution 3, Heuristic mechanism]~\\
	The proposal, design, and evaluation of an adaptive heuristic-based mechanism to enhance video transmissions over wireless networks is one of the contributions of this thesis. 
	The proposed mechanism uses the human experience and knowledge of the tackled issues to define a strategy to improve the video quality without adding unnecessary redundancy, thus saving the wireless resources.
	The main objective of heuristic methods is not to provide a perfect solution, but rather a practical process of dealing with a problem with satisfactory results.
	This contribution is studied in Section~\ref{sec:viewfec}.
	
	\item[Contribution 4, Random neural networks mechanism]~\\
	A different mechanism was proposed using the random neural networks as means to both define the video content category related to the motion intensity and to define in real-time an adequate amount of redundancy.
	In order to do that, first, the mechanisms have to be trained prior to the execution. 
	This involves feeding the neural network with a large enough number of video sequences to perform a numeric categorisation of the frames in terms of motion intensity.
	After trained and validated it is ready to perform the video content characterisation and to attribute an adaptive redundancy.
	Section~\ref{sec:neuralFEC} presents this contribution.
	
	\item[Contribution 5, Ant colony optimisation mechanism]~\\
	The conceptual definition, implementation, and assessment of mechanisms that uses ant colony optimisation to overcome the network-related problems regarding the video transmission is another contribution of this thesis. 
	These mechanisms rely on ant colony metaheuristic to define a precise amount of redundancy according to the degree of motion intensity, as well as other video characteristics.
	This contribution is evidenced in Section~\ref{sec:PredictiveAnts}.
	
	\item[Contribution 6, Fuzzy logic mechanism]~\\
	Another contribution of this thesis is the design and evaluation of several mechanisms that employ fuzzy logic to decide how much redundancy is needed for each type of video sequence.
	The definition of the fuzzy logic rules and sets is an offline process that depends on the feedback obtained from Contribution~2. 
	However, once defined, they are ready to be used in real-time.
	In addition, hierarchical fuzzy logic structures are also designed and assessed. 
	This allows building even-more-complex systems at lower computation costs.
	This contribution can be found in several mechanisms in Chapters~\ref{ch:UAV} and~\ref{ch:VANET}.
	
\end{description}
\section{Thesis Outline}
\label{sec:intro:thesisOutline}

This thesis is organised as follows. 

\begin{description}

	\item[Chapter~\ref{ch:networkStuff} - Network Conceptualization and Feature Highlights]~\\
	Introduces the related network concepts and main definitions about several types of networks that are used in the experiments, namely wireless mesh networks, flying ad-hoc networks, and vehicular ad-hoc networks.
	The usage of the wireless channels and the routing protocol characteristics are also investigated in this chapter, as well as the ways to implement a cross-layer design.
	In addition, the dynamic nature and the time-varying channels conditions of these networks are explored because they can have a significant impact on delay-sensitive applications such as the video services.

	\item[Chapter~\ref{ch:videoStuff} - Video Coding Design and Quality of Experience]~\\
	Addresses the video-related components and details, such as compression techniques, different types of frames, the group of pictures format, as well as the concept of macroblocks and blocks.
	In addition, it is also examined how the video sequences are distressed by impairments, the importance of the different types of frames, and the impact of the motion intensity on the quality. 
	At the end of the chapter, quality of experience assessment methods are discussed.

	\item[Chapter~\ref{ch:rw} - Advances on Improved Video Transmissions]~\\
	Discusses the concepts surrounding error-correcting codes.
	Two types are investigated, namely the automatic repeat request and the forward error correction, along with the advantages and disadvantages of both.
	This chapter also shows the literature review on mechanisms to enhance the video transmission quality. At the end, the summary and open issues are presented. 

	\item[Chapter~\ref{ch:MESH} - Mechanisms for Resilient Video Transmission over WMNs]~\\
	Proposes and assesses three mechanisms to shield video transmissions over wireless mesh networks.
	All the proposed mechanisms adjust the redundancy amount in real-time according to several video characteristics.
	The first mechanism~(ViewFEC) uses a heuristic method to adjust the redundancy. 
	The second mechanism~(neuralFEC) adopts random neural networks to both classify the motion intensity and in the decision-making process.
	The last mechanism~(PredictiveAnts) is based on a random neural network, to categorise motion intensity, and uses ant colony optimisation for the dynamic redundancy allocation.

	\item[Chapter~\ref{ch:UAV} - Mechanisms for Resilient Video Transmission over FANETs]~\\
	Proposes and assesses two adaptive video-aware mechanisms to safeguard unmanned aerial vehicle real-time video transmissions against packet loss.
	The first mechanism~(uavFEC) uses motion vectors to define the spatial video complexity and fuzzy logic to adjust the redundancy amount.
	The second mechanism~(MINT-FEC) follows the same procedure, but differs by using additional parameters, such as the temporal video intensity and the ability to cope with videos of arbitrary size.

	\item[Chapter~\ref{ch:VANET} - Mechanisms for Resilient Video Transmission over VANETs]~\\
	Proposes and assesses two adjustable mechanisms to increase the video transmission resiliency over vehicular ad-hoc networks.
	The former mechanism~(CORVETTE) uses a hierarchical fuzzy logic model to define a specific amount of redundancy according to the video characteristics and the network conditions.
	The latter mechanism~(SHIELD) adopts a similar procedure for the video details, however, a much more comprehensive method is used to diagnose and evaluate the network conditions.

	\item[Chapter~\ref{ch:Conclusions} - Conclusions and Future Work]~\\
	Presents the final remarks and conclusions, as well as the outline of future research to further advance this work.

\end{description}
 \cleardoublepage

\setcounter{mtc}{9}
\chapter{Network Conceptualization and Feature Highlights }
\chaptermark{Network concepts and Feature Highlights}
\label{ch:networkStuff}

\renewcommand*{\dictumwidth}{.43\textwidth}
\renewcommand*{\dictumauthorformat}[1]{({#1})\bigskip}
\dictum[George Orwell, Nineteen eighty-four]{\ Big Brother is Watching You !}

\minitoc

\lettrine[lines=3]{\color{gray}\bf{T}}{} he optimisation of video transmission over wireless networks is a complex task which involves a number of concepts and definitions. 
In this chapter, several of these components will be discussed, providing a general understanding on this subject. 
First, the cross-layer design will be described and after that, the characteristics of wireless networks will be discussed.

\section{Introduction}
\label{sec:net:concepts}

Wireless networks have become a cost-effective way to distribute variate contents. 
Another advantage is the untethered convenience allowing a good range of mobility to the end-users, as well as the self-management and multi-hopping features.
Because of that, the widespread deployment of wireless networks, as well as the usage of mobile devices, has increased significantly in recent years. 
These networks are well known for their dynamic nature along with their time-varying channel conditions. 

Apart from that, several new services have stringent requirements and work in real-time. 
This situation is a cause of great concern when it comes to video transmission, which can be considered to be one of the most demanding services.
Environments that do not provide the necessary conditions will have a poor performance in this type of service and the video impairments will be very noticeable to the end-users.
In this case, the intrinsic network details must be exploited in order to understand how to provide the necessary conditions to better serve these resource-demanding services.

The remainder of this chapter presents the concepts of cross-layer design in Section~\ref{sec:net:cross} and the main characteristics of the wireless channels in Section~\ref{sec:net:wirelesschannels}. 
This is followed by Section~\ref{sec:net:wanets} which describes wireless ad-hoc networks. 
The chapter's summary is given in Section~\ref{sec:net:summary}.

\section{Cross-layer Design}
\label{sec:net:cross}

The purpose of a layering principle is to reduce the constraints and the complexity of the design.
The Open Systems Interconnection~(OSI) model~\cite{Day1983} proposes seven conceptual layers, where each one is only able to communicate and provide information to the adjacent layers.
This means that non-adjacent layers are not allowed to exchange information. 
The ever-increasing network traffic, along with the emergence of novel services, e.g., video transmission which has stringent requirements, means that it is a very challenging task to meet the end-to-end requirements without interaction between non-adjacent protocol layers~\cite{Setton2005}.

The cross-layer design was proposed to overcome this limitation. 
It can be defined as the violation of the referenced layered communication architecture through direct communication or the sharing of information between non-adjacent layers. 
There are several ways that cross-layer design can be carried out, for example, by creating new layer interfaces and allowing direct communication between non-adjacent layers. 
An alternative approach could be to perform a vertical calibration across layers, where the values to set parameters could span across non-adjacent layers. 
Merging adjacent layers using a joint design can also be considered to be cross-layer~\cite{Srivastava2005}.

\begin{figure}[!htb]
  \begin{center}
      \includegraphics[width=100mm]{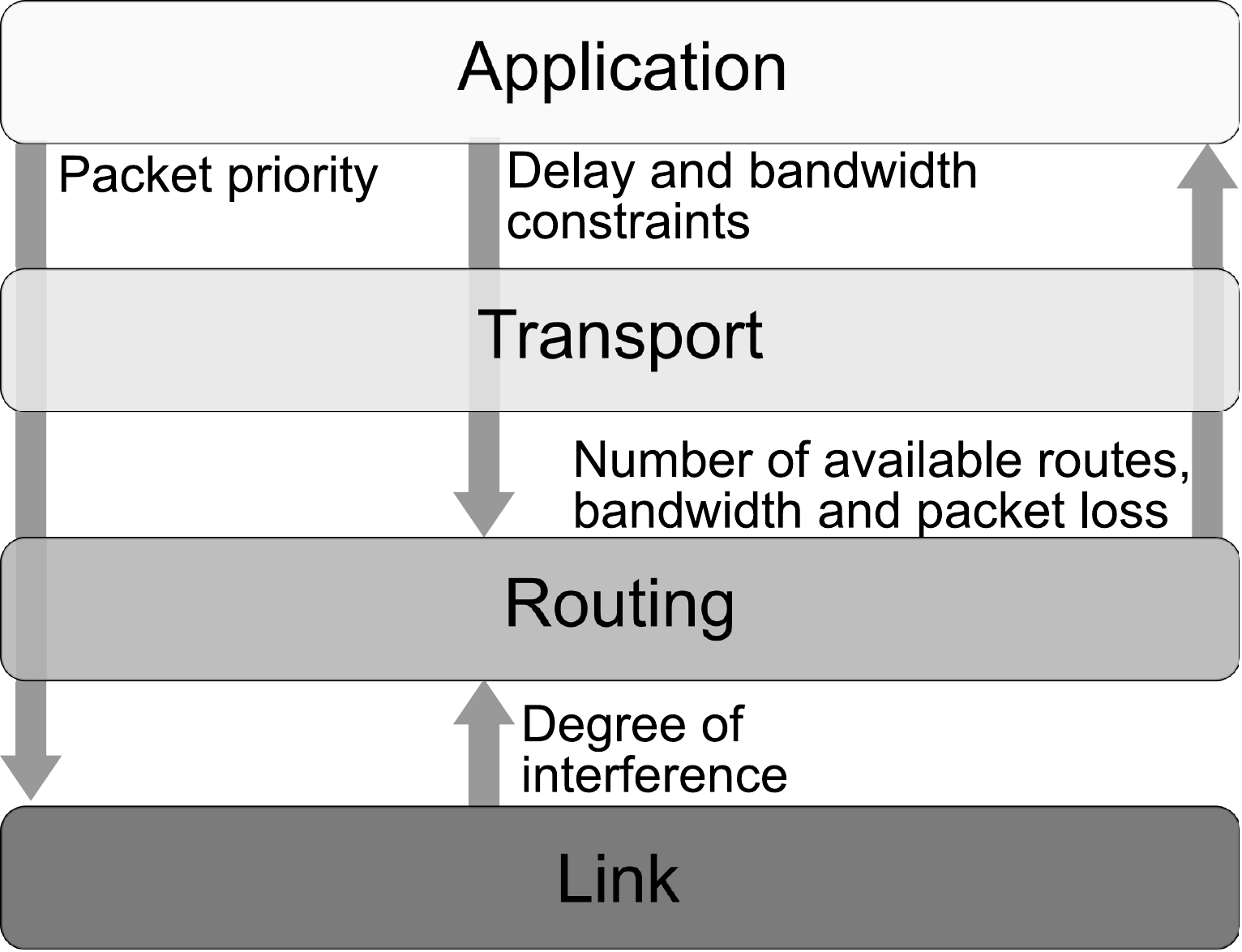}
  \end{center}
  \caption{Cross-layer information exchanges}
  \label{fig:net:cross-layer}
\end{figure}

Figure~\ref{fig:net:cross-layer}~(adapted from~\cite{Lindeberg2011}) depicts several examples of information exchange using a cross-layer design. 
For instance, from the application layer, it is possible to set packet priorities directly to the link layer or to define the delay and bandwidth requirements in the routing layer. 
Furthermore, as well as defining information, it is also possible to have access to other's layers data. 
For example, the application layer can have information about the offered load from the transport layer or the number of available routers, bandwidth, and packet losses from the routing layer.

Since the use of this technique is widespread, several studies have been proposed for addressing the question of cross-layer design, and some of them have achieved promising results in improving the system performance of wireless networks~\cite{Wang2010,Oh2010,Liu2008,Andreopoulos2006}. 
However, a number of problems may arise, e.g. incompatibility with existing protocols, unclear protocol-layer abstractions, increased design complexity, difficulty in maintenance and management, among others.

Overall, the cross-layer techniques are a flexible means of both accessing and defining information in non-adjacent protocol layers. 
This is especially important when assessing the end-user perception of video quality because the use of this information provides a better chance of interpreting the general user perception of the service correctly. 
Nevertheless, it is important to take into account the relevance of compatibility and standardisation, as otherwise the interoperability of the solution will be impaired.

\section{Wireless Channels}
\label{sec:net:wirelesschannels}

A channel can be considered a medium to convey a data signal from the sender to the receiver.
While the nodes in a wired network transmit information using links that have static and well-known characteristics, wireless nodes communicate via a much more dynamic and unpredictable environment. 
A radio transmits electromagnetic waves through the wireless channels over a non-uniform and unknown medium, where the properties of each scenario might be radically different and oscillate over time~\cite{Adlakha2007}.
These properties can be affected by surrounding objects, communication between other nodes and even themselves, as well as radio frequency~(RF) interference from other equipment. 
This means that, as the signal travels, it is subjected to a wide range of physical effects that can cause a deterioration of the quality.

These abnormalities sometimes occur as a result of events like multipath fading and shadowing which lead to a decrease of signal quality and thus tend to increase packet error rates. 
Multipath fading is the consequence of more than one interfering reflections of the same signal. 
Shadowing occurs when large objects block~(or partially block) the propagation path, causing severe signal attenuation~\cite{Lindeberg2011}.

In addition to the fact that a shared medium is used, the radio antennas are unable to sense and transmit simultaneously, which makes matters worse. 
Another known issue is the hidden terminal problem, which happens when a node is visible from one Access Point~(AP), although not from the other nodes connected to the same AP.
It is also possible to name the exposed node issue that occurs whenever a node is kept from transmitting packets due to transmissions from other neighbouring nodes.
These constraints increase the packet loss rates even further, as a result of a higher number of collisions. 
This situation becomes even worse in high node density networks. 
Moreover, as the distance between the nodes increases, the signal strength gradually decreases.

Thus, owing to the time-varying link characteristics and the unpredictable nature of wireless channels, mechanisms are needed to enhance the transmission.
This is especially true in view of the stringent delay, jitter, and packet loss requirements imposed by video transmission services.

\section{Wireless Ad-Hoc Networks~(WANETs)}
\label{sec:net:wanets}

A WANET is composed of a collection of wireless nodes which do not rely on a pre-existing infrastructure.
This is a self-organized and decentralised type of network where the communications are performed by multi-hop wireless paths.
Additionally, these networks can be spontaneously created as the nodes dynamically choose where to forward the data according to the network connectivity. 
On top of that, due to its quick deployment and a reduced configuration, they are very convenient for several types of applications.

The ascendance of video services, however, has created the need for a stable and trustworthy communication link between the nodes in order to guarantee a high level of QoS and QoE support.
Considering that WANETs are highly dynamic, have near-unpredictable node mobility, and suffer from time-varying network conditions, the QoE and QoS provisioning is very complex as well as challenging. 
Because of that, new methods and mechanisms have to be proposed to address these issues, being this one of the main goals of this thesis.

According to its application, WANETs can be further divided into more specialised categories. The ones related to this work are presented below.

\subsection{Wireless Mesh Networks~(WMNs)}
\label{sec:net:wmns}

WMNs have been used as a possible means of extending the reach of wireless APs, proving to be a viable solution for spreading reliable wireless Internet across wide areas~\cite{Pal2011}. 
This type of network is formed by a set of mesh routers, generally static, which use multi-hop transmissions to create a backbone network, allowing, in this way, communication between mesh clients~\cite{Akyildiz2008}. 
A WMN does not have a fixed structure or a defined and immutable path to convey the information because these networks rely on dynamic paths which can be changed and discovered through the routing protocol in real-time. %
In doing that, it is possible to assume that every router is responsible for maintaining the information flow between other routers in the neighbourhood. 

The mesh client, in these networks, can be any type of device that has some kind of wireless connectivity, namely a notebook, a smartphone, or a tablet. 
Usually, one or more mesh routers are connected to the Internet acting as gateways to the clients. 
In general, WMNs incur low installation and maintenance costs and offer a reliable service due to their dynamic self-configuration and self-organization capabilities~\cite{Akyildiz2005}. 
This type of network has been adopted for many industrial and academic deployments, in small and large offices, as well as in home applications.
This means that this technology can be used in any place that does not have a wired local area network~(LAN) or has high-deployment costs, e.g., instant surveillance systems, and back-haul service for large-scale wireless sensor networks~\cite{Kim2006}.

Recently, these multi-hop wireless networks have come to be regarded as the most important wireless technology to provide last mile access in future wireless multimedia networks. 
WMNs will allow thousands of fixed and mobile users to access, produce and share multimedia content in an ubiquitous way. 
These multi-hop communications are expected to meet quality requirements, namely QoS and QoE, to achieve this goal. 
They are also expected to ensure easy deployment, flexibility, and high reliability~\cite{Bruno2005,Zwinkels2004}.

\subsubsection*{Routing Protocols and Metrics}
\label{sec:routingprotocols}

Due to the dynamic nature of the WMNs, a routing protocol is needed to find and maintain connectivity between the nodes, allowing it to achieve the best possible overall performance.
In other words, the routing protocol can be described as the process to determine the end-to-end path that connects the source node to the destination node~\cite{Waharte2006}. 
Another role of this type of protocol is to share the routing information with the network nodes as a means of providing the knowledge required to build the network topology.

There is a wide range of routing protocols, but generally, the approaches can be divided into proactive, reactive, or hybrid categories. 
In proactive routing protocols, there is a periodic exchange of routing messages to keep the routing tables updated, even when there is no network traffic, and hence a waste of resources. 
Reactive or on-demand routing protocols were developed to serve as an alternative to the constant message exchanges. 
These protocols discover routes upon request avoiding the waste of resources. 
The problem arising from this approach is that there is a higher initial delay when retrieving a routing path, and depending on the service type, this may not be acceptable. 
Lastly, there are hybrid routing protocols which exploit the advantages of both proactive and reactive protocols~\cite{Ancillotti2011}.

The routing protocols can use several metrics to find the best end-to-end path. 
Hop count is the simplest metric, where only the number of hops between the source and destination are considered. 
A different performance metric is known as Expected Transmission Count~(ETX)~\cite{DeCouto2005}, which takes into account the data loss and the number of retransmissions. 
Another metric is the Expected Transmission Time~(ETT)~\cite{Draves2004}, which is an improvement of the ETX metric that includes the link bandwidth to calculate the route path~\cite{Parissidis2011}. 
A few more examples of metrics used in other studies include the location-dependent contention, such as the Contention-Aware Transmission Time~(CATT)~\cite{Genetzakis2008}, Metric of Interference and Channel-switching~(MIC)~\cite{Yang2006}, the number of per-link admitted flows e.g., Load Aware ETT~(LAETT)~\cite{Aiache2008}, interference and load through passive monitoring in Metric for INterference and channel Diversity~(MIND)~\cite{Borges2009}, as well as the load-dependent cost, such as Weighted Cumulative Expected Transmission Time with Load Balancing~(WCETT-LB)~\cite{Ma2007}.

Furthermore, QoS-based routing is also desirable due to the recent improvements in mobile devices as well as the emergence of real-time video services. 
In this approach, the end-to-end paths are determined by using information about the resource availability in the network in conjunction with the QoS requirements of the flows. 
Additionally, the routing protocol is able to find a path that satisfies a specific flow requirement, increasing the network resource utilisation.
An important feature of this approach is that it has a collision detection mechanism which enables it to avoid disruptive interferences~\cite{luo2012new,Bouhouch2007}.

Another important feature of some WMN routing protocols is the support for multipath routing. 
These routing algorithms take advantage of the multiple connections between the mesh nodes to provide more than one path between source-destination pairs~\cite{Lindeberg2011}. 
One advantage of multipath routing approaches is that they can provide alternative paths according to the required QoS, without a frequent route discovery~\cite{Yang2006a}. 
A further benefit is the ability to transmit information, from the source node, through a number of disjoint paths, to the destination node, by enabling bandwidth aggregation, load balancing, and reduced delay as well as improving fault tolerance. 
The fault tolerance feature is especially important for real-time transmission because when a high degree of multipath diversity is achieved, it is less feasible that a link disruption will occur in more than one path at the same time, thus making the chance of a packet loss less likely~\cite{Tsirigos2004}.
Therefore, if this happens, it will reduce the impact of a single link failure on the video quality.

\subsection{Vehicular Ad-Hoc Networks~(VANETs)}
\label{sec:net:vanets}

VANETs enable vehicles to create and maintain wireless communication links between them in the absence of a central base station. 
One of the main goals of this technology is to provide ubiquitous connectivity as well as to allow efficient vehicle-to-vehicle communications, which is necessary to implement an Intelligent Transportation Systems~(ITS)~\cite{Al-Sultan2014}.
This network uses the IEEE 802.11p Wireless Access for Vehicular Environments~(WAVE)~\cite{Jiang2008} standard and is envisioned to be used in a multitude of environments, such as in highways as well as in rural and urban scenarios, including support for many applications.

A typical VANET is composed of several components, as described below:

\begin{description}

	\item[On board unit~(OBU)] This is an onboard device inside the vehicles to provide connectivity with other OBUs or Roadside units~(RSUs). 
	Usually, it is equipped with a specialised interface for ad-hoc communication and short-range wireless capabilities using IEEE WAVE. 
	The OBU is also responsible for providing data forward to and from both others OBUs or RSUs.

	\item[Roadside unit~(RSU)] This component is a WAVE-enabled device typically fixed that is placed along the roadside or in predetermined locations.
	The main goal of this device is to extend de transmission range of other RSUs or OBUs. 
	In order to do that, this device is usually equipped with a WAVE radio and more often than not with another communication link directly to the infrastructural network.
	Because of that, it is also possible to provide Internet connectivity to passing by OBUs.

	\item[Application unit~(AU)] This element stores and acts as interface for the applications and/or services supported by the provider.
	This element can be a dedicated hardware or be implemented using the same physical components of the OBU.
	This means that the difference between the OBU and AU could be only software-based, i.e., applications and interface.
	Nevertheless, the network functions are still the OBU responsibility.
	
\end{description}

The above-described components allow distinct types of communication setup as briefly described below:

\begin{description}

	\item[Vehicle-to-Vehicle~(V2V)] This is an ad-hoc network allowing the direct communication between the vehicles, e.g., OBU to OBU. 
	In this format, there is no need for a fixed or pre-existing infrastructure. 
	The main applications of V2V are safety information, warning functions, security, co-operative assistance, as well as the dissemination of application-related data.

	\item[Vehicle-to-Infrastructure~(V2I)] In this type of communication the vehicles exchange information directly with the roadside infrastructure, e.g., OBU to RSU.
	This model can offer a one-hop connection to the Internet as well as a high bandwidth link.

	\item[Hybrid architecture] This design allows both V2V and V2I communication.
	In doing this, it enables multi-hop transmissions between vehicles and/or the roadside infrastructure.
	For example, vehicles that are far away from each other can use one or multiples RSUs to maintain contact. 
	Another possible scenario to be considered is vehicles distant from an RSU that can use another vehicle's OBU as a hop to extend the delivery range, thus reaching the closest RSU.

	\item[Vehicle-to-Everything~(V2X)] It adopts a broader approach than Hybrid architectures by allowing the communication between the vehicles and any other entity, such as V2V, V2I, Vehicle-to-Grid~(V2G), Vehicle-to-Pedestrian~(V2P), and Vehicle-to-Device~(V2D).
	In other words, it comprises each and every message exchange in the network.
	It works in ad-hoc mode and therefore it does not require a pre-existing infrastructure, which is an advantage in little developed or remote areas.
	
\end{description}

There are some key characteristics of this model that are worth mentioning.
For example, due to the high mobility of the nodes, this network environment has a very dynamic topology.
Another characteristic related to the high node mobility is the frequent disconnections. 
These characteristics can be explained by the speed of the vehicles, which can be high, especially on highways. 
In addition, the vehicles may also be moving in opposite directions.
This means that as quickly as they can join a network they can also leave, leading to the mentioned dynamic topology and the frequent disconnections.

Other aspects of VANETs are the predictable mobility patterns and the absence of power constraints. 
Despite the very dynamic topology, due to the mobility patterns, it is feasible to predict the future position of the vehicles as they move on pre-defined roadways. 
Furthermore, they have to respect the speed limit, traffic lights and signals, as well as being restricted to traffic conditions.
An additional feature that can bring advantages is the power supply. 
Vehicles are equipped with long-life batteries which can be recharged as they move. 
This allows processing more complex and computationally intensive applications.

Together with useful services and applications, VANETs also present new challenges that have to be addressed to ensure a good QoE.
For example, in the ad-hoc mode, the lack of an infrastructure transfers the responsibilities to the vehicles.
In doing that, they are responsible not only for their communication but also for the message forwarding of other vehicles or entities associated with that network.
It is also possible to highlight several other challenges such as signal fading and the small effective diameter of the network, incurring in a weak connectivity between the nodes, as well as bandwidth limitations.

\subsubsection*{Routing Protocols and Metrics}

Due to the dynamic nature of VANETs, which have highly mobile nodes and suffer from rapid topology changes, the design of an efficient routing protocol is very challenging. 
This protocol should allow delivering the packets in the minimum period of time with the smallest packet loss rate~(PLR) possible.
They are also responsible for avoiding transmission conflicts and reducing the degree of interference.
The routing protocols can be classified into several categories, such as topology-driven routing, position-based routing, cluster-based routing, and geocast routing, just to name a few.

The topology-driven or ad-hoc routing uses several known mobile ad-hoc network~(MANET) protocols to create and maintain the delivery paths, such as Ad-hoc On-demand Distance Vector~(AODV)~\cite{Perkins2003} and Dynamic Source Routing~(DSR)~\cite{Johnson1996}.
However, these protocols are designed for general purpose ad-hoc networks and tend to have poor performance in VANETs~\cite{Liu2004}.

A different approach is taken by the position-based routing, 
where information about the neighbouring environment and the vehicles well-defined mobility patterns can be employed allowing the selection of better routing paths~\cite{Fonseca2013}.
The Geographic Source Routing~(GSR)~\cite{Lochert2003} is one example of a position-based protocol.
It uses the city topology, with aid of a pre-loaded map, and geographic routing to determine the best nodes to receive and forward the messages.
Another proposal is the Anchor-based Street and Traffic Aware Routing~(A-STAR)~\cite{Seet2004}.
This protocol is similar to GSR but differs by adding traffic awareness, through statistically and dynamically rated maps, in the path selection.
In doing this, it tries to solve the issue of unevenly distributed vehicles which tend to concentrate more on some roads than other.

It is also possible to use cluster-based routing in VANETs.
The same method used in MANETs is adopted here, where the cluster members can perform intra-cluster communication and the cluster head is responsible for both intra- and inter-cluster message exchange.
However, due to the fast changing topology and high-speed vehicles, the clustering techniques used in MANETs are not appropriated for VANETs.
To address this issue, the Trust dependent Ant Colony Routing~(TACR)~\cite{Sahoo2012} uses the real-time position and an ant colony routing technique based on a trust value for each vehicle.
In addition, it also considers the direction and the relative speed in this process.
This allows it to better define the cluster members, providing higher scalability and fewer transmissions failures. 

A further example of VANETs routing methods is the geocast protocols.
The aim of these protocols is to deliver the packets in a form of a location-based multicast~\cite{Singh2014,Kaiwartya2014}.
This means that all nodes within a specific geographical region are able to receive these messages.
One example is the Guaranteed Geocast Routing (GGR) protocol~\cite{Kaiwartya2015}.
It uses acknowledgements to ensure the one-hop deliveries as well as a heuristic procedure to select the next hop vehicle.

\subsection{Flying Ad-Hoc Networks~(FANETs)}
\label{sec:net:fanets}

This network model can be defined as a form of MANET, and a subset of VANET, where the nodes are UAVs~\cite{Bekmezci2013}.
In the same ways as in the VANETs, the communications links can be delineated as UAV-to-UAV, UAV-to-Ground, and hybrid.
However, there are several differences between this network and other ad-hoc networks.
Firstly, the node mobility has very specific characteristics, for example, it is much faster than in VANETs. 
In addition, it also has a higher degree of movement, which means that the node positions are far less predictable.
Even in autonomous mode, the UAVs are constantly correcting and adjusting the predefined flight plan~\cite{Sahingoz2014}.
Another difference is that these nodes are capable of making fast and sharp turns, changing direction very easily, which has to be considered in the mobility models.
Apart from that, the node density is lower than in other ad-hoc networks. 
This happens because the UAVs are scattered in the sky and can be positioned far away from each other.

Besides the node mobility and density, the radio propagation model is also peculiar to these networks.
In VANETs the nodes are just above the ground and in several cases, there is no line-of-sight communication between them.
Because of that, the radio signals are disturbed by the terrain features and geographical structures.
On the other hand, in FANETs the nodes tend to be far away from the ground allowing line-of-sight communications, not being directly impacted by the ground structures and the terrain.
However, this creates a different source of interference, namely the ground reflection effect.
Another relevant issue in UAVs is the environmental weather conditions.

Furthermore, the UAV platform offers small energy constraints and the possibility of high computation power.
Modern UAVs are equipped with high-capacity batteries that can provide energy for an extended period of time.
However, this can be an issue with small form factor UAVs.
In addition, UAVs are outfitted with state-of-the-art processors which offer a considerable high computation power.
Both of these features are only constrained by the size and weight of the desired hardware.
Lighter UAVs have extended fly range and additional payload capacity, which can be used to deploy supplementary sensors or any other necessary peripherals.

\subsubsection*{Routing Protocols and Metrics}

Several routing protocols have been proposed for MANETs and VANETs networks, however, due to the specific issues that have to be addressed these routing strategies are not ideal.
Because of the fast pace and high mobility nodes in FANETs, a resourceful routing protocol is needed to provide the best possible communication channels between the UAVs.

One example of a position-based routing protocol for FANETs is the Geographic Position Mobility Oriented Routing~(GPMOR)~\cite{Lin2012a}. 
This protocol uses the movement prediction of UAVs to choose the next hop and adjust the routing tables.
Another example is the Directional Optimized Link State Routing Protocol~(DOLSR)~\cite{Alshabtat2010}.
It uses OLSR as base, modifying, however, the procedure of choosing the multipoint relay nodes through the use of directional antennas, reducing the end-to-end latency.

Another example of FANET routing protocols is the Greedy Perimeter Stateless Routing~(GPSR)~\cite{Karp2000}. 
This protocol uses only the position of routers and the packet destination to construct the routing tables.
In doing that, it provides very good scalability in densely deployed networks.
Furthermore, hierarchical protocols also tackle scalability issues. 
The UAVs are divided into groups and each cluster head communicates directly with an upper layer UAV or a satellite. 
One of the main issues with this approach is how to predict the mobility patterns of the UAVs and the clusters.
The Mobility Prediction Clustering Algorithm~(MPCA)~\cite{Zang2011} uses a dictionary based on a digital tree algorithm and the link expiration timeout to minimise this issue.

A different approach is adopted by Data-centric routing algorithms.
These models can be adapted to FANETs and usually have a publish-subscribe model type as communication architecture~\cite{Erman2008}.
The UAVs that are producing data~(publishers) are connected directly with other UAVs or the ground station that wants to consume the data~(subscribers). 
Because of that, the routing algorithm creates the tables based on the data content. 
The use of data aggregation algorithms is desirable to provide energy-efficient content dissemination.
The Autonomous self-deploying and operation of Wireless sensor-actuator networks cooperating with AeRial objEcts~(AWARE)~\cite{Gil2007} is one example of this approach.

\section{Summary}
\label{sec:net:summary}

In this chapter important components of wireless networks are discussed, including cross-layer design, wireless channels, wireless ad-hoc networks, and routing protocol characteristics, 
First of all, the concept of cross-layer implementation was presented in Section~\ref{sec:net:cross}. 
This technique can be employed to overcome the restricted form of communication of the network layers.
Because of that, it can be considered a violation of the referenced layered communication to provide more flexible mechanisms to improve the network transmissions.

Section~\ref{sec:net:wirelesschannels} outlined the wireless channels characteristics, such as the time-varying link properties and the unpredictable communication quality. 
In addition, the common issues found in these networks, such as shadowing, multipath fading, and signal attenuation were also pointed out.

Furthermore, Section~\ref{sec:net:wanets} detailed the WANETs characteristics and presented the specialised network categories related to this work, namely WMNs, VANETs, and FANETs.
The WMNs are commonly used to extend the reach of wireless access points by creating a multi-hop network. 
There is no fixed structure in these networks, therefore, they are easy to expand and reliable, as they do not have a single point of failure.
VANETs enable ubiquitous connectivity between the network members.
The main applications are related to transport efficiency and information/entertainment services. 
The last network category, FANETs, are responsible for providing reliable communication channels for flying nodes or UAVs.
They are a subset of VANETs having several similar characteristics but differ in regards to the node mobility and density, as well as in the radio propagation model.
 \cleardoublepage

\setcounter{mtc}{10}
\chapter{Video Coding Design and Quality of Experience}
\chaptermark{Video Coding Design and QoE}
\label{ch:videoStuff}

\renewcommand*{\dictumwidth}{.45\textwidth}
\renewcommand*{\dictumauthorformat}[1]{({#1})\bigskip}
\dictum[George Orwell, Nineteen eighty-four]{His thin dark face had become animated, his eyes had lost their mocking expression and grown almost dreamy. - It's a beautiful thing, the destruction of words.}

\minitoc

\lettrine[lines=3]{\color{gray}\bf{R}}{} eal-time video services are becoming a large part of the daily routines of people all over the world. 
They have been used to spread information ranging from education to entertainment content.
In addition, this technology has been used by many companies as a part of a business drive 
as well as by non-professional users, resulting in a considerable growth of online videos consumption.
In this chapter, several details about the video format and the quality of experience will be explored to provide an overall awareness on this subject.

\section{Introduction}

The rapid growth of real-time video services is evident in recent years, particularly from wireless mobile devices~\cite{Adobe2016}.
This leap is related to the technological advances in broader connectivity and the widespread adoption of smart devices, as well as the rise in popularity of this type of content. 
Many companies have used video streaming as part of a business drive to increase productivity, improve collaboration, reduce costs, and streamline and optimise business operations. Following the same trend, non-professional users are producing, sharing and accessing thousands of videos by using both wired and wireless systems~\cite{Cisco2016}. 

The video quality of these services can be impacted by several factors, including the video characteristics.
The type of codec, the streaming bitrate, the frame type, as well as the format and the length of the Group of Pictures~(GoP), can alter how the video quality is perceived.
The video content also plays an important role on the perceived quality. Videos with a small degree of movement and few details tend to be more resilient to packet loss, keeping the quality high. In contrast, videos with high levels of details and movement are more susceptible to losses and the flaws will be more noticeable~\cite{Khan2010}.

In order to assess the video integrity, Quality of Experience~(QoE) methods are desirable.
These metrics evaluate the level of the video impairments and can be defined in terms of how users perceive the quality of an application or service~\cite{Piamrat2009}. 
In other words, the QoE metrics assess the video quality considering the end-users point-of-view.

The remainder of this chapter features the main concepts and definitions needed to understand the characteristics of video formats~(Section~\ref{sec:videoFormat}).
This is followed by the examination of how video is distressed by impairments in Section~\ref{sec:videoImpairments}. 
Section~\ref{sec:videoQoE} presents an analysis of the metrics and standards used for video quality assessment. 
At the end, Section~\ref{sec:videoSummary} gives the summary.

\section{The Video Format}
\label{sec:videoFormat}

Video technology is used to record, display, and broadcast multimedia content through electronic media. In a simple way, a video flow is composed of a sequence of temporal-related images arranged to create one fluid moving picture. 
Video coding formats, or standards, aim to create efficient digital video archives. These documents describe how the pictures have to be stored and transmitted. 
Examples of video standards include Theora~\cite{Xiph.Org2011}, MPEG-4 Part 10~(H.264)~\cite{ITU-T2005}, and High Efficiency Video Coding~(HEVC)~\cite{Sullivan2012}.
On the other hand, the implementation of these formats is called codec, such as x264, OpenH264, and Xvid.
\subsection{Video Compression}

In the video technology, the compression or coding is one of the main processes~\cite{Richardson2004}. It is responsible for converting the video signal so it takes less storage space and consequently uses less bandwidth during the transmission.
The compression process can be in both lossy and lossless format.
In the lossless compression, the original data can be perfectly reconstructed, which does not allow high compression rates. 
On the other hand, in the lossy compression, several video details can be suppressed, achieving higher compression rates. 
However, with the increase in the compression ratio, there is a decrease in the video quality.

The video compression is performed by excluding some types of redundant information. 
Usually, compression algorithms exploit three main scenarios, namely spatial~(intra-frame), temporal~(inter-frame), and perceptual redundancy. 
The first two are also called statistical redundancy and can be used in lossless compression because all the information is only reorganised and not dropped. 
The spatial redundancy compression takes advantage of the significant correlation between neighbouring pixels in a given area of the frame.
All the pixels with the same characteristics are represented by a pointer, reducing the amount of information that needs to be stored.
The temporal redundancy compression capitalises on the statistical correlation among pixels in subsequent video frames.
In the same way as in the spatial compression, a pointer can be used to represent the characteristics of a block of pixels that has motion attributes, however, with static stored values. 

The last one~(perceptual redundancy) is only used in lossy compression because it requires dropping non-essential details.
This compression is based on the fact that the Human Visual System~(HVS) does not perceive all the optical information equally. 
Just to give one example, the human eye has a better perception of luminance variation than to colours deviations.
These small details, which cannot be clearly recognised, can be neglected without incurring in a perceptual alteration of the video quality.

\subsection{Coding Elements}
\begin{figure}[!htb]
	\begin{center}
		\ifBW \includegraphics[width=143mm]{./MPEG_structure_gray-eps-converted-to.pdf}
		\else \includegraphics[width=143mm]{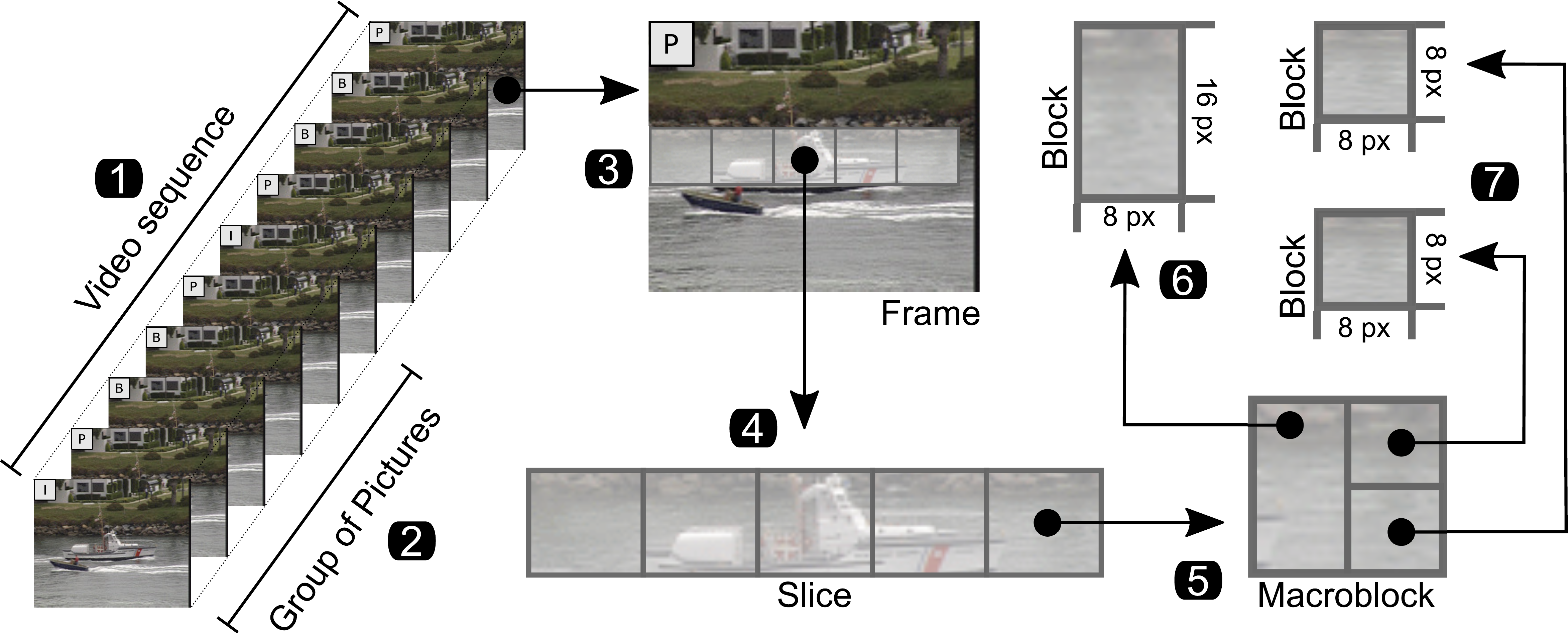}
		\fi
	\end{center}
	\caption{H.264 standard video structures}
	\label{fig:video:MPEG}
\end{figure}

The coding format is another component of the video technology. 
The H.264 is, to date, the most commonly used format of video content~\cite{Ma2016}.
Figure~\ref{fig:video:MPEG} depicts an overview of the H.264 structure and components.
The video sequence is composed of frames with different types~(1), which are, in turn, divided in the GoP~(2). Each frame~(3) has its own type and can be divided into slices~(4). The slices can hold one or more macroblocks~(5) that can be broken down in blocks~(6). These components are explained below.

A video frame can be defined as an amount of data after compression.
There are three major types of frames: Intra-coded frames~(I-Frames), Predictive-coded frames~(P-Frames), and Bi-directionally predictive-coded frames~(B-Frames). The compression in I-Frames is carried out by reducing the spatial redundancy only in the current raw frame.
Such compression takes advantage of the human eye inability to detect certain minor changes in the image.
This frame is like a conventional static image and because of this, these frames are reconstructed independently. 
Because of that, as soon as the information that this frame holds is needed, it can be displayed without having to consult other frames.
On the other hand, P- and B-Frames only enclose segments of the image information, reduce the coding data, and thus improve the video compression rates. 
On the negative side, these frames need other I- or P-Frames to be decoded, respectively.

A P-Frame only contains the changes of the actual image when compared with the previous I- or P-Frame. With this type of prediction, less coding data is required~($\approx$ 50\% less in contrast to I-Frame size). B-Frames also use prediction to reduce the amount of coding data, but unlike the P-Frame, they use both the previous and following frames to determine the content. 
The size of a B-Frame is usually $\approx$ 25\% smaller than a P-Frame~\cite{Wiegand2003}. 
In both types of frame the data is expressed as motion vectors and transforms coefficients which are used in prediction correction and motion compensation. 

\begin{figure}[!htb]
	\begin{center}
		\ifBW \includegraphics[width=100mm]{./GoP_realvideo_gray-eps-converted-to.pdf}
		\else \includegraphics[width=100mm]{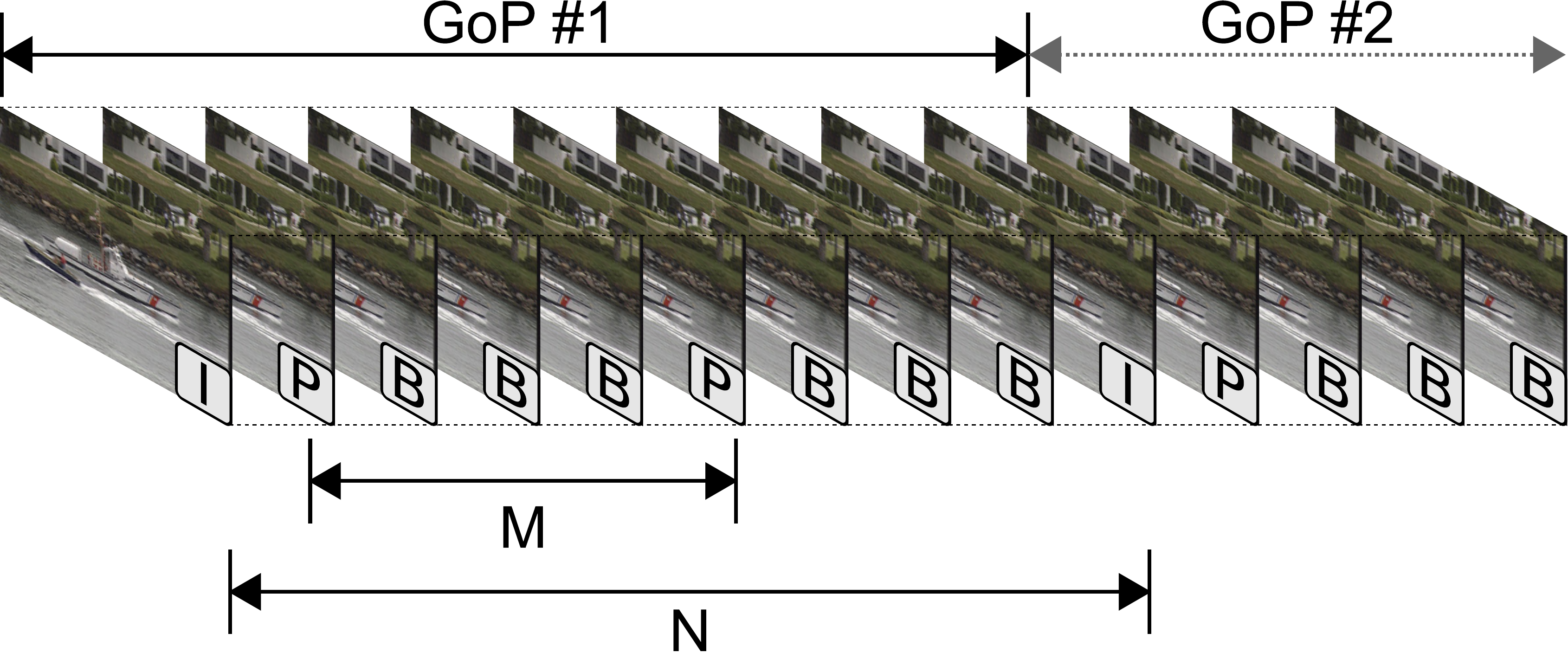}
		\fi
	\end{center}
	\caption{GoP structure}
	\label{fig:video:gop}
\end{figure}

GoP are sequences of frames grouped together.
The MPEG standard uses a hierarchically-structured GoP. This structure is a full sequence, which means that it contains all the information that is needed to decode the video images, within that period of time. Therefore, it enables random access into that portion of the video.
Figure~\ref{fig:video:gop} shows an example of a GoP structure, where $M$ represents the distance between successive P-Frames and $N$ defines the separation between the adjacent I-Frames. The GoP frame ratio is $N:M$, in the example, the ratio is 9:3. This means that it contains one I-Frame at every 9 frames and one P-Frame after each three B-Frames.
This structure is flexible, and both the frame types and their location within a GoP are adjusted in accordance with the encoding format desired.

A slice is another MPEG structure, which are special regions defined during the coding process.
Each slice represents a spatially distinct region inside a frame. 
In view of this, everything within the slice's area will be encoded independently from the remaining of the frame. 
This gives the opportunity to the codec to decide to use a distinct amount of compression inside the same frame.
In other words, a slice is composed of an arbitrary and successive number of macroblocks. 

The macroblocks are fundamental processing units for motion compensation. Each one can be associated with a unique motion vector. 
Due to the hierarchical MPEG structure, each frame type has a predefined way to handle motion compensation. 
For example, I-Frames can only have intra-coded macroblocks. 
This means that it has to be self-contained. On the other hand, P-frames can have both intra-coded or predicted motion compensation. 
Finally, B-Frames can have intra-coded, predicted, and/or bi-predicted motion compensation macroblocks.

Blocks are the smallest coding unit defined by the MPEG format. This is the most basic element in frame coding. Usually the frames are segmented into blocks of 8x8 pixels, however, the H.264 standard allows blocks with 4x4, 16x8, and 16x16 pixels, which can be defined per-macroblock basis~\cite{ITU-T2005}.

\section{Video Impairments}
\label{sec:videoImpairments}

In general, video impairments tend to be transient, thus it is difficult to detect and act upon them.
Consequently, this is a hard problem to solve in a satisfactory way, especially in wireless networks. 
The situation is aggravated by the fact that video transmission is more sensitive to impairments than data distribution. 
This can lead to a serious perceived quality degradation even with a low rate of packet loss.
Apart from this, not all packets have an equal impact on the perceptual video quality. 
Each packet, or group of packets, can carry different types of information about the frames that are being transmitted, and some are more important than others. 
This originates from the video frame arrangement carried out by the encoding/compressing process.
Besides that, subjective factors also play an important role; for example, the motion and complexity levels, as well as the amount of detail in the video sequence can determine how the packet loss is perceived.

Additionally, the encoding and compression techniques performed by the codec will have an influence on the video quality as well.
The basic principle of the compression techniques is to reduce the amount of information that is stored and/or transmitted. 
There are several ways to achieve this reduction, such as by limiting the redundant data, discarding less important information, and using prediction schemes~\cite{Seeling2004}. 
With higher compression, less data needs to be sent through the transmission channel, however, any packet loss that may happen will have a greater impact on the video quality.

Another important trait to be considered is the video content.
Studies have shown that videos with slight movement have better packet loss resilience, whereas videos with rapid movement are less resilient and the flaws generated are more noticeable.
Generally, the video content can be classified into three categories according to its motion intensity~\cite{Lotfallah2006}. 
The first one is low intensity, which includes sequences with a small moving region of interest, generally someone's face and traditionally on a static background.
This type of video sequence can handle up to 20\% of packet loss and still have an acceptable video quality.  
The second classification, medium intensity, contains sequences where adjacent scenes are modified. This category usually has frames with higher image details. 
The packet loss natively supported is up to 10\%.
Finally the last classification, high motion intensity, includes wide-angle image sequence and a lot of movement, as usually the entire picture is moving.
This classification is more sensitive to packet losses, thus it can handle up to 6\%.
Table~\ref{tab:video:lossImpact} summarises the relationship between the video content, motion intensity, and natural resilience to packet loss~\cite{Khan2009}.

\begin{table}[!ht]
\caption{Impact of packet loss on different motion intensities}
\begin{center}
\begin{tabular}{c|l|l}
\hline \textbf{Motion intensity} & \textbf{Video characteristic} & \textbf{Packet loss} \\ 
\hline
\hline Low & small moving region & up to 20\% \\ 
\hline Medium & higher movements and image details & up to 10\% \\ 
\hline High & wide-angles and non-static backgrounds & up to 6\% \\
\hline 
\end{tabular}
\end{center}
\label{tab:video:lossImpact}
\end{table} 

\begin{figure}[!htb]
  \begin{center}
    \ifBW \includegraphics[width=100mm]{./packet_loss_realvideo_gray-eps-converted-to.pdf}
    \else \includegraphics[width=100mm]{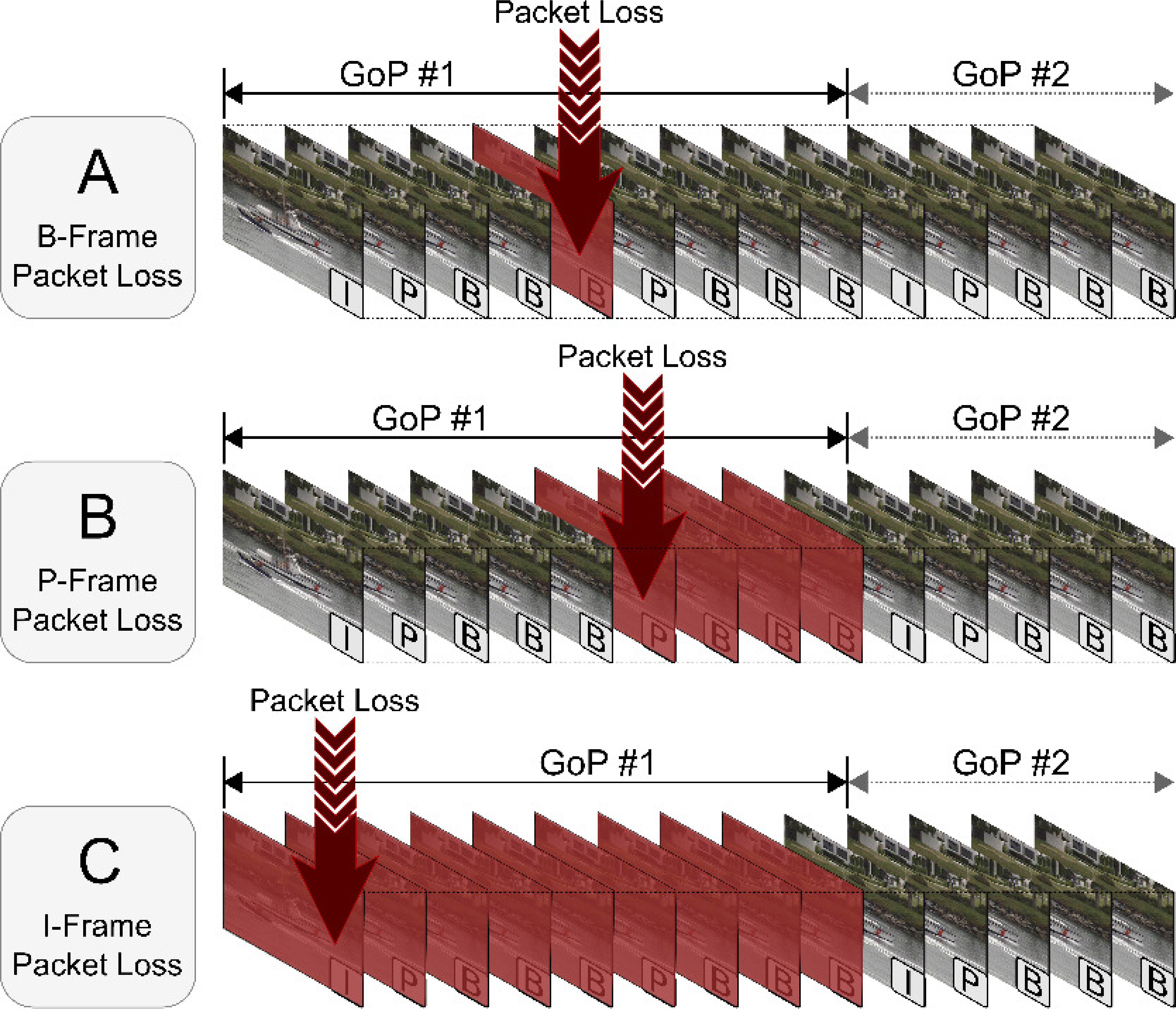}
    \fi
  \end{center}
  \caption{Packet loss repercussion on the GoP}
  \label{fig:video:packet_loss}
\end{figure}

Furthermore, the MPEG's hierarchically-structured GoP also has an influence on how the packet loss will be perceived. 
Because of its ranked order used to organise the frames, which allows reducing the file size, some pieces of information are more important than others.
As a result, the packet losses may affect the video quality in different ways, depending on the information that was lost. 
Figure~\ref{fig:video:packet_loss} depicts three hypothetical scenarios. 
In Scenario~(A), the network drops packets within a B-Frame. 
In this case, just this particular frame will be damaged, because it needs information from other frames, but none of the others needs its information. 
Scenario~(B) describes dropped packets in a P-Frame. 
This will produce impairments that are extended through the rest of the GoP. 
This means that the error will only be corrected with the arrival of an I-Frame. 
Finally, Scenario~(C) shows one I-Frame being damaged by packet losses. 
In this case, the impairment will spread throughout the remainder of the GoP. 
This happens because all other frames need the information carried by the I-Frame, and thus the video quality will only improve when the decoder receives the next I-Frame.

\section{Video Quality Assessment}
\label{sec:videoQoE}

The video quality measurement can be defined as an estimation of the video degradation. 
Several types of artifacts and/or distortions can be introduced during the encoding, decoding, and transmission process of video content, which can negatively impact the end-user perception of quality.
It is not difficult for human subjects to point out video defects, especially if they are happening over an extended period of time. 
On the other hand, it is not a straightforward task to quantify these impairments. 
Therefore, QoE metrics are used to provide support for this task.

Quality of Experience~(QoE) can be defined as a measure of customer's personal observation of a service, product, or application.
QoE is related to Quality of Service~(QoS) metrics, which measure the network operation conditions, such as crosstalk, noise, lost or dropped packets, however differs from it because adopts a holistic approach assessing the end-to-end connection as well as the final user requirements or expectations.
It is possible to say that QoE extends the concept of QoS by encompassing additional factors, aiming to provide an appraisal of human expectations in respect to a particular activity.

In regards to the QoE, there are two types of metrics that can be used to quantify video impairments: subjective and objective. 
Objective metrics use mathematical models with a set of indicators that are correlated with the user's perception of quality. 
These metrics obviate the need for human intervention and can have instant results.
Subjective evaluations use human individuals to assess the video quality. 
These metrics are able to capture all the details that might affect the user experience, however, they need long rating process, are expensive, and the results are not reproducible.

\subsection{Subjective metrics}
One of the most widely used approaches for subjective video evaluation is the Mean Opinion Score~(MOS)~\cite{P.910,ITURec500-13}. 
Using this technique, it is possible to capture the full degree of subjectivity that end-users adopt to perceive the impairments in the video sequence. 
MOS is recommended by the ITU-R and uses a group of people voting in video sequences, according to a predefined quality scale, to rate the quality. The MOS scale goes from 1 to 5, where 5 is the best possible score, as shown in Table~\ref{tab:video:mosscale}.

\begin{table}[!hbt]
	\caption{Mean Opinion Score scale }
	\begin{center}
		\begin{tabular}{c|l|l}
			\hline
			\textbf{MOS} & \textbf{ACR} & \textbf{Impairment}\\
			\hline 
			\hline 5 & Excellent & Imperceptible\\
			\hline 4 & Good & Perceptible but not annoying\\
			\hline 3 & Fair & Slightly annoying \\
			\hline 2 & Poor & Annoying \\
			\hline 1 & Bad & Very annoying\\
			\hline
		\end{tabular}
		\label{tab:video:mosscale}
	\end{center}
\end{table}

The recommendation ITU-R BT.500-13~\cite{ITURec500-13} describes several subjective quality test methods, such as Absolute Category Rating~(ACR), Degradation Category Rating~(DCR), and Double-Stimulus Continuous Quality-Scale~(DSCQS). In the ACR method, the video sequences to be ranked are presented one by one. 
The viewers watch the sequences only once and then rate the quality using the scale shown in Table~\ref{tab:video:mosscale}.
The ACR scale is related to the MOS scale.
This is considered one of the most suitable methods for conducting the quality assessment of video applications~\cite{Seshadrinathan2010}.

In the DCR method, the video sequences are displayed in pairs. 
The first one is the original video, and the second is the impaired sequence which is a result of the experiments. 
In this procedure, voters are asked to rank the impairment of the second video using as reference the first one. 
Finally, in the DSCQS method, the video sequences are showed in pairs as well, similarly to DCR, one is the impaired and another is the reference. However, the subjects do not know which one is the reference and are asked to score the quality of both videos. The resulting scale is present in pairs containing both the scores for the reference and the impaired videos.

The subjective methods provide good results, however, they are fairly expensive to set up, need the collaboration of a considerable number of humans voters, and it takes a long time to get the results~\cite{Winkler2008}.
Taking this into consideration, objective metrics are desirable. 
They are fast and intend to be unbiased. 
In addition, they are computed through mathematical calculations, and thus, measurable and verifiable.

\subsection{Objective metrics}
One of the most common objective metrics is the Peak Signal to Noise Ratio~(PSNR)~\cite{Ma2011}. 
This is a non-perception-based metric that is widely used to evaluate the performance of several optimisation techniques. 
This metric gives the distortion measurements between two video images, which are averaged over the video sequence. 
Through this approach, a single well-defined metric is provided and it has proven useful to compare different optimisation mechanisms using a single scale. 
However, the absolute PSNR values of different video content may not be directly related~\cite{Huynh-Thu2008}. 
Additionally, due to the use of averages, this metric is not able to differentiate between individual loss occurrences over a period of time.

Equation~\ref{eq:video:psnr} defines the PSNR calculation, where the output result is a logarithmic decibel value. 
This scale is used because of the wide dynamic range of several signals. 
The PSNR calculation is based on the Mean Squared Error~(MSE), defined in Equation~\ref{eq:video:mse}. 
Table~\ref{tab:video:psnrnotation} summarises the notations adopted.

\begin{table}[!hbt]
\caption{PSNR Notation}
\begin{center}
\begin{tabular}{c|l}
\hline \textbf{Parameter} & \textbf{Meaning} \\
\hline
\hline $MAX$ & Upper pixel value limit of the input image\\
\hline $A$ and $B$ & Assessed images \\
\hline $m$  & Image width in pixels \\
\hline $n$  & Image height in pixels \\
\hline $m \times n$ & Number of pixels\\
\hline
\end{tabular}
\label{tab:video:psnrnotation}
\end{center}
\end{table}
\begin{equation}
\mathit{MSE}=\frac{1}{m\times{n}} \sum_{i=1}^{m}\sum_{j=1}^{n} (|A_{(i,j)}-B_{(i,j)}|)^{2}
\label{eq:video:mse}
\end{equation}
\begin{equation}
\mathit{PSNR(dB)}=10 \times \log_{10} \left ( {\frac{\mathit{MAX}^{2}}{\mathit{MSE}}} \right )
\label{eq:video:psnr}
\end{equation}

At the end, the PSNR is a common objective metric to assess data fidelity. 
However, it is based on a byte-by-byte comparison disregarding what the information actually represents. 
Additionally, PSNR does not recognise the pixel structure in the image nor the spatial relationship between the pixels, thus, it does not consider the visual importance of each pixel~\cite{HuynhThu2008a}.
Since the results obtained from PSNR do not correlate with subjective human perception, other metrics have been proposed. Two of the most widely adopted objective QoE metrics that correlate with subjective perception are Structural Similarity Metric~(SSIM) and Video Quality Metric~(VQM)~\cite{Chikkerur2011}.

The SSIM~\cite{Wang2004} assesses video quality through frame-to-frame measurements. 
This metric employs the video structural distortion as an estimate of the visual quality.
This provides a good correlation with the perceptual quality because the human visual system is very effective in the extraction of structural information, thus, any impairment in this information will be proportional to the perceived errors.

SSIM indexes are composed of three basic components, namely, luminance, contrast and structural similarity. 
The index values are calculated only for carefully selected blocks inside each frame, and not the whole frame. 
This allows saving computation resources while providing precise video quality assessments.
At the end, an average of all SSIM indexes will give the frame quality.
This information is used together with the local luminance to define a final perceptual quality.

Finally, the general video quality is found by the weighted sum of all frames indexes. 
The weighting value is based on each frame's motion intensity.
The overall index value is a decimal number between 0 and 1, where 1 stands for precisely the same video. 
A simplified SSIM calculation is defined through Equation~\ref{eq:video:ssim} and Table~\ref{tab:video:ssimnotation} specifies the parameters used.

\begin{table}[!hbt]
\caption{SSIM Notation}
\begin{center}
\begin{tabular}{c|l}
\hline \textbf{Parameter} & \textbf{Meaning} \\
\hline
\hline $c$ & Contrast correlation \\
\hline $l$ & Luminance correlation \\
\hline $s$ & Structure correlation \\
\hline $x$ and $y$ & Measurement between two windows of equal size\\
\hline $\alpha, \beta, \gamma$ & Significance of each component \\
\hline
\end{tabular}
\label{tab:video:ssimnotation}
\end{center}
\end{table}
\begin{equation}
\mathit{SSIM(x,y)} = [c(x,y)]^{\alpha} \times [l(x,y)]^{\beta} \times [s(x,y)]^{\gamma}
\label{eq:video:ssim}
\end{equation}

VQM~\cite{Pinson2004} is a modified discrete cosine transform-based metric to assess video quality. Values closer to 0 correspond to the best video quality. 
In order to calculate these values, this metric uses the same features as the human eye to perceive the video quality, including colour and block distortion, blurring and global noise. 
More specifically, this model employs a linear combination with seven independent parameters. 
Four parameters are extracted from spatial gradients~($SI_{loss}$, $SI_{gain}$, $HV_{loss}$, $HV_{gain}$), two are obtained from a chrominance vector~($CHROMA_{spread}$, $CHROMA_{extreme}$), and the last one is derived from absolute temporal and contrast details~($CT\_ATI_{gain}$). These parameters are defined as flows:

\begin{itemize}
	\renewcommand{\labelitemi}{$\bullet$}
	\item \textsc{Parameter $SI_{loss}$} - It is responsible for detecting differences in the spatial information. It uses a 13-pixel spatial information filter~(SI13) that evaluates any meaningful visual edge impairments~(e.g., blurring);
	
	\item \textsc{Parameter $SI_{gain}$} - This parameter measures the opposite situation of $SI_{loss}$. This means that any enhancement in the video quality provided by any edge sharpening technique will be reflected here;
	
	\item \textsc{Parameter $HV_{loss}$} - It is in charge of identifying and measuring any edge fluctuation from horizontal/vertical position to a diagonal orientation. In doing this, it is able to catch horizontal/vertical edges that suffer more blurring than diagonal edges;
	
	\item \textsc{Parameter $HV_{gain}$} - This parameter computes the reversed edge movement of $hv_loss$. In other words, it detects diagonal-oriented edges that had tilted to horizontal/vertical-oriented edges~(e.g., blocking artifacts);
	
	\item \textsc{Parameter $CHROMA_{spread}$} - It identifies the variation of two-dimensional colour samples. In the same way as in the other parameters, the colour variation is only reported if it is over a large area which can impact significantly on the perceived quality;
	
	\item \textsc{Parameter $CHROMA_{extreme}$} - As the parameter's name suggest it is also responsible for detecting colour variations. Its operation is similar to the $CHROMA_{spread}$, but differs by exploring acute colour impairments in a restricted area~(e.g., common impairment produced by digital transmission errors);
	
	\item \textsc{Parameter $CT\_ATI_{gain}$} - This parameter infers the motion intensity of a defined region by computing the product of temporal and spatial intensity. This information is important to grade how much the impairments will impact on the perceived video quality. A high motion intensity region will have more perceptual artifacts than a low motion intensity region~(e.g., noise and error blocks).
\end{itemize}

After the definition of all parameters, the VQM applies a linear combination to score the video quality.
This metric is optimised to obtain a good correlation between objective and subjective assessments.

\section{Summary}
\label{sec:videoSummary}

This chapter described several components and details related to video technology.
In Section~\ref{sec:videoFormat} video compression techniques were discussed, along with the description of several methods that allow reducing the redundant video information.
These algorithms include statistical compression~(spatial and temporal), as well as perceptual redundancy compression. 
Additionally, the H.264 structure and components were described, including the GoP format, the different types of frames, macroblocks, and blocks.

Section~\ref{sec:videoImpairments} presented how the video sequences are impacted by packet loss. 
It was evidenced that not all packets have equal importance on how the quality is perceived. 
This means that some packets~(or group of packets) are more important than others.
Following the same idea, I-Frames are more important than, P-Frames, which in turn, are more important than B-Frames. 
Furthermore, the position inside the hierarchically-structured GoP is equally important. 
Frames closer to the beginning of the GoP are more important than frames closer to the end.
Additionally, the motion intensity of the videos also plays an important role. 
Videos with low motion intensity are more resilient to packet loss than a video with high motion intensity.

The video quality assessment was examined in Section~\ref{sec:videoQoE}. 
The concept of QoE was defined as a measure of customer's personal perception of a service or application. 
This metric is related to QoS but differs by encompassing additional factors providing an end-to-end assessment. 
The methods used to score the QoE can be objective or subjective. 
The former uses mathematical models and the latter requires human voters to quantify the video quality.
 \cleardoublepage

\setcounter{mtc}{11}
\chapter{Advances on Improved Video Transmission}
\label{ch:rw}

\dictum[George Orwell, Nineteen eighty-four]{And even technological progress only happens when its products can in some way be used for the diminution of human liberty.}

\minitoc

\lettrine[lines=3]{\color{gray}\bf{T}}{} he design of techniques to improve the video transmission quality is a complex task which involves a number of concepts and definitions, especially error-correcting codes. 
This chapter discusses several of these details to provide a general understanding on this subject. 
It also presents a literature review of mechanisms that aim to enhance the video transmissions with and without error-correcting codes.

\section{Introduction}

Video services are very demanding applications, as they require a steady and continuous flow of packets.
Taking this into account, both network and application-level technologies must be adopted to enable the delivery of a video stream with satisfying quality.
At the network level, QoS techniques can be used to safeguard video transmissions. 
However, this does not guarantee that the perceived quality by the end-user is at the best possible.
Thus, other mechanisms or techniques are required to provide the end-users with high perception video.

Video transmissions over wireless channels introduce a set of challenges. 
The omnidirectional signal has the tendency to endure time-varying physical effects which can result in a degraded video quality.
The outcome of shadowing, multipath fading, hidden terminal, and the antenna range may originate errors in packets leading them to be dropped.
Additionally, the strict bandwidth, delay requirements, and node overpopulation can also aggravate the situation creating channels with high error rates.

To mitigate these issues, error correction techniques have been successfully used to protect the transmission of real-time video services~\cite{Nafaa2008}. 
These techniques provide robust video transmission through redundant information that is sent along with the original data set~(FEC-based), or by resending the lost packets~(ARQ-based).
However, despite the latest developments, there is still a shortage of adaptive QoE-driven mechanisms to improve real-time video transmissions~\cite{Jiang2012,Bellalta2014}.
These mechanisms need to be able to perform on unforeseen situations to protect the most QoE-sensitive data, while not adding unnecessary network overhead.
In order to do that, several aspects of the video details, along with the network characteristics and condition need to be considered.
As a result, it is possible to assign an optimal amount of redundant data only to QoE-sensitive data, or carefully choose which information should be sent again in case of losses.
These are important features in any mechanism, especially if it will be used in highly dynamic networks.

To address the aforementioned issues, this chapter introduces and discusses error correction techniques in Section~\ref{sec:rw:EC}. 
After that, several important studies about the optimisation of video transmission are presented and reviewed in Section~\ref{sec:rw:mechanisms}, followed by the summary and open issues in Section~\ref{sec:rw:summary}.

\section{Error-Correcting Codes (ECC)}
\label{sec:rw:EC}

ECC or error correction~(EC) techniques are used to enhance the communication over unreliable channels. 
This type of approach is especially needed in wireless channels which are subject to a series of issues.
Using EC techniques it is possible to reconstruct the original error-free data if losses occur during the transmission. 
Generally speaking, the error correction can be carried out with two distinct methods, namely Forward Error Correction~(FEC) and Automatic Repeat reQuest~(ARQ). 
FEC-based schemes send redundant information~(parity bits) along with the original data set, which can be used to recover the original data in case of loss. 
On the other hand, ARQ-based schemes use error-detection codes along with positive and negative acknowledgements~(ACK).
If a negative ACK is received~(or a time-out happens), the sender performs a retransmission to correct erroneous or lost data.
Both FEC and ARQ methods are discussed below.

\subsection{Automatic Repeat reQuest (ARQ)}

ARQ is an error-control~(or error-correction) method based on the retransmission of the packets that fail to reach their destination and/or arrive in a damaged state. 
This retransmission process is generally triggered by two conditions:~(1) packets that are not acknowledged on time, in the case of packet loss~(delay-constrained retransmission), and~(2) through data request, when erroneous packets arrive~(parity retransmission). 
Taking this into consideration, the use of a reliable feedback channel is of critical importance to achieve a good performance.
This is a closed-loop mechanism also known as backwards error correction, and is commonly adopted in unicast protocols~\cite{Rizzo1997}. 
These techniques are particularly helpful if the end-to-end network statistics are unknown~\cite{Setton2008}. 

There are basically three types of ARQ, namely Stop-and-wait, Go-Back-N, and Selective Repeat ARQ~(ARQ-SR). 
The first, Stop-and-wait ARQ, is the most naive method because it sends one frame at a time, and then, waits for a positive ACK, a negative acknowledgement~(NACK), or a time-out incident. 
In the case of an ACK, the next frame is transmitted; however, in the case of a NACK or time-out, the same frame is transmitted again, until an ACK is received. 
This method ensures that the data is not lost, as well as correct packet order. 
It is clear that this method raises several real-life implementation issues. 
For instance, if an ACK is lost or damaged, the same frame will be re-sent; the latency can significantly increase; the channel throughput is just a fraction of what is expected. 
This issue is addressed in the other ARQ methods through the addition of a larger sequence number~\cite{Zhang2012}. 
In this way, it is possible to send several packets at the same time, and then wait for an ACK to acknowledge the whole set.

The second method, Go-Back-N ARQ, sends the number of frames defined by a sliding window, before receiving any ACK from the receiver. 
To keep track of the sequence number, the receiver process always acknowledges what has been sent with the number of the last frame received correctly, and in the right sequence. 
If this number matches some point in the sliding window, it slides to that point, and the transmission or retransmission of the missing frames is started. 
On the other hand, if the number does not match any point, all the frames inside must be retransmitted. 
This process continues until there are no more packets. 
It is a more efficient method because there is no need to wait for an ACK for each packet, and hence, time is not spent waiting for another packet to be sent~\cite{Chen2012}. 
The drawback is that this method can involve having to send duplicate frames because if one frame in the window is lost or damaged, every subsequent frame has to be retransmitted.

The last method, ARQ-SR, also continues to send the frames inside the sliding window even after losses.
However, it differs from the Go-Back-N as it only resends the lost frames.
On the one hand, this makes it more bandwidth efficient, on the other hand, it needs a more complex process to receive and acknowledge the frames. 
This means that it has to keep track of the earliest frame that has not been received, and reports to the sender the ACKs and NACKs.
As a result, the sender will use the ACKs to move the sliding window and the NACKs will be the only ones that are retransmitted~\cite{Cai2008}.

These error correction methods are appropriate as a means of recovering from a few errors when the network is temporarily congested or undergoing some type of interference. 
If the conditions of the network have deteriorated, even the ARQ-SR scheme will produce a large number of packets that need to be retransmitted, generating a feedback implosion problem, which can degrade the network even further~\cite{Li2007}.
Another drawback is that, although ARQ does not consume unnecessary bandwidth, it can increase the round-trip time~(RTT) and the delay~\cite{Cai2008}. 
In this case, it is only helpful to retransmit packets if they have a chance to arrive on time, or in other works, before its playout time.
Additionally, several enhancement techniques have been proposed for the ARQ methods~\cite{Tsai2009,Han2010,Hassan2010}, including some in combination with FEC-based schemes. 
Further details can be found in Section~\ref{sec:rw:ec_based}.

\subsection{Forward Error Correction (FEC)}

FEC are methods that, if correctly employed, can enhance the video quality in noisy or lossy networks.
They are open-loop mechanisms that send redundant data along with the original set. 
This means that if some original information is lost, the error-free data can be reconstructed using the redundant information, without the need for retransmissions.
One advantage of this technique lies on the fact that no further interaction between the sender and receiver is required.
This happens because the receiver is able to reconstruct the lost information solely through the correctly received packets.

One advantage of this technique is to ensure a good performance when applied to multiple receivers in an error-prone environment. 
In this type of scenario, the video impairments tend to occur due to uncorrelated and time-varying losses.
A good application of this technique, for example, is to provide reliable multicast on the Internet~\cite{Rizzo1997}.

This section presents three categories of FEC codes as shown below. 
Although there are several other implementations of EC techniques, our objective here is not to make a survey of these mechanisms but just to name a few examples. 

\begin{itemize}
	\renewcommand{\labelitemi}{$\bullet$}

	\item \textbf{Block codes}: These ECCs work on a fixed-size block of $k$ input bits~(or symbols) generating $n$ bits in the output~$(n,k)$. 
	This means that the sender builds blocks of predetermined size~($k$), adds a predefined amount of redundancy~(parity bits), and transmits~($n$) to the receiver. 
	The receiver, in turn, uses a decoding mechanism to rebuild the error-free data.
	The overall performance and success of the transmissions rely upon the parameters chosen for the block code configuration. 
	Examples of block codes are the Hamming(7,4)~\cite{Hamming1950}, 
	the Reed-Solomon~(RS) code~\cite{Reed1960}, and the Low-density parity-check~(LDPC) codes~\cite{Gallager1962}.
	
	\item \textbf{Convolutional codes}: It can be characterised as continuous or codes with arbitrary block length.
	These codes differ from the previous, as their size is not settled by any algebraic properties.
	The generation of the parity bits~(or symbols) is generally applied using a sliding operation of a function with boolean polynomial proprieties. 
	Examples of convolutional codes are the Turbo code~\cite{Berrou1996} and the Serial concatenated convolutional codes~(SCCC)~\cite{Benedetto1998}.
	
	\item \textbf{Fountain codes}: This technique produces an endless supply of encoded symbols.
	The original error-free data can be recovered from any subset of encoded symbols that is larger or at least equal in size to the source set.
	These codes are considered rateless because they do not have a fixed code rate.
	Examples of fountain codes are the LT codes~\cite{Luby2002} and the Raptor codes~\cite{Shokrollahi2006}.
	
\end{itemize}

All FEC techniques add some type of redundant information which is sent along with the original data set. 
In order to provide a good result at the receiver side, a suitable amount of redundancy has to be added.
One disadvantage of this scheme is the increased computational effort required to compose the redundancy. 
There are however FEC schemes which provide good performance and can be used in real-time video transmission, RS codes, which will be detailed below.
Another drawback is the network overhead, as it needs a larger bandwidth to add a larger amount of redundant information for all the data~\cite{Neckebroek2010,Abboud2011}.

This problem can be overcome by employing coding optimisations that include content-aware models with a UEP scheme~\cite{Nafaa2008}. 
The UEP scheme can take advantage of the knowledge about video details obtained through the cross-layer models. 
In this way, these methods can provide different amounts of redundancy to the video packets that are more important, and will thus cause a bigger impact if they are lost. 
In doing this, packets with more QoE-sensitive data are effectively protected, and this reduces the impairments in the video sequence, while saving network resources.

The RS code~\cite{Reed1960} is one of the most widely used FEC-based methods.
RS code belongs to the linear block codes category and it is a non-binary cyclic ECC that uses univariate polynomials over blocks with a predetermined size.
This code can be use to provide error control in a great variety of digital systems, such as CDs, DVDs, Blu-ray discs, barcodes, digital television, ADSL, xDSL, satellite data links, as well as in wireless and mobile communications.

\begin{figure}[!htb]
	\begin{center}
		\includegraphics[width=4.0in]{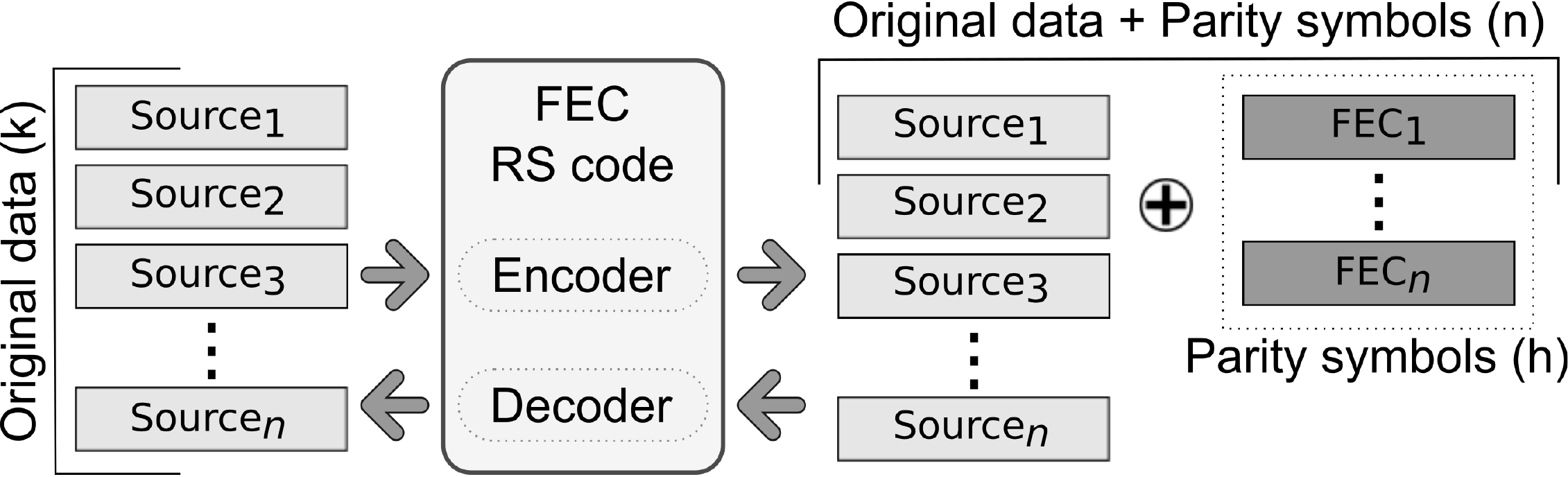}
	\end{center}
	\caption{Reed-Solomon code}
	\label{fig:rw:fec}
\end{figure}

This ECC method can detect and correct multiple errors in each block. In order to do that, the encoder receives the source blocks and adds a determined amount of redundancy. This information is sent to the receiver and the decoder will attempt to detect any possible error and use the redundant data if needed to rebuild the original error-free data, as depicted in Figure~\ref{fig:rw:fec}.

Table~\ref{tab:rw:rsnotation} shows the RS notation. 
The code can be specified as $RS(n, k)$.
This means that the total block size, including the redundancy data symbols, is represented by~$n$.
The parameter~$k$ indicates the original data set size, resulting in the~$(n, k)$ parity code. 
The last parameter is~$h$, and it defines the amount of redundancy~(or parity symbols), which is the same as~$h = ( n - k )$.
In order to recover the entire original data set $k$, at least~$( n - h )$ packets have to arrive successfully. 
The robustness to losses is determined by the size of $h$, and the error recovery against an average packet loss rate can be expressed as~$h / n$ or~$( n - k ) / n$.

\begin{table}[!hbt]
	\caption{Reed-Solomon Notation }
	\begin{center}
		\begin{tabular}{c|l}
			\hline
			\textbf{Notation} & \textbf{Meaning} \\
			\hline 
			\hline 
			$k$ & Original data set \\
			\hline
			$n$ & Original data set $+$ parity symbols \\
			\hline
			$h$ & Redundancy amount~(or parity symbols) \\
			\hline
			$h / n$ & Error recovery rate\\
			\hline
		\end{tabular}
		\label{tab:rw:rsnotation}
	\end{center}
\end{table}

The RS erasure code has low complexity and therefore offers a better performance for real-time services~\cite{Neckebroek2010}. 
This is important when considering its adoption in services that include video delivery as they demand a continuous flow of packets as well as low delay.

\section{Video Transmission Optimisation}
\label{sec:rw:mechanisms}

Owing to the wide range of solutions, the mechanisms presented in this section were divided into Non-EC and EC-based methods. 
It is possible to enhance video transmission with several techniques such as packet prioritisation, link adaptation, and video bitrate adjustment. 
However, without using an EC-based scheme this improvement will only be attainable to a limited degree. 
At the end, an error correction mechanism is needed to ensure a high video quality as perceived by end-users, whatever network adversity may occur.

\subsection{Non-EC based Optimisation Mechanisms}

Several techniques have been proposed to enhance the quality of video transmissions without using EC-based methods.
One approach is the combination of an adaptive algorithm in the link layer and a scalable video codec in the application layer~\cite{Haratcherev2006} to improve the quality of real-time video streams over wireless channels. 
Its architecture is based on a cross-layer technique that continuously adapts the video encoding rate~(through the use of scalable video codec) at the application layer, in accordance with its link layer status.
Another advantage of this approach is that it employs a link adaptation algorithm. 
This algorithm is able to efficiently manage the modifications in the wireless channel conditions by using both SNR and the wireless channel statistics to automatically adjust the radio and link layer parameters, with the aim of obtaining the best possible packet transmission quality. 

One of the weaknesses of this architecture is that it does not take into account the video content that is being delivered. 
It is known, for instance, that videos with lower spatial or temporal resolution are more loss resilient and in this case, the encoding rate should not be changed, since it leads to unnecessary modifications, and consumes time and resources. 
Another problem is that the encoding characteristics of the scalable video have to be chosen prior to the running time. 
If these parameters are not carefully chosen, there may be some network conditions that do not fit any of the existing encoding systems, and this will lead to the selection of non-optimal coding.

Another cross-layer proposal allows the video packets to be mapped, from the application layer to the appropriate Access Categories~(ACs), at the link layer~\cite{Chilamkurti2009,Lin2009}.
This classification is performed depending on the importance of the video information that the packets are carrying.
In doing this, it enables a content-aware categorization to prioritise the video packets in terms of their frame type.
This means that I-frames will have higher priority, followed by P-frames, and B-frames will have lower priority. 
In addition, downward probabilities are employed.
This means that, if the high priority AC queue exceeds a pre-determined threshold, the less important packets are mapped to lower priority ACs.

This solution provides good results when the network is not heavily loaded. 
However, when packet losses are caused by network congestion~(e.g., concurrent transmissions from other nodes), it is not sufficient to prioritise the packets.
In this case, some type of error recovery technique must be applied to improve~(or even maintain) the video quality. 
A further weakness is the lack of assessment of the motion activity, as frame types have different degrees of importance depending on the motion activity of each frame.
Finally, the position of each frame inside the GoP also plays an important role. 
It should be taken into account that frames close to the beginning of the GoP produce larger video impairments if lost than frames close to the end of the GoP~\cite{Greengrass2009a}.

Another proposed solution is a technique to improve the quality of the video transmission through optimising the network resource utilisation in accordance with the video content classification~\cite{Khan2010}.
The optimisation is achieved by determining the initial quality and then adapting the Video Sender Bitrate and the Frame Rate to the users QoE requirements. 
This allows the impact of the QoS parameters to be mapped to the end-to-end perception of video quality. 
The need of preprocessed videos reduce the applicability of this solution and also require a lot of processing power to carry out the content classification since they are not suitable for real-time services. 
Additionally, the video content which could improve the video performance, even more, is not taken into account.
Another approach is employ a border proxy~(located at the edge of the WMNs) to improve the video traffic delivery and provide two services~\cite{Qiu2010}.
The first service is a route selection algorithm, called Minimum Interference Route Selection~(MIROSE), which is able to choose routes based on the amount of interference. 
In doing this, the route with minimal interference can be chosen and thus allow a better usage of network resources. 
The second service is responsible for establishing the optimal video streaming rate according to the dynamic and unpredictable nature of the wireless channel conditions and is called Network State Dependent Video Compression Rate~(NSDVCR)~\cite{Qiu2009}. 
In carrying this out, the NSDVCR algorithm can temporarily buffer the video stream to optimise the compression rates so that it suits the network conditions of the end-to-end path. 
This re-compression is based on a heuristic that determines the optimal compression rate that corresponds to the current network status.

This technique has several weaknesses; for example, re-coding is a processor-intensive activity which raises serious scalability issues, and the same applies to the usage of a buffer in the gateway. 
As a result, it would have to include a comprehensive storage space to accommodate a higher number of concurrent streams. 
This architecture has a single point of failure, which means that if the gateway is down, the video enhancement will not work. 
Some of these issues can be remedied with a high availability and load-balancing system, although this increases deployment and maintenance costs.
A different method was proposed for VANETs. 
It adopts an adaptive multi-objective Medium Access Control~(MAC) retransmission limit strategy~\cite{Asefi2012}. 
At the Road Side Units~(RSUs), channel statistics and packet transmission rate are used as input to the optimisation framework in order to tune the MAC retransmission limit. 
Although this optimisation improves the performance of video transmission, it only aims to minimise the playback freezes and reduce the start-up delay. 
These are important characteristics, however, QoE metrics should be used to assess the image quality. 
This evaluation would provide a more comprehensive assessment of the proposed mechanism. 
Additionally, the authors only took into account the use of RSUs and two-hop communications. 
It is known that the major advantages of VANETs come from the communication directly between the vehicles, without the need for a fixed infrastructure. 
This severely restricts the application of the mechanism.

Another VANET proposal relies on routing protocol adaptations, such as the QoE-based routing protocol for video streaming over VANETs~(QOV)~\cite{Pham2014}. 
In QOV, the perceptual quality of the videos is assessed in real-time, at the receivers, using the Pseudo-Subjective Quality Assessment~(PSQA)~\cite{Rubino2005} metric. 
After that, the results are announced to the neighbours throughout \texttt{Hello} packets. 
This allows the routing protocol to choose the best paths to deliver the video sequences. 
Nevertheless, VANETs are very dynamic networks and because of that, the proposed mechanism would have to update very quickly the PSQA result announcement, overloading the network with \texttt{Hello} packets. 
Another weakness of this proposal is that it does not include any type of EC. 
As aforementioned, the video quality can be maintained only up to a certain level without using EC, however, if the network has a high packet loss rate the quality will decrease.

The mechanisms presented in this section are able to improve the performance of the video transmissions, however, there is a lack of any type of EC on them.
As previously mentioned, without using an EC technique the video quality can only be sustained up to some degree. 
After that, an impact on the visual video quality will be noticed if the number of errors overcomes the natural video resilience to packet loss.

\subsection{EC-based Optimisation Mechanisms}
\label{sec:rw:ec_based}
Several EC-based techniques have been proposed in the past years to improve the perceived video quality.
One example is the Adaptive Multi-Hop FEC~(AM-FEC) protection scheme~\cite{Tsai2009} to improve the quality of video streaming data. 
The proposal seeks to reduce the end-to-end delay and FEC computational costs at the same time. 
Through a heuristic algorithm, the AM-FEC protection minimises the amount of redundant information sent with the original data set. 
The video frame rate is also used to dynamically adjust the FEC scheme on each link. 

One disadvantage of this approach is that it does not use QoE metrics to improve the performance.
The QoE methods are only used in the assessment.
Other important parameters that are missing in the AM-FEC protections are codec type and motion complexity, which have proved to be efficient in this kind of scheme.
The Adaptive Cross-Layer FEC~(ACFEC) mechanism uses packet-level error correction~\cite{Han2010}. Through a cross-layer design, these packets are monitored by the mechanism at MAC layer, and when a loss occurs, a failure counter is increased. 
The information held by the failure counter determines whether the number of FEC recovery packets is increased or decreased. 
In this way, when the counter is zero, it means that there is no packet loss and the wireless connection is good, and thus recovery packets are not generated which results in less redundant traffic. 
However, no assessment of the network overhead is conducted, which is very important, especially in a wireless environment. 

As mentioned earlier, this type of network generally does not have a fair bandwidth allocation which, means that increasing the overhead of one node may affect the communication between the others. 
In addition, the approach discussed above does not consider the video content, and, as is well-known, this information has a direct influence on how the video is resilient to packet loss~\cite{Aguiar2012}. 
Although the ACFEC mechanism seems to be a good solution when the network is healthy and there is sporadic packet loss, when network congestion occurs, this mechanism will start to generate more and more FEC redundancy packets, which will increase the congestion.

Another technique to enhance the quality of the video transmission employs a forward error correction and retransmission-based adaptive source-channel rate control~\cite{Hassan2010}. 
This scheme uses real-time monitoring of the decoder buffer occupancy and the channel state, to calculate the optimal parameters for FEC redundancy. 
This information is regularly returned, through a feedback channel or out-of-band using Real Time Streaming Protocol~(RTSP)~\cite{Schulzrinne1998}, to the video encoder at the server site, which proceeds to adapt it to its own transmission parameters. 

The authors claim that there has been an improvement in the QoE for end-users, however, the main objective of this scheme is to ensure the continuity of video playback with unpredictable channel variations and avoid unnecessary FEC redundancy. 
Moreover, information such as video content and frame type were not considered in the definition of the proposal~\cite{Aguiar2012}. 
This approach does not assess QoE metrics, as it relies on packet loss values to predict QoE levels, and it does not measure the overhead that has been introduced.

A distinct proposal to enhance video transmission over wireless local area networks are based on a method which adapts in real-time the amount of FEC redundancy and the transmission rate~\cite{Alay2010}. 
In order to adjust the FEC redundancy and the transmission rate, the receivers periodically send the packet error rate information to the Access Point~(AP). 
Using this information, the AP can identify the worst channel's condition and then adjust the transmission rate and FEC. 
The application level FEC redundancy adaptation is done by multiple pre-encoded videos with different bit rates and FEC rate, so, in order to adapt theses parameters, the system has to switch to a different bit stream.
The bit-stream switching is always initiated when the next frame is an I-frame because it is independently decodable.  

One of the downsides of this proposal is the need of a pre-processed video which reduces the applicability of this solution. 
It also demands high processing power and storage space, since there is the need to encode multiple times the same video with different bit rates and FEC redundancy. 
Moreover, the FEC overhead amount introduced by this mechanism was 48\%~(without taking into account the feedback messages overhead), which is higher than the proposed mechanisms as shown in Chapters~\ref{ch:MESH}, \ref{ch:UAV}, and~\ref{ch:VANET}.
The Adaptive Hybrid Error Correction Model~(AHECM) solution adopts a dynamic FEC block length~\cite{Tsai2011}. 
This FEC block can be adjusted in real-time depending on Markov models to estimate the PLR and the number of continuous losses, to boost video transmissions. 
This mechanism is heavily based on network parameters leaving out important QoE-sensitive information, such as frame size and type, as well as the motion intensity. 
Furthermore, the mechanism uses a buffer to cope with the impact of packet disordering. 
This should increase delay and lead to the discarding of the packets by the encoder due to playback time out.
Following the same pattern as the studies outlined above, no attempt is made to measure the network overhead.

An alternative mechanism is the Adaptive Packet and Block length FEC~(APB-FEC)~\cite{Tsai2011a}. 
This mechanism uses smaller packet lengths in order to increase the size of the FEC block. 
A feedback channel is used to receive packet loss information in order to adapt the video sequences to the network characteristics. 
Additionally, the mechanism also adapts the buffer size according to the network conditions. 
The use of buffers is not optimal and can increase the delay. 
Also, relying on information from the receiver can be problematic due to the fact that if the communication is hindered the feedback information may not reach the sender.
In addition, the use of past PLR can lead to an inaccurate characterization of the network due to outdated information. 
The performance assessment is based on the effective packet loss rate, the network overhead, and PSNR metric. 
However, the PSNR metric is known to not correlate well with how the QoE is perceived by end-users.

A FEC-based mechanism was proposed to enhance the quality of video streaming using video-aware techniques~\cite{Diaz2011}. 
In this approach, some parameters are used to define the importance of a data set, namely the frame relevance, channel state, and bitrate constraints. 
With this information, the proposed algorithm is able to select the most suitable frames to give protection in real-time.
This mechanism also employs an unequal loss protection technique, providing the possibility to select the most suitable frames to add redundancy, and consequently, securing the delivery of the most important information.

One of the weaknesses of this solution is that it only uses the frame type in order to define the relevance of the data and does not take into consideration the motion activity levels, which can have a considerable impact in the perceived impairments~\cite{Greengrass2009a}.
It also requires the addition of extra tags in RTP encapsulation, which means that the videos have to be reprocessed.

The Cross-Layer Mapping Unequal Error Protection~(CLM-UEP) assigns a different level of redundancy according to the frame type of the video sequences and the packet loss rate~\cite{Lin2012}. 
Moreover, this mechanism has an adaptive mapping algorithm to direct the video data and redundant packets to the suitable Access Category~(AC) queues. 
This operation also takes into consideration the frame type and the packet loss rate, as well as the AC queue occupancy to avoid congestion-induced packet losses.
The amount of redundancy is defined through the analysis of the frame type and the past PLR. 

The mechanism was assessed using the Playable Frame Ratio~(PFR) and the PSNR score. 
Nonetheless, the past PLR may not repeat in the near future leading to a mischaracterization of the network. 
Moreover, the average PLR will not capture fast time-varying changes in the wireless network channels. 
Another major drawback of this mechanism is the lack of use of important video characteristics, such as levels of motion activity and position of the frames within the GoP. 
As evidenced before, this information plays a substantial role to define the best amount of redundancy, allowing the system to save important network resources.
An additional work proposes the use of a 2-state Markov hierarchical model to predict the short-term losses and hidden Markov models~(HMM) to forecast the longer-term network losses~\cite{Silveira2012}. 
In doing this, both the PLR and burstiness are categorised and used as input to configure the amount of redundancy added by the FEC scheme. 
The assessment is performed with the Perceptual Evaluation of Speech Quality~(PESQ) and MOS metrics. 
This proposal does not consider the video characteristics in the decision-making process. 
These characteristics are known to have a direct impact on the video resiliency to packet loss and consequently, on the QoE for the end-users.

The ``Transport Audiovisuel avec Protection In\'{e}gale des Objets et Contr\^{o}le d'Admission''~(TAPIOCA) mechanism divides each GoP by layers assigning different priority values to each one~\cite{Lecuire2012}.
This allows the protection of the most important layers. 
The assessments of the mechanism were performed by the Decodable Frame Rate~(DFR) and the Protection System Efficiency~(PSE) metrics. 
These assessment techniques are so unique that they do not provide much information about the QoE performance of the scheme.
The process of dividing each GoP into layers is both computationally heavy and time-dependent, making this scheme unsuitable for real-time use.
Additionally, the mechanism does not take into consideration the motion intensity of the video sequences, which can have a considerable influence in the perceived impairments.

The Optimised Cross-Layer FEC~(OCLFEC)~\cite{Talari2013} computes priority values based on the mean squared error of each frame. 
Two error correction codes are used, namely Luby Transform~(LT) and Rate-compatible Parity Check~(RCPC). 
The former is used to encode the data and the latter to add check bits. 
Performance optimisations are made on both for specific situations.
The GoP information is the obtained using a cross-layer technique and then assigned different priorities according to the commutative mean squared error of the entire GoP. 
The GoPs are encoded and cyclic redundancy check bits are added to detect coding errors. 
Afterwards, the FEC codes are optimised with different parameters for different situations. 
This mechanism has several optimisation phases, for each frame, that are very time-consuming, increasing the delay and degrading the QoE.

The mechanism is assessed in terms of QoS performance, which does not guarantee a good QoE for end-users. 
The only metric used to assess the performance of the mechanism is the PSNR, which by itself does not tell much information about QoE. 
Besides that, the OCLFEC does not take into account the motion intensity and the network state, leaving out important characteristics that should be considered to protect video sequences.

A different mechanism to improve the video quality over wireless networks compares the efficiency of Random Linear Coding~(RLC) and XOR-based coding~\cite{Rezende2013}. 
The benchmark results show that both erasure codes are able to improve the video quality by increasing the number of successfully received packets over error-prone networks. 
The results also show that XOR-based coding outperforms the RLC scheme. 

In addition, the proposed mechanism finds the optimal packet block size, which allows adding a more precise amount of redundancy. 
While this is true, important network and video characteristics are not taken into consideration.
Some of these features, such as packet loss, video content, and codec, are important in the optimisation process to provide a way to compute a precise amount of redundancy leading to both high video quality and low network overhead.

The Hybrid Video Dissemination Protocol~(HIVE)~\cite{Naeimipoor2014} uses a multi-layer strategy to improve the video quality. 
The HIVE multi-layer strategy is based on the joint use of traffic congestion control scheme, node selection method, and application layer erasure coding technique. 
This allows higher packet delivery ratio while keeping latency and packet collisions low. 

The results show improvement in the PSNR assessment, leading the authors to claim that they improved the QoE for end-users. 
However, relying on only one metric is not enough to prove that, especially considering that the PSNR results do not correlate well with the human vision system~\cite{HuynhThu2008a}. 
Another issue is the lack of video characteristics assessment. 
It is known that these video details have a considerable impact on how resilient a video sequence is when experiencing packet loss.

Table~\ref{tab:rw:mesh_related} compiles the main characteristics of all above-named works. The parameters are defined as follows:

\begin{itemize}
	\renewcommand{\labelitemi}{$\bullet$}
	
	\item \textbf{FEC-based}: accounts for mechanisms that employ FEC;
	
	\item \textbf{ARQ-based}: mark mechanisms that use ARQ;
	
	\item \textbf{QoE-sensitive data}: this parameter demonstrates mechanisms that identity and/or considerate the video content to define the EC policy;
	
	\item \textbf{Video-aware}: check mark is given to mechanisms that use any video characteristics to define the amount of redundancy and/or retransmission;
	
	\item \textbf{High-quality video}: it is marked if the mechanisms are using videos equal or higher than 720p~(HD ready);
	
	\item \textbf{Network status}: this parameter defines if the mechanisms use the information about the network healthy to define the redundant data; 
	
	\item \textbf{UEP-enabled}: means that different amounts of redundancy are being added to distinct portions of the video.
	
	\item \textbf{QoE assessment}: used if the mechanism's assessment is performed by subjective and/or objective QoE metric;
	
	\item \textbf{Network assessment}: this parameter reflects the network-related evaluation, especially the network overhead caused by the redundancy;
	
\end{itemize}
\newcommand{\V}{\raisebox{-0.5mm}\CheckmarkBold}%
\newcommand{\X}{\raisebox{-0.5mm}{\scriptsize\XSolidBrush}}%
\newcommand{\1}{\tnote{1}}
\newcommand{\2}{\tnote{2}}
\newcommand{\3}{\tnote{3}}
\begin{table}[!htb]
	
	{\scriptsize
		\caption{State of the Art of EC-based optimisation mechanism}
		\begin{center}
			\begin{threeparttable}
			\renewcommand{\TPTminimum}{\linewidth}%
			\resizebox{\columnwidth}{!}{%
				\setlength{\tabcolsep}{6pt} %
				\renewcommand{\arraystretch}{1.1} %
				\rowcolors{2}{gray!10}{white}
				\begin{tabular}{l|c|c|c|c|c|c|c|c|c}
					{Proposals} / {Parameters} & 
					\rotatebox{90}{FEC-based} &
					\rotatebox{90}{ARQ-based} &
					\rotatebox{90}{QoE-sensitive data} & 
					\rotatebox{90}{Video-aware} & 
					\rotatebox{90}{High-quality video} & 
					\rotatebox{90}{Network status} & 
					\rotatebox{90}{UEP-enabled} & 
					\rotatebox{90}{QoE assessment} &
					\rotatebox{90}{Network assessment} \\
					\hline
					\hline
\cite{Tsai2009}		& \V & \V & \X & \X & \X & \X		& \X & \V/\X\1	& \X \\
\hline
\cite{Alay2010} 	& \V & \X & \X & \V & \X & \V 		& \X & \V/\X\1	& \V \\
\hline
\cite{Han2010} 		& \V & \V & \X & \X & \X &\V/\X\2	& \X & \V/\X\1	& \X \\
\hline
\cite{Hassan2010} 	& \V & \V & \V & \X & \V &\V/\X\3	& \V & \X 	& \X \\
\hline
\cite{Tsai2011} 	& \V & \X & \X & \X & \X & \V 		& \X & \X 	& \V \\
\hline
\cite{Tsai2011a} 	& \V & \X & \X & \X & \X & \V		& \X & \X 	& \V \\
\hline
\cite{Diaz2011} 	& \V & \X & \X & \V & \X & \V 		& \V & \V/\X\1	& \X \\
\hline
\cite{Lin2012} 		& \V & \X & \X & \V & \X & \V 		& \V & \V/\X\1	& \V \\
\hline
\cite{Silveira2012}	& \V & \X & \X & \X & \X & \V		& \X & \V	& \V \\
\hline
\cite{Lecuire2012}	& \V & \X & \X & \V & \X & \X 		& \V & \X	& \V \\
\hline
\cite{Talari2013} 	& \V & \X & \V & \V & \X & \X 		& \V & \V/\X\1	& \X \\
\hline
\cite{Rezende2013} 	& \V & \X & \X & \X & \X & \X 		& \X & \X	& \V \\
\hline
\cite{Naeimipoor2014}	& \V & \X & \X & \X & \X & \X		& \X & \V/\X\1	& \V \\
\hline

				\end{tabular}
			}
			\begin{tablenotes}
				\item[1] Uses only PSNR, which does not correlate well with how humans see video impairments
				\item[2] Based only in the last PLR
				\item[3] Uses the probability of errors, no real measure nor actual feedback
			\end{tablenotes}
		\end{threeparttable}
			\label{tab:rw:mesh_related}
		\end{center}
	}
\end{table}

As evidenced, these works fell short to produce an holistic mechanism to provide resilient video transmission without imposing unnecessary network overhead. 
Some of the proposed works do not provide a QoE-driven procedure to compute the necessary amount of redundancy. 
This is an important step towards an efficient mechanism, as the most QoE-sensitive information has to be better protected than other less important data.
Without this, an unnecessary network overhead can be generated.
This leads to another problem; several studies do not provide a clear account of this matter, and this hampers any comparisons from being made. 
Other proposals fail to take into consideration the video-aware data.
It is known that this content plays an important role in the video transmission process~\cite{Khan2010}, and without this information, the proposed mechanism cannot achieve the most advantageous performance.

\section{Summary and Open Issues}
\label{sec:rw:summary}

The widespread deployment of wireless networks, as well as the usage of mobile devices, has increased significantly in recent years. 
These networks are well known for their dynamic nature along with their time-varying channel conditions. 
In addition, several new services have stringent requirements and need to be operated in real-time. 
This is especially true when dealing with video transmissions which can be considered one of the most demanding services in terms of delay, packet loss rate, and bandwidth.
Environments that do not provide the necessary conditions will have poor performance in this type of service and the video impairments will be very noticeable to the end-users.

Section~\ref{sec:rw:EC} presented several concepts about error-correcting codes. 
The first method discussed was the ARQ, where packets are retransmitted if lost or after a determined time-out time. 
The advantage of this technique is that it adds very little network overhead, i.e., only small messages on the feedback channel.
The disadvantage is that it can add a delay in the packets delivery.
If the delay is high, the packet may be received after its play-out time, meaning that it can no longer be used.
This trait is not appropriate for delay-sensitive applications with real-time transmissions, such as the video services.

The FEC method, instead of retransmitting the lost information, adds an amount of redundant data~(or parity bits), allowing the receiver rebuild the original error-free data if something is lost or damaged.
These types of error correction are widely diffused and have been used to improve the reliability of storage data, as well as to shield the communication in error-prone channels.
There are several FEC codes with different characteristics and applications.
One of the most used is the Reed-Solomon code because of its simplicity and real-time features. 

This chapter also presented a state of the art revision about the mechanisms that improve the video transmission quality~(Section~\ref{sec:rw:mechanisms}). 
Two types of mechanisms were discussed, non-EC based and EC-based.
The former does not use any type of error correction, being limited the amount of improvement that it can provide, especially in high-error rate networks.
The latter adopts error correction, which can be ARQ or FEC and a combination between both as well.

As evidenced in the review there is a lack of QoE-driven, Video-aware, and high-resolution ready mechanisms.
This can be explained by a large number of interfering components~(e.g., video characteristics, network status, how the human vision perceives video quality) and the complex relationships between them.
Taking everything into consideration, these EC methods have to be well configured to improve the video quality without adding unnecessary redundancy or delay.

In the case of FEC-based mechanisms, a number of open issues arose. 
To start off, one challenge is how to provide a reliable way to calculate a proper FEC block size. 
Smaller blocks are more efficient in slower upload links and are more resilient to burst packet losses.
The downside is that they lead to a lower FEC encoding efficiency. 
On the other hand, bigger blocks, have better encoding efficiency but they are more susceptible to burst losses.

Once decided the FEC block sizes, it is necessary to configure the encoding FEC algorithm which will generate the redundant data.
This is a challenging process because it involves the selection of the appropriate encoding rates to ensure that the missing data can be reconstructed.
If a low level of redundancy is added, a small amount of network overhead is produced, however, it may not be possible to reconstruct the original error-free block due to the limited information. 
Conversely, if a high level of redundancy is chosen, these FEC blocks will be more resilient to packet loss, however, a larger network overhead occurs.

On top of that, different encoding rates can be used to protect distinct categories of data in determined network conditions.
For example, high rates can be used to protect the most QoE-sensitive data when the network presents high PLR. 
On the other hand, low rates can be used if the network is healthy.
In addition, the mid-level data on the scale of importance can be protected with low encoding rates even if the network has a high-errors incidence as they will not have great impact if lost.


\setcounter{mtc}{12}
\chapter{Mechanisms for Resilient Video Transmission over WMNs}
\chaptermark{Resilient Video Transmission over WMNs}
\label{ch:MESH}

\dictum[Aldous Huxley, Brave New World]{You can't consume much if you sit still and read books.}

\minitoc

\lettrine[lines=3]{\color{gray}\bf{T}}{} he video delivery over wireless networks is now a part of the daily life of users since it is a solution that delivers a wide range of information. 
Despite the issues previously outlined, WMN provides a cost-efficient way of distributing broadband Internet access.
Another advantage is its flexibility and reliability for a large set of applications in a wide coverage area~\cite{Zhu2011,Akyildiz2005}. 
Meanwhile, several difficulties can impair the success of the transmission, such as limited network resources and high error rates, as well a fluctuating signal strength that may lead to variable bandwidth. 
The use of these error-prone networks unveils the need for an adaptive mechanism to improve the video transmission.
Adaptive FEC-based techniques that are able to assure a high QoE for end-users are a convenient means of delivering video data to wireless users in this case.
This chapter describes and assesses three proposed FEC-based adaptive mechanisms to shield video transmissions over WMN.

\section{Introduction}

The usage of online video services has been increasing rapidly in recent years, particularly from wireless mobile devices~\cite{Adobe2016}. 
This upswing is related to several technological improvements in mobile devices, as well as the rise in popularity of this type of service. 
For example, several companies are using live video services to reduce costs and increase both collaboration and productivity. 

Following the same trend, the number of non-professional users creating, sharing, and consuming online videos is growing apace. 
Cisco predicted that over 82\% of all Internet traffic by 2020 will be some sort of video content~\cite{Cisco2016}. 
As an example, roughly a million minutes of IP video will be crossing the network every second.
This figure means that a single individual would have to spend over 5 million years to watch the entire video content transmitted in only one month.
Because of the video traffic ascendance, the probability of errors arising from interference and network congestion increases.
This unveils the need for an adaptive mechanism to shield the video delivery, otherwise, the above factors will impact on the video quality, degrading the QoE for the end-users.

Taking this into consideration, new mechanisms for increasing the transmission quality are required to support the growth of video traffic, as the video quality may be affected by several factors.
Some of them are owing to the video characteristics, such as codec type, bitrate, format, and the length of the GoP, as well as, the content/genre of the video~\cite{Yuan2006}. 

In addition, when packet losses occur, the perceptual quality is not harmed in the same way by all the packets.
This happens because there is a link between the packet content and the impact it has on the user's perception of video quality. 
When this is taken into account, the most important information should be best protected and thus encouraging the use of UEP-based mechanisms. 
In addition, the video content also plays an important role during the transmission. 
Videos with a small degree of movement and fewer details tend to be more resilient to packet loss. 
In contrast, videos with higher levels of detail and movement are more susceptible to losses and the impairments will be more noticeable~\cite{Khan2010}.

Provided that most of the video services are real-time applications, they need a steady and continuous flow of packets.
This constraint can be affected by a number of factors, especially in wireless environments. 
The channel conditions in these networks can suddenly change over time due to noise, co-channel interference, multipath fading, and also, the mobile host movement~\cite{Lindeberg2011}. 
Nevertheless, as afore-stated one of the major challenges in WMN is how to distribute the available bandwidth fairly among the requesting nodes to support real-time video traffic~\cite{Liu2009}. 
For this reason, it is important to optimise the resource usage and thus avoid congestion periods and a high packet loss rate, particularly in resource-consuming applications, such as video streaming.

The adoption of an adjustable data protection is of crucial importance to enhance video transmission, providing both well-perceived video quality and low network overhead. 
FEC-based schemes have been successfully used in real-time systems~\cite{Nafaa2008}. 
FEC allows robust video transmission through redundant data, which is sent along with the original set. 
As a result, if some information is lost, the original data can be reconstructed through the redundant information~\cite{Lee2011}. 
However, the resources might be limited and unfairly distributed as well. 
An adjustable FEC-based mechanism must use UEP schemes to reduce the volume of redundant information. 
In the UEP approach, the amount of redundancy is chosen in accordance with the relevance of the protected data and thus giving better protection to the most important video details.

Several techniques can be adopted to enable the use of UEP. These techniques can vary from heuristic-based mechanisms to sophisticated schemes using Random Neural Networks~(RNN)~\cite{Abraham2005} and Ant Colony Optimization~(ACO)~\cite{Dorigo1996}. 
In heuristic-based mechanisms, the steps to solve a problem usually come from the knowledge of past problems with similar characteristics. 
The main objective is not to guarantee an optimal or perfect solution, but a practical and/or quick method that offers a satisfactory outcome to the proposed problem~\cite{Pearl1984}.

RNNs can also be used to the same end. They are a type of Neural Network~(NN) that provides an information-processing paradigm based on the central nervous system. 
Through a learning process, based on pattern reading and connection weight adjustment, it is possible to configure these techniques for a specific application, being widely used in pattern recognition and data classification~\cite{Mohamed2002,Aguiar2012a}. 

Another option is the use of ACO, which is a probabilistic algorithm, based on the behaviour of ants. 
This strategy is used to dynamically solve computational problems by finding the best path in a graph. 
In this method, ants span through the paths between the nodes to find a solution. 
In every path followed, a pheromone marker is deposited. 
At the end, the paths with the greater amount of pheromone represent the best-fitted solutions~\cite{Dorigo2006}.

Considering the aforementioned issues, this chapter describes the design and evaluates three adaptive FEC-based mechanisms. 
The first one is the adaptive cross-layer \textsc{VI}d\textsc{E}o-a\textsc{W}are FEC-based Mechanism with Unequal Error Protection scheme~(ViewFEC) in Section~\ref{sec:viewfec}. 
The second proposed mechanism is the adaptive Video-aware Random Neural Networks~(RNN) based mechanism~(neuralFEC), in Section~\ref{sec:neuralFEC}. 
The last one is the QoE-driven motion- and video-aware mechanism~(PredictiveAnts), presented in Section~\ref{sec:PredictiveAnts}.

\section{Adaptive Video-aware FEC-based Mechanism (ViewFEC)}
\label{sec:viewfec}

Motivated by the open issues afore-identified, this section proposes and validates the adaptive cross-layer \textsc{VI}d\textsc{E}o-a\textsc{W}are FEC-based Mechanism with Unequal Error Protection scheme~(ViewFEC).

\subsection{ViewFEC Overview}

The aim of ViewFEC is to strengthen video transmissions while increasing user satisfaction and improving the usage of wireless resources. 
Owing to these factors, the use of video-aware FEC-based mechanisms is suitable to transmit videos with better quality, although it needs additional bandwidth to send the redundant information data. 
ViewFEC is an adaptive mechanism that overcomes this problem by dynamically configuring itself according to the video characteristics and user perception of quality. 
Using this process, only the more sensitive data sets will carry an unequal amount of redundant information, thus maintaining a good video quality and saving resources. 

In the ViewFEC mechanism, decisions are made at the network layer resorting to two modules, the \textsc{C}luster \textsc{A}nalysis \textsc{K}nowledge bas\textsc{E}~(CAKE) and the \textsc{C}ross-\textsc{LA}yer infor\textsc{M}ation~(CLAM). 
The decision-making process at the network layer provides deployment flexibility and allows the ViewFEC mechanism to be implemented in access points, routers, or video servers. 
The analysis of the information obtained from these two modules enables the proposed mechanism to estimate the most advantageous redundancy ratio necessary to sustain a good video quality, without adding unnecessary network overhead.

\begin{figure*}[!htb]
	\begin{center}
		\ifBW \includegraphics[width=143mm]{./f33_novel_revisado_vfec_gray-eps-converted-to.pdf}
		\else \includegraphics[width=143mm]{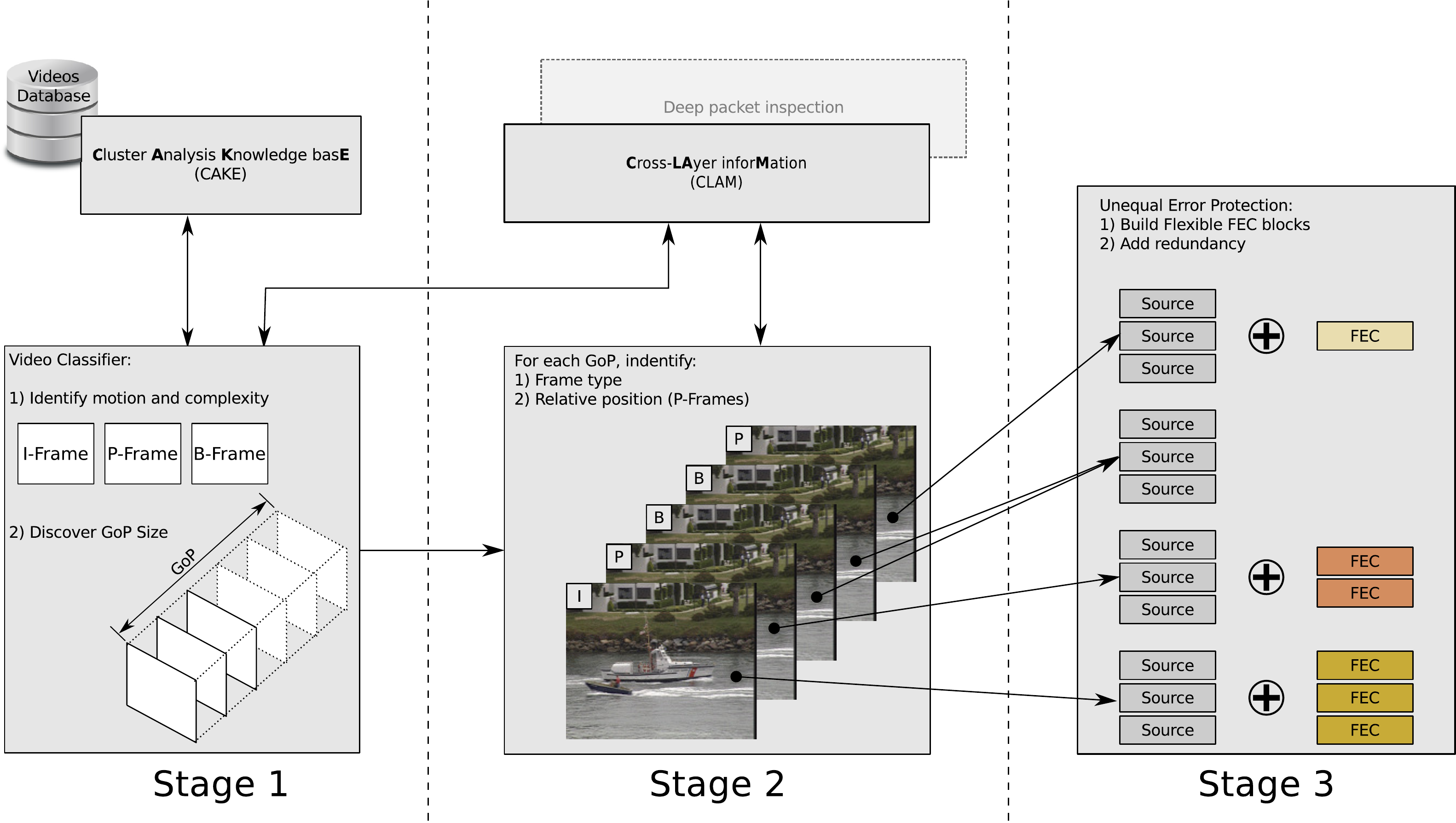}
		\fi 
	\end{center}
	\caption{ViewFEC stages}
	\label{fig:vfec:overall}
\end{figure*}

Figure~\ref{fig:vfec:overall} depicts the ViewFEC mechanism. There are three distinct stages. In the first stage, the proposed mechanism identifies several key video characteristics, such as motion and complexity levels, as well as the GoP length. 
In the second stage, further details about the video sequence are gathered, namely the type and relative position of the frames within its GoP. By the aid of these details, the ViewFEC mechanism will be able to correctly identify the video characteristics needed to configure the amount of redundancy in the next stage. 
The offline process is important because it allows a fast and more accurate real-time execution since few variables need to be handled. 
The construction of the FEC blocks and the UEP redundancy assignment takes place in the third stage. Further details of each module are described later.

\subsection{Towards the design of ViewFEC}

First of all, an exploratory analysis using hierarchical clustering is performed to conceive a knowledge database in the CAKE module. 
The hierarchical clustering is a statistical method of partitioning data into groups that are as homogeneous as possible~\cite{Revelle1979}. 
The video sequences are clustered according to the size of the I-, P- and B-Frames because they tend to have similar motion activity.
This operation only has to be performed once during the setup phase of the mechanism. 
Afterwards, when the mechanism is running, the relationship between the database information and the videos that are being transmitted in real-time is used to determine a couple of video characteristics, namely motion activity and complexity levels.
This directory also stores information about the relation between video characteristics and their impact on video quality.

The video sequences of the experiments were chosen in compliance with the recommendations of the VQEG~\cite{Staelens2011} and ITU~\cite{ITUTJ2008}. 
A total of 20 different videos were assessed. 
Ten of them were used to assemble the database and another set of ten was used to evaluate the ViewFEC mechanism. 
While remaining in compliance with the recommendations, the videos cover different distortions and content, since they are representative of regular viewing material. 
The video sequences also contain distinct temporal and spatial details, luminance stress, and still and cut scenes.

\begin{figure}[!htb]
	\vspace{-0.0cm}
	\begin{center}
		\ifBW \includegraphics[width=74mm]{./dendrogram_gray-eps-converted-to.pdf}\\
		\else \includegraphics[width=74mm]{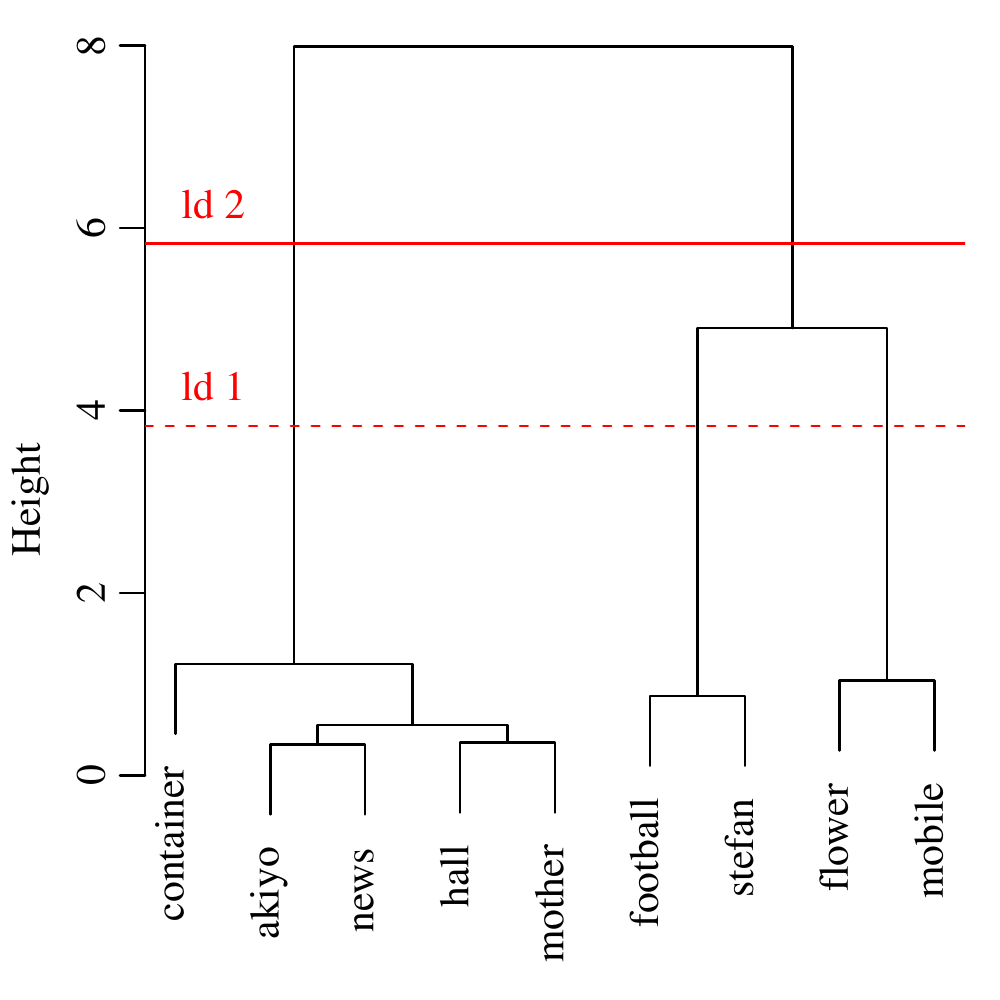}\\
		\fi
		\vspace{-0.35cm}\hspace{-2.8cm}\small{\textit{ld - linkage distance}}
	\end{center}
	\vspace{-0.0cm}
	\caption{Cluster Dendrogram}
	\label{fig:vfec:dendrogram}
	\vspace*{-0.0cm}
\end{figure}

Video motion and complexity are commonly classified into three categories, namely low, medium, and high~\cite{Khan2010,Aguiar2012}~(see Figure~\ref{fig:vfec:dendrogram} at linkage distance~(ld) 1). 
Nevertheless, throughout the experiments, videos with both medium and high complexities behaved roughly the same. 
Therefore, the linkage distance of the cluster analysis algorithm was chosen to only produce two clusters~(Figure~\ref{fig:vfec:dendrogram} at ld 2). 
The linkage criterion is responsible for setting the observation distance between the nodes of the cluster.
In other words, this is how the elements will be grouped together.
The proposed mechanism also employs the Ward method which seeks to reduce the sum of squares between the samples inside the cluster, this better reflects the experiment's findings.

\begin{figure}[!htb]
	\begin{center}
			\ifBW \includegraphics[width=100mm]{./size_qoe_ssim_mobile_Thesis_gray-eps-converted-to.pdf}
			\else \includegraphics[width=100mm]{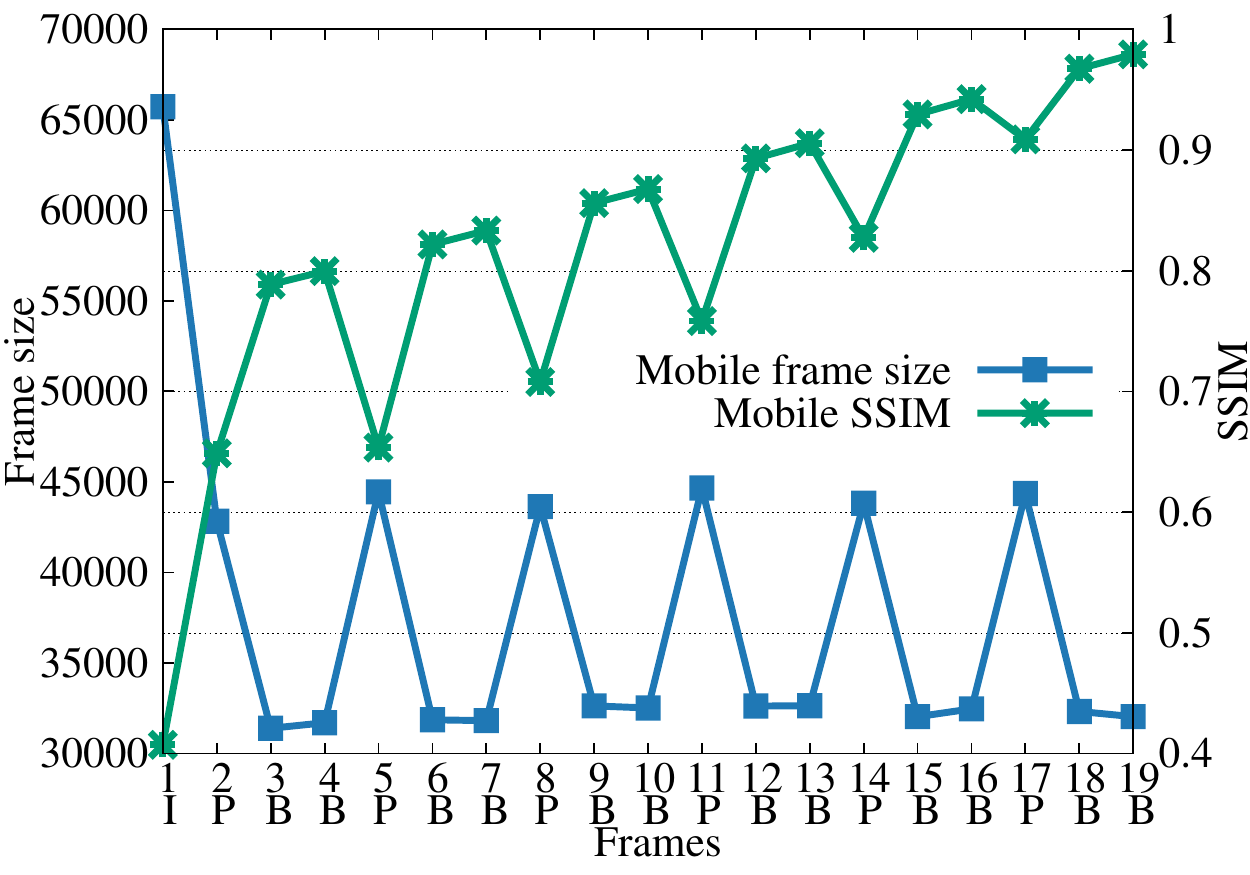}
			\fi
	\end{center}
	\caption{Frame size x QoE (SSIM) - Mobile video sequence}
	\label{fig:vfec:sizexqoeA}
\end{figure}

\begin{figure}[!htb]
	\begin{center}
			\ifBW \includegraphics[width=100mm]{./size_qoe_ssim_akiyo_Thesis_gray-eps-converted-to.pdf}
			\else \includegraphics[width=100mm]{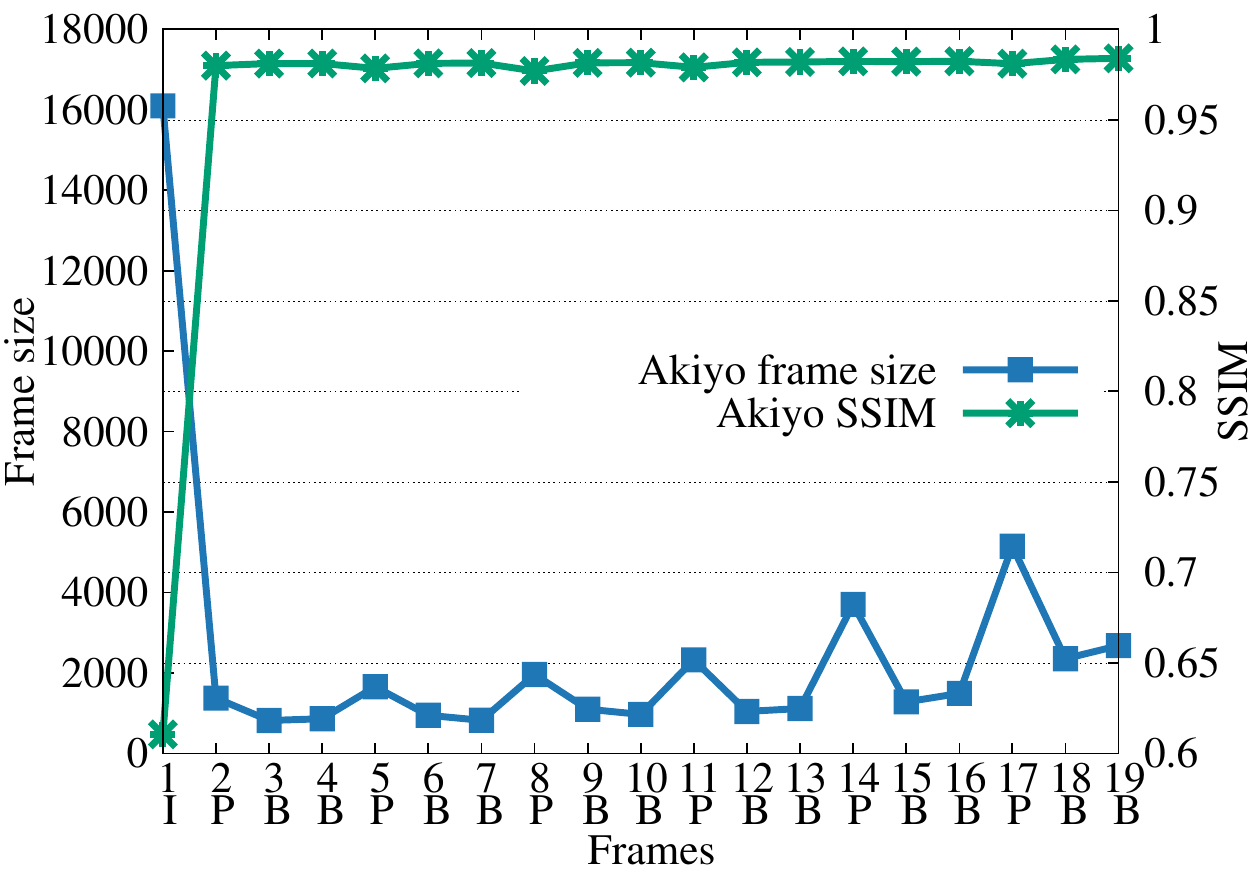}
			\fi
	\end{center}
	\caption{Frame size x QoE (SSIM) - Akiyo video sequence}
	\label{fig:vfec:sizexqoeB}
\end{figure}

The relationship between motion and complexity levels, as well as frame size~(in bytes) and frame position, is shown in Figures~\ref{fig:vfec:sizexqoeA} and~\ref{fig:vfec:sizexqoeB}.
The former depicts the Mobile video sequence and the latter the Akiyo video sequence, one from each motion cluster.
Only the first GoP of each video was considered, which made it easier to visualise the results. 
The Mobile sequence has uninterrupted scene modification and a wide-angle camera, and thus, high motion and complexity levels. 
For this reason, the video has larger frames and also a greater difference, in terms of size, between P- and B-Frames, as shown in Figure~\ref{fig:vfec:sizexqoeA}.
In contrast, the Akiyo video sequence has only a small region of interest that is moving, which is concentrated around the face and shoulders, and also a static background. 
As a result, it has a low motion and complexity levels, leading to a smaller difference in size between P- and B-Frames, as depicted in Figure~\ref{fig:vfec:sizexqoeB}.


Additionally, both figures show the SSIM scores with frames been deliberately discarded. 
The measurement of this metric is fairly simple, even though it is consistent with the human visual system, and yields good scores~\cite{Wang2004}. 
The SSIM results were acquired by removing the frame which occupied that position, i.e. the first SSIM value was calculated without the first frame, and the same is applied to all other frames.
In the Mobile video results~(Figure~\ref{fig:vfec:sizexqoeA}), it is clear that I- and P-frames have greater significance, and also that the frames closest to the beginning of the GoP have more impact on the video quality when discarded. 
As expected, the Akiyo sequence behaves differently~(Figure~\ref{fig:vfec:sizexqoeB}). 
It has a lower motion and complexity levels being more resilient to packet loss and achieving higher SSIM values~\cite{Khan2010}. 
The CAKE module is aware of these video characteristics and can determine the motion activity and complexity levels of each GoP that is being transmitted. 
This is done because it is possible to have a different motion and complexity levels inside the same video sequence, as expected for Internet videos. 

\subsubsection{Video-aware Information Module}

The next step~(Stage 2) is the CLAM module. 
The functions of this module are implemented using cross-layer techniques.
This allows accessing information from the application layer, such as the video characteristics, to the network layer, where the module is deployed.
It has three basic functions. 
The first one is to identify the GoP length. 
As previously discussed, when the GoP length is larger, the packet loss has a greater influence on the video impairments. 
This happens partially because a new I-Frame that is needed to fix the error, will take longer to arrive.

The role of the second function is to identify the frame type. 
Different frame types need distinct amounts of redundancy. 
For example, the loss of an I-Frame will cause more impairments than the loss of a P-Frame, and hence, the loss of a P-Frame will be worse than the loss of a B-Frame.
Another important remark is that video sequences with a high level of spatial complexity tend to have larger I-Frames in relation to P- and B-Frames. 
On the other hand, videos with higher temporal activities tend to have larger P- and B-Frames. 
Larger frames mean that more network packets will be required to carry their data, increasing the chance of these packets being lost. 
Thus, these packets need more redundancy. 
Incidentally, as the GoP length increases, the size of both B-, and especially P-Frames, also increases. 

The task of the last function is to identify and compute the relative position of P-Frames inside the GoP. 
P-Frames that are closer to the beginning of the GoP have more impact if lost than those close to the end, and as a result, need more redundancy packets.
The combined use of these functions enables ViewFEC to enhance the video quality transmission without adding unnecessary network overhead, and thus, support a higher number of simultaneous users sharing the same wireless resource.

In addition, the ViewFEC mechanism has a flexible structure, making possible to swap the modules to obtain the desired behaviour. 
When it is not feasible to use a cross-layer design to obtain application layer information, e.g. in a router device, the CLAM module can be exchanged for one that uses another technique to obtain the desired information, for instance, packet and deep packet inspection. 
By analysing the packet header of some of the protocols, such as UDP, RTP and TS, it is possible to discover information about codec type and coding parameters, among others~\cite{Schulzrinne2003a}. 
On the other hand, the video content information can only be accessed through deep packet inspection scheme. 

\subsubsection{Defining an Adaptive Redundancy Amount}

In the last step~(Stage 3), the amount of redundancy needed is calculated in accordance with the details obtained from the previous stages. 
This tailored amount of redundancy is used to optimally adjust the FEC scheme. 
The RS code was adopted because it offers less complexity, and consequently achieves a better performance for real-time services~\cite{Neckebroek2010}. 
Nevertheless, any other alternative scheme could be used if needed.
A RS code consists of $n$, $s$, and $h$ elements. 
The total block size, including the redundancy data, is represented by $n$, and $s$ indicates the original data set size, therefore the parity code is~$(n,s)$. 
Finally, the parameter $h$ defines the amount of redundancy, which could also be represented as $h=n-s$. 
Before the original data set $s$ can be restored, at least~$(n-h)$ packets have to arrive successfully. 
The recovery rate can be expressed as $h/n$ or~$(n-s)/n$, which means that the robustness to losses is given by the size of $h$.

The ViewFEC mechanism settles the parity code in real-time. 
In other words, both $n$ and $h$ parameters are adjusted at Stage 3. 
This is done based on video characteristics found at Stages 1 and 2, obtained from the CAKE and CLAM modules, respectively. 
The first parameter of the parity code, $n$, is used to build the Flexible FEC Block~(FFBlock) scheme. 
This scheme involves dividing the I- and P-Frames into groups of packets, allowing each group to have an individual redundancy data size. 
This unique size is defined by the second parameter, $h$, and provides a tailored amount of redundancy for each FFBlock. 
Hence, rather than using a single redundancy amount to all the frames and video sequences, the ViewFEC mechanism uses an adjustable amount. 
Consequently, it is capable of yielding good results in different network conditions and also supporting a wide range of video characteristics.

The adjustable amount of redundancy data assigned by ViewFEC is the outcome of the joint evaluation of the frame type and position inside the GoP as well as the video motion and complexity levels. 
By adopting this procedure, we are able to infer the spatio-temporal video characteristics and, as a result, to choose the most beneficial redundancy amount, $h$, for each FFBlock. 
Owing to this, the ViewFEC mechanism is able to achieve better video quality and has the further advantage of reducing the amount of data that needs to be sent through the network, decreasing the overhead and providing a reasonable usage of wireless resources.
 
The reduction of the network overhead is important, because as the network grows larger, the number of concurrent transmissions increases, and this may cause serious interference problems. 
The situation gets worse if more overhead is added due to redundant information. 
This means that, if the overhead is reduced, a larger number of users will be able to receive more videos with better quality, thus boosting the overall capabilities of the system.

Algorithm~\ref{algo:vfec:pseudocode} shows a pseudo-code of the ViewFEC operation.
It illustrates how the GoP length and motion detection are performed, and also, the steps taken to assign a tailored amount of redundancy.
The algorithm has two nested loops which are responsible for going through the GoP and also the frames within each GoP.
For each frame, several conditions are assessed.
The information retrieval from CAKE and CLAM modules occurs through lines 2, 3, 5, and 11. 
Since the redundancy amount of P-Frames also depends on their relative position inside the GoP, it has to be treated differently from the I-Frames; this difference is noticeable at line 11.
At the end, is only added an adjusted amount of redundancy according to each frame's characteristic. 

\iflatextortf
\else
{%
	\SetKwProg{Fn}{}{}{}
	\begin{algorithm}[!htb]
		{\small
			\Fn{\textbf{for} each GoP \textbf{do}}{
				CAKE.getGopMotion(GoP)\;
				CLAM.getGopLength(GoP)\;
				\Fn{\textbf{for} each frame \textbf{do}}{
					\Fn{\textbf{case} (CLAM.getFrameType(frame)) \textbf{do}}{
						\Fn{I-Frame:}{
							buildFFBlock(frame)\;
							addRedundancy(frame)\;
							sendFrame(frame)\;
						}
						\Fn{P-Frame:}{ 
							CLAM.getRelativePosition(frame)\;
							buildFFBlock(frame)\;
							addRedundancy(frame)\;
							sendFrame(frame)\;
						}
						\Fn{B-Frame:}{
							sendFrame(frame)\;
						}
						
					}\textbf{\textit{end case}}
				}\textbf{\textit{end for}}
			}\textbf{\textit{end for}}
		}
		\caption{ViewFEC pseudo-code}\label{algo:vfec:pseudocode}
	\end{algorithm}
}
\fi

Equation~\ref{eq:vfec:redundancy} shows the amount of redundancy added by the ViewFEC mechanism to each GoP ($R_{GoP}$). 
$FS_i$ describes the number of packets of the frame that are being transmitted and $FT_i$ holds the frame type, as shown in Equation~\ref{eq:vfec:frametype}. 
If $\gamma > 0$, some level of redundant information will be added to the frame under discussion. 
If we have the vector $(\gamma{I},\gamma{P},\gamma{B})$ with elements $(1,1,0)$, for example, only I- and P-Frames will receive redundant information. 
The notation used in the equations is shown in Table~\ref{tab:vfec:notation}.

\begin{table}[!hbt]
	{\small
		\caption{Adopted Notation}
		\begin{center}
			\begin{tabular}{c|l}
				\hline \textbf{Notation} & \textbf{Meaning} \\
				\hline
				\hline $R_{GoP}$ & ViewFEC redundancy amount per GoP \\
				\hline $FS_i$ & Frame size in packets of number $i_{th}$ frame \\
				\hline $FT_i$ & Frame type of number $i_{th}$ frame \\
				\hline $C_{GoP}$ & GoP motion and complexity level \\
				\hline $RP_i$ & Relative position of number $i_{th}$ P-Frame \\
				\hline $N_{GoP}$ & Number of GoPs in the video sequence \\
				\hline 
			\end{tabular}
			\label{tab:vfec:notation}
		\end{center}
	}
	\vspace*{-0.0cm}
\end{table}

\begin{minipage}{\linewidth} %
\begin{equation}
	R_{GoP} = \sum_{i=0}^{GoPLength}\left [ FS_i \times FT_i \times C_{GoP} \times \frac{1}{RP_i}\right ]
	\label{eq:vfec:redundancy}
\end{equation}
\begin{equation}
	FT_i = \left\{ \begin{array}{rl}
		\gamma > 0 &\mbox{, \textit{send frame with redundancy}} \\
		0 &\mbox{, \textit{frame without redundancy }}
	\end{array} \right.
	\label{eq:vfec:frametype}
\end{equation}
\end{minipage}

The parameter $C_{GoP}$ in Equation~\ref{eq:vfec:gopcomplexity} describes the motion and complexity levels. 
If the mechanism is using two distinct video clusters, it is possible to define the vectors $(\alpha{High/Medium},\alpha{Low})$, with elements $(1,0.5)$, for example.
This means that the cluster with high motion and complexity levels would receive twice the amount of redundancy than the cluster with low levels. 
If more redundancy levels are needed, the vectors could be defined as $(\alpha{High},\alpha{Medium},\alpha{Low})$, with elements $(1,0.5,0.25)$, for example.
In this configuration, three levels of motion intensity will be addressed, high, medium and low, respectively. 

\begin{equation}
	C_{GoP} = \left\{ \begin{array}{rl}
		1 &\mbox{, \textit{high motion and complexity}} \\
		0 \leq\alpha < 1  &\mbox{, \textit{otherwise}}
	\end{array} \right.
	\label{eq:vfec:gopcomplexity}
\end{equation}

$RP_i$ is the last parameter in Equation~\ref{eq:vfec:redundancy}, which defines the relative distance of the P-Frames inside the GoP. 
As previously mentioned, frames closer to the end of the GoP are likely to receive less redundant information because the impact of packet loss will be smaller than a loss near the beginning of the GoP, especially in video sequences with larger GoP length.

The total amount of redundant information within a video sequence can be computed by the sum of all the redundant information of each GoP, which is given by $R_{GoP}$. 
On the other hand, the average amount of redundant data, $\bar{R}$, can be found using Equation~\ref{eq:vfec:avgredundancy}.

\begin{equation}
	\bar{R} = \frac{1}{N_{GoP}} \sum_{i=0}^{N_{GoP}} R_{GoP(i)} 
	\label{eq:vfec:avgredundancy}
\end{equation}
\subsection{ViewFEC Performance Evaluation and Results}
\label{sec:vfec:evaluation}

The main objective of the ViewFEC mechanism is to reduce the network overhead introduced by FEC-based schemes while maintaining videos with an acceptable level of quality.

\subsubsection{Experiment settings}

The assessment of the benefits and impact of ViewFEC on WMNs were carried out by using NS-3.
The evaluation scenario comprises six nodes distributed in a grid form~(3x2); each node is 90 meters away from its closest neighbour~\cite{Oh2010}. 
Optimized Link State Routing Protocol~(OLSR)~\cite{Clausen2003} was used as the routing protocol, although any other protocol can be used, such as the Hybrid Wireless Mesh Protocol~(HWMP)~\cite{Bahr2007}. 
A Constant Bit Rate~(CBR) was set as background traffic at 800 kbps and ten video sequences were used in the evaluation scenario~\cite{traces}, with Common Intermediate Format~(CIF) size~(352x288), H.264 codec. 
The GoP size was set at a 19:2 ratio, which means that it contains one I-Frame at every 19 frames and one P-Frame after each two B-Frames.
A Frame-copy technique is used as error concealment method, which means that the decoder will replace each lost frame with the last good one received.
Table~\ref{tab:vfec:parameters} shows the simulation parameters.

\begin{table}[!ht]
	{ \small
		\caption{ViewFEC Simulation parameters}
		\begin{center}
			\begin{tabular}{l|l}
				\hline \textbf{Parameters} & \textbf{Value} \\ 
				\hline
				\hline Display size & CIF - (352 x 288)\\
				\hline Frame rate mode & Constant\\
				\hline Frame rate & 29.970 fps\\
				\hline GoP & 19:2 \\ 
				\hline Video format & H.264\\
				\hline Codec & x264 \\ 
				\hline Container & MP4 \\
				
				\hline Error concealment method & Frame-copy \\
				\hline Wireless standard & IEEE 802.11g \\
				\hline Propagation model & FriisPropagationLossModel \\
				\hline Background traffic & 800 kbps CBR\\
				\hline Routing Protocol & OLSR \\
				\hline Number of nodes & 6 nodes (grid of 3x2) \\
				\hline Error model & Gilbert-Elliot\\
				\hline
			\end{tabular}
			\label{tab:vfec:parameters}
		\end{center}
	}
\end{table}

\begin{figure}[!htb]
	\vspace{-0.0cm}
	\begin{center}
		\includegraphics[width=2.5in]{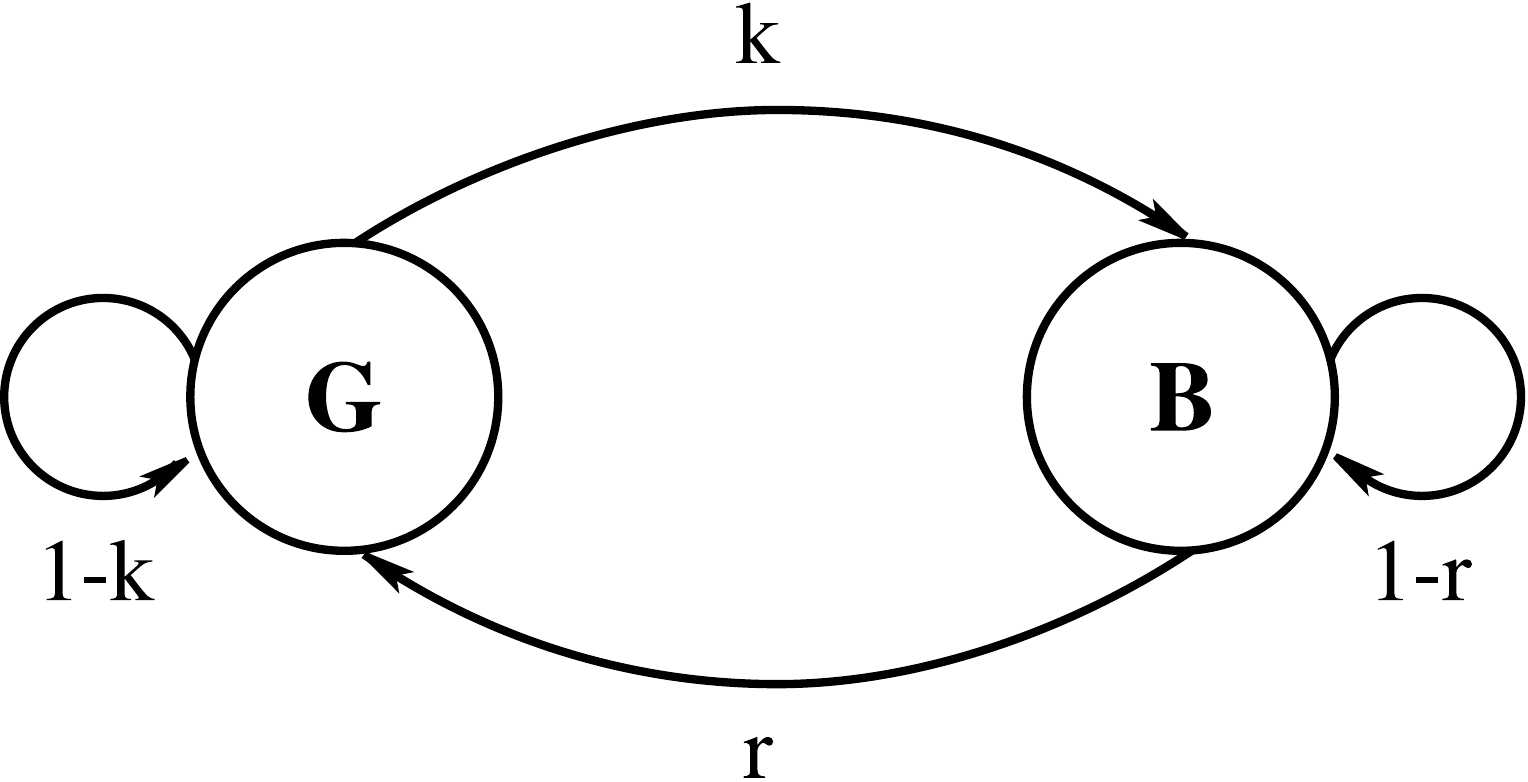}
	\end{center}
	\caption{Gilbert-Elliot loss model}
	\label{fig:vfec:ge}
\end{figure}

Apart from the background traffic, a two-state Markov chain model was implemented to better reflect the network environments in practice. 
This model is also known as the Gilbert-Elliot loss model. 
This is considered a realistic model to simulate network losses because of its burst loss pattern representation, which is commonly found in wireless channels~\cite{Wilhelmsson1999}.
Figure~\ref{fig:vfec:ge} shows this model which is composed of two states~($G$, $B$) and has four parameters~($PG$, $PB$, $r$, $k$).
In state $G$, the highest number of packets must be received correctly. 
The $PG$ parameter holds a probability that indicates when a packet has successfully arrived without any kind of errors. 
The opposite is true for state $B$, and the parameter $PB$ indicates the probability of a packet being lost or damaged. 
The state transition is given by $k$ and $r$, where $k$ represents the probability of transition from state $G$ to $B$, and $r$ represents the probability of transition from state $B$ to $G$.

It is possible to compute the steady-state probability for the $G$ and $B$ states, $\varphi{G}$ and $\varphi{B}$ respectively, for $0<k, r<1$ with Equations~\ref{eq:vfec:sg} and~\ref{eq:vfec:sb}. 

\begin{equation}
\varphi{G} = \frac{r}{(r+k)}
\label{eq:vfec:sg}
\end{equation}
\begin{equation}
\varphi{B} = \frac{k}{(r+k)}
\label{eq:vfec:sb}
\end{equation}

It is also possible to calculate the average packet loss probability, $\bar{P}_{loss}$, using Equation~\ref{eq:vfec:pavg}.

\begin{equation}
\bar{P}_{loss} = PG \times \varphi{G} + PB \times \varphi{B}
\label{eq:vfec:pavg}
\end{equation}

Three scenarios with different mechanisms were assessed. 
The first experiment, which served as a baseline, was carried out without any enhancement~(Without FEC). 
The second was implemented with a non-adaptive Video-aware FEC approach~(Video-aware FEC), where a fixed amount of data redundancy~(80\%) was statically added to both I- and P-frames. 
This amount of redundancy was selected according to an extensive set of experiments, which showed the best video quality situation taking into consideration the characteristics of the scenario defined for the experiment.
Finally, the last scenario adopted the proposed adaptive approach with unequal error protection~(ViewFEC). 
Each one of these three experiments was simulated 20 times with different packet loss patterns due to distinct initial seeds for random number generation~\cite{Salyers2008} used by the Gilbert-Elliot model. 
The average loss was approximately 20\%.

The video quality obtained in the different evaluation scenarios was assessed through objective and subjective measurements. 
The objective metrics used to assess the video quality were SSIM and VQM since there was a lack of correlation in PSNR values according to the subjective human perception.
Both SSIM and VQM are among some of the most widely used to this end~\cite{Chikkerur2011}. 
The SSIM analyses the structural similarity, contrast and luminance of the transmitted images to rank it according to the likeness from the original data. 
Values closer to one represent better video quality. 
VQM uses a discrete cosine transform to assess the spatial-temporal property of the human visual system, allowing it to evaluate the image distortion. 
Values closer to zero represent better video quality. 
The objective quality assessment of the video sequences was carried out with EvalVid~\cite{Klaue2003} and MSU Video Quality Measurement Tool (VQMT)~\cite{DmitriyVatolin2011}. 

Additionally, a subjective evaluation was performed based on the MOS with single-stimulus and ACR.
The subjective experiments were conducted using a Desktop PC with Intel Core i5, 4GB RAM and a 21" LCD monitor, with an application that displays the video sequences and collects the user scores. 
All the sequences are played in a random order in the middle of the screen. 
A neutral grey background is displayed to avoid distracting the attention of the observer. 
25 observers participated in the experiments; they all had normal vision and their ages ranged from 18 to 45. 
The observers included undergraduates, postgraduate students, and university staff.

\subsubsection{QoE assessments}

\begin{figure}[!htb]
	\begin{center}
		\includegraphics[width=116mm]{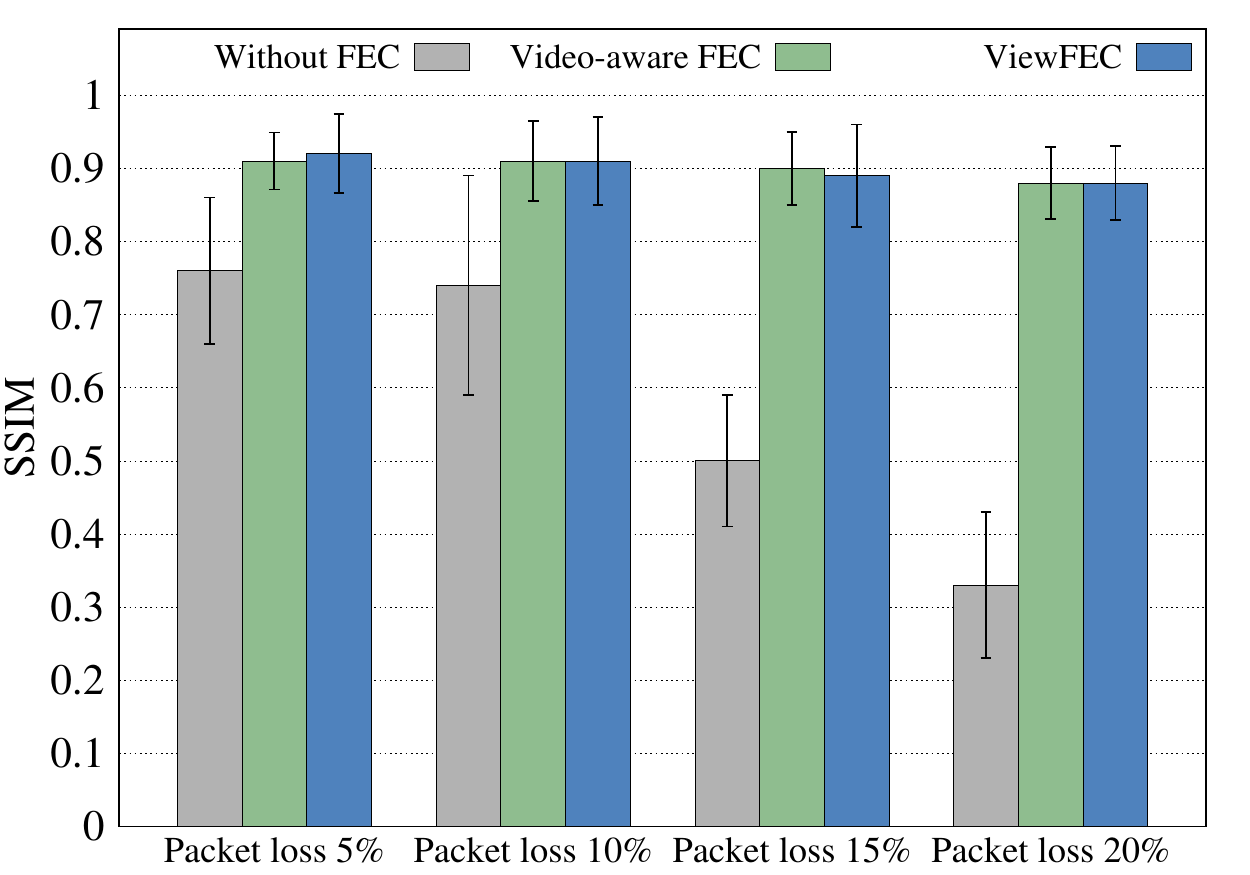} 
	\end{center}
	\caption{\small Average SSIM values for all the video sequences}
	\label{fig:vfec:ssim_cost}
\end{figure}

Figure~\ref{fig:vfec:ssim_cost} shows the average SSIM scores of all the video sequences. 
In the SSIM metric, values closer to one indicate better video quality. 
It is possible to notice that when there is an increase in packet loss rate, there is a sharp decrease in the video quality of sequences that are being transmitted without any type of protection mechanism. At the same time, video sequences that are using either type of FEC-based mechanisms, are able to maintain a good quality. Another important aspect that is worth highlighting, is that with 5\% and 10\% of packet loss rates, the video quality of sequences without FEC are, on average, virtually the same. This can be explained by the natural video resilience to a certain amount of packet loss. Generally speaking, video sequences with low spatial and temporal complexities are more resilient to losses, achieving better results in the QoE assessment. Other sequences, with high spatial and temporal complexity, had poorer results and despite the similar average, the standard deviation was higher with a packet loss rate of 10\%. 
In other words, the QoE assessment values that were obtained are more distant from each other.

\begin{figure}[!htb]
	\begin{center}
		\includegraphics[width=116mm]{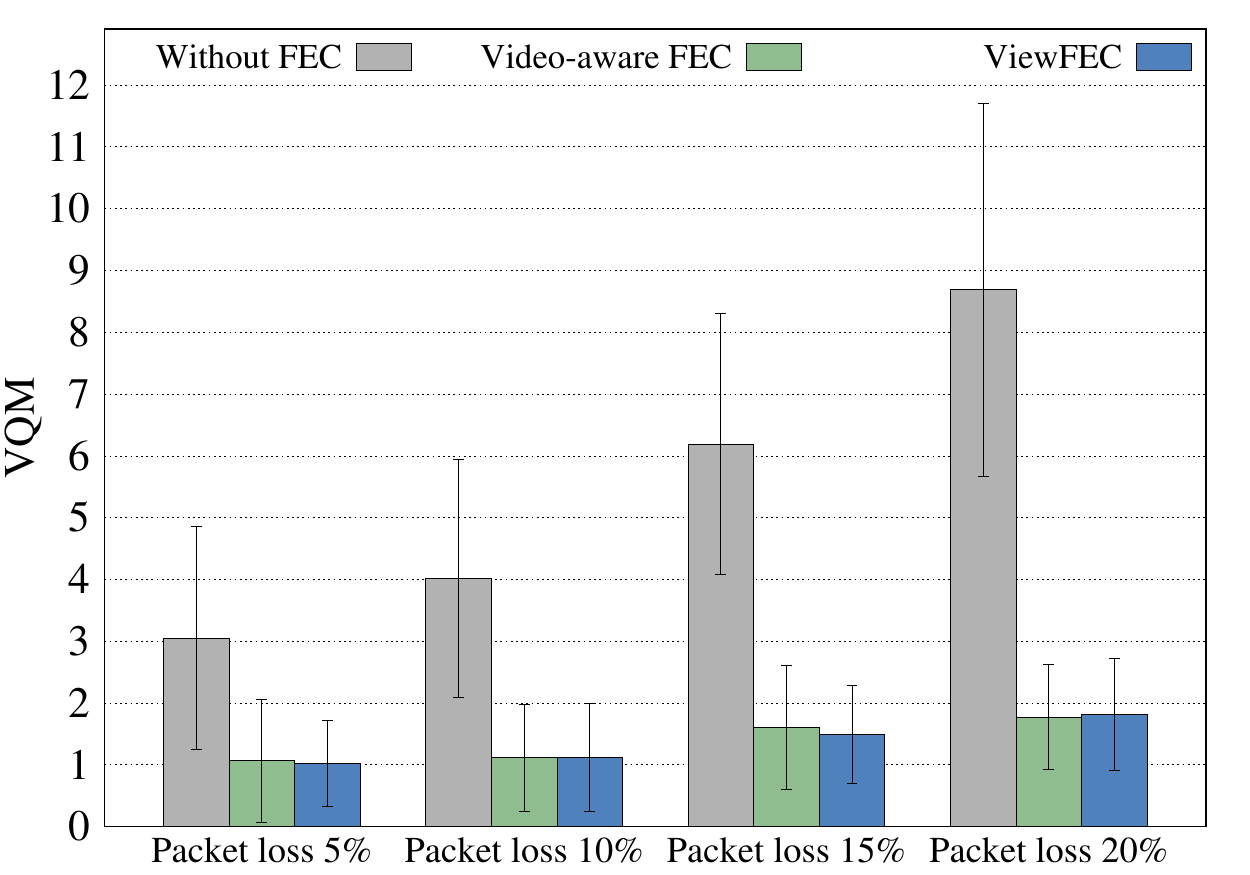}
	\end{center}
	\caption{\small Average VQM values for all the video sequences}
	\label{fig:vfec:vqm_cost}
\end{figure}

Almost the same pattern is discernible in the VQM values showed by Figure~\ref{fig:vfec:vqm_cost}. In this metric, videos with a better quality score close to zero. With a packet loss rate of 5\% or 10\%, the VQM values are also very close to each other. This is not so evident as in the SSIM metric because VQM tends to be more rigid with regard to video impairments, because of that, videos with fewer flaws have poorer results. For the same reason, the standard deviation of this metric tends to be higher than the SSIM metric.

Figure~\ref{fig:vfec:vqm_cost} also shows a large standard deviation for the scenario without FEC. 
The reason for this is that the VQM metric tends to rapidly increase the negative scores of the impaired videos. 
If no error correction technique is employed, some videos will have more defects than others. 
Therefore, because this metric has the tendency to assign really low scores to videos with high degrees of impairments, the sequences transmitted without error correction will receive fairly bad scores, resulting in a larger standard deviation.

\begin{figure}[!htb]
	\begin{center}
		\includegraphics[width=116mm]{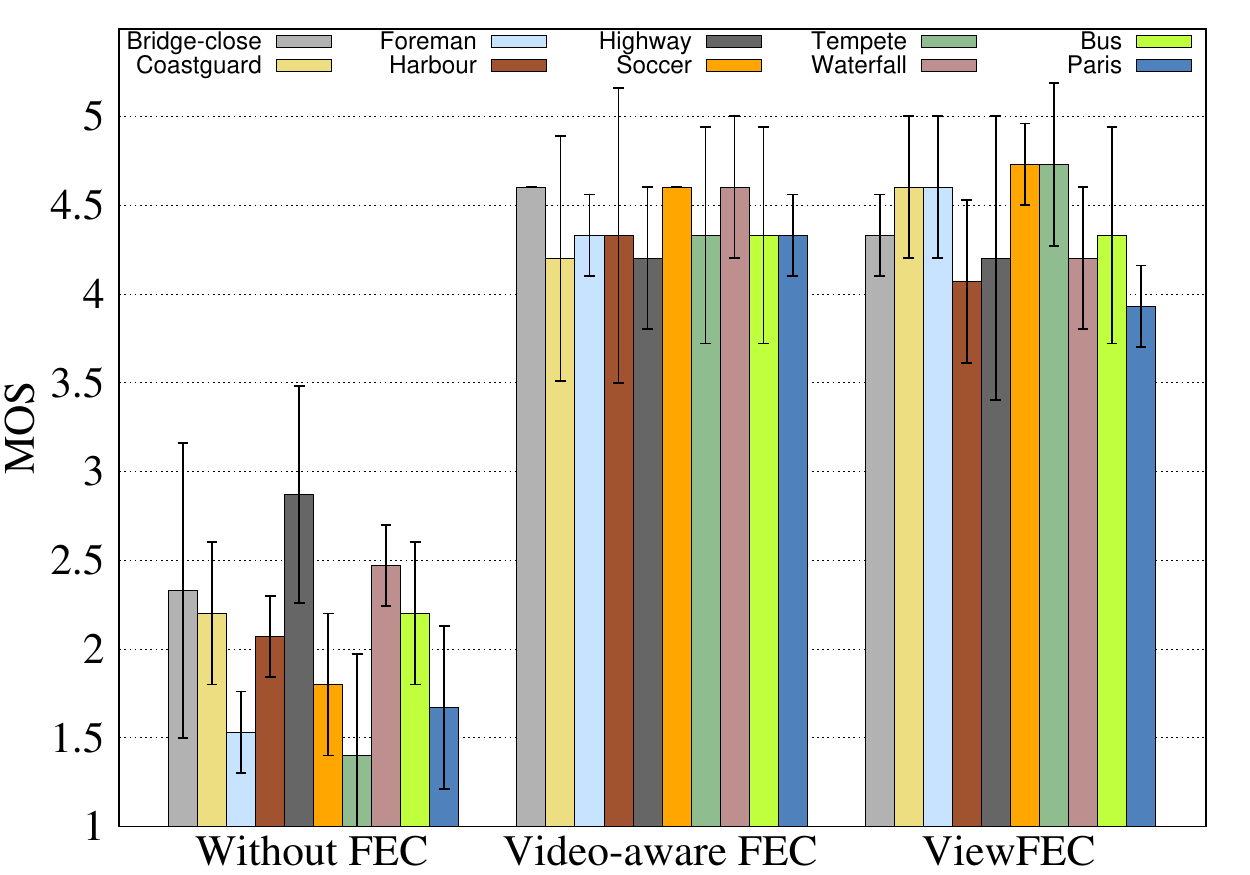}
	\end{center}
	\caption{\small Average MOS per video sequence}
	\label{fig:vfec:mos}
\end{figure}

A detailed analysis was conducted with the scenario that the proposed mechanism achieved the best results~(20\% packet loss rate).
Figure~\ref{fig:vfec:mos} shows the results of the subjective experiments. 
Without using a FEC-based scheme to protect the transmission, the average MOS was 2.05, which is considered poor video quality with annoying impairments. 
When the non-adaptive Video-aware FEC and ViewFEC mechanisms were employed, the MOS average values were 4.39 and 4.37, respectively. 
These values are between good and excellent quality, with perceptible but not annoying impairments. 
The results showed that one of the objectives of ViewFEC mechanism had been attained, which was to maintain the video quality.

\begin{figure}[!htb]
	\vspace{-0.0cm}
	\begin{center}
		\includegraphics[width=116mm]{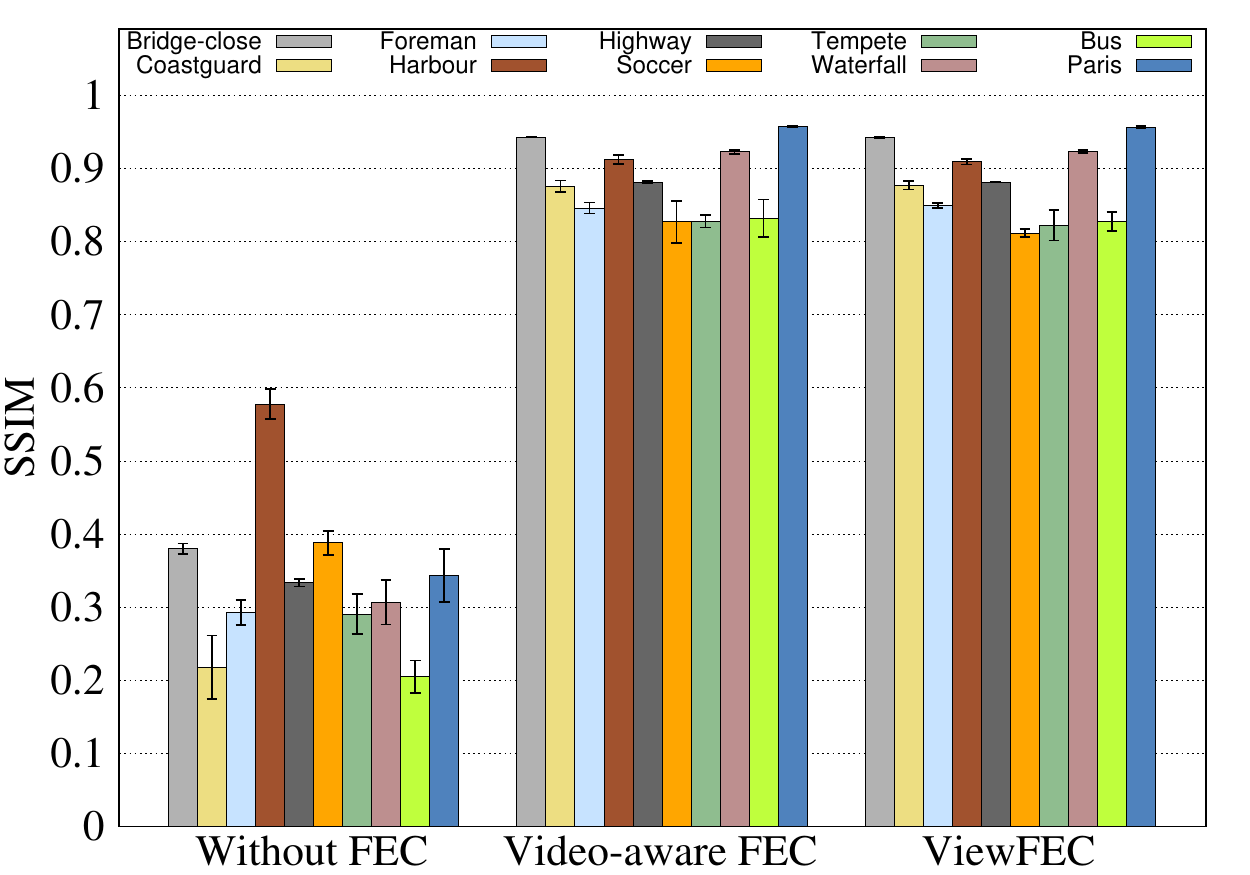}
	\end{center}
	\caption{\small Average SSIM per video sequence}
	\label{fig:vfec:ssim}
\end{figure}

\begin{figure}[!htb]
	\vspace{-0.0cm}
	\begin{center}
		\includegraphics[width=116mm]{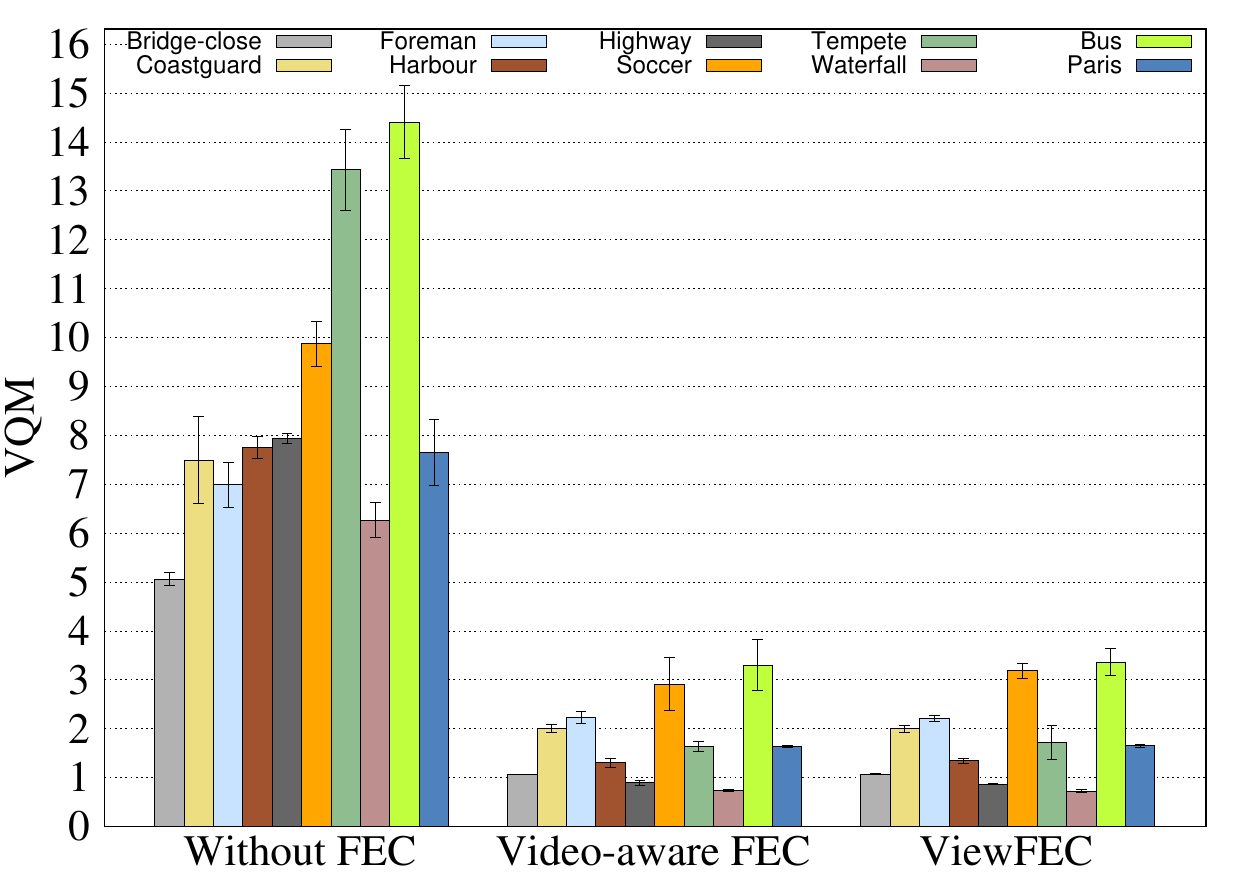}
	\end{center}
	\caption{\small Average VQM per video sequence}
	\label{fig:vfec:vqm}
\end{figure}

Figure~\ref{fig:vfec:ssim} and~\ref{fig:vfec:vqm} show the SSIM and VQM values of each video sequence used in the tests. 
The SSIM average value~(when the FEC schemes were not used) was 0.33 and the VQM value was 8.68, representing low-quality levels and confirming what was found in the subjective assessment. 
On the other hand, the SSIM average of the non-adaptive Video-aware FEC and ViewFEC mechanism was 0.88, and the VQM values were 1.81 and 1.77, respectively. 
These scores show a good video quality, once again, corroborating the subjective findings. 
The distinct QoE scores reached by the different video sequences in the experiments are due to the unique characteristics of each video. 
As mentioned before, small differences in motion and complexity levels can influence the obtained values. 
In view of this, it is important to make use of several types of videos when conducting the experiments.

\subsubsection{Network footprint analysis}

\begin{figure}[!htb]
	\begin{center}
		\includegraphics[width=116mm]{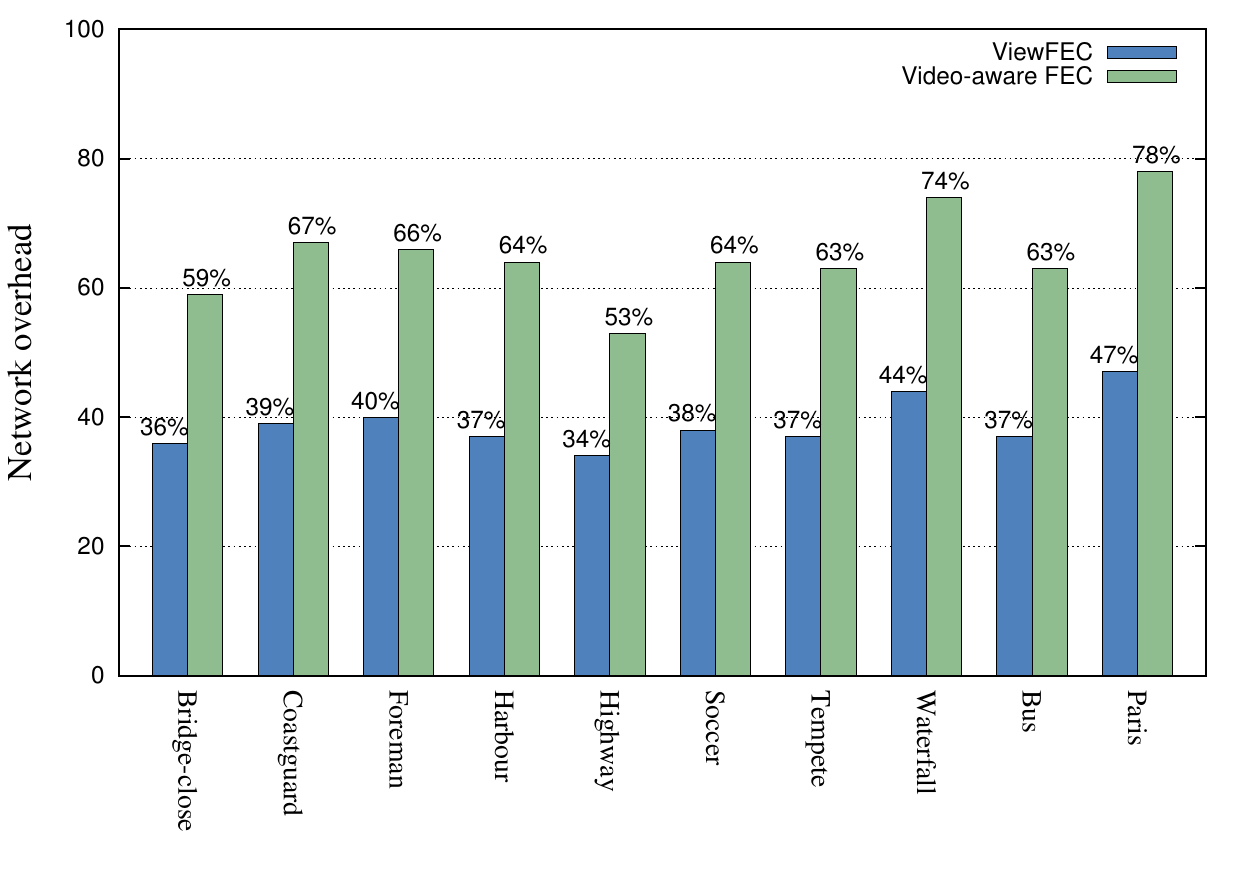}
	\end{center}
	\caption{\small Network overhead (\%)}
	\label{fig:vfec:overhead}
\end{figure}

All the QoE assessments demonstrated that the ViewFEC mechanism was able to maintain a good video quality. Nevertheless, the main goal of this mechanism was to reduce the network overhead. This is important in wireless networks, due to the limited channel resources, uneven bandwidth distribution, and interference caused by concurrent transmissions. 
In the experiments, the network overhead is the sum of all video frames transmitted, which includes the redundant data, minus the original frame size. 
The non-adaptive Video-aware FEC mechanism was able to deliver videos with a network overhead between 53\% and 78\%, as shown in Figure~\ref{fig:vfec:overhead}.
On the other hand, when the ViewFEC mechanism was used, the network overhead remained between 34\% and 47\%. The ViewFEC mechanism imposes, on average, 40\% less network overhead than the non-adaptive Video-aware FEC, with equal or slightly better video quality, as shown in Figure~\ref{fig:vfec:mos},~\ref{fig:vfec:ssim} and~\ref{fig:vfec:vqm}.

\begin{figure}[!htb]
	\begin{center}
		\includegraphics[width=116mm]{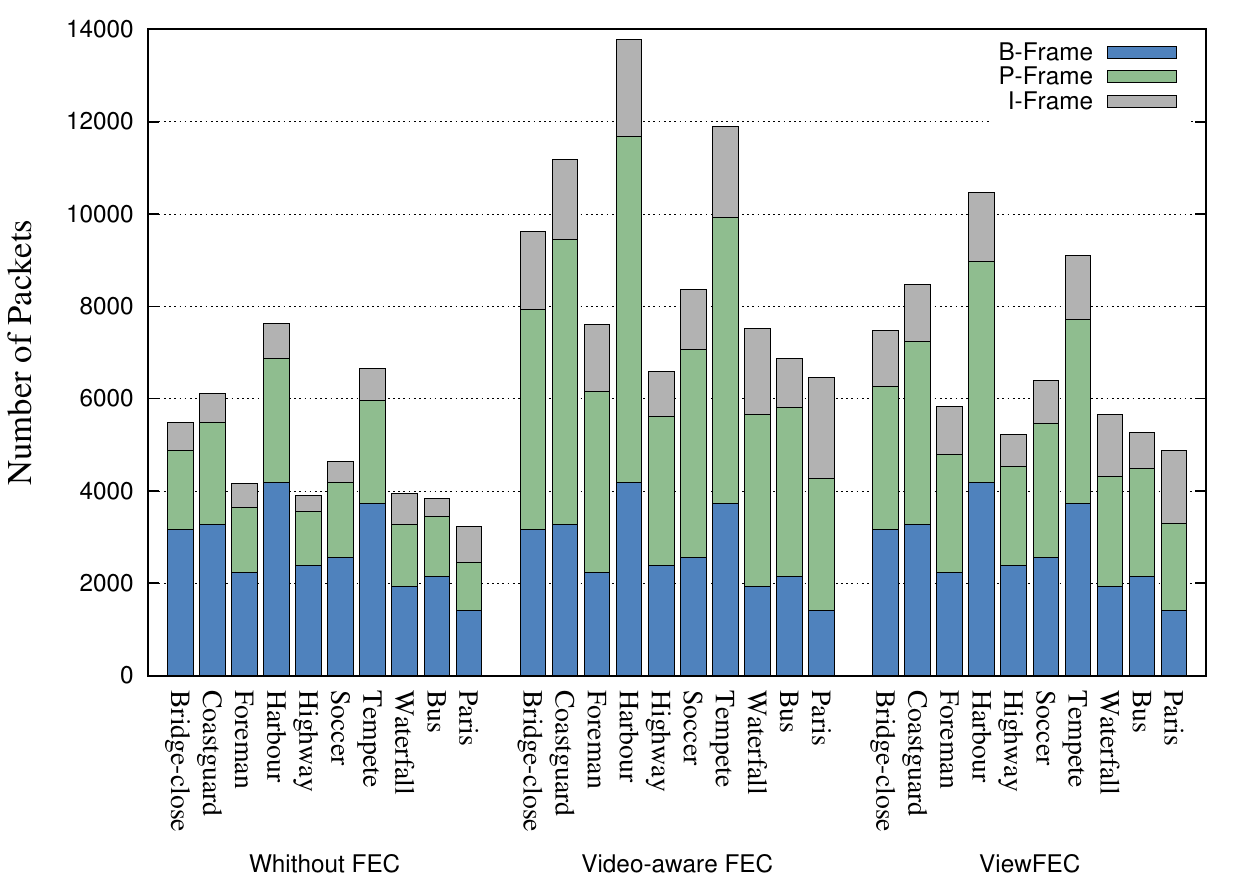}
	\end{center}
	\caption{\small Network overhead in packets}
	\label{fig:vfec:pktsize}
\end{figure}

The video sequence Highway had the smallest reduction in network overhead~(36\%) and the Coastguard sequence had the largest~(42\%). This can partially be explained by the size of the I-, P- and B-Frames. Figure~\ref{fig:vfec:pktsize} shows the size of the frames of all videos. 
The y-axis shows the number of sent packets, including the redundant information.
When the Highway values are analysed, one can notice that over 61\% of the packets belong to B-Frames, which are not considered in either the non-adaptive Video-aware FEC or in ViewFEC, because they lead to minor impairments if lost. This means that less than 39\% of the packets are optimised by the ViewFEC mechanism, and this result in a smaller reduction in the overhead. Conversely, the Coastguard sequence has more than 46\% of the packets in I- or P-Frames, which can be optimised, resulting in a greater reduction of network overhead.

\subsubsection{Overall results}

Table~\ref{tab:vfec:allpktloss} summarises the results and shows the improved video quality~(in percentage) in each scenario for both mechanisms. 
In the VQM metric, the more negative the percentage showed in the table is the better the result~(e.g., $\downarrow$79.60\%). 
On the other hand, the opposite is true for the SSIM metric. 
As expected, both FEC-based mechanisms produce more valuable results when the network has a higher PLR.
When the average PLR was 20\%, for example, it was possible to achieve a reduction of over 79\% in VQM values; this means $\approx 4.9$ times smaller scores. 
With the SSIM metric, there was an increase of over 166\% in the results, meaning $\approx 2.66$ times higher values.

\begin{table}[!ht]
	{\scriptsize
		\caption{ViewFEC QoE values and improvement}
		\begin{center}
			\resizebox{\columnwidth}{!}{%
			\begin{tabular}{c||c|c|c|c|c|c}
				\hline
				\raisebox{-1.5ex}{Packet loss}& \raisebox{-1.5ex}{QoE} & \raisebox{-1.5ex}{Without} &  & \raisebox{-1.5ex}{Video-aware FEC}  &  & \raisebox{-1.5ex}{ViewFEC} 	 \\
				\raisebox{1.0ex}{rate} & \raisebox{1.0ex}{Metric} & \raisebox{1.0ex}{FEC} & \raisebox{1.5ex}{Video-aware FEC} & \raisebox{1.0ex}{Improvement} & \raisebox{1.5ex}{ViewFEC} &  \raisebox{1.0ex}{Improvement}	 \\
				\hline
				\hline
				\multirow{2}{*}{Packet loss 5\%}	&VQM	&3.05	&1.06	&$\downarrow$65.14\%	&1.02	&$\downarrow$66.48\%	\\
				& SSIM	&0.76	&0.91	&$\uparrow$19.74\%	&0.92	&$\uparrow$21.05\%	\\
				\hline 
				\multirow{2}{*}{Packet loss 10\%}	&VQM	&4.01	&1.11	&$\downarrow$72.36\%	&1.12	&$\downarrow$72.09\%	\\
				& SSIM	&0.74	&0.91	&$\uparrow$22.97\%	&0.91	&$\uparrow$22.97\%	\\
				\hline 
				\multirow{2}{*}{Packet loss 15\%}	&VQM	&6.19 & 1.60	& $\downarrow$74.09\% &	1.49	& $\downarrow$75.87\%	\\
				& SSIM	&0.50	&0.90	&$\uparrow$80.00\%	&0.89	&$\uparrow$78.00\%	\\
				\hline 
				\multirow{2}{*}{Packet loss 20\%}	&VQM	&8.68 &	1.77	&$\downarrow$79.60\%	&1.81	&$\downarrow$79.14\%	\\
				& SSIM	&0.33	&0.88 &$\uparrow$166.67\%	&0.88	&$\uparrow$166.67\%	\\
				\hline
			\end{tabular}
			}
			\label{tab:vfec:allpktloss}
		\end{center}
	}
\end{table}

Taking into consideration the results of the experiments, it is possible to say that the proposed ViewFEC mechanism showed good performance. 
Additionally, it also highlights the fact that it is feasible to improve the QoE of video sequences delivered over wireless networks.

\section{Video-aware and RNN-based mechanism (neuralFEC)}
\label{sec:neuralFEC}

The mechanism described in Section~\ref{sec:viewfec} provided positive results, however, several issues were identified during the experiments. Most of these are related to the heuristic used in the process to classify the motion intensity. A more accurate technique to perform this task could improve the mechanism. 

Using more information about the video sequences on the classification method can also be beneficial. 
An efficient way to quantify the pace of action is through motion vectors. These vectors play a key part in the video compression process, allowing the system to store changes from adjacent frames, including both previous and future frames. 
Therefore it is possible to quantify the motion intensity of a given frame using the information inside its vectors.
Taking these issues into account, this section proposes the adaptive Video-aware Random Neural Networks~(RNN) based mechanism~(neuralFEC).
This mechanism was proposed in collaboration with MSc student Pedro Borges, which performed his work under my co-supervison.

\subsection{neuralFEC Overview}

The neuralFEC mechanism aims at overcoming the limitations of non-adaptive schemes, such as the inability to take into consideration a precise classification of the video's motion intensity, which is crucial to a high QoE. 
The mechanism proposed in this section mitigates these problems by adaptively selecting the amount of redundancy given to individual frames, showing better results than the previously proposed mechanism. 
The adaptive redundancy is chosen according to the analysis of the frame type and the motion characteristics using a RNN~\cite{Abraham2005}. 
Neural Networks are computational models inspired by biological central nervous systems, which are able to go through the process of machine learning and pattern recognition. 
The NN can be trained by feeding to it learning patterns and letting it change the weights according to some learning rule.
Random Neural Networks are known for they success in pattern recognition and classification problems~\cite{Mohamed2002}, making this type of NN suitable for the proposed mechanism.

\begin{figure*}[!htb]
	\begin{center}
		\ifBW \includegraphics[width=143mm]{./rnnFEC_v3_Thesis_gray-eps-converted-to.pdf}
		\else \includegraphics[width=143mm]{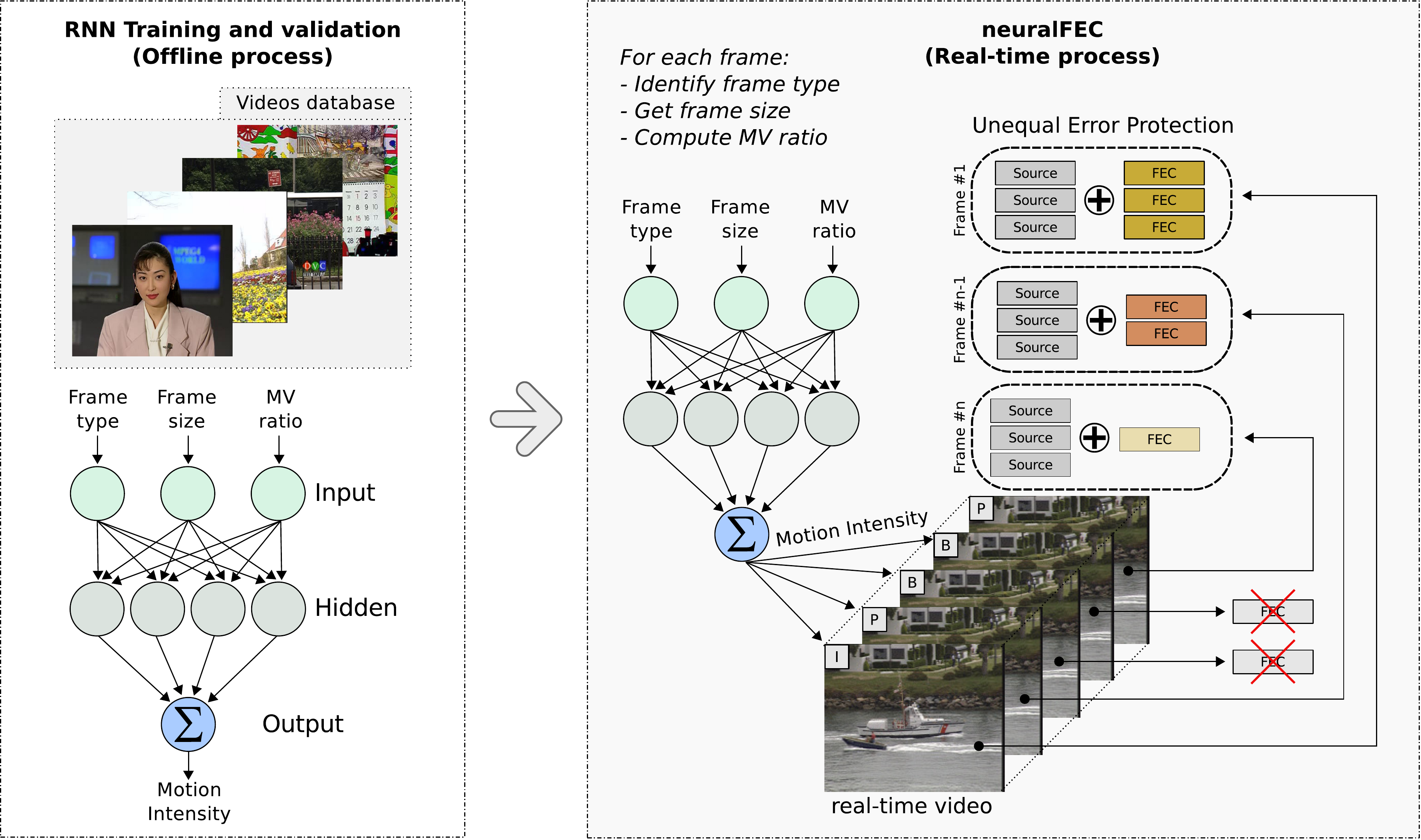}
		\fi
	\end{center}
	\caption{neuralFEC mechanism}
	\label{fig:neuralFEC:design}
\end{figure*}

Figure~\ref{fig:neuralFEC:design} depicts the overall operation of the neuralFEC mechanism. 
First of all, the same procedure adopted by ViewFEC has been used again in the offline process. 
This means that an exploratory analysis using hierarchical clustering was carried out, however, here the results are used to train the RNN. 
The RNN was validated through the human experience about the intrinsic video characteristics and several simulation experiments.
After this, the RNN can be used in real-time. 
The decision-making process conducted by the RNN determines a specific amount of redundancy needed by each frame. This allows the neuralFEC to shield only the QoE-sensitive data against packet loss, resulting in better video quality as perceived by the end-users, while saving network resources. A detailed explanation of the proposed mechanism is presented afterwards. 

\subsection{Towards the design of neuralFEC}

In order to perform the classification of each frame according to its motion intensity, a RNN was employed. 
As other NNs, this method has the capability for learning and generalisation, providing particularly good results in pattern recognition and classification problems~\cite{Mohamed2002}. 
By training the network successfully with an adequate range of video samples, it can be used in real-time to classify a given video sequence according to the intensity of the movement. 
This is achieved by attributing a specific value to determine frames with different motion intensities. 
After that, the neuralFEC is able to select, in real-time, the appropriate amount of redundancy to be transmitted so that the network overhead is minimised and QoE is maximised. 

The RNN structure consists of three input nodes, seven hidden layer nodes and one output. The three input nodes represent each frame's characteristics specifically frame size, frame type, and motion vectors ratio~(the total number of vectors divided by the distance described by them). 
The ratio was used because a certain frame can have several vectors pointing to a close distance while other frames can have fewer vectors pointing further away, however, and consequently defining a situation of higher motion intensity. Finally, the output node provides the motion intensity classification value, computed by the network from the given inputs. Through these parameters, it is possible to characterise the video motion intensity and choose the optimal amount of redundancy on a frame-by-frame basis.

To adequately train the RNN, an exploratory hierarchical cluster analysis using Ward's method~\cite{WardJr1963} was performed to categorise selected video sequences which represent different types of movement. 
As always, the video sequences were selected according to the recommendations of the VQEG~\cite{Staelens2011} and ITU~\cite{ITUTJ2008}.
A set of 15 videos was selected to perform the hierarchical cluster analysis. 
Each video was broken down into three parameters, namely about the frame size and type, and motion vectors ratio.

Using the exploratory analysis results, the video sequences were classified into three categories of motion intensity, namely low, medium and high intensity. 
Afterwards, two videos of each motion intensity category were randomly selected to train the RNN. 
The training of the RNN consisted in feeding the information of this set of selected videos to the inputs of the network for about 600 iterations which were the point at which the Minimum Mean Squared Error~(MMSE) stabilised. 
After the training period it was validated with a different set of video sequences, the remaining 9 videos from the exploratory analysis, which also cover all three motion intensity characteristics. 

After the training and validation phases, the RNN can be used in the real-time process. Using cross-layer techniques, the neuralFEC mechanism is able to obtain important information about several video characteristics, namely frame type and size, as well as the number of motion vectors and the Euclidean distance pointed by them. All these details are fed to the RNN which in turn provides, in real-time, an accurate motion intensity value for each frame. 

Once the video frame is classified, it is encoded with the amount of redundancy selected by the RNN. 
Through this procedure, a precise UEP amount can be assigned to each frame, where only the QoE-sensitive data will be protected. 
The result of this is better video quality while reducing the amount of redundancy data needed.
Therefore, not adding unnecessary redundancy will allow more users to access services, improving the overall system performance.

Algorithm~\ref{algo:neuralFEC:pseudo} shows the pseudo-code portraying the neuralFEC real-time operation. All procedures are performed inside a for-loop, at line 01, which will go through all the frames in the video sequence. At line 02, the frame type is identified to be used in the selection control mechanism~(if statement) at line 03. This allows the change in the control flow according to neuralFEC needs, which is to assign a tailored redundancy amount to I- and P-Frames, and send B-Frames without additional data. At lines 04, 05, 06, and 07 it is possible to observe the identification of the frame size, the computation of the motion vectors ratio, the classification of the video frame motion intensity using the RNN, and the assignment of an unequal amount of redundancy to the most QoE-sensitive data, respectively. Lines 08 and 10 are responsible for sending the frame with or without redundancy.

\begin{algorithm}[!htb]
	\For{each Frame}{
		FT $\leftarrow$ \textsc{getFrameType}($Frame$)\;
		\eIf{(FT equal (I- or P-Frame))}{
			FS $\leftarrow$ \textsc{getFrameSize}($Frame$)\;
			MVratio $\leftarrow$ \textsc{calculateRatio}(getMV($Frame$))\;
			MotionIntensity $\leftarrow$ \textsc{RNN}($FT, FS, MVratio$)\;
			\textsc{addRedundancy}(\textsc{RS}($MotionIntensity$))\;
			\textsc{sendFrame}($Frame+Redundancy$)
		}{
		\textsc{sendFrame}($Frame$)
		}
	}
\caption{neuralFEC pseudo-code}\label{algo:neuralFEC:pseudo}
\end{algorithm}
\subsection{neuralFEC Performance Evaluation and Results}
\label{sec:neuralFEC:evaluation}

The neuralFEC goal is to use an optimised UEP scheme according to the video frame's motion intensity characteristics, making it possible to better protect the most QoE-sensitive frames. Therefore,  providing both the reduction in the impact of packet loss on the video quality and the use of only the necessary amount of redundancy. Through this, it is possible to distribute the redundancy in a way that will improve the QoE for the end user, while sparing network resources.

\subsubsection{Experiment settings}

In order to assess the performance of the proposed mechanism in wireless networks, several experiments were performed by using NS-3. The scenario for evaluation is comprised of 25 nodes in a grid disposition~(5x5), separated by 50 meters. The OLSR was used as the routing protocol. Ten video sequences were used in this scenario, namely Bowing, Coastguard, Container, Crew, Foreman, Hall, Harbour, Mother and Daughter, News and Soccer. These particular sequences were selected in order to have a great variety of motion intensities. They are in CIF with a resolution of~352x288 and coded with the H.264 codec. The GoP size was set to 19:2 thus after each I- or P-frames, come two B-frames. 
Table~\ref{tab:neuralFEC:parameters} shows the simulation parameters.

\begin{table}[!ht]
	{ \small
		\caption{neuralFEC Simulation parameters}
		\begin{center}
			\begin{tabular}{l|l}
				\hline \textbf{Parameters} & \textbf{Value} \\ 
				\hline
				\hline Display size & CIF - (352 x 288)\\
				\hline Frame rate mode & Constant\\
				\hline Frame rate & 29.970 fps\\
				\hline GoP & 19:2 \\ 
				\hline Video format & H.264\\
				\hline Codec & x264 \\ 
				\hline Container & MP4 \\
				
				\hline Error concealment method & Frame-copy \\
				\hline Wireless standard & IEEE 802.11g \\
				\hline Propagation model & FriisPropagationLossModel \\
				\hline Background traffic & 800 kbps CBR\\
				\hline Routing Protocol & OLSR \\
				\hline Number of nodes & 25 nodes (grid of 5x5) \\
				\hline Error model & simplified Gilbert-Elliot\\
				\hline
			\end{tabular}
			\label{tab:neuralFEC:parameters}
		\end{center}
	}
\end{table}

A two-state discrete-time Markov chain model was implemented following a simplified Gilbert-Elliot packet-loss model~\cite{Razavi2009}, which approximates the behaviour of a wireless network. 
It produces simulation results which are closely related to those of burst loss patterns of wireless channels~\cite{Wilhelmsson1999}. 
This model differs from the Gilbert-Elliot used in Section~\ref{sec:viewfec} because the probability of packet loss in the Good state~(G) is set at 0, which means no losses. 
On the other hand, the probability of packet loss in the Bad state~(B) is set at 1, where all packets are lost. 
The PLR can be obtained by Equation~\ref{eq:neuralFEC:burst_loss}, where $P_{BG}$ represents the probability of transitioning from the Bad state to the Good state and vice-versa with $P_{GB}$.

\begin{equation}
	PLR = \frac{P_{BG}}{P_{BG} + P_{GB}}
	\label{eq:neuralFEC:burst_loss}
\end{equation}

To validate and compare the results, three experiments with different mechanisms were performed. The first experiment serves as a baseline as there was no FEC mechanism in use. The second experiment was performed with a non-adaptive video-aware FEC mechanism~(Video-aware FEC), where a fixed amount of 38\% of redundancy was added only to I- and P-frames. 
This amount of redundant data was chosen after several experiments and represents the best video quality in the considered scenarios.
The last experiment is the proposed adaptive mechanism with RNN and UEP~(neuralFEC). Each of the three experiments was simulated 10 times with an error rate of 20\% representing an average loss~\cite{Immich2013} obtained through a simplified Gilbert-Elliot packet loss model.

The video quality of each evaluation scenario was assessed through an objective measurement, namely the SSIM. The objective quality assessment of the video sequences was performed with EvalVid and the MSU VQMT.

\subsubsection{Network footprint analysis}

\begin{figure}[!htb]
	\begin{center}
		\includegraphics[width=116mm]{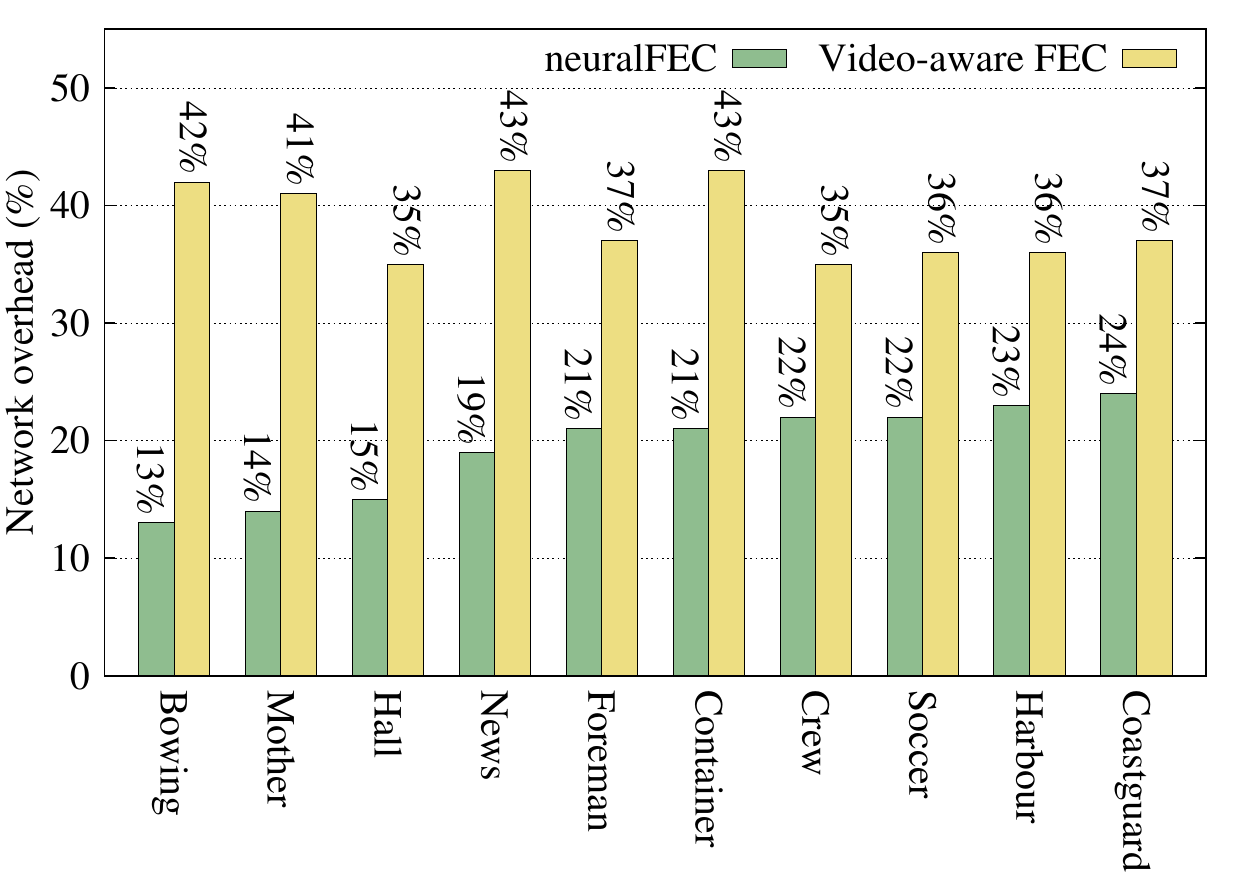}
	\end{center}
	\caption{Network Overhead}
	\label{fig:neuralFEC:overhead}
\end{figure}

Figure~\ref{fig:neuralFEC:overhead} shows the results in terms of network overhead. 
The non-adaptive Video-aware FEC mechanism had an overhead between 35\% and 43\%. 
On the other hand, when the neuralFEC mechanism was employed, the amount of overhead remained between 13\% and 24\%. 
This means that the average redundancy added by the non-adaptive mechanism was around 38\% in contrast to only 19\% added by neuralFEC. 
It is also clear that the proposed mechanism can assess the importance of frames according to motion intensity. This assessment is performed by the RNN, which attributes a higher classification for frames with a great amount of movement, and a lower classification for frames with less amount of movement. In doing that, a greater amount of redundancy was attributed to video sequences such as Crew, Soccer, Harbour and Coastguard. On the contrary, video sequences which are classified as being of lesser motion intensity, such as Bowing, Mother and Hall are given less redundancy. These results show that the neuralFEC mechanism performs better than the non-adaptive Video-aware FEC mechanism in terms of overhead, by reducing in average a half of the redundancy needed to protect the data.

\subsubsection{QoE assessments}

\begin{figure}[!htb]
	\begin{center}
		\includegraphics[width=116mm]{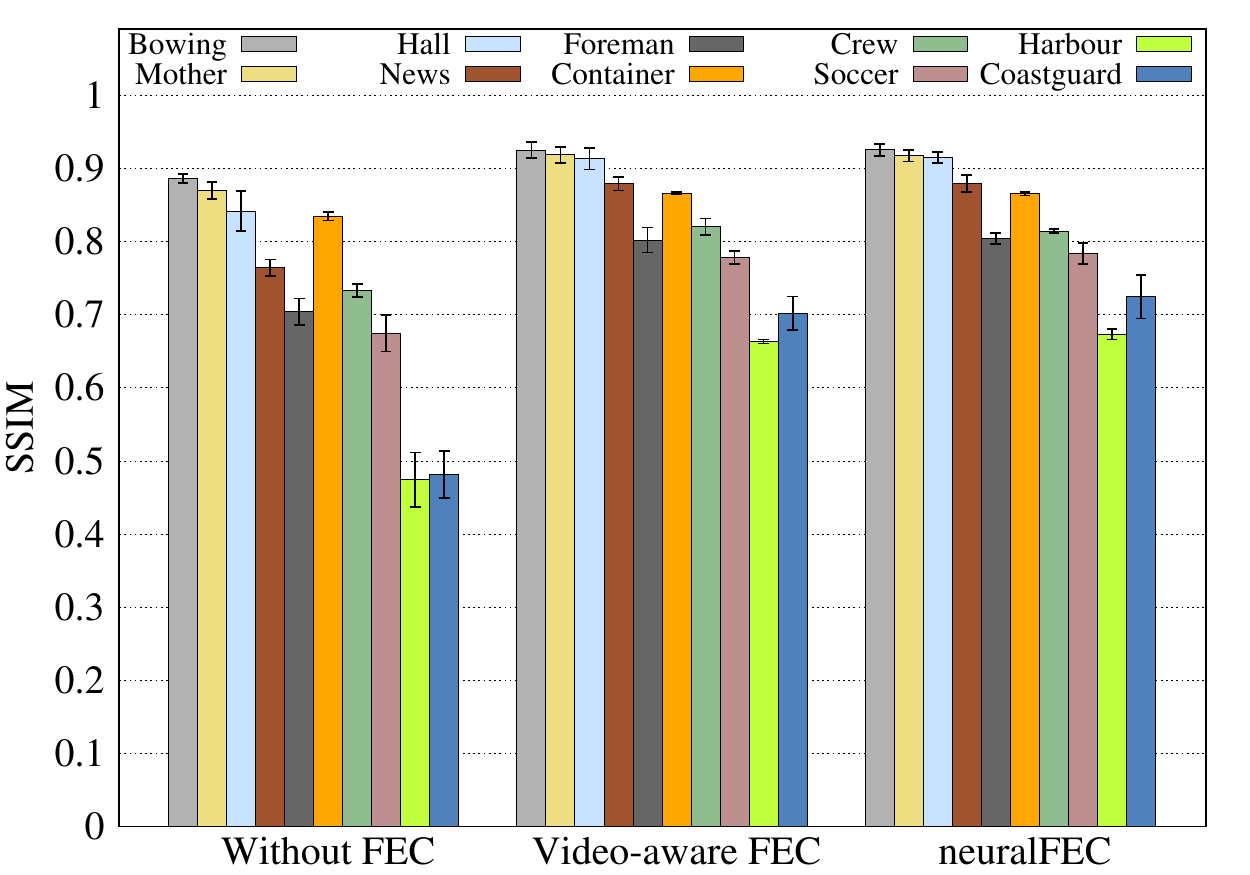}
	\end{center}
	\caption{neuralFEC Objective QoE assessment (SSIM)}
	\label{fig:neuralFEC:ssim}
\end{figure}

Besides saving the already scarce network resources by not adding unnecessary redundancy, it is also important to provide good video quality. In order to verify this situation, a set of assessments was performed using the SSIM metric. Figure~\ref{fig:neuralFEC:ssim} depicts the SSIM values for each video sequence while using the three aforementioned protection schemes. The results show the neuralFEC mechanism obtained an average SSIM value of 0,831 against a value of 0,819 for the video-aware FEC mechanism and 0,726 for the mechanism that did not use any type of protection.
This represents a slight improvement of around 1,5\%, on average, in terms of SSIM value for the adaptive neuralFEC mechanism in comparison to the non-adaptive video-aware FEC mechanism. In further detail, the SSIM score achieved by neuralFEC for the Harbour video sequence was of 0,675 against 0,662 for the video-aware mechanism and 0,485 for the mechanism without FEC. Although all videos were transmitted with the same PLR, the SSIM score obtained by the same three mechanisms for the Bowing sequence was of 0,915, 0,914, and 0,920, respectively. 
This can be explained by the different characteristics of these two sequences. 
The Harbour video sequence has a greater amount of motion compared to the Bowing sequence, meaning that packet loss has a greater effect on this type of sequences. 
This results in lower SSIM scores for sequences with a higher degree of motion intensity and also shows that videos with a lower degree of motion intensity have greater resilience to packet loss. 
Due to this, it is important to employ adaptive FEC mechanisms, such as neuralFEC to protect the contents of the video taking into account its motion intensity characteristics.

\subsubsection{Overall results}

Table~\ref{tab:neuralFEC:summ} summarises the results presenting the average SSIM and network overhead for all video sequences. It demonstrates that the proposed neuralFEC mechanism had a slightly improved video quality. Most importantly, it was able to do so while drastically reducing the network overhead by not adding unnecessary redundancy. This is of great importance in wireless networks, due to the limited nature of the channel resources, which can be aggravated by packet loss due to interference from concurrent transmissions and network congestion.

\begin{table}[h]
	\caption{Average SSIM and network overhead of neuralFEC}
	\begin{center}
		\begin{tabular}{l|c|c|c}
			\hline
			& \multicolumn{1}{|l}{\textbf{neuralFEC}} & \multicolumn{1}{|l}{\textbf{Video-aware FEC}} & \multicolumn{1}{|l}{\textbf{Without FEC}} \\
			\hline
			\hline
			{SSIM}     & 0,831                                  & 0,819                                        & 0,726                                 \\
			\hline
			{Overhead} & 19,334\%                               & 38,460\%                                     & --                                  \\   
			\hline
		\end{tabular}
	\end{center}
	\label{tab:neuralFEC:summ}
\end{table}

The results showed that the neuralFEC mechanism, through an accurate motion intensity classification of video sequences with distinct characteristics, is able to add a precise amount of protection. In doing that, it can offer less overhead during transmission in a wireless mesh network setting while providing as good video quality as non-adaptive FEC mechanisms.

\section{QoE-driven Motion- and Video-aware Mechanism (PredictiveAnts)}
\label{sec:PredictiveAnts}

The aforementioned mechanism was able to provide a notable outcome including a slightly better video quality as well as the reduction of the unnecessary network overhead. 
Nevertheless, there is still room for improvement in the classification method, by using a higher number of video characteristics.
Additionally, both ViewFEC and neuralFEC lack the capability to assess the network conditions to adjust the redundancy amount, which can have a significant impact. 

Considering the open issues aforementioned, particularly the absence of motion-aware mechanisms that use a broad amount of video characteristics together with a network assessment, this section presents the proposed QoE-driven motion- and video-aware mechanism~(PredictiveAnts). 
This mechanism was proposed in collaboration with MSc student Pedro Borges.

\subsection{PredictiveAnts Overview}

The adaptive FEC-based mechanism proposed in this section uses several video characteristics and packet loss rate prediction to shield real-time video transmission over wireless mesh networks, improving both the user experience and the usage of resources. 
This is possible through a combination of a RNN, to categorise motion intensity of the videos, and an ACO scheme, for dynamic redundancy allocation, allowing the protection of the most important information.

\begin{figure}[!htb]
	\begin{center}
		\ifBW \includegraphics[width=143mm]{./figure1_pAnts_gray-eps-converted-to.pdf}
		\else \includegraphics[width=143mm]{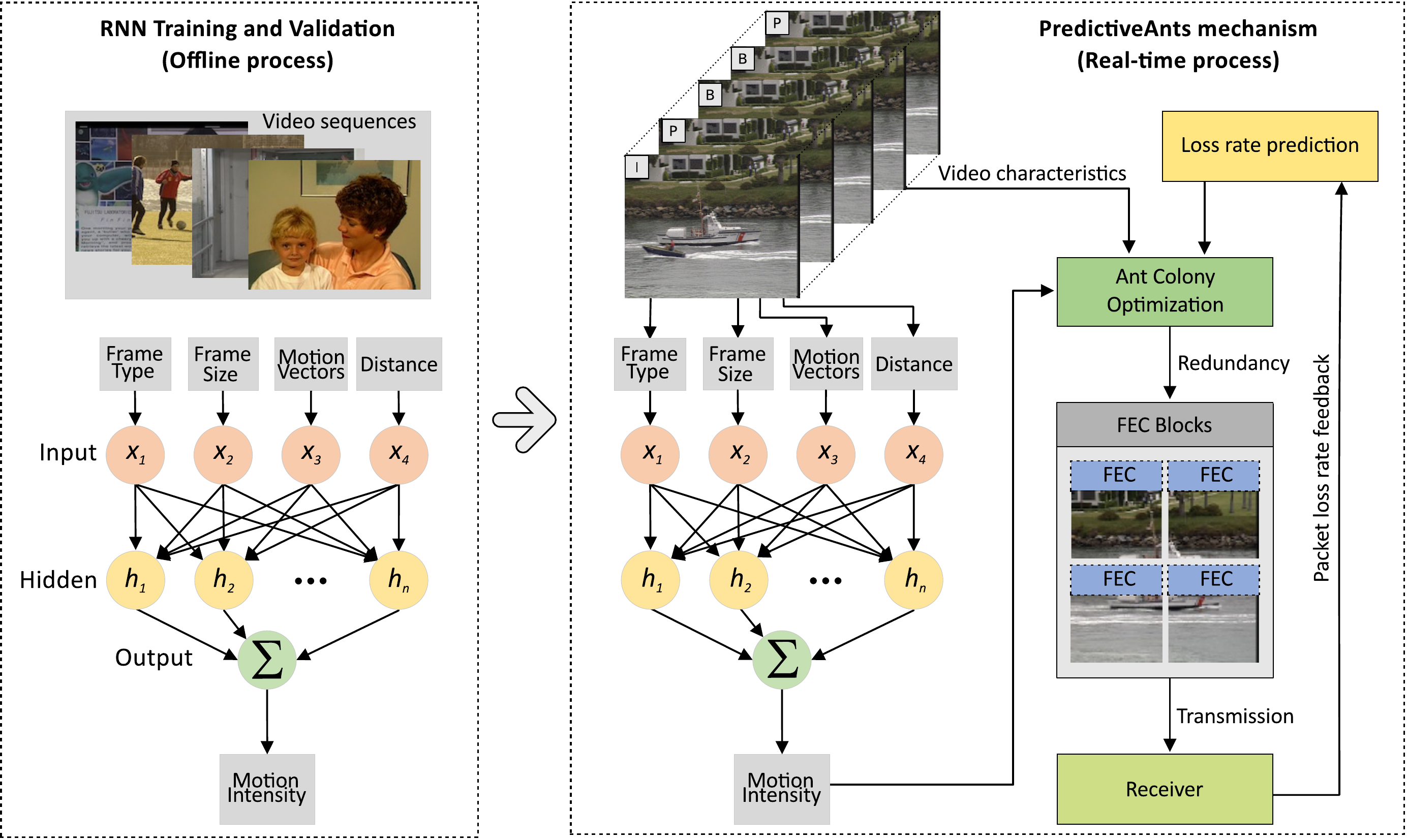}
		\fi
	\end{center}
	\vspace{-0.0in}
	\caption{PredictiveAnts mechanism}
	\label{fig:pAnts:mechanismFEC}
\end{figure}

Figure~\ref{fig:pAnts:mechanismFEC} depicts the PredictiveAnts mechanism. It is composed of two processes, one is performed offline and the other one in real-time. 
In the same way as in the other mechanisms, the Offline process is responsible for the RNN training and validation steps. The main objective of the RNN is to characterise the motion intensity of video sequences according to several inputs, such as the frame type and size, the number of motion vectors, and the Euclidean distance pointed by these vectors. Since it is an offline process, it needs to be executed only once. After that, the RNN can be used in real-time. The offline process is important because it allows a fast and more accurate real-time execution since few variables need to be handled. 

The real-time process consists of several modules, each one having very peculiar tasks, as follows:

\begin{itemize}
	\renewcommand{\labelitemi}{$\bullet$}
	\item \textsc{Motion Intensity} - The motion intensity characterization is performed by the RNN in real-time since it was already trained and validated in the offline process;
	\item \textsc{Feedback Receiver} - The feedback mechanism is responsible for the retrieval of loss statistics. The information is collected by the receiver and sent to the transmitter;
	\item \textsc{Loss Rate Prediction} - Using the feedback statistics, the properties of the error probability are estimated on the server side; 
	\item \textsc{Video Characteristics} - This module fetches information from the video sequences that are being transmitted to identify video characteristics such as the frame type and size, as well as the motion vectors;
	\item \textsc{Ant Colony Optimization} - The ACO is responsible for making a joint analysis of all the information gathered by the other modules, establishing the most suitable amount of redundancy to each FEC block;
	\item \textsc{FEC Blocks} - The FEC blocks are built and a specific amount of redundancy designed by the ACO is assigned to each one.
\end{itemize}

The ACO is a probabilistic algorithm, based on the behaviour of ants, used to dynamically solve computational problems by finding the best path in a graph. In this solution, ants span through the paths between the nodes to find a solution. In every path followed, a pheromone marker is deposited. At the end, the paths with a greater amount of pheromone represent the best-fitted solutions~\cite{Dorigo1996}.

\subsection{Towards the design of PredictiveAnts}

As previously mentioned, the PredictiveAnts mechanism comprises several processes and modules that are going to be detailed in this section. First of all, in the same way as in the previous mechanism, it is necessary to train and validate the RNN for the motion intensity categorization. 
The RNN is composed of four input nodes, seven hidden nodes and one output node. The input nodes are the frame size, the frame type, the number of motion vectors and the Euclidean distance described by the vectors. The hidden nodes are generated through a stochastic process whereas the output node gives the motion intensity value. The RNN was trained using a set that comprises distinct motion scenarios and validated with a different set. The selection of the sets was performed through an exploratory hierarchical cluster analysis to group the video sequences according to the motion intensity. Several video characteristics are used in this analysis, such as frame type, frame size and motion vectors. The results are well-defined clusters that can be used in the RNN. As aforementioned, the offline process needs to be executed only once and after that, the RNN can be used in real-time.

After the offline process, all further computations are done in real-time. 
One of the main improvements of this proposed solution is the design and use of a simple error prediction scheme. This is performed, instead of just using the instantaneous network loss rate, to attribute a customised amount of redundancy to the video sequence being transmitted. The use of a loss prediction scheme enables a further reduction of added overhead, by balancing the allocation of redundancy data between the network's good and bad states.

\begin{figure}[!htb]
	\begin{center}
		\ifBW \includegraphics[width=4in]{./figure2_pAnts_gray-eps-converted-to.pdf}
		\else \includegraphics[width=4in]{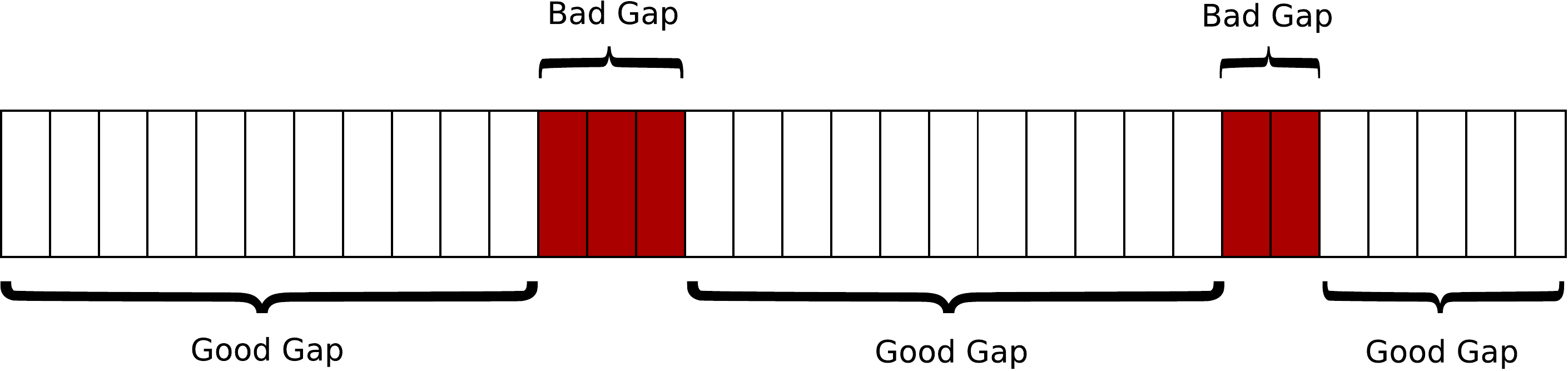}
		\fi
	\end{center}
	\caption{Packet gaps during transmission}
	\label{fig:pAnts:gbgaps}
\end{figure}

Forecasting future events is a very important task, as the predicted data can be used as input for the decision-making process. There are several proposals to forecast the PLR using, for example, time-series, sparse basis models, and hidden Markov models. 
However, as it is needed a fast mechanism, which should be able to run in real-time, a simpler model was designed. Thereby, the error prediction scheme was developed based on the concept of good and bad gaps~\cite{Karner2007}. In order to do that, a feedback mechanism was implemented to enable the retrieval of loss statistics. 
This feedback information comprises the distribution of good and bad gaps during transmission. As shown in Figure~\ref{fig:pAnts:gbgaps}, a good gap is defined as the interval of packets that were successfully received between two bad gaps~(white squares). A bad gap is the interval of packets during which a burst of errors is occurring~(red squares). The feedback information is collected by the receiver and sent to the transmitter in the form of a vector containing the size of every gap of each type. From the measured statistics, the characteristics of the error probability can be computed on the server side. This information is used as a predicting value of a higher probability of the occurrence of an error in the next block of packets to be transmitted. Therefore, the error prediction scheme has an influence on ACO, leading to an adjustment in redundancy based on the prediction of the occurrence/non-occurrence of an error.

Another enhancement in PredictiveAnts is the use of FEC blocks. This means that, instead of performing redundancy allocation on a frame-by-frame basis, the mechanism was improved to support the transmission of each frame on blocks of several packets. The size of the blocks can be adjusted according to video and/or network characteristics. Particularly in this case, the size of the blocks was set to 10 network packets. This value was selected through extensive experimentation to provide more flexibility of the mechanism while dealing with the error correcting code, and better control over the data. Furthermore, this provides enough granularity to comply with the sizes of the gaps. Therefore, the selected packet block size was that which allowed for the block to be isolated (i.e., whole block inside a good gap) from an error gap through error prediction in all packet loss rate conditions.

Once the information about the motion intensity, video characteristics and the packet loss prediction is gathered, the ACO mechanism can be defined. It also has an offline process and after that, through the aforementioned information, the exact amount of redundancy needed for each FEC block is computed in real-time. The ACO metaheuristic that needs to be defined are the construction graph, the creation of the candidate list, the delimitation of the heuristic information, and the pheromone trails definition.

\subsubsection{The ``Construction Graph'' details}

The construction graph is one of the principal elements of the ACO metaheuristic. It is used to map the problem under consideration onto a graph~\cite{Dorigo2006}, so the feasible solutions are encoded as walks on the graph. 
This means that, as the ants traverse the construction graph, they construct a solution to the problem. 
In other words, the result value of the objective function in each walk corresponds to a viable solution to the original problem~\cite{Gutjahr2000}.

A hierarchical graph was used to meet the needs of the mechanism.
This means that, once the ants start to walk they can only go to the next layer and always forward.
Since it is only possible to move forward and from the previous to the next layer, the construction graph is not fully connected and the number of vertices is equal to the number of layers. 
Additionally, the distance between the nodes is directly proportional to the amount of redundancy required to improve, or at least maintain, a superior QoE. By taking this into consideration, the construction graph is built to better reflect this condition, enabling PredictiveAnts to find the best possible solution for each scenario.

The resulting construction graph for this problem is described as a connection graph $G_c = (C, L)$, where nodes $C$ are the components and $L$ represents the set of partially connected components $C$, also called connections.
The problem constraints are given by the function $\Omega$ and follow these conditions:

\begin{enumerate}[{(1)}]
	\item In $G_c$, there is only one start node and it is located at the first layer;
	\item Let $Q$ be the set of tours~(complete walks) $q$ in $G_c$ which satisfy the conditions below:
	\begin{enumerate}[{(i)}]
		\item $q$ always starts at the start node of $G_c$ in the first layer;
		\item $q$ contains exactly one node of each layer of $G_c$;
		\item The last node on $q$ belongs to the last layer of $G_c$;
	\end{enumerate}
\end{enumerate}

Then $\Omega$ maps the set $Q$ onto the collection of attainable solutions for this specific problem instance. Following this definition, the construction graph $(G_c, \Omega)$ gives the set of all feasible solutions. The construction graph refers to the association of a set of QoE- and network-related parameters~(e.g. motion intensity nodes, frame type and size nodes, as well as the packet loss rate nodes) with a set of vertices in the graph, meaning the connections between the nodes.
Using the results of the exploratory data analysis it is possible to create an efficient construction graph.

\subsubsection{The ``Heuristic Information'' details}

The heuristic information, also called heuristic value, provides the ability to exploit problem-dependent knowledge obtained prior to the execution or at run-time if retrieved from a different source other than the ants. 
This information will guide the ants' probabilistic solution, meaning that the ants have to take into consideration fewer options to decide how to move on the graph. 
In doing that, it will strongly reduce the local search spectrum and consequently improve the solutions.
Owing to this fact, the ACO algorithm is able to provide good performance in real-time.

Using the results of the exploratory data analysis, together with the knowledge database and human expertise the heuristic information $I_h$ is defined. 
This information is composed of the length $d_{ij}$ of the arc connecting the nodes $i$ and $j$. Therefore, it is possible to define the heuristic information as $I_h = 1 / d_{ij}$. 
As mentioned before, the length of the $arc$ is directly proportional to the amount of redundancy required to improve or at least maintain a superior QoE. 
For this reason, the longer the tour the ants are walking the higher the redundancy amount needed.
In the PredictiveAnts mechanism, all the $I_h$ is pre-computed once at the bootstrap time, a table with all the possible values is generated, and it remains unchanged during the whole mechanism's run.

\subsubsection{The ``Candidate List'' details}

The candidate list is used to reduce the number of possible choices that have to be considered at every construction step.
In order to accomplish such task, this list holds a small number of promising choices of next stop.
The static lists are built utilising prior knowledge of the problem, however, they can also be generated dynamically with information gather on-the-fly. 
Since the proposed mechanism uses a hierarchical graph and it does not change over time, it can use a static candidate list composed of all the nodes of the next layer.

Let $L_c$ be the candidate list of any specific node, an arc $(i,j)$ is included in this list if the following conditions are met:

\begin{enumerate}[{(1)}]
	\item The arc $(i,j)$ it is not already included in $L_c$;
	\item The arc $(i,j)$ establishes a connection from the origin layer to a higher one, which means that it does not create cycles or backwards links;
	\item The arc $(i,j)$ holds that $I_{h(ij)} > 0$, which implies that this connection needs to add some useful heuristic information;
\end{enumerate}

The adoption of candidate lists is twofold. Fist of all, it restricts the walking path of the ants to certain conditions. 
This is of primordial importance to PredictiveAnts due to its hierarchical graph design, and thus not allowing the ants to walk horizontally inside the same layer, but just between the layers. Secondly, it strongly reduces the dimension of the search space of each ant, improving the real-time performance and therefore speeding up the solution process.

\subsubsection{The ``Pheromone Trails'' details}

The pheromone trails in the ACO metaheuristic help to guide the ants to make probabilistic decisions, and thus, construct possible solutions for the problem that is being solved. 
These trails are composed of numerical information distributed in the paths along the graph. 
During the algorithm's execution, the ants adapt the pheromone value to express their search knowledge.

\subsubsection{PredictiveAnts ACO structure}

Figure~\ref{fig:pAnts:acograph} shows PredictiveAnts ACO graph. It has fourteen nodes characterising video and network details. These nodes were chosen because they represent a combination of factors that directly affect the video quality, as follows:

\begin{figure}[!htb]
	\vspace{0cm}
	\begin{center}
		\includegraphics[width=4in]{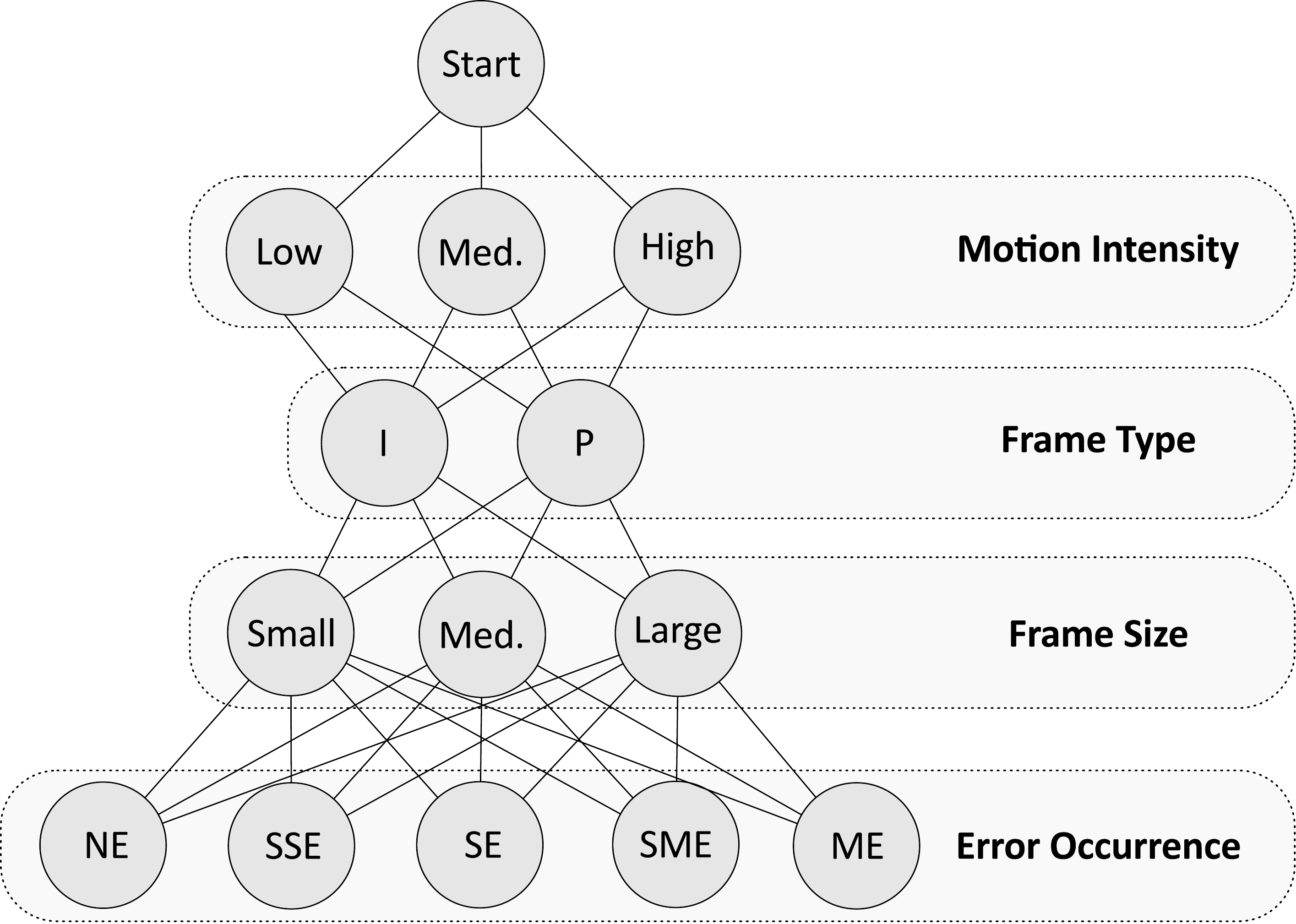}
	\end{center}
	\vspace{0cm}
	\caption{Graph used in the ACO mechanism}
	\label{fig:pAnts:acograph}
	\vspace*{0cm}
\end{figure}
\begin{itemize}
	\renewcommand{\labelitemi}{$\bullet$}
	\renewcommand{\labelitemii}{$\circ$}
	
	\item \textsc{Start} - The first node is just the starting point. As the total amount of redundancy is given by the travelled path all ants must start at the same point;
	
	\item \textsc{Motion Intensity} - These three nodes feature the RNN classification in terms of motion intensity, which can be low, medium and high motion;
	
	\item \textsc{Frame type} - The frame type, I- or P-frame, is represented by these two nodes. These are the most important frames in the MPEG standard. The loss of one I- or P-Frame will be more noticeable by the end-user because the error will only be corrected when another I-Frame arrives, in other words, in the beginning of the next GoP. Thus, those frames need to be protected with redundant information;
	
	\item \textsc{Frame size} - These three nodes characterise the frame size, which can be small, medium, and large;
	
	\item \textsc{Error occurrence} - The last layer of nodes represents the possibility of occurrence of an error instead of the instantaneous packet loss rate. The five nodes represent five different scenarios which can occur: 
	
	\begin{itemize}
		\item \textsc{NE (No Error)} - It is a scenario where no error is accounted for the current FEC block; 
		
		\item \textsc{SSE (Shared Single Error)} - In this scenario, a single error is predicted, which will be shared by this FEC block and the next; 
		
		\item \textsc{SE (Single Error)} - A single error is predicted only for the current FEC block; 
		
		\item \textsc{SME (Shared Multiple Errors)} - It is a scenario where the occurrence of two or more error gaps is predicted in the current FEC block continuing to the next one; 
		
		\item \textsc{ME (Multiple Errors)} - In this scenario, two or more blocks of errors occur only on the current FEC block.
		
	\end{itemize}
	
\end{itemize}

A simple ACO model was used with both the number of iterations and ants set to 10. The values were reached through extensive experimentation to obtain a solution which did not worsen the delay of the PredictiveAnts mechanism. In run-time, the ants search the graph while leaving pheromone in the travelled path, this reinforces the best solutions for the problem which can be re-used for similar conditions. The value computed by the ACO mechanism will be used to configure the amount of redundancy in the RS algorithm~\cite{Reed1960}. This algorithm is of low complexity being suitable for real-time use. By adding a tailored amount of redundancy to each FEC block, it is possible to better protect the most QoE-sensitive data, maximising the video quality and, at the same time, minimising the network overhead.

Algorithm~\ref{algo:pAnts:pseudo} shows the PredictiveAnts mechanism pseudo-code. All operations are repeated for each frame~(01). The first step is to get the frame type~(02). Since only I- and P-Frames are protected it is necessary to check this condition~(03). If false, the frame is sent immediately~(13). If true, other information about the frame is required, such as frame size~(04), the number of motion vectors~(05) and the Euclidean distance of the motion vectors~(06). All these are input information to the RNN for motion intensity categorization~(07). Once this value is found, it can be fed to the ACO mechanism~(08), together with the frame type, frame size and the loss rate prediction, to compute the amount of redundancy~(09). Afterwards, the FEC blocks are built using the original frame and the redundancy~(10) and then the blocks are sent~(11). The feedback information about the packet loss is received~(15) and the loss rate prediction is calculated~(16).

\begin{algorithm}[!htb]
	\For{each Frame}{
		FT $\leftarrow$ \textsc{getFrameType}($Frame$)\;
		\eIf{(FT equal (I- or P-Frame))}{
			FS $\leftarrow$ \textsc{getFrameSize}($Frame$)\;
			MV $\leftarrow$ \textsc{getMotionVectors}($Frame$)\;
			MVDist $\leftarrow$ \textsc{computeDistance}($MV$)\;
			MotionIntensity $\leftarrow$ \textsc{RNN}($FT, FS, MV, MVDist$)\;
			RSpar $\leftarrow$ \textsc{ACO}($MotionIntensity, FT, FS, LP$)\;
			Redundancy $\leftarrow$ \textsc{RS}($RSpar$)\;
			\textsc{buildFECBlocks}($Frame+Redundancy$)\;
			\textsc{sendFECBlocks}($FB1, FB2, ..., FBn$)\;
		}{
		\textsc{sendFrame}($Frame$)\;
	}
	LOSS $\leftarrow$ \textsc{receiverFeedback}()\;
	LP $\leftarrow$ \textsc{calculateLossRatePrediction}($LOSS$)\;
}

\caption{PredictiveAnts pseudo-code}
\label{algo:pAnts:pseudo}
\end{algorithm}
\subsubsection{Computational complexity of PredictiveAnts}

The computational complexity of the main PredictiveAnts components is as follows. Equation~\ref{eq:ACOcomplexity} represents the ACO complexity of finding a solution with an expected number of iterations in a graph with $n$ nodes, $m$ edges and where $\rho$ is the evaporation rate of the pheromone used by the ants~\cite{Attiratanasunthron2008}.

\begin{equation}
O(~\frac{1}{\rho}\,n^2\,m\,log\,n)
\label{eq:ACOcomplexity}
\end{equation}

Concerning Reed-Solomon, the encoding computational complexity is comprised of two steps, namely the pre-computing of the Generator Matrix~(GM) of the code, followed by the multiplication of the source vector by the GM. Equation~\ref{eq:RSEcomplexity} represents the total computational complexity of the encoding per element~\cite{Lacan2009}, where $k$ represents the rows and $n$ represents the columns of the GM matrix.

\begin{equation}
O(~\frac{k}{(n-k)}*(log\,k)^2 + log\,k)
\label{eq:RSEcomplexity}
\end{equation}

The decoding steps of the Reed-Solomon code involve the computation of the $k*k$ sub-matrix of the GM. Afterwards, this matrix is inverted and multiplied by the received vector in order to recover the original vector. The computational complexity~\cite{Lacan2009} per element of these steps is represented by Equation~\ref{eq:RSDcomplexity} where $k$ is the number of received elements.

\begin{equation}
O((log\,k)^2)
\label{eq:RSDcomplexity}
\end{equation}

Overall, the PredictiveAnts mechanism has the capability to be used in real-time. Moreover, due to the accurate categorization of the motion intensity in the video sequences and the PLR prediction, the adaptive PredictiveAnts mechanism can downsize the network overhead, reducing the video delivery footprint while improving the video quality.

\subsection{PredictiveAnts Performance Evaluation and Results}
\label{sec:pAnts:evaluation}

The PredictiveAnts mechanism aims to improve the usage of wireless network resources by reducing the overhead while assuring a good perceived video quality. The performance evaluation goal is to show that the PredictiveAnts mechanism can effectively decrease the network overhead while still providing high QoE. 

\subsubsection{Experiment settings}

The evaluation experiments were carried out by using the NS-3. The scenario is composed of a grid of 25 static nodes~(5x5), 90 meters apart from each other. The OLSR was used as the routing protocol. A data set of ten video sequences in CIF format, GoP length of 19:2, and H.264 codec was used. The selected video sequences are different from those used to train the RNN. These videos cover different distortions and subjects, which represent content usually found in on-line video services. The Frame-Copy error concealment method was used. Table~\ref{tab:pAnts:parameters} shows the simulation parameters.

\begin{table}[!ht]
	{ \small
		\caption{PredictiveAnts Simulation parameters}
		\begin{center}
			\begin{tabular}{l|l}
				\hline 
				\textsc{\textbf{Parameters}} & \textsc{\textbf{Value}} \\ 
				\hline
				\hline Display size & CIF - (352 x 288)\\
				\hline Frame rate mode & Constant\\
				\hline Frame rate & 29.970 fps\\
				\hline GoP & 19:2 \\ 
				\hline Video format & H.264\\
				\hline Codec & x264 \\ 
				\hline Container & MP4 \\
				\hline Error concealment method & Frame-copy \\
				\hline Wireless standard & IEEE 802.11g \\
				\hline Propagation model & FriisPropagationLossModel \\
				\hline Routing Protocol & OLSR \\
				\hline Number of nodes & 25 nodes (grid of 5x5) \\
				\hline Error model & Simplified Gilbert-Elliot\\
				\hline
				
			\end{tabular}
			\label{tab:pAnts:parameters}
		\end{center}
	}
\end{table}

In order to simulate the burst loss patterns found in wireless networks~\cite{Wilhelmsson1999}, a simplified two-state discrete-time Markov chain scheme following the Gilbert-Elliot~(GE) packet-loss model~\cite{Yu2005,Razavi2009} was implemented. 
By adjusting these probabilities, it is possible to generate different error patterns, which can be translated to specific PLR values. In the experiments, the values were set to 5\%, 10\%, 15\%, and 20\%, which are commonly present in wireless networks.

Five different cases were simulated as follows: (1) without any type of FEC. This case will serve as baseline to compare with the others; (2) Video-aware Equal Error Protection~(VaEEP)~(where both I- and P-Frames are equally protected) with a pre-defined amount of redundancy set to 38\%; (3) Video-aware UEP~(VaUEP), here again both I- and P-Frames are protected, this time, however, with a different amount of redundancy depending on the type. An average of 30\% redundancy amount is added. It is important to notice that the protection of only I- and P- frames is a common practice in the video transmission industry. The redundancy amounts used by both VaEEP and VaUEP mechanisms were attained after a thorough set of simulation studies. They showed, in average, a good tradeoff between video quality and network overhead under the different PLR; the next case is the AntMind mechanism which uses a combination of an RNN and ACO for Unequal Error Protection~\cite{Immich2014a}; Finally, the last case adopts the mechanism proposed in this section, the PredictiveAnts mechanism.

\subsubsection{QoE assessments}

Two main QoE metrics were employed to carry out the video quality assessment, namely SSIM and VQM. 
The objective quality assessment was conducted using Evalvid and MSU VQMT.

\begin{figure}[!htb]
	\vspace{0cm}
	\begin{center}
		\includegraphics[width=116mm]{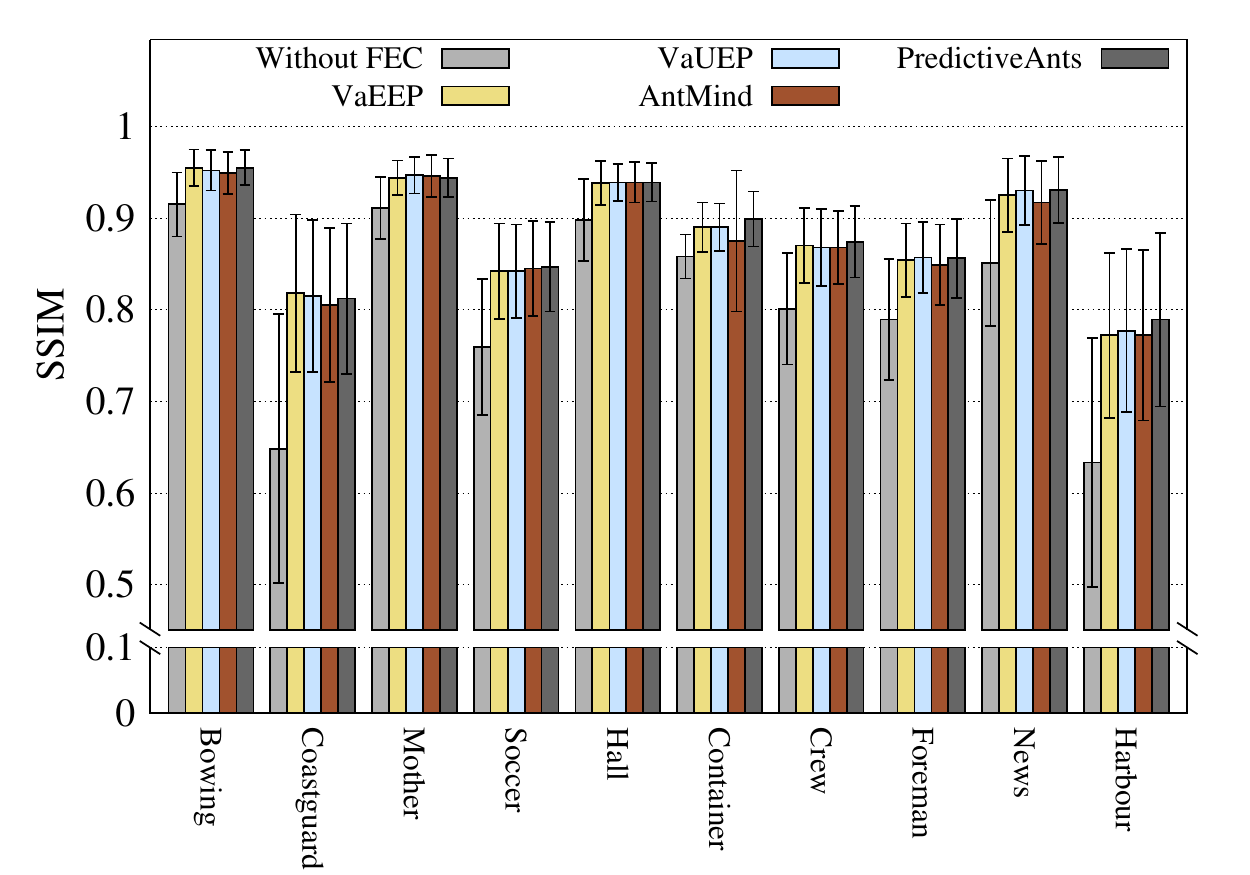}
	\end{center}
	\vspace{0cm}
	\caption{Objective QoE assessment (SSIM)}
	\label{fig:pAnts:ssim}
	\vspace*{0cm}
\end{figure}

Figure~\ref{fig:pAnts:ssim} shows the SSIM assessment for all of the video sequences in each of the five schemes. The values are an average of all PLRs for each video. The scheme without FEC averaged a value of 0,806.
The VaEEP mechanism averaged a value of 0,880 and the VaUEP obtained 0,881. The AntMind mechanism had an average of 0,876 and the PredictiveAnts score 0,884, which was the highest average value. The distinct values for the different video sequences are due to the unique characteristics of each video, this highlights the need for a motion- and video-aware mechanism. These results show therefore that the PredictiveAnts mechanism offers a better video quality than its competitors. Taking this into account and the reduction of the added overhead, it is possible to say that it provides a more precise protection scheme.

\begin{figure}[!htb]
	\vspace{0cm}
	\begin{center}
		\includegraphics[width=116mm]{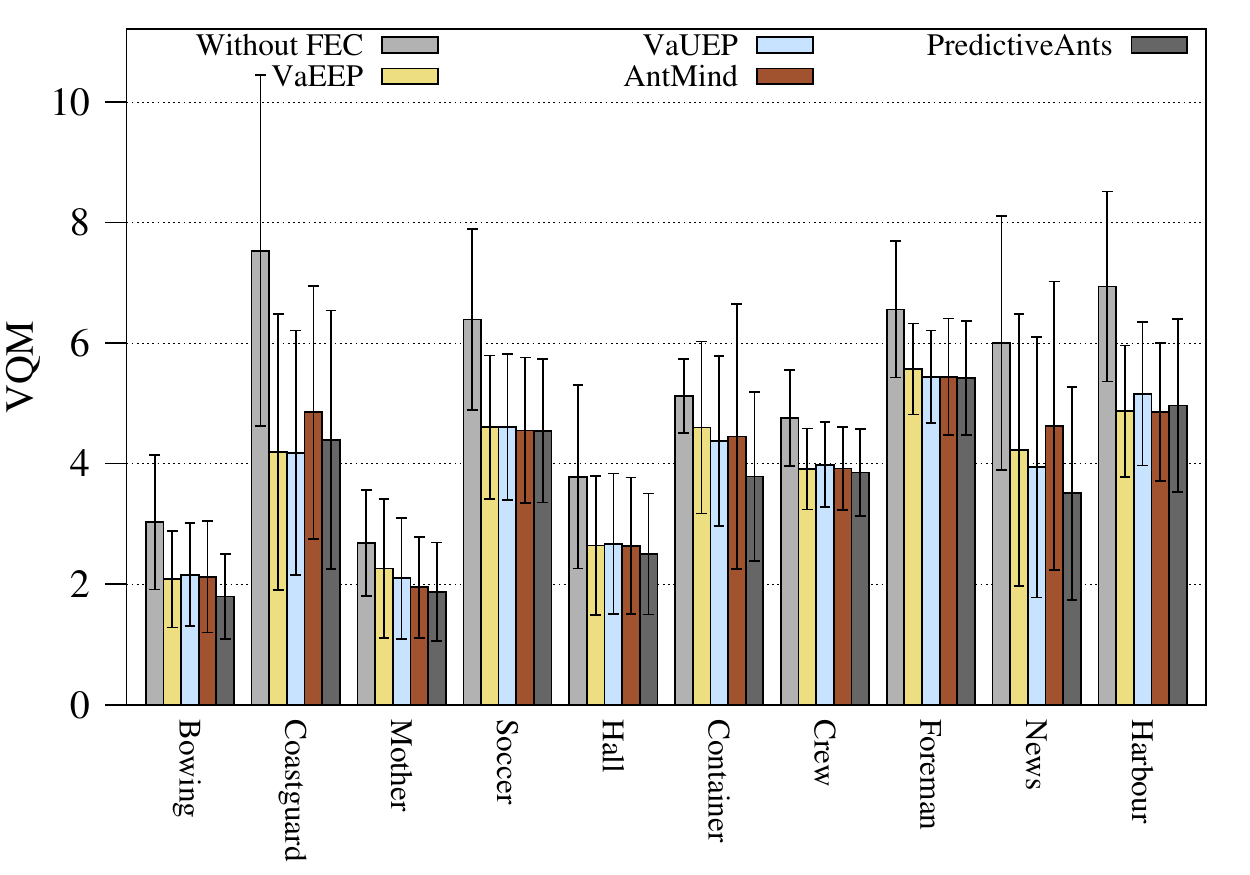}
	\end{center}
	\vspace{0cm}
	\caption{Objective QoE assessment (VQM)}
	\label{fig:pAnts:vqm}
	\vspace*{0cm}
\end{figure}

Figure~\ref{fig:pAnts:vqm} presents the VQM scores. The scheme without FEC averaged a value of 5,277. The VaEEP and VaUEP mechanisms had an average of 3,895 and 3,860, respectively. The AntMind mechanism achieved an average of 3,940 and the PredictiveAnts scores 3,664. The same way as in the SSIM assessment, the PredictiveAnts had the better video quality. This proves once again that the improvements made to the PredictiveAnts were able to reduce the network overhead, while improving the video quality.

\subsubsection{Network footprint analysis}

\begin{figure}[!htb]
	\vspace{0cm}
	\begin{center}
		\includegraphics[width=116mm]{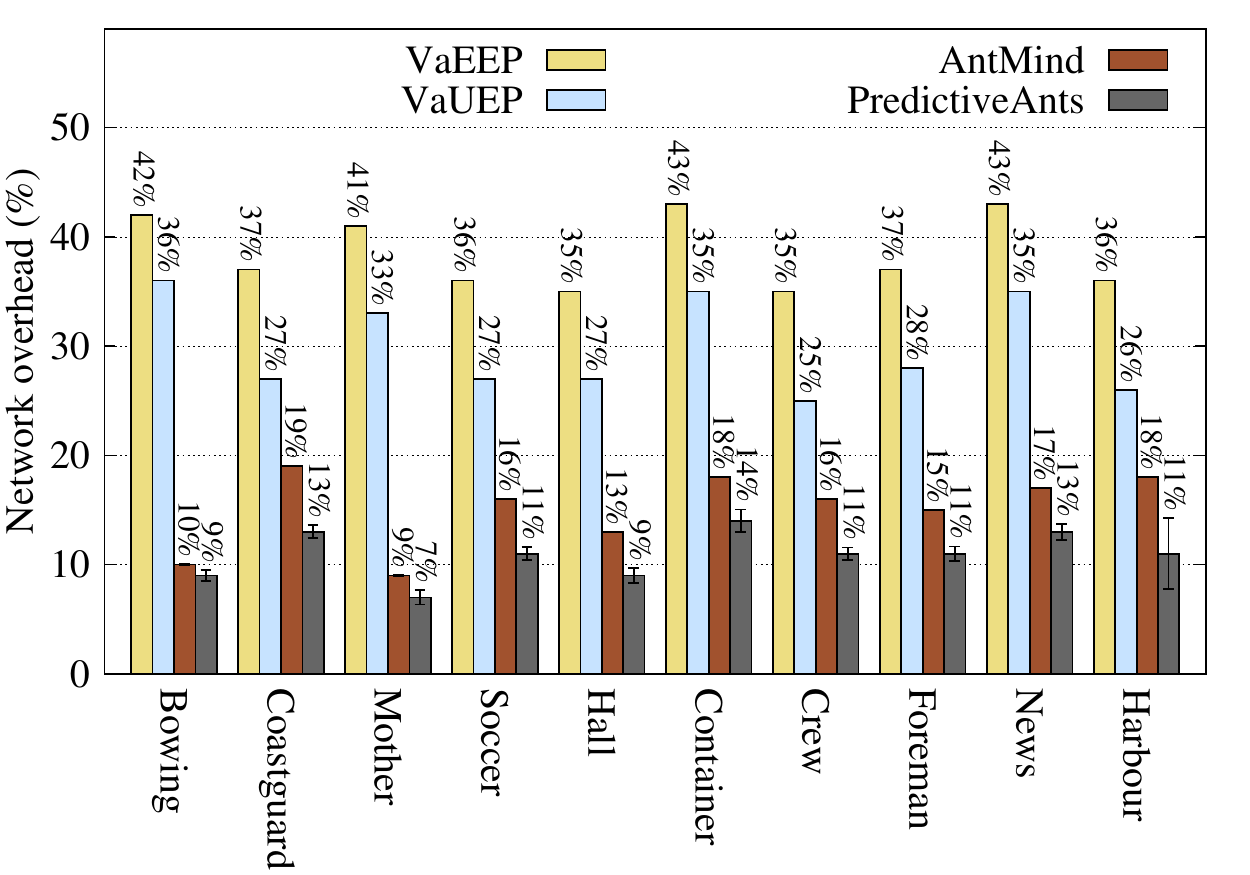}
	\end{center}
	\vspace{0cm}
	\caption{PredictiveAnts Network Overhead}
	\label{fig:pAnts:overhead}
	\vspace*{0cm}
\end{figure}

Figure~\ref{fig:pAnts:overhead} shows the network overhead results of all PLRs using the four FEC schemes. The first scheme, without FEC, is not shown because it does not produce overhead. VaEEP's average overhead was 38\% with values ranging from 35\% to 43\%, and VaUEP's average overhead was 30\% with values ranging from 25\% to 36\%. The AntMind mechanism had an average overhead of 15\%, with values between 9\% and 19\%. This is a notable result, with an overall reduction of more than 50\% in the redundancy amount~(60\% over VaEEP and 50\% over VaUEP). The proposed PredictiveAnts was able to produce even better results, providing an average overhead of 11\%, with values ranging from 7\% to 13\%. This represents a further improvement of on average over 27\% less redundancy. This means that far less redundancy data is used by the PredictiveAnts opposed to VaEEP, VaUEP, and AntMind.

Additionally, it is worth pointing out that PredictiveAnts correctly characterises the importance of the frames according to their motion intensity details. 
In all video sequences, PredictiveAnts outperforms the AntMind mechanism. However, the biggest reductions in the network overhead were achieved on the video sequences which have greater amounts of motion intensity, such as Harbour~(36\%) and Coastguard~(33\%). The lowest reductions in the network overhead were found on the videos that are opposite to these two in terms of motion intensity, specifically Bowing~(11\%) and Mother~(17\%). 
Since the PredictiveAnts mechanism is an enhancement of AntMind, the greater gain in the overhead reduction has already been achieved. Therefore, this explains why the video sequences with lower intensities of motion had a slight reduction in the network overhead.

\subsubsection{Overall results}

Table~\ref{tab:pAnts:RNNACOfec} summarises the SSIM, VQM, and network overhead results. It demonstrates that the proposed mechanism was able to considerably cut down the network overhead by not adding unnecessary redundancy. The PredictiveAnts achieved an average of 67\% of redundancy reduction over the non-adaptive schemes~(71\% less redundancy than VaEEP and 63\% less than VaUEP).  It also provides a very good result over the AntMind mechanism, saving more than 27\% in the network overhead. Additionally, the PredictiveAnts mechanism managed to achieve the best results in terms of video quality. 
These outcomes are very important in wireless environments due to the already scarce network resources.

\begin{table}[h]
	{\small
		\caption{PredictiveAnts Average SSIM, VQM and network overhead}
		\begin{center}
			\resizebox{\columnwidth}{!}{%
			\begin{tabular}{l|c|c|c|c|c}
				\hline
				& \multicolumn{1}{|l}{\textbf{PredictiveAnts}} & \multicolumn{1}{|l}{\textbf{AntMind}} & \multicolumn{1}{|l}{\textbf{VaEEP}} & \multicolumn{1}{|l}{\textbf{VaUEP}} & \multicolumn{1}{|l}{\textbf{Without FEC}} \\
				\hline
				\hline
				{SSIM}     & 0,884 & 0,876                                  & 0,881                                     & 0,880         & 0,806 \\
				\hline
				{VQM}      & 3,664 & 3,940                                  & 3,895                                     & 3,860         & 5,277                         \\   
				\hline
				{Overhead} & 10,970\% & 14,898\%                               & 38,460\%                                  & 29.827\%               & --                                  \\   
				\hline
			\end{tabular}
			}
			\label{tab:pAnts:RNNACOfec}
		\end{center}
	}
\end{table}

On the basis of the results referred to above, the PredictiveAnts mechanism showed that it is able to considerably reduce the network overhead. This is only possible due to an accurate categorization of the motion intensity details. Additionally, the packet loss prediction scheme allows anticipating the amount of redundancy that will be needed before the transmission. In doing that, it shields the video delivery against losses by ensuring an adequate protection to any kind of video sequence. This leads to an improved QoE for end-users.

\section{Summary}

This chapter described and assessed three proposed mechanisms to safeguard video transmission resiliency over WMN. 
The ViewFEC mechanism proposed in Section~\ref{sec:viewfec} adjusts the redundancy amount according to several video characteristics. 
This proactive adjustment occurs in real-time and all the information needed is gathered through cross-layer techniques. 
It also benefits from a motion activity database which is built before the execution of the mechanism.
The main goal of this mechanism was to achieve the best video quality possible without adding unnecessary network overhead, due to the redundancy packets.
This ability was confirmed through the simulation results where the ViewFEC mechanism outperformed non-adaptive FEC-based schemes in both the video quality and the network overhead. 
Generally speaking, the ViewFEC mechanism presented good results, it had, however, several issues in regards to the heuristic used in video classification procedure. 
In this case, a more precise technique could be used to better classify the motion intensity of the video sequences.

The neuralFEC mechanism~(detailed in Section~\ref{sec:neuralFEC}) improves on the above issues by using RNN in both the classification process and the decision-making steps.
In doing that, it provides the possibility to shield the video transmission in wireless networks, protecting only the most QoE-sensitive data, maximising the video quality, while saving network resources by not sending unnecessary redundancy. 
The experimental simulation results showed that neuralFEC was able to reduce the amount of network overhead by 50\% while maintaining or even improving the QoE for the end user. 
This is a considerable enhancement over non-adaptive FEC mechanisms and also reinforces the importance of using adaptive FEC-based scheme which takes into account motion intensity when protecting a video stream with varying characteristics. 
Altogether, both the ViewFEC and the neuralFEC mechanisms provided very good results, especially in reducing unnecessary network overhead. However, considering the video quality, the mechanisms still presented some deficiency that can be improved upon. Additionally, both mechanisms do not assess the network state to adjust the redundancy, which has an impact on the final result.

In the light of the aforementioned issues, the PredictiveAnts mechanism, described in Section~\ref{sec:PredictiveAnts}, uses several video characteristics and packet loss rate prediction to shield real-time video transmission over static wireless mesh networks.
This allows the improvement of both the user experience and the usage of resources. 
The proposed mechanism is based on a combination of a RNN, to categorise motion intensity of the videos, and an ACO scheme, for dynamic redundancy allocation. 
The experiment results evidenced that PredictiveAnts was able to enhance the video quality without adding an unnecessary amount of redundancy. In comparison to the non-adaptive mechanism, it has reduced the network overhead by 67\% on average. When compared to the adaptive mechanism, it provides a further 27\% savings in the network overhead. This is a great enhancement over both non-adaptive and adaptive FEC mechanisms. It only reinforces the relevance of using adaptive FEC mechanisms, which take into consideration the motion intensity and packet loss prediction to protect a video streaming with fluctuating characteristics. 

The literature review and the proposed mechanisms described in this chapter culminated in the following publications:

\bigbreak
\textbf{Journal papers:}
\begin{itemize}

	\item {Immich}, R. and Borges, P. and Cerqueira, E. and Curado, M., ``\textbf{QoE-driven video delivery improvement using packet loss prediction}'', International Journal of Parallel, Emergent and Distributed Systems, Volume 30, Issue 6, pp 478-493, Taylor \& Francis, 2015
	
	\item Ros\'{a}rio, D. and Cerqueira, E. and Neto, A. and Riker, A. and {Immich}, R. and Curado, M., ``\textbf{A QoE handover architecture for converged heterogeneous wireless networks}'', The Journal of Mobile Communication, Computation and Information, Wireless Networks, Volume 19, Issue 8, pp 2005–2020, Springer, 2013

\end{itemize}

\bigbreak

\textbf{Book chapter:}
\begin{itemize}

	\item {Immich}, R. and Cerqueira, E. and Curado, M., ``\textbf{Cross-layer FEC-based Mechanism for Packet Loss Resilient Video Transmission}'', in Data Traffic Monitoring and Analysis: From measurement, classification and anomaly detection to Quality of experience, Volume 7754, pp 320-336, Springer LNCS, 2013

\end{itemize}

\bigbreak

\textbf{Conference papers:}
\begin{itemize}

	\item {Immich}, R. and Borges, P. and Cerqueira, E. and Curado, M., ``\textbf{Adaptive Motion-aware FEC-based Mechanism to Ensure Video Transmission}'', in the 19th IEEE Symposium on Computers and Communications (ISCC), 2014
	
	\item {Immich}, R. and Borges, P. and Cerqueira, E. and Curado,M., ``\textbf{Ensuring QoE in Wireless Networks with Adaptive FEC and Fuzzy Logic-based Mechanisms}'', in the IEEE International Conference on Communications (ICC), 2014
	
	\item {Immich}, R. and Borges, P. and Cerqueira, E. and Curado, M., ``\textbf{AntMind: Enhancing Error Protection for Video Streaming in Wireless Networks}'', in the 5th IEEE International Conference on Smart Communications in Network Technologies (SaCoNET), 2014
		
	\item {Immich}, R. and Cerqueira, E. and Curado, M., ``\textbf{Adaptive Video-Aware FEC-based Mechanism with Unequal Error Protection Scheme}'', in the 28th Annual ACM Symposium on Applied Computing (SAC), 2013
	
	\item Zhao, Z. and Braun, T. and Ros\'{a}rio, D. and Cerqueira, E. and {Immich}, R. and Curado, M., ``\textbf{QoE-aware FEC Mechanism for Intrusion Detection in Multi-tier Wireless Multimedia Sensor Networks}'', in the International Workshop on Wireless Multimedia Sensor Networks (WMSN), 2012

\end{itemize}

\bigbreak

\textbf{Special session presentations:}
\begin{itemize}
	
	\item Borges, P. and {Immich}, R. and Cerqueira, E. and Curado, M., ``\textbf{Mechanisms for resilient video transmission in wireless networks: Adaptive FEC mechanism with random neural network classification and ant colony optimization}''. 18 Semin\'{a}rio Rede Tem\'{a}tica de Comunica\c{c}\~{o}es M\'{o}veis (RTCM), 2014
	
	\item  {Immich}, R., Cerqueira, E. and Curado, M., ``\textbf{Cross-layer FEC-based Mechanism to ensure Quality of Experience in Video Transmission}''. 15 Semin\'{a}rio Rede Tem\'{a}tica de Comunica\c{c}\~{o}es M\'{o}veis (RTCM), 2012
	
\end{itemize}
 \cleardoublepage

\setcounter{mtc}{13}
\chapter{Mechanisms for Resilient Video Transmission over FANETs}
\chaptermark{Resilient Video Transmission over FANETs}
\label{ch:UAV}

\dictum[George Orwell, Nineteen eighty-four]{Who controls the past controls the future; who controls the present controls the past}

\minitoc

\lettrine[lines=3]{\color{gray}\bf{U}}{} AV are rising in popularity together with video applications for both military and civilian use. 
This unveils the need for an adaptive video-aware mechanism capable of overcoming a number of challenges related to the scarce network resources, device movement, as well as high error rates, to ensure a good video quality delivery.
Adaptive FEC-based techniques are known to be suitable to enhance the QoE of video transmitted over error-prone wireless networks with high mobility, which is an intrinsic characteristic of FANETs using UAV-to-Ground connection model.
Additionally, as mentioned before, the unique characteristics of each video sequence, such as the spatial complexity and the temporal intensity, strongly affect how the QoE will be impacted by the packet loss.
This chapter describes two adaptive video-aware mechanisms to safeguard UAV real-time video transmissions against packet loss, providing a better user experience, while saving resources.

\section{Introduction}

The rapid growth of both, autonomous and nonautonomous UAV~\cite{Kumar2001}, with the objective of video surveillance, exploitation, and reconnaissance is evident in the last years. 
The deployment of these vehicles is no longer exclusive of military and special operation applications, as the civilian use of small UAVs has also increased due to ease operation, robust, and cost-effective wireless networking technologies, such as 4th Generation Networks~(4G) Long-Term Evolution~(LTE).

The adoption of UAVs can be helpful in a broad range of situations, more often than not replacing fixed video cameras due to their mobility and low-cost operation in contrast to manned systems. 
Some examples of UAV applications are in traffic surveillance, sports events, festivals, public parades, or at any place that has the potential of gathering plenty of people~\cite{Puri2005,Bekmezci2013}. 
It is also worth highlighting the use for monitoring and inspection of critical infrastructures, such as harbours, large industrial areas, railways, long pipelines, power plants, as well as to cover large areas with lack of infrastructure, such as interior border control, countryside properties or even in natural disaster sites and rescue missions~\cite{Bernard2011}.

The benefits of UAV with video capability are clear, however, even with proper equipment, robust data integration and visualisation tools, poor-quality video streaming can compromise the usability of the system. These video streams are watched by humans, and a good quality is essential to, for example, identify faces, damaged power lines or pipes, as well as track conditions. 
However, as mentioned before, real-time video transmissions with ensured QoE are resource-demanding services, especially over wireless networks. 

Incidentally, FANETs tend to have poor connectivity quality~\cite{Frew2009}. In addition, channel conditions can quickly fluctuate over time owing to the high mobility of the nodes and terrain structures, as well as other wireless communication issues like noise, multipath fading, and channel interference~\cite{Lindeberg2011}.

Another challenge is to fairly use the available bandwidth~\cite{Liu2009}. It is critical to make an efficient use of resources preventing the induction of network congestion and a high packet loss rate. This is especially important in resource-consuming services like video transmission. 
Therefore, an optimised distribution of live video streams with QoE support is one of the main challenges in highly dynamic wireless environments, such as FANETs with UAV-to-Ground model. Choosing the proper adaptive redundancy control mechanism with QoE and network-awareness is decisive for an efficient use of resources while increasing the video quality as perceived by end-users.

This chapter describes two video-aware mechanisms that use motion vectors details, FEC, and Fuzzy logic to improve the resilience of UAV video transmission with both UEP and QoE-awareness. The first one is the cross-layer adaptive video-aware FEC mechanism~(uavFEC), in Section~\ref{sec:uavFEC}, and the second is the \textsc{M}otion \textsc{INT}ensity and video-aware mechanism~(MINT-FEC), in Section~\ref{sec:MINT-FEC}.

\section{Adaptive Video-aware Fuzzy Logic Mechanism~(uavFEC)}
\label{sec:uavFEC}

Considering the open issues aforementioned this section describes and evaluates the proposed cross-layer adaptive video-aware FEC mechanism~(uavFEC).
The proposed mechanism aims to supply an alternative considering the lack of QoE- and video-aware proposals which include clear indicators of motion intensity, as well as, the network conditions. 
The main goal of the proposed mechanism is to enhance the video transmission of small UAV.

\subsection{uavFEC Overview}

The uavFEC mechanism uses motion vectors details, FEC, and Fuzzy logic to improve the resilience of UAV video transmission with both UEP and QoE-awareness. 
The uavFEC mechanism dynamically configures itself, using fuzzy logic, to send redundant information of only the most important data. 
This improves the human experience when watching live video flows while providing users and authorities~(e.g., firefighters and paramedics) with a high perception of videos. 
This is important because it allows reducing the human reaction times in case of an emergency.

\begin{figure*}[!htb]
	\begin{center}
		\ifBW \includegraphics[width=4in]{./uavFEC_mech_v1_gray2-eps-converted-to.pdf}
		\else \includegraphics[width=4in]{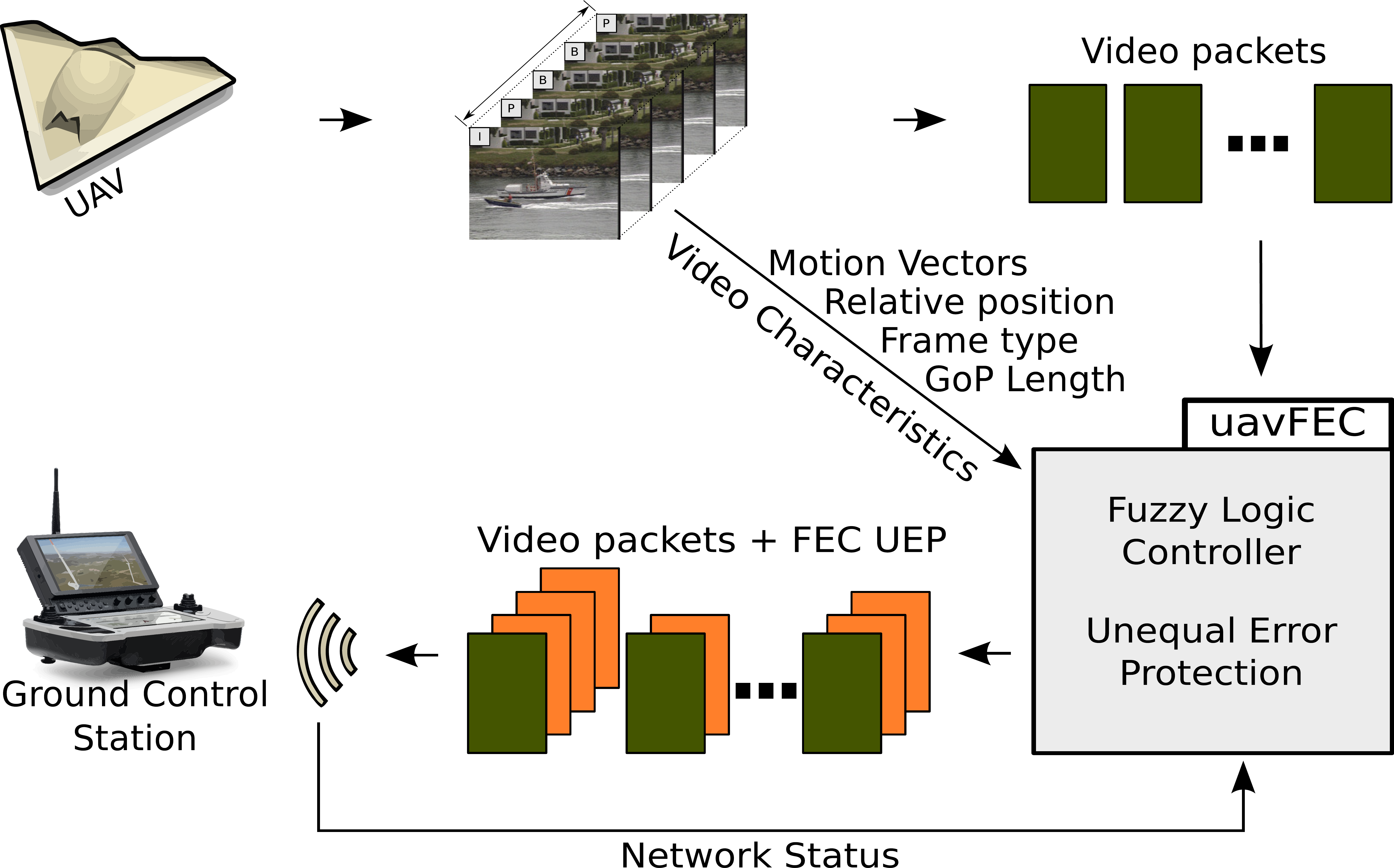}
		\fi
	\end{center}
	\caption{General view of the uavFEC mechanism}
	\label{fig:uavFEC}
\end{figure*}

Figure~\ref{fig:uavFEC} depicts the overall operation of the proposed mechanism. First of all, the video is captured, packetized, and delivered to uavFEC. After that, the mechanism will gather information about the video characteristics, such as the distance pointed by the motion vectors, frame type, GoP length, and relative position of P-Frames. This information is obtained through cross-layer techniques and loaded on the fuzzy interface engine to compute a suitable redundancy amount. Another important feature of uavFEC is the use of the network status to improve even further the amount of redundancy, which allows enhancing the video quality without adding unnecessary network overhead.

\subsection{Towards the design of uavFEC}
\label{sec:uavFEC:design}

In order to conceive uavFEC, a knowledge database needs to be created. In order to do that, an exploratory analysis using hierarchical clustering was adopted. This database stores information about the relation between several video characteristics and their impact on the quality of the videos. %
The combined use of this knowledge database and human expertise allows the definition of several fuzzy rules and sets. The offline process needs to be executed only once. 
Following this analysis, the information is loaded into the fuzzy interface engine and can be used in the real-time decision-making process. This is an important step since the real-time mechanism can be faster and more accurate, as fewer variables need to be handled.

uavFEC also uses the network state as one of the inputs to the adaptive mechanism. This information is jointly employed with the GoP length, motion vectors distance, frame type, and the relative position, to determine a suitable amount of redundancy. After that, using an improved UEP technique, a proper amount of redundancy will be added, sparing resources while increasing the video quality. A detailed description of the adaptive mechanism is presented below.

The use of fuzzy logic in the proposed mechanism allows it to be more comprehensive and dynamic because it can take into consideration a larger number of video and network details and still be fast enough to operate in real-time schemes as expected in a highly dynamic UAV network. Additionally, fuzzy logic can be considered a problem-solving methodology that aims to define what the system should do rather than attempting to fully understand its operation. It adopts a simple approach to provide definitive conclusions relying on imprecise, ambiguous, or vague information.

In order to use fuzzy logic, it is necessary to define several components, such as rules, sets, and membership functions. The rules define how the system behaves. The fuzzy sets, in contrast to classical sets that an element either belongs or does not belong to, are capable of having a degree of membership. 
Finally, the membership functions are designed to represent the significance of each element in the fuzzy set. 

The process of designing the fuzzy logic components that will be used in the uavFEC mechanism enfolds a series of exploratory analysis to define the behaviour and value of each one of them. The first step is to quantify the motion intensity. In order to do that, an exploratory analysis using hierarchical clustering with Euclidean distance was conducted. This is a statistical method of partitioning data into groups that are as homogeneous as possible~\cite{Revelle1979}. Motion vectors data is used to create these clusters. As mentioned before, the idea of motion vectors was obtained from classical mechanics and their vector-oriented model of motion. This model describes the movement of objects as simple as the sequence of small translations on a plane~\cite{LeGall1991}. 

To produce a comprehensive database, the motion vectors of several UAVs video sequences were extracted. Then, through Euclidean distance, it was computed how far each vector is pointing and summed together with all others in the same frame. 
This was used instead of just counting them because one frame can have several vectors pointing to a close distance where another frame can have fewer vectors, but pointing much farther away, thus presenting higher motion intensity. 

Figures~\ref{fig:uavFEC:mv21} and~\ref{fig:uavFEC:mv34} depict an example of the aforementioned situation. 
At frame~\#21~(Figure~\ref{fig:uavFEC:mv21}) the UAV is turning right, thus, it is possible to see that the motion vectors are longer than in frame~\#34~(Figure~\ref{fig:uavFEC:mv34}). 
In the frame~\#34 the UAV finished the turn and starts hovering, therefore, the motion vectors are smaller.
Incidentally, several of them are so small that they are represented by dots instead of arrows.
The Euclidean distance sums of all motion vectors in these frames are 109300 and 14117, for frame~\#21 and~\#34 respectively. 
However, the total number of vectors in each frame is 4959~(frame~\#21) and 4963~(frame~\#34). 
This means that even tough frame~\#34 has more vectors, they are describing less motion than those stored at frame~\#21, more precisely, they are 7.74 times smaller.

{%
\begin{figure}[!htb]
	\begin{center}
		\includegraphics[width=103mm]{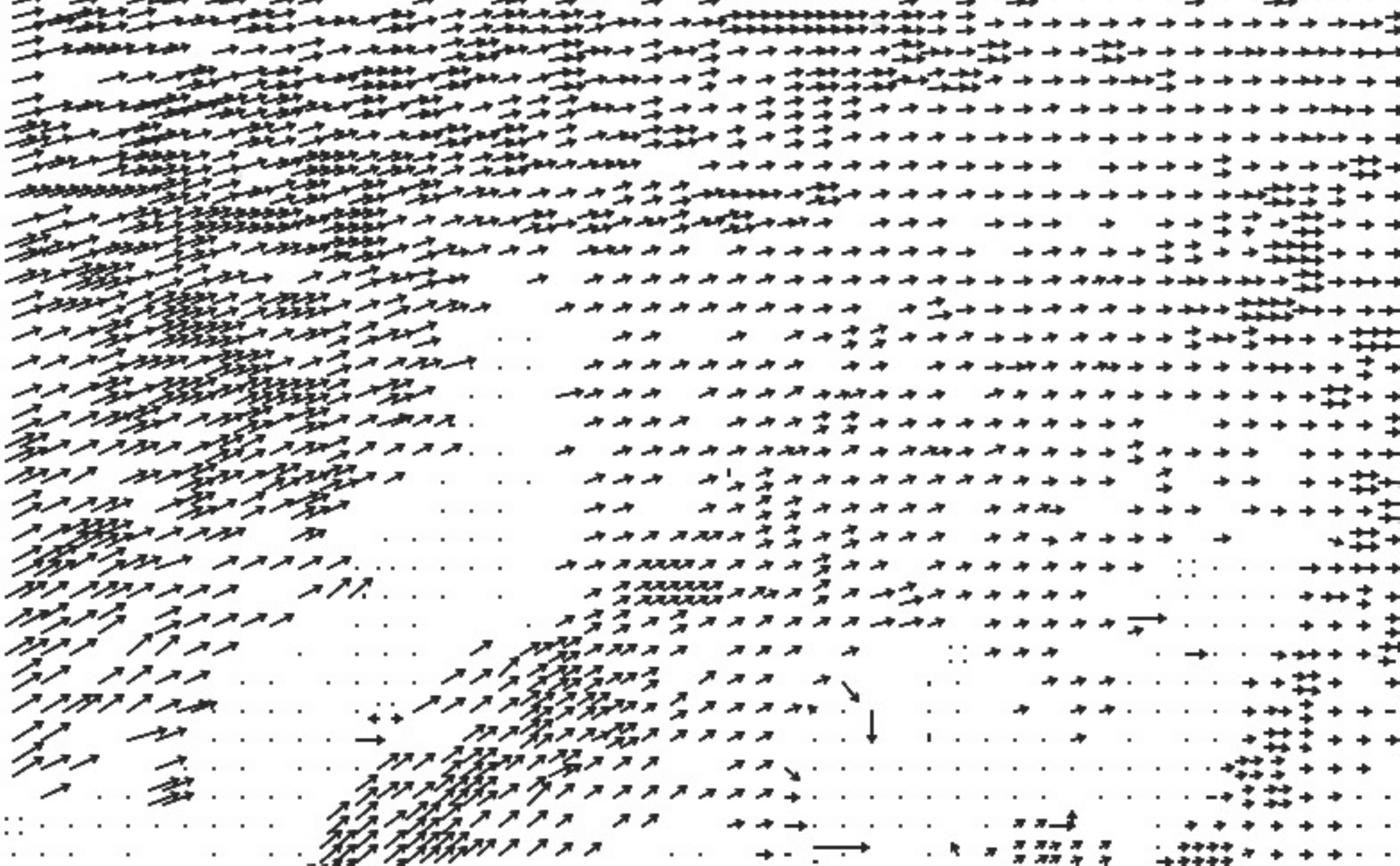}
	\end{center}
	\caption{Motion vectors of frame \#21}
	\label{fig:uavFEC:mv21}
	\begin{center}
		\includegraphics[width=100mm]{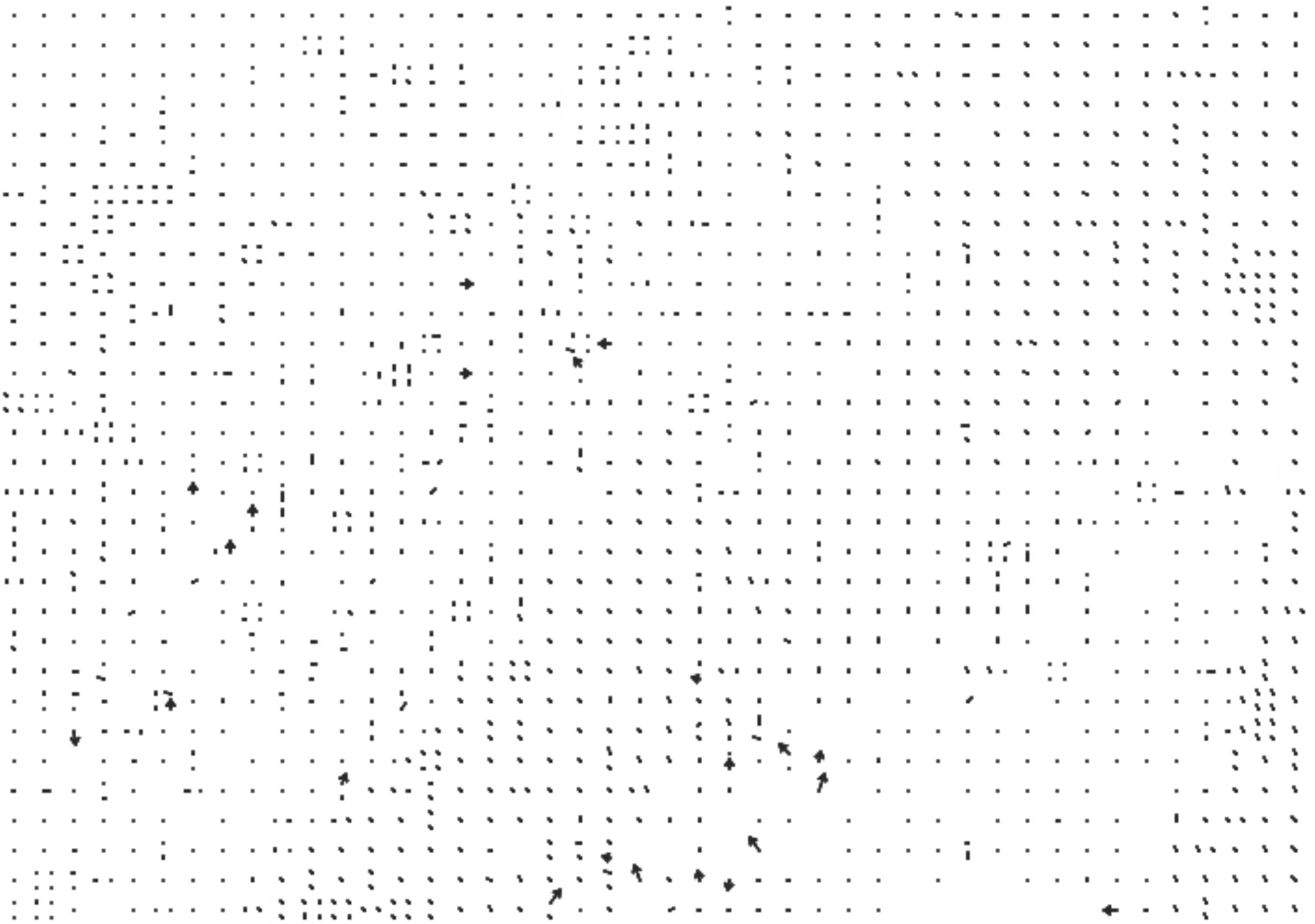}
	\end{center}
	\caption{Motion vectors of frame \#34}
	\label{fig:uavFEC:mv34}
\end{figure}
}

Initially, the distance described by all the motion vectors in all frames is computed. After that, the frames are clustered together according to the motion intensity. Based on the linkage distance between the clusters, the motion intensity was divided into three groups, namely ``small'', ``medium'', and ``high''. Using this information, the motion intensity set can be defined as presented in Algorithm~\ref{algo:uavFEC:motionIntensity}.

\iflatextortf
\else
{\LinesNumberedHidden \SetAlgoVlined
\SetKwProg{Fn}{}{}{}
\begin{algorithm}[!htb]
	{\small
		\Fn{InputLVar* \textbf{Motion} = new InputLVar(``\textbf{MotionIntensity}'');}{
			Motion $\rightarrow$ addTerm( ShoulderTerm(``\textit{LOW}'', 10000, 30000, true))\;
			Motion $\rightarrow$ addTerm( TriangularTerm(``\textit{MEDIUM}'', 21000, 80000))\;
			Motion $\rightarrow$ addTerm( ShoulderTerm(``\textit{HIGH}'', 60000, 130000, true))\;
		}
		engine.addInputLVar(\textbf{Motion})\;
	}
	\caption{Motion Intensity input set}\label{algo:uavFEC:motionIntensity}
\end{algorithm}
}
\fi

\begin{figure}[!htb]
	\vspace*{-0.0cm}
	\begin{center}
		\includegraphics[width=8.5cm]{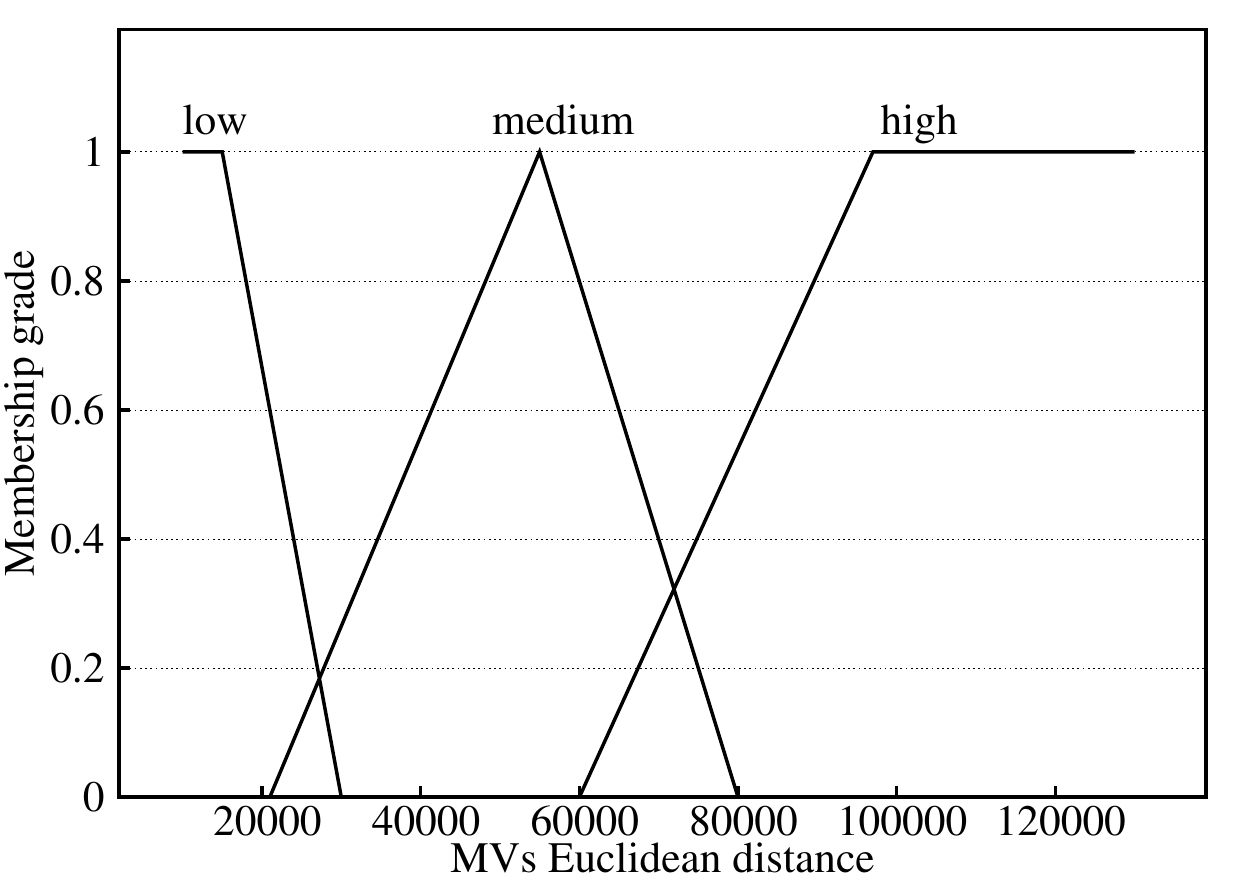}
	\end{center}
	\vspace{-0.0cm}
	\caption{Motion intensity membership function}
	\label{fig:uavFEC:MVmembership}
	\vspace*{-0.0cm}
\end{figure}

After defining the sets, it is necessary to set up the membership functions. This definition is a complex and problem-dependent task. Taking this into account, it is preferable to use piecewise linear functions~(formed by straight-line sections), because they are simple and more efficient with respect to computability and resource requirements. Figure~\ref{fig:uavFEC:MVmembership} shows the graphical representation of the membership functions.

After delineating the motion intensity, the packet loss rate set must be defined. The aim of this activity is to quantify the packet loss rate against the video quality in terms of QoE. In other words, a loss rate of 10\% can be considered low in the proposed approach however, it might be unacceptable in other applications, such as a voice over IP call. To define this set, a number of network simulations with several packet loss rates as well as a broad collection of UAV video sequences were carried out. On average, the video quality was considered good when the network losses were between 0\% and 10\%. Between 5\% and 20\%, a tolerable video quality was perceived, but over 15\% the quality quickly decreased, soon becoming unacceptable. In view of this, three categories were defined, namely ``low'', ``medium'', and ``high'', as showed in Algorithm~\ref{algo:uavFEC:networkErrorInput}.

\iflatextortf
\else
{\LinesNumberedHidden \SetAlgoVlined
	\SetKwProg{Fn}{}{}{}
	\begin{algorithm}[!htb]
		{\small
			\Fn{InputLVar* \textbf{PLR} = new InputLVar(``\textbf{PacketLossRate}'');}{
				PLR $\rightarrow$ addTerm( TriangularTerm(``\textit{LOW}'', 0, 15))\;
				PLR $\rightarrow$ addTerm( TriangularTerm(``\textit{MEDIUM}'', 5, 30))\;
				PLR $\rightarrow$ addTerm( TriangularTerm(``\textit{HIGH}'', 20, 100))\;
			}
			engine.addInputLVar(\textbf{PacketLossRate})\;
		}
		\caption{Packet loss rate input set for strict video resolutions}\label{algo:uavFEC:networkErrorInput}
	\end{algorithm}
}
\fi

\begin{figure}[!htb]
	\vspace*{-0.0cm}
	\begin{center}
		\includegraphics[width=8.5cm]{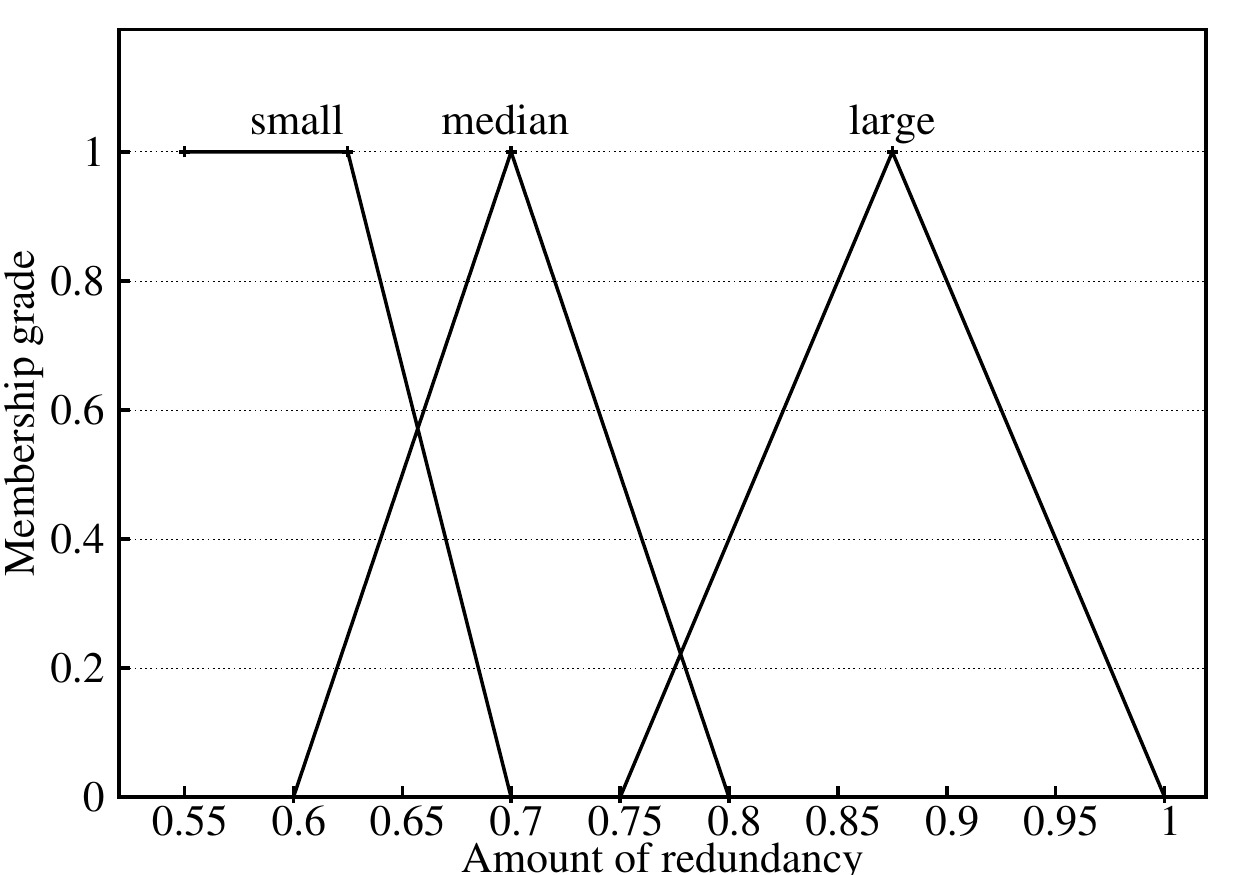}
	\end{center}
	\vspace{-0.0cm}
	\caption{Redundancy amount membership function}
	\label{fig:uavFEC:redundancyMembership}
	\vspace*{-0.0cm}
\end{figure}

Another stage is to delineate the redundancy set. The main goal of this set is to establish the output value which will be used to add the redundancy. Here again, a combination of experiments and human knowledge in the field was used to specify what could be considered a ``small'', ``medium'', and ``large'' amount of redundancy. The values obtained and the graphical representations of the membership functions are displayed in Algorithm~\ref{algo:uavFEC:redudancyOutput} and Figure~\ref{fig:uavFEC:redundancyMembership}, respectively.

\iflatextortf
\else
{\LinesNumberedHidden \SetAlgoVlined
	\SetKwProg{Fn}{}{}{}
	\begin{algorithm}[!htb]
		{\small
			\Fn{OutputLVar* \textbf{Redundancy} = new OutputLVar(``\textbf{RedundancyAmount}'');}{
				Redundancy $\rightarrow$ addTerm( ShoulderTerm(``\textit{SMALL}'', 0.55, 0.70, true))\;
				Redundancy $\rightarrow$ addTerm( TriangularTerm(``\textit{MEDIUM}'', 0.60, 0.80))\;
				Redundancy $\rightarrow$ addTerm( TriangularTerm(``\textit{LARGE}'', 0.75, 1))\;
			}
			engine.addOutputLVar(\textbf{RedundancyAmount})\;
		}
		\caption{Motion activity output set}\label{algo:uavFEC:redudancyOutput}
	\end{algorithm}
}
\fi

After defining all the fuzzy sets, the IF-THEN structure must be created. This is a straightforward procedure, because if the transmitted video has low levels of motion activity~(according to the motion vectors) and the packet loss rate is low as well, then the uavFEC will attribute also a low redundancy. The same procedure is valid for ``medium'' and ``high'' motion activities and packet loss rate as partially depicted in Algorithm~\ref{algo:uavFEC:packetLossRules}.

\iflatextortf
\else
{\LinesNumberedHidden \SetAlgoVlined
	\SetKwProg{Fn}{}{}{}
	\begin{algorithm}[!htb]
		{\small
			\textit{RuleBlock* \textbf{block} = new RuleBlock();}
				\BlankLine
				\Fn{block $\rightarrow$ addRule( \textit{new} MamdaniRule(``}{
					\Fn{\textbf{if} (Motion is LOW and PacketLossRate is LOW) \textbf{then}}{
						RedundancyAmount is SMALL\;}
					'', engine))\;}
					
				\BlankLine
				\Fn{block $\rightarrow$ addRule( \textit{new} MamdaniRule(``}{
					\Fn{\textbf{if} (Motion is MEDIUM and PacketLossRate is MEDIUM) \textbf{then}}{
						RedundancyAmount is MEDIUM\;}
					'', engine))\;}
				\BlankLine
				\Fn{block $\rightarrow$ addRule( \textit{new} MamdaniRule(``}{
					\Fn{\textbf{if} (Motion is HIGH and PacketLossRate is HIGH) \textbf{then}}{
						RedundancyAmount is LARGE\;}
					'', engine))\;}	
			}
		\caption{Packet loss x redundancy amount rules}\label{algo:uavFEC:packetLossRules}
	\end{algorithm}
}
\fi

After defining the rules and sets, they need to be loaded in the fuzzy logic controller. This activity has to be performed just once, during the system setup period~(bootstrap). After the definition, the controller will calculate the degree of membership of each input information, resulting in a precise amount of redundancy on-the-fly.

This is important because video transmission is delay-sensitive, meaning that if a frame is received after its decode deadline it cannot be displayed. Moreover, unlike neural networks or genetic algorithms, the fuzzy logic controller does not need a period of online training or convergence, making it a proper tool for real-time control. Additionally, the calculations can be very simple, especially when triangular or trapezoidal membership functions are adopted~\cite{Pedrycz1994}, and even further reduce to a simple operation through fuzzy control surface.

\subsection{uavFEC Performance Evaluation and Results}
\label{sec:uavFEC:performance}

The main goal of the uavFEC mechanism is to improve the perceived video quality without adding unnecessary network overhead, thus saving wireless network resources.

\subsubsection{Experiment settings}

The evaluation scenario is composed of up to four UAVs, equipped with a 4G LTE radio at 800MHz. These UAVs can be operated in autonomous or nonautonomous mode. In a surveillance scenario, for example, it is possible to have a human operating the UAV. This allows having an instant change of direction and speed during the pursuit of a suspect. Therefore, the mobility model was defined as random waypoint~\cite{Bouachir2013}.

All UAVs are in line-of-sight and communicate directly with an ad-hoc connection to the ground control station which was equipped with a portable base station and antenna. To simulate the video transmission, a set of twenty real UAV video sequences in high definition~(720p), GoP length of 19:2, and H.264 codec was used.
Due to the ad-hoc communication and the high definition videos, the flying range is limited to a radius of 900 meters from the base station. 

A Frame-Copy error concealment method is active, this means that lost frames are replaced by the last good one received. 
The PLR varies according to the movement of the UAVs, namely distance from the portable base station and velocity, and also due to concurrent transmissions of others UAVs. Owing to the aforementioned details, the PLR can range from 0\% to 45\%. Table~\ref{tab:uavFEC:parameters} shows the simulation parameters.

\begin{table*}[!ht]
	{ %
		\caption{uavFEC Simulation parameters}
		\begin{center}
			\begin{tabular}{l|l}
				\hline 
				\textbf{Parameters} & \textbf{Value} \\ 
				\hline
				\hline Display size & 1280 x 720\\
				\hline Display aspect ratio & 16:9\\
				\hline Frame rate mode & Constant\\
				\hline Frame rate & 29.970 fps\\
				\hline GoP & 19:2 \\ 
				\hline Video format & H.264\\
				\hline Codec & x264 \\ 
				\hline Container & MP4 \\
				\hline Propagation model & FriisPropagationLossModel \\
				\hline UAV velocity & 45-65 km/h (28-40 mph) \\
				\hline LTE Frequency band & 800MHz \\
				\hline LTE Mode & FDD \\
				\hline LTE Bandwidth & 5 MHz \\
				\hline eNodeB Operating Power & 22 dBm \\
				\hline Antenna Gain & 16 dBi \\
				\hline
			\end{tabular}
			\label{tab:uavFEC:parameters}
		\end{center}
	}
\end{table*}

In order to compare the results, five different cases were simulated. 
The first is without FEC, serving as a baseline to compare with the others. The second case is a non-adaptive video-aware FEC-based approach. In this case, only I- and P-Frames are protected with an equal amount of redundancy, which was set at 65\%. This amount was chosen because it provides a good tradeoff between video quality and network overhead under several PLRs. 
The next case is the previously proposed mechanism in Section~\ref{sec:viewfec}, which uses a simple adaptive unequal error protection~(ViewFEC).
Another case is an implementation of the Cross-Layer Mapping Unequal Error Protection~(CLM-UEP)~\cite{Lin2012}. 
The last case is the proposed uavFEC mechanism. 
The simulation setup is composed of 20 real UAVs video sequences and 5 cases, each one was simulated 30 times with each video, resulting in 3.000 simulations in total.
\subsubsection{QoE assessments}

\begin{figure}[!htb]
	\vspace*{-0.0cm}
	\begin{center}
		\includegraphics[width=4.3in]{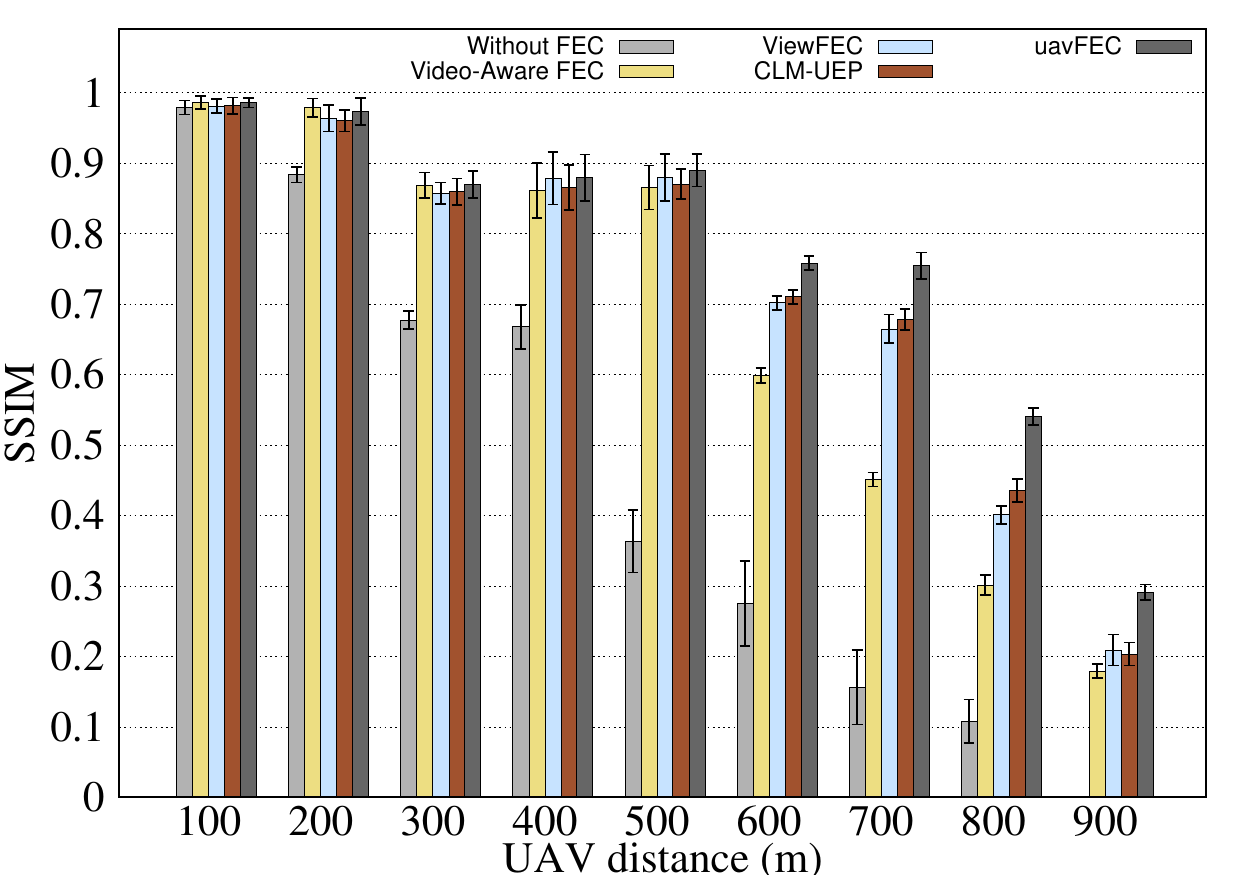}
	\end{center}
	\vspace{0.0cm}
	\caption{SSIM for all scenarios with one UAV}
	\label{fig:uavFEC:qoe_ssim_uav1}
	\vspace*{-0.0cm}
\end{figure}

Figure~\ref{fig:uavFEC:qoe_ssim_uav1} depicts the average SSIM for all video sequences when only one UAV is transmitting. 
The measurement of this metric is fairly simple however, it is consistent with the human visual system, given good results~\cite{Wang2004}. 
In SSIM, values closer to one indicate a better video quality. As expected, when the UAV is far away from the ground control station there is a decline in the video quality. In the baseline case, without FEC, a sharp decline in the video quality after 400m is perceived. 
Conversely, the UAVs using a FEC-based mechanism are able to sustain a better video quality longer, and it is only noticeable after 500m for case 2, and after 700m for cases 3 to 5. 

\begin{figure}[!htb]
	\vspace*{-0.0cm}
	\begin{center}
		\includegraphics[width=4.3in]{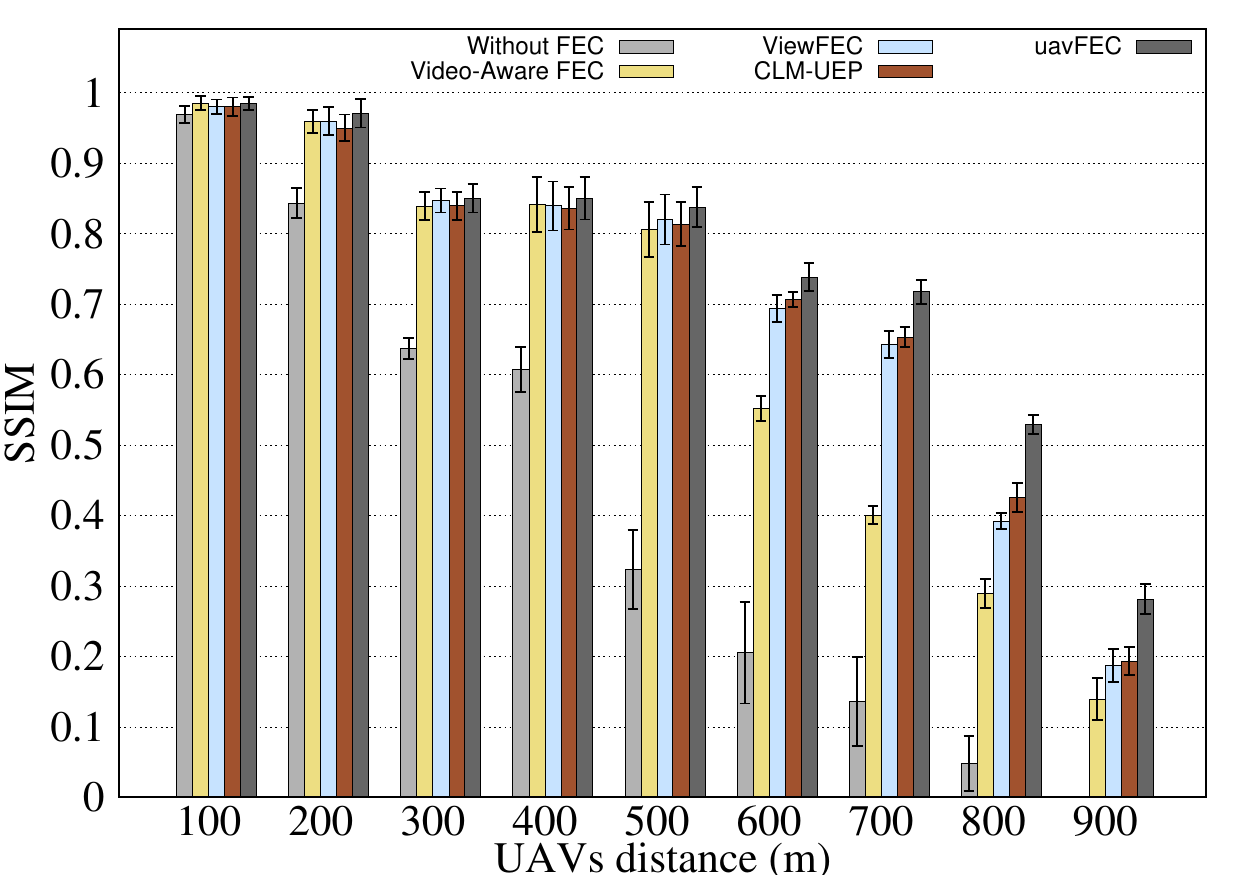}
	\end{center}
	\vspace{-0.0cm}
	\caption{SSIM for all scenarios with two UAVs}
	\label{fig:uavFEC:qoe_ssim_uav2}
	\vspace*{-0.0cm}
\end{figure}

Almost the same behaviour is shown in Figure~\ref{fig:uavFEC:qoe_ssim_uav2} which demonstrates the results for 2 UAVs transmitting simultaneously. 
One clear difference between these two scenarios is the increase in the standard deviation of the baseline case. This can be explained by the natural resiliency of some videos to packet loss due to different video characteristics. Video sequences with low motion intensity are more resilient to losses, and generally, have better results in the QoE-aware assessment. On the other hand, videos with high motion intensity tend to have poor results. 

\begin{figure}[!htb]
	\vspace*{-0.0cm}
	\begin{center}
		\includegraphics[width=4.3in]{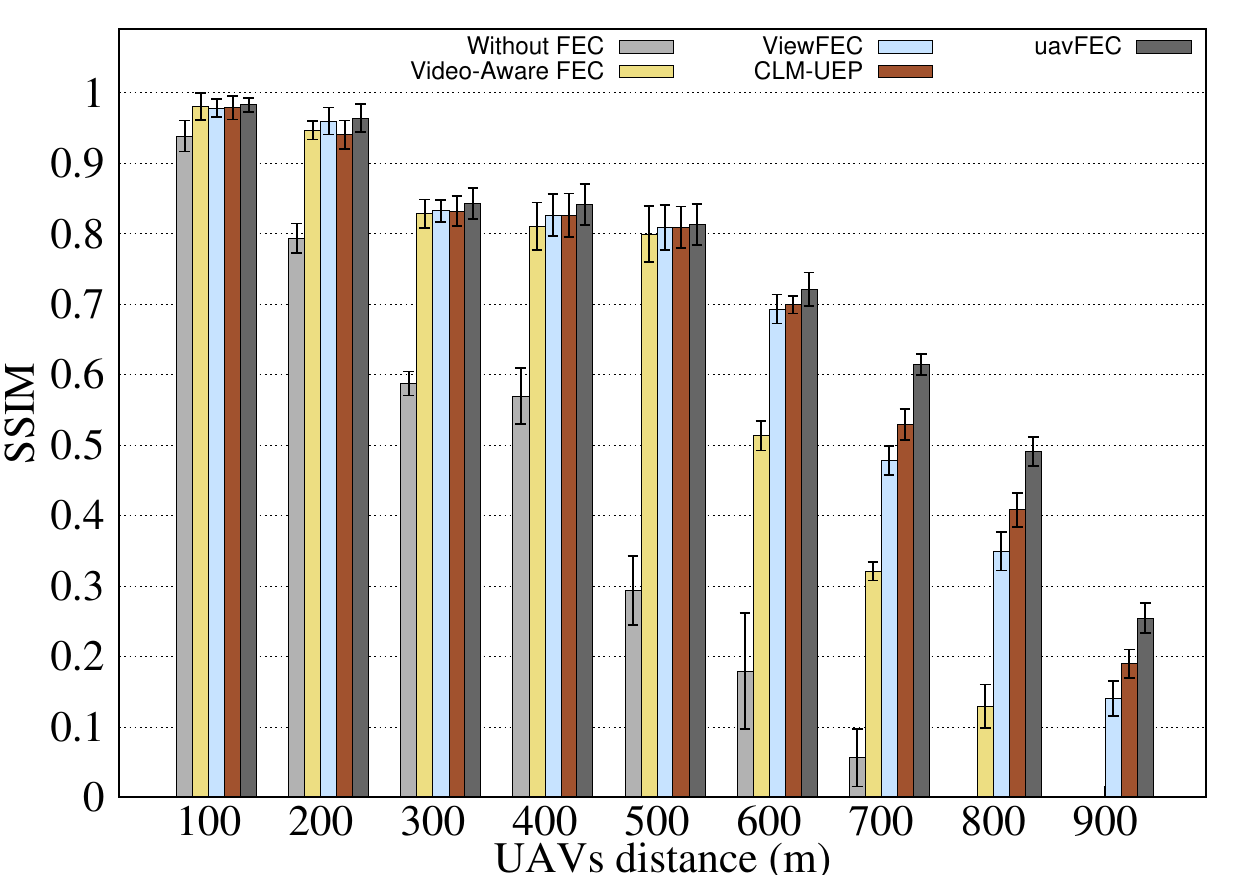}
	\end{center}
	\vspace{-0.0cm}
	\caption{SSIM for all scenarios with three UAVs}
	\label{fig:uavFEC:qoe_ssim_uav3}
	\vspace*{-0.0cm}
\end{figure}

\begin{figure}[!htb]
	\vspace*{-0.0cm}
	\begin{center}
		\includegraphics[width=4.3in]{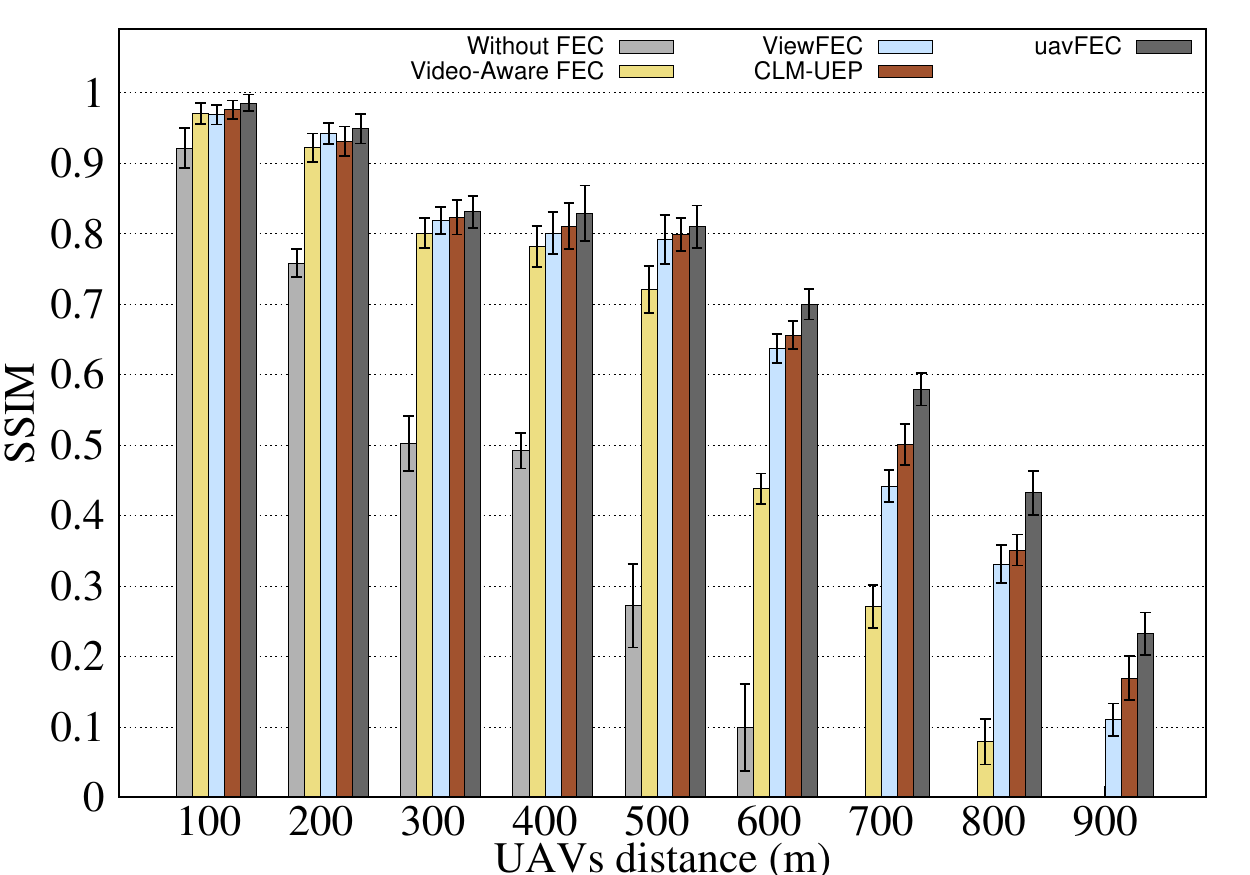}
	\end{center}
	\vspace{-0.0cm}
	\caption{SSIM for all scenarios with four UAVs}
	\label{fig:uavFEC:qoe_ssim_uav4}
	\vspace*{-0.0cm}
\end{figure}

As the number of video sequence flows begins to increase, the quality of the transmitted video starts to decrease sooner than before. 
Figures~\ref{fig:uavFEC:qoe_ssim_uav3} and~\ref{fig:uavFEC:qoe_ssim_uav4} depict this tendency. In the first two scenarios~(with one and two UAVs), the uavFEC managed to keep the SSIM above 0.7 up to 700m~(other approaches only up to 600m). 
However, with three and four UAVs, the uavFEC was able to maintain the SSIM over 0.7 only up to 600m, after that, there is a sharp decline in the video quality in all of the assessed mechanisms. This can be attributed to a more congested network due to several transmissions together with the distance from the ground control station.

\subsubsection{Network footprint analysis}

Throughout the QoE assessment was demonstrated that the uavFEC mechanism enhances the video quality over several scenarios, having particularly good results over higher distances and with increased network traffic. 
Besides the video quality, the uavFEC was also designed to add as little as possible redundancy, to maintain a low overhead and thus saving resources. 

\begin{figure}[!htb]
	\vspace*{-0.0cm}
	\begin{center}
		\includegraphics[width=4.3in]{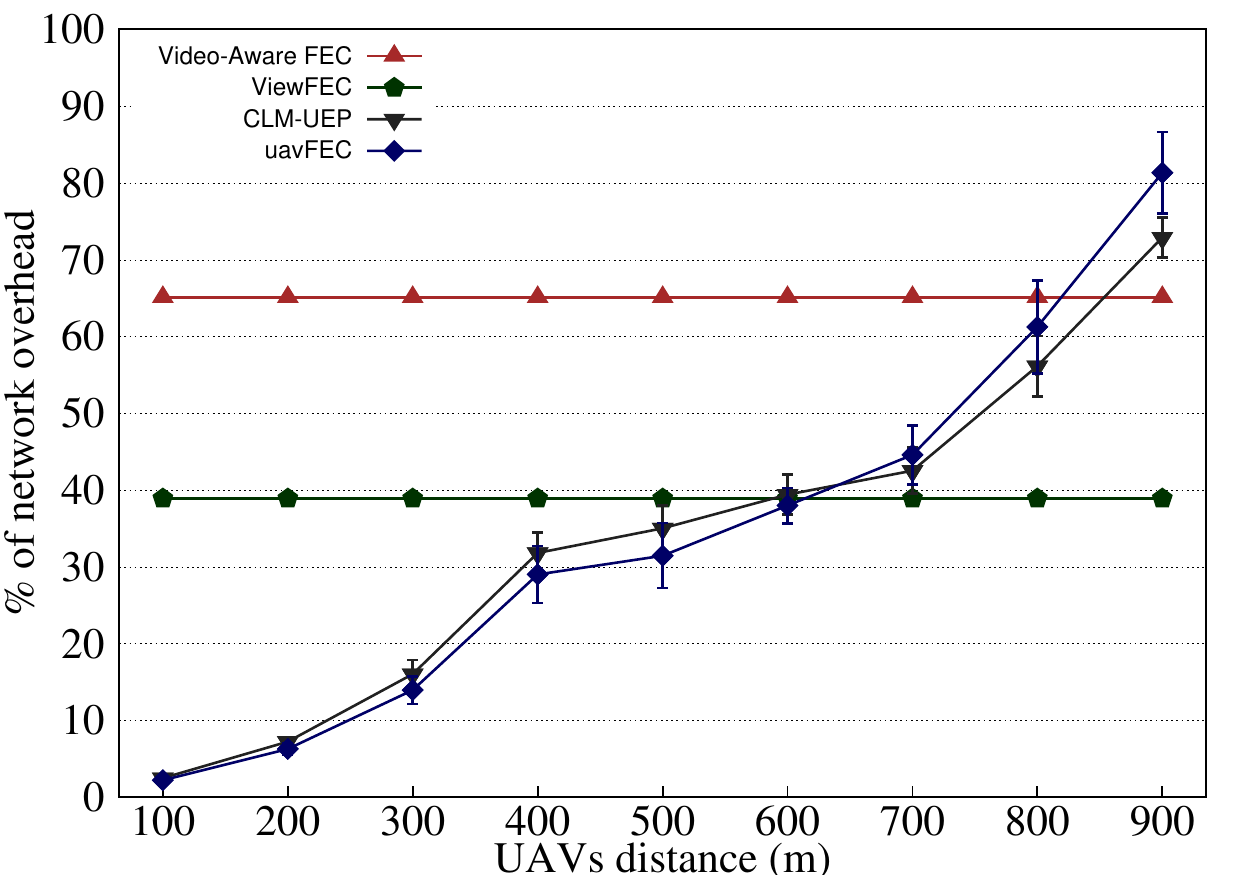}
	\end{center}
	\vspace{-0.0cm}
	\caption{\small Network overhead for all scenarios}
	\label{fig:uavFEC:net_overhead}
	\vspace*{-0.0cm}
\end{figure}

The network overhead was computed by summing the size of all video frames transmitted by each mechanism. This means that, if the original frame size is subtracted, it is possible to find the specific amount of redundancy added only by the approaches. Two mechanisms assessed are non-adaptive, video-aware FEC and ViewFEC, and because of that, they have the same network overhead in all distances, which was 65.10\% and 38.90\%, respectively, as showed in Figure~\ref{fig:uavFEC:net_overhead}. These mechanisms are not appropriate because even when the UAVs are close to the ground control station they add a considerable amount of redundancy, wasting resources. The same figure depicts the results for uavFEC and CLM-UEP. Both mechanisms perform close to each other up to 600m, but in average the uavFEC has lower network overhead. Over 600m, the uavFEC starts to add more redundancy, increasing the network overhead, however, providing better video quality.

\subsubsection{Overall results}

\begin{figure}[!htb]
	\vspace*{-0.0cm}
	\begin{center}
		\hspace*{-0.0cm}
		\includegraphics[width=4.3in]{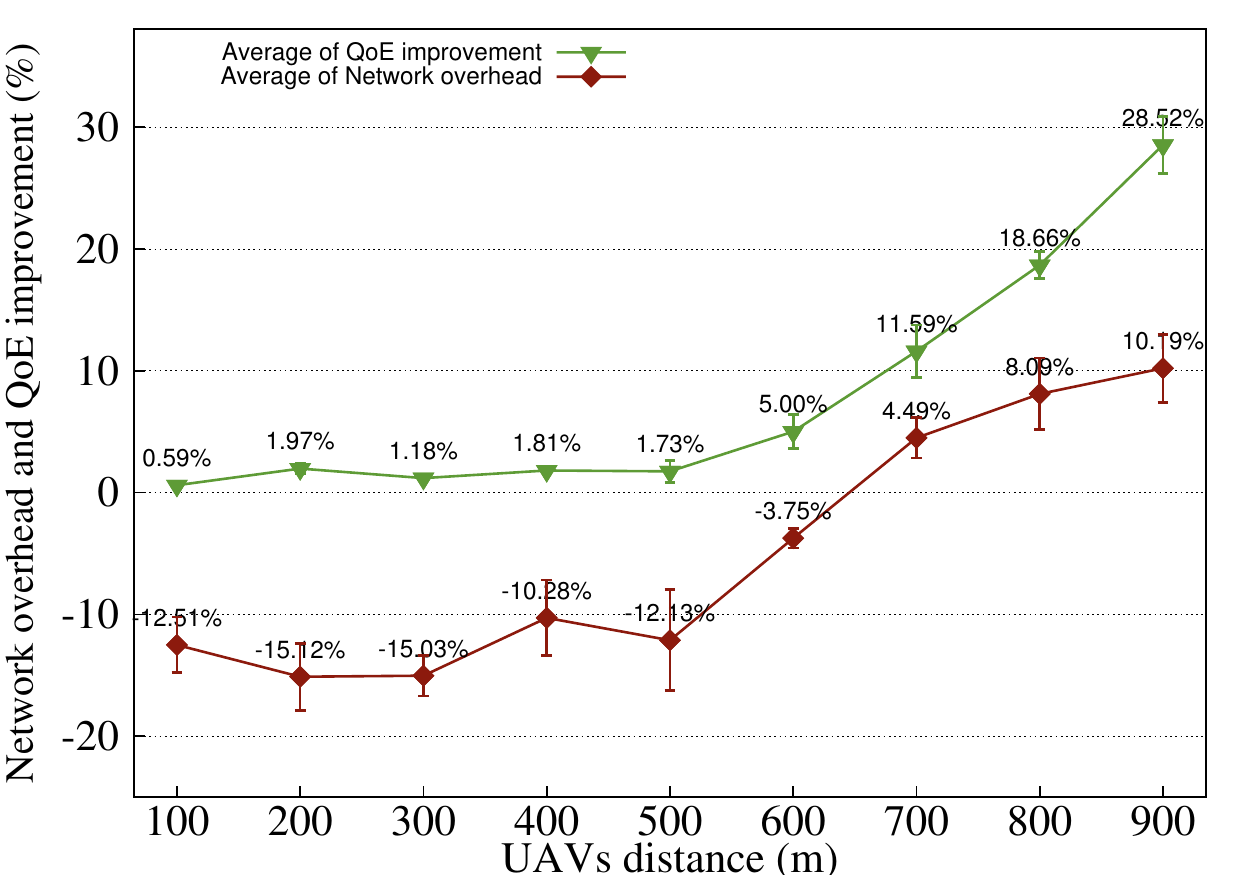}
	\end{center}
	\vspace{-0.0cm}
	\caption{\small QoE and Redundancy against UAV distance}
	\label{fig:uavFEC:qoe_redund_dist}
	\vspace*{-0.0cm}
\end{figure}

Figure~\ref{fig:uavFEC:qoe_redund_dist} depicts the comparison of uavFEC and the related work~(CLM-UEP)~\cite{Lin2012}. The graph shows the average percentage of QoE improvement against the amount of redundancy added by the mechanisms in all scenarios~(from 1 to 4 UAVs). 
A positive percentage means that uavFEC had better QoE results than CLM-UEP. In all four scenarios, uavFEC presented a slightly better video quality until 600m, on average between 0.59\% and 5.00\% better. 
The real advantage of uavFEC is noticeable after the 700m when it enhances even further the video quality. 
The uavFEC was able to achieve improvements, on average, between 11.59\% and 28.52\%, better than CLM-UEP. Taking this into consideration, it is clear that the proposed mechanism performs better in a higher distance, where the PLR is also higher. 
This gives uavFEC the capability to operate in wide coverage areas.

It is also shown in Figure~\ref{fig:uavFEC:qoe_redund_dist} the comparison of the amount of redundancy added by both, CLM-UEP and uavFEC. 
A negative percentage means that the proposed mechanism adds less redundancy than CLM-UEP. 
In all four scenarios, uavFEC added less redundancy until 600m, which was around 3.75\% and 15.12\% less on average and still managed to transmit the videos with higher QoE. This means that uavFEC was able to improve the video quality and at the same time save resources. 
After 700m, the uavFEC mechanism begins to increase the redundancy. This happens because the uavFEC was developed to enhance the video quality over higher distances, which make the networks more susceptible to errors. Considering this, the mechanism will have to increase the protection of the most important video data, adding more overhead. For example, at 700m the uavFEC mechanism added on average 4.49\% more redundancy, and at 900m added 10.19\%. 
Increasing the redundancy is an expected response of the proposed mechanism to further improve the video quality, which can be confirmed through the QoE assessment in the same figure. In summary, uavFEC provides a good tradeoff between video quality and network overhead.

A further analysis of Figure~\ref{fig:uavFEC:qoe_redund_dist} shows that up to 500m both mechanisms had similar QoE results, with uavFEC having a modest higher video quality. The major difference was the considerably smaller network overhead, this means that uavFEC, through its QoE- and Video-aware techniques, was able to add redundancy to the most important video data only. 
At 600m, the uavFEC mechanism still adds less redundancy than the related work, but it is already showing better results, with an improvement of 5.00\% on QoE. After this threshold, considering the increasing distance and in order to improve the video quality, the proposed mechanism starts to add a larger amount of redundancy. 
The result of this approach are videos transmitted on average with more than 28\% of better quality than CLM-UEP while adding less than 11\% of redundancy.

The uavFEC mechanism achieved good results making the video transmission more resilient to packet loss and thus, enabling a longer video transmission range for the UAVs. 
The results are particularly beneficial in a higher distance with several UAVs, providing a better video quality of live video flows, allowing end-users, such as civilians and/or authorities, to have a high-quality perception of videos and thus reducing reaction times.

\section{Adaptive Motion Intensity Awareness mechanism~(MINT-FEC)}
\label{sec:MINT-FEC}

The mechanism presented in Section~\ref{sec:uavFEC} obtained favourable results however, it presented some drawbacks in regard to the motion intensity classification and the strict video resolution dependence.
To improve on these issues, the adaptive FEC-based mechanism with motion intensity awareness~(MINT-FEC) was proposed.

\subsection{MINT-FEC Overview}

The MINT-FEC mechanism aims to enhance the resilience of UAV real-time video transmissions. 
One of the major weaknesses in the mechanisms found in the literature, including the uavFEC, is the use of unnecessary redundancy.
To tackle this issue, the MINT-FEC dynamically adapts itself by using fuzzy logic, to add a precise amount of redundancy to only the most QoE-sensitive data, while ensuring high-quality video and downsizing the usage of scarce wireless resources. 

In the MINT-FEC mechanism, the motion intensity is now given by combining the spatial complexity and temporal intensity. 
Spatial complexity is how distinct one frame is from another, as well as the colour and luminance saturation. The temporal intensity can be defined as how fast and how much the image is changing frame-by-frame. 
Their joint use provides a more accurate motion classification to be used for adding a adequate amount of redundancy. Another improvement from the work described in Section~\ref{sec:uavFEC} is the video resolution independence. By normalising the values of all the video characteristics, as well as using the motion vector distance and macroblock size, it is possible to add, on-the-fly, an adaptive amount of redundancy to videos with arbitrary resolution. A detailed description of all the novel components is given next.

Following the same core structure as uavFEC, the MINT-FEC mechanism also depends on Fuzzy logic and thus several fuzzy components need to be defined, such as the sets, membership functions, and rules.
The offline process needs to be executed only once. After that, all the generated information can be loaded into the fuzzy interface engine to be used in real-time.

\subsection{Towards the design of MINT-FEC}

The design process of MINT-FEC starts with the definition of the fuzzy components. 
In a similar fashion to the uavFEC mechanism, the first step is to quantify the spatial complexity. This component represents how much static information a frame is carrying compared to the previous one. The most common way to compute this difference is using the Sum of Absolute Differences~(SAD)~\cite{Vanne2006}. It is not a complex operation however, it is very time-consuming because it compares each pixel from both frames, making this impractical in real-time. 

Another way to locate this information is through the frame sizes. The problem of using the frame size is that several video characteristics can impact on it, such as different resolutions~(picture size), content, as well as temporal intensity. To be able to compare the frame sizes among different videos, it is necessary to normalise all the information. Using Equation~(\ref{eq:MINT:avgFrameI}) the average frame size is calculated, and the same operation is also executed for P- and B-frames. After that, through Equation~(\ref{eq:MINT:normFrameI}), all frame sizes are normalised, as before, this is also done for P- and B-frames. This process is performed for each video sequence separately. Table~\ref{tab:MINT:notation} shows the adopted notation.

\begin{equation}
\mu I_{s} = \frac{1}{nF}\sum_{i=0}^{nF-1} I_{s(i)} 
\label{eq:MINT:avgFrameI}
\end{equation}

\begin{equation}
\hat{\mu}I_{s} = \frac{\mu I_{s}}{\mu I_{s} + \mu P_{s} + \mu B_{s}}
\label{eq:MINT:normFrameI}
\end{equation}

\vspace*{-0.0in}
\begin{table}[!hbt]
	\caption{MINT-FEC Adopted Notation}
	\vspace{-0.0in}
	\begin{small}
		\begin{center}
			\begin{tabular}{c|l}
				\hline \textsc{\textbf{Notation}} & \textsc{\textbf{Meaning}} \\
				\hline
				\hline $\mu I_{s}, \mu P_{s}, \mu B_{s}$ & Frame size average \\
				\hline $\hat{\mu}I_{s}, \hat{\mu}P_{s}, \hat{\mu}B_{s}$ & Normalised frame size average \\
				\hline $I_{s(i)}, P_{s(i)}, B_{s(i)}$ & Frame size of the $i_{th}$ frame \\
				\hline $nF$ & Number of frames in the video sequence \\
				\hline $\left | MV \right |$ & Euclidean distance of a motion vector \\
				\hline $\left | MV_{(i)} \right | $ & Euclidean distance of the $i_{th}$ motion vector \\
				\hline $MB_{h}$ & Macroblock height \\
				\hline $MB_{w}$ & Macroblock width \\
				\hline $aMB$ & Macroblock area \\
				\hline $aMB_{(i)}$ & Area of the $i_{th}$ macroblock \\
				\hline $nMB$ & Number of macroblock in the frame \\
				\hline $TI_{\Delta t}$ & Temporal intensity \\ 
				\hline 
			\end{tabular}
			\label{tab:MINT:notation}
		\end{center}
	\end{small}
	\vspace*{-0.0in}
\end{table}

\begin{figure}[!htb]
	\begin{center}
		\includegraphics[width=3.5in]{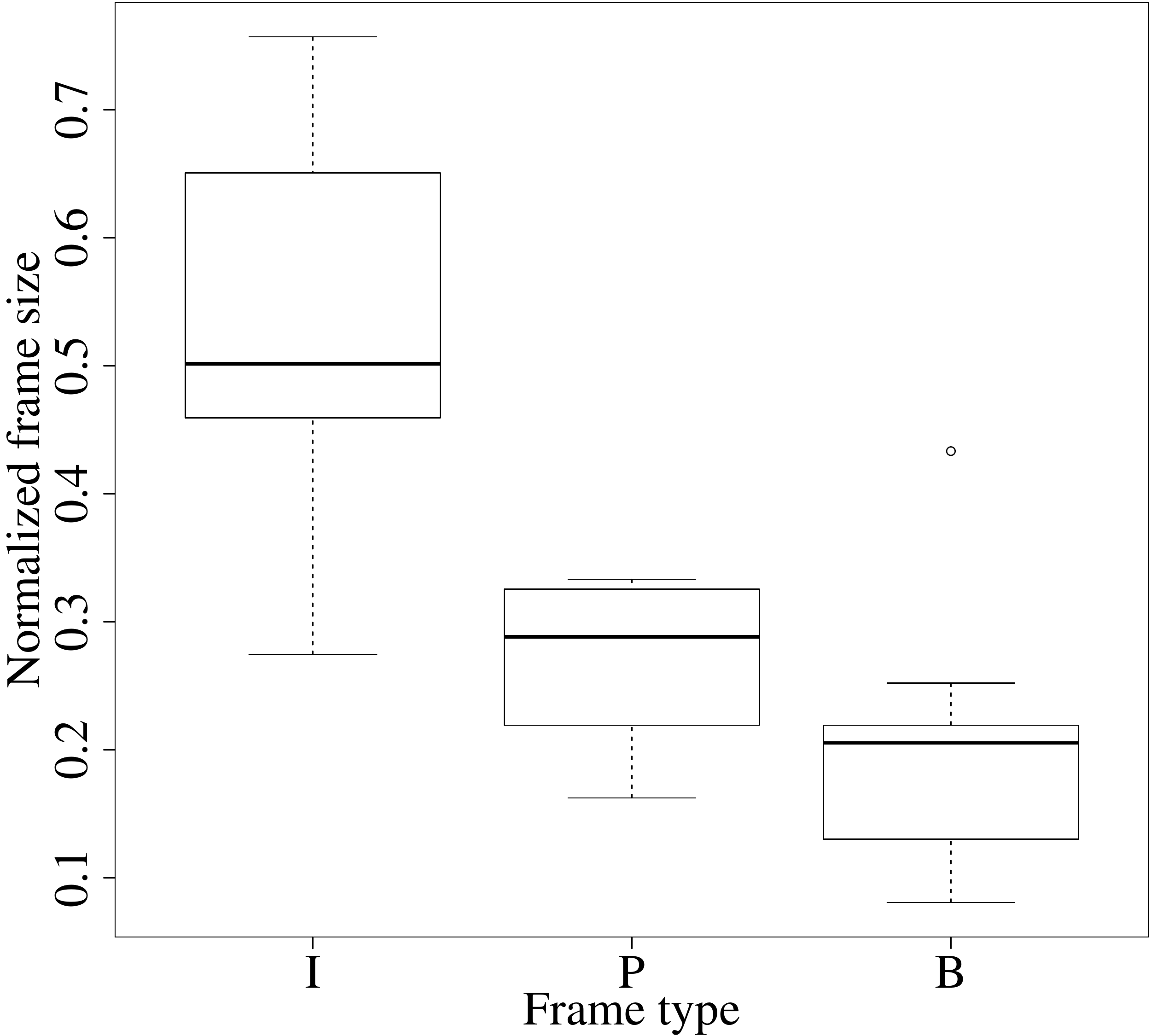}
	\end{center}
	\vspace{-0.0in}
	\caption{\small Spatial Complexity }
	\label{fig:MINT:spatialInt}
\end{figure}

Once all the frame sizes are normalised, it is possible to perform an exploratory analysis to cluster all frames of all video sequences together according to their sizes. Based on the linkage distance of the clusters it was possible to divide them into three distinct groups, namely ``small'', ``medium'', and ``large''. After defining the clusters, a boxplot was used to summarise and display the distribution of the data. This is an important tool in the exploratory analysis because it displays the shape of the distribution of each cluster along with the central value and the variability. Figure~\ref{fig:MINT:spatialInt} shows the boxplot for the spatial complexity. The fuzzy sets for spatial complexity were defined using the information displayed by the boxplot, as showed by Algorithm~\ref{algo:MINT:spatialIntSets}.

\iflatextortf
\else
{\LinesNumberedHidden \SetAlgoVlined
	\SetKwProg{Fn}{}{}{}
	\begin{algorithm}[!htb]
		{\small
			FuzzyOperator\& \textbf{op} = FuzzyOperator::DefaultFuzzyOperator()\;
			FuzzyEngine engine(``complex-mamdani'', op)\;
			\BlankLine
			
			\Fn{InputLVar* \textbf{Isz} = new InputLVar(``\textbf{I-size}'');}{
				Isz $\rightarrow$ addTerm( ShoulderTerm(``\textit{SMALL}'', 0.274, 0.459, true))\;
				Isz $\rightarrow$ addTerm( TriangularTerm(``\textit{MEDIUM}'', 0.274, 0.651))\;
				Isz $\rightarrow$ addTerm( ShoulderTerm(``\textit{LARGE}'', 0.502, 0.757, true))\;
			}
			engine.addInputLVar(\textbf{Isz})\;
			\BlankLine
			
			\Fn{InputLVar* \textbf{Psz} = new InputLVar(``\textbf{P-size}'');}{
				Psz $\rightarrow$ addTerm( ShoulderTerm(``\textit{SMALL}'', 0.162, 0.219, true))\;
				Psz $\rightarrow$ addTerm( TriangularTerm(``\textit{MEDIUM}'', 0.162, 0.325))\;
				Psz $\rightarrow$ addTerm( ShoulderTerm(``\textit{LARGE}'', 0.288, 0.333, true))\;
			}
			engine.addInputLVar(\textbf{Psz})\;
			\BlankLine

			\Fn{InputLVar* \textbf{Bsz} = new InputLVar(``\textbf{B-size}'');}{
				Bsz $\rightarrow$ addTerm( ShoulderTerm(``\textit{SMALL}'', 0.081, 0.13, true))\;
				Bsz $\rightarrow$ addTerm( TriangularTerm(``\textit{MEDIUM}'', 0.081, 0.219))\;
				Bsz $\rightarrow$ addTerm( ShoulderTerm(``\textit{LARGE}'', 0.205, 0.252, true))\;
			}
			engine.addInputLVar(\textbf{Bsz})\;
			\BlankLine		
		}
		\caption{Spatial complexity (Frame size sets)}\label{algo:MINT:spatialIntSets}
	\end{algorithm}
}
\fi

\begin{figure}[!htb]
	\begin{center}
		\includegraphics[width=3.5in]{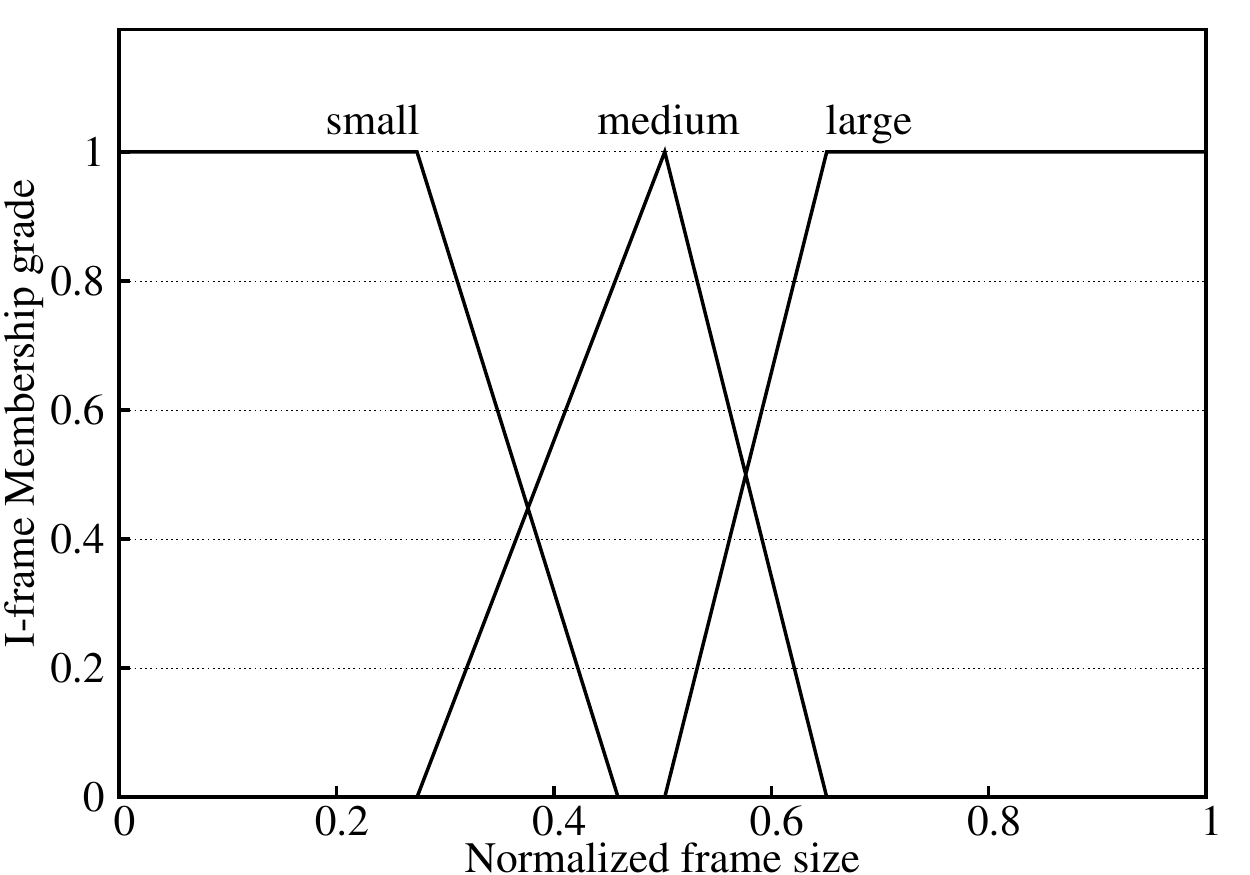}\\
		\includegraphics[width=3.5in]{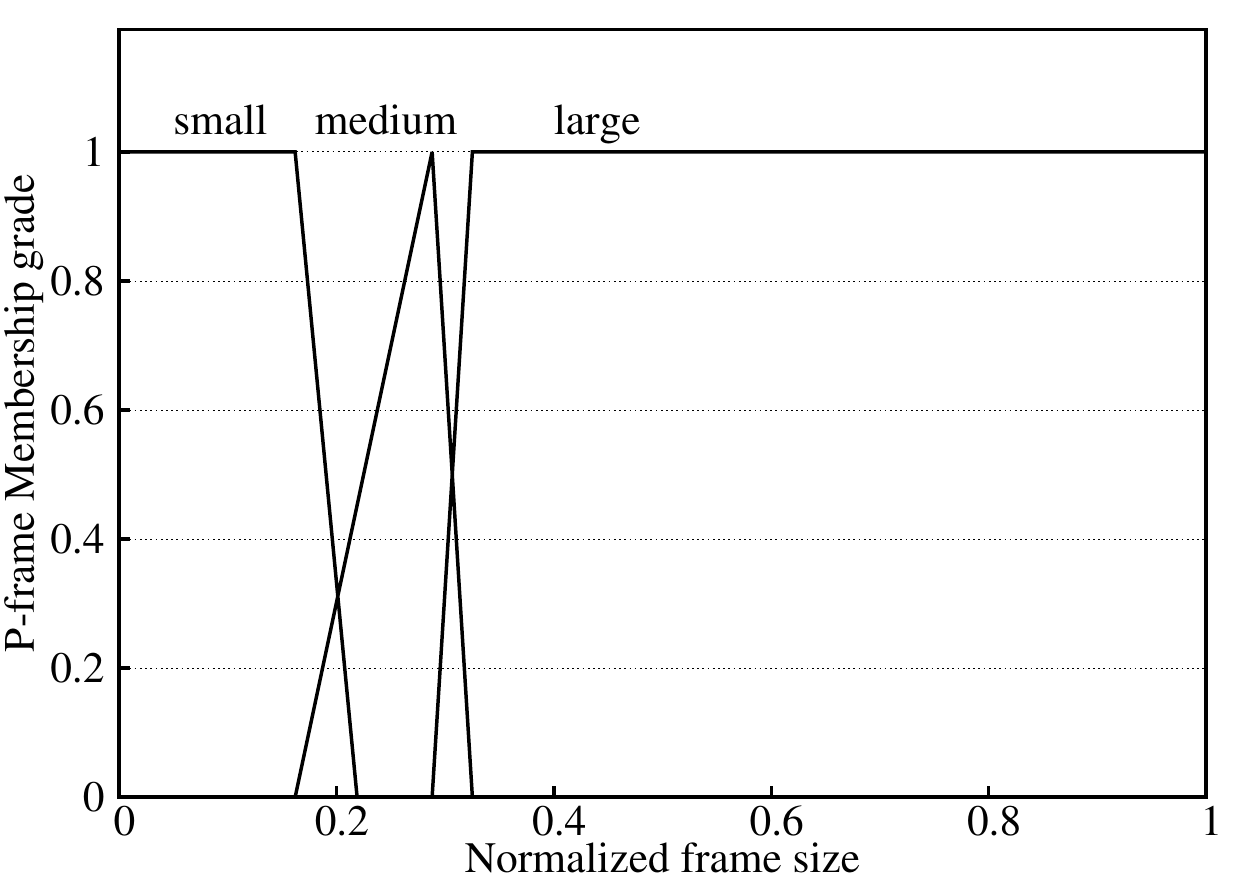}\\
		\includegraphics[width=3.5in]{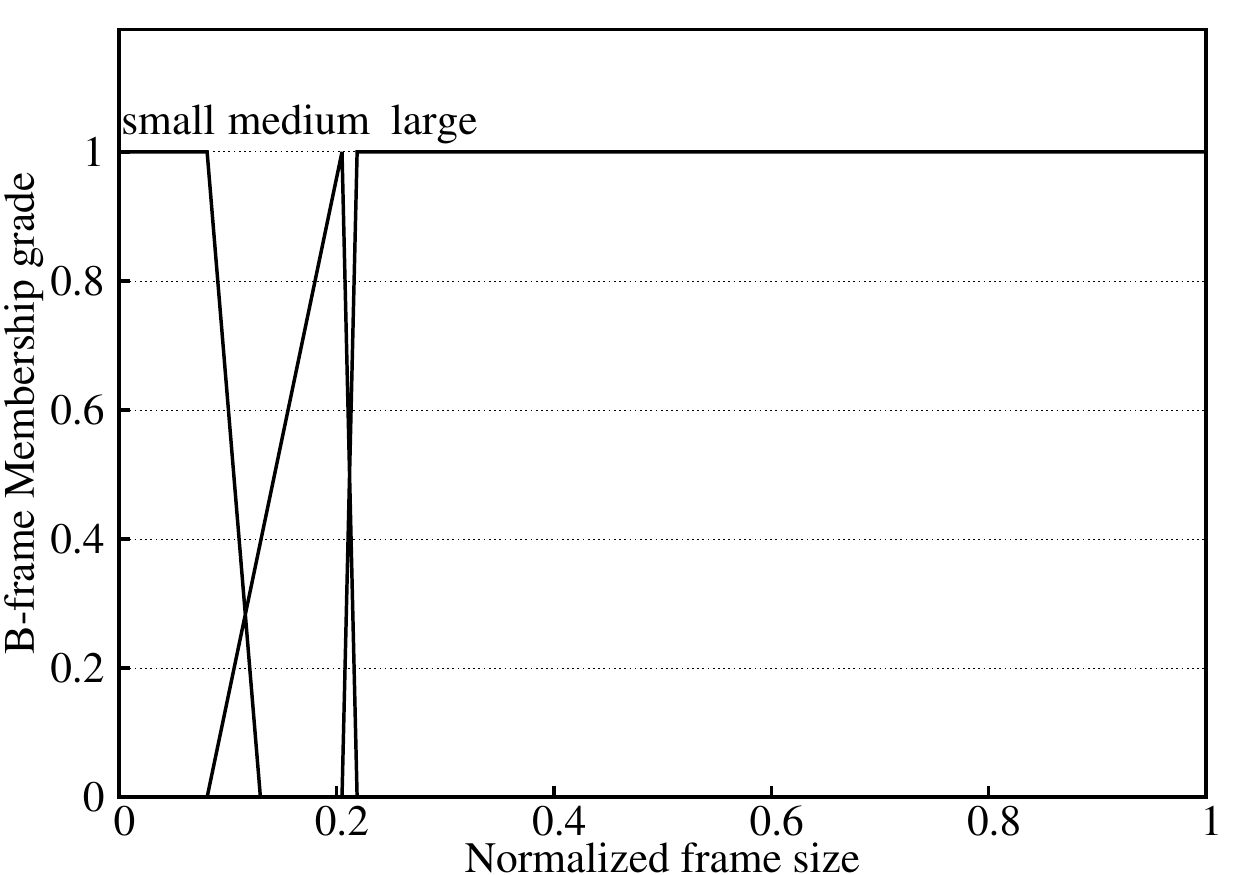}
	\end{center}
	\vspace{-0.15in}
	\caption{\small Frame size membership function }
	\label{fig:MINT:membershipSize}
\end{figure}

After defining the set, the membership functions need to be outlined. This is problem-dependent, as well as a complex task being difficult to find an optimal solution~\cite{Wong2005}. Considering that, it is better to use piecewise linear functions~(formed of straight-line sections). These functions are both simpler and more efficient regarding computability, leading to lesser resource requirements. Figure~\ref{fig:MINT:membershipSize} shows the graphical representation of the chosen membership functions for the frame sizes.

Apart from the spatial complexity, the fuzzy components for the motion intensity also need to be created. The analysis of this criterion is performed through motion vectors details. 
In order to better infer the amount of movement described by the vectors, instead of counting them, it is computed how far each one is pointing using the Euclidean distance.
This represents the motion intensity better because it is possible to have one frame with several vectors pointing to a close distance meanwhile, another frame with fewer vectors, pointing farther away although, and thereby having higher motion intensity.

As defined in the MPEG standard, the motion vectors describe the movement of macroblocks from some position in one frame to another position in another frame. It is important to note that not all MB have the same size, as well as videos with higher resolution will have more macroblocks than videos with lower resolution. 
To be able to compare video sequences with different macroblocks sizes and resolutions, the MB area is used, given by Equation~(\ref{eq:MINT:mbArea}), together with the motion vectors. 
Additionally, using Equation~(\ref{eq:MINT:motionIntensityFrame}) it is possible to calculate for each macroblock how many pixels have been moved and how far away, which can be translated as temporal intensity.

\begin{equation}
aMB = MB_{h} \times MB_{w}
\label{eq:MINT:mbArea}
\end{equation}
\begin{equation}
TI_{\Delta t} = \frac{1}{nMB} \sum_{i=0}^{nMB-1} aMB_{(i)} \times \left | MV_{(i)} \right | 
\label{eq:MINT:motionIntensityFrame}
\end{equation}

\begin{figure}[!htb]
	\begin{center}
		\includegraphics[width=3.5in]{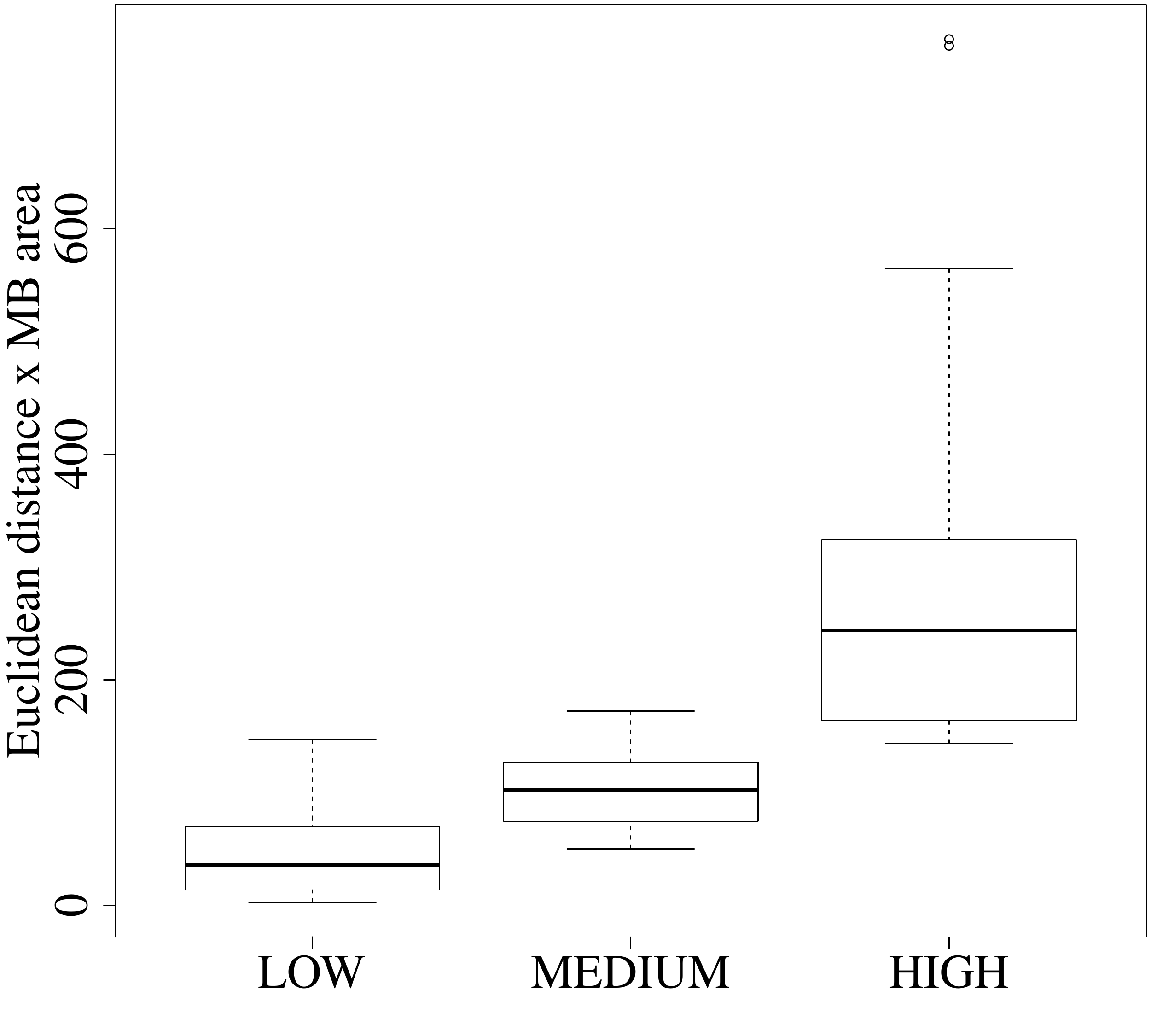}
	\end{center}
	\vspace{-0.0in}
	\caption{\small Temporal Intensity}
	\label{fig:MINT:temporalInt}
\end{figure}

Using the aforementioned details, another exploratory analysis was performed to classify the video sequences in terms of temporal intensity. This time, instead of breaking the video sequences in frames, the whole video was analysed. The values found through Equation~(\ref{eq:MINT:motionIntensityFrame}) were used to cluster the videos into three distinct groups, namely ``low'', ``medium'', and ``high'' temporal intensity. Additionally, in the same way as before, a boxplot was used to summarise and display the data distribution, as well as to create the sets, as presented in Figure~\ref{fig:MINT:temporalInt}.

Another important step is to define the PLR. The primary objective is to find out the influence of different PLR in the QoE for a set of videos.
The same procedure adopted by uavFEC was applied to MINT-FEC however, different values were found.
To find the PLR that best represents the video quality, a number of network simulations using a broad collection of UAV video sequences were carried out. As the MINT-FEC can handle arbitrary video resolution, several of them were used in the experiments.
For PLR between 0\% and 10\%, the video quality was good, in comparison to the uavFEC, where the good range was between 0\% and 15\%.
A tolerable video quality was observed for most of the videos between 5\% and 20\% of PLR~(uavFEC had values between 5\% and 30\%). Over 15\% of PLR the quality quickly decreased in videos with higher motion intensity, and over 25\% it became unacceptable. In general, the PLR input set is lower in MINF-FEC than in uavFEC because videos with higher resolution tend to consume more network resources.
Based on the results, the PLR set was defined, as shown in Algorithm~\ref{algo:MINT:PLRSet}. 

\iflatextortf
\else
{\LinesNumberedHidden \SetAlgoVlined
	\SetKwProg{Fn}{}{}{}
	\begin{algorithm}[!htb]
		{\small
			\Fn{OutputLVar* \textbf{PLR} = new OutputLVar(``\textbf{PacketLossRate}'');}{
				PLR $\rightarrow$ addTerm( TriangularTerm(``\textit{LOW}'', 0, 10))\;
				PLR $\rightarrow$ addTerm( TriangularTerm(``\textit{MEDIUM}'', 5, 20))\;
				PLR $\rightarrow$ addTerm( TriangularTerm(``\textit{HIGH}'', 15, 100))\;
			}
			engine.addOutputLVar(\textbf{PLR})\;
		}
		\caption{Packet loss rate input set for arbitrary video resolutions}\label{algo:MINT:PLRSet}
	\end{algorithm}
}
\fi

The last fuzzy set is the redundancy set. The MINT-FEC mechanism uses the same set as uavFEC, which was explained in Section~\ref{sec:uavFEC:design} and Algorithm~\ref{algo:uavFEC:redudancyOutput} defines it. With all sets delineated, it is necessary to create the rules. 
This activity also involves human knowledge about the video characteristics, namely spatial complexity and temporal intensity, as well as the frame type, and the PLR. As mentioned before, in videos with high spatial complexity the I-Frame needs a greater amount of protection, because it holds a large amount of information. 
On the other hand, in videos with high temporal intensity, the I-Frame also needs to be protected.
The P-Frame also plays an important role because it holds the temporal information about that sequence, and needs to have almost the same protection as the I-Frame.
Algorithm~\ref{algo:MINT:lossVideoRules} shows two rules that represent this case.

\iflatextortf
\else
{\LinesNumberedHidden \SetAlgoVlined
	\SetKwProg{Fn}{}{}{}
	\begin{algorithm}[!htb]
		{\small
			\textit{RuleBlock* \textbf{block} = new RuleBlock();}
				\BlankLine
				\Fn{block $\rightarrow$ addRule( \textit{new} MamdaniRule(``}{
					\Fn{\textbf{if} (SpatialComplexity is HIGH and\\
						\Indp 
						PacketLossRate is HIGH and \\
						FrameType is I)\\
						\textbf{then}}{
						RedundancyAmount is HIGH\;}
					'', engine))\;}
				
				\BlankLine
				\Fn{block $\rightarrow$ addRule( \textit{new} MamdaniRule(``}{
					\Fn{\textbf{if} (TemporalIntensity is HIGH and \\
						\Indp
						PacketLossRate is HIGH and \\
						FrameType is I or P) \\
						\textbf{then}}{
						RedundancyAmount is HIGH\;}
					'', engine))\;}
							
			}
		
		\caption{Packet loss x video characteristics rules}\label{algo:MINT:lossVideoRules}
	\end{algorithm}
}
\fi

MINT-FEC utilises the same core structure of uavFEC, so once all the fuzzy rules and sets are defined, they are employed in real-time in the fuzzy logic controller. In the same way as before, the offline process needs to be performed just once, after that the controller will be able to compute a suitable amount of QoE-aware redundancy on-the-fly. 

\subsection{MINT-FEC Performance Evaluation and Results}
\label{sec:MINT-FEC:preformance}

The MINT-FEC goal is to improve on uavFEC~(Section~\ref{sec:uavFEC}) to ensure an even higher perceived QoE for end-users, while avoiding unnecessary network overhead.

\subsubsection{Experiment settings}

\begin{figure}[!htb]
	\begin{center}
		\includegraphics[width=3.6in]{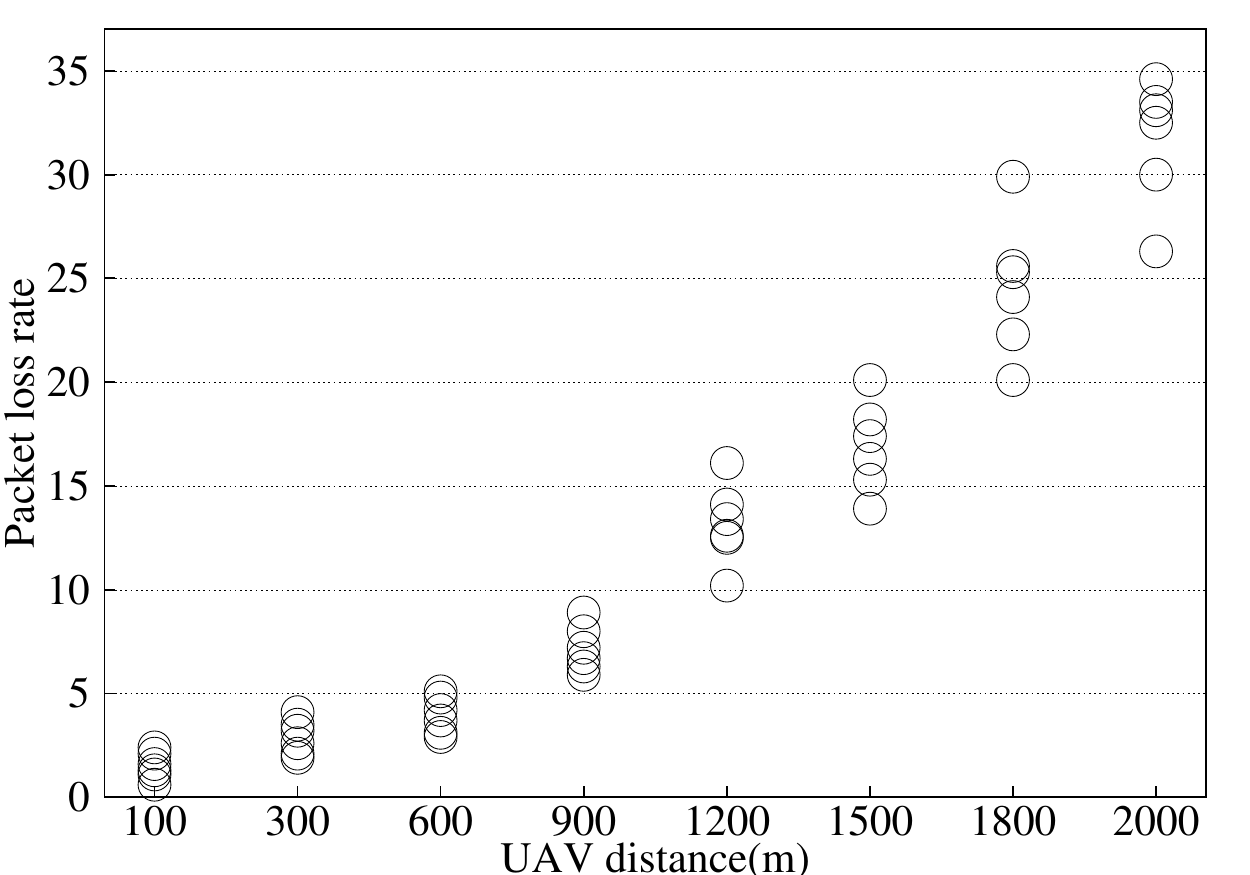}
	\end{center}
	\vspace{-0.0in}
	\caption{\small MINT-FEC's experiment PLR distribution}
	\label{fig:MINT:lossDist}
\end{figure}

The assessment scenario consists of up to four UAVs operating in autonomous mode, with 4G LTE radio at 800MHz. To better reflect a UAV scenario, the Gauss-Markov distribution mobility model was used. 
This model provides a uniform spatial distribution of the nodes and also simulates inertia in the movements, which are a characteristic of UAVs in autonomous mode. The ground control station is equipped with a portable base station and antenna. All UAVs are in line-of-sight and communicating in ad-hoc mode. 
Only real UAV video sequences were used in the experiments. To be more precise, twenty of each video resolution~(1080p, 720p, and SVGA), giving a total of sixty video sequences. All of them were encoded with both same GoP length of 19:2 and same H.264 codec. Considering the portable base station power, the ad-hoc communication, and the very demanding high definition videos, the flying range was limited to a radius of 2000 meters from the base station. Due to the harsh environment and the low-gain antenna, the PLR can range from 0\% to 35\%. Figure~\ref{fig:MINT:lossDist} shows the packet loss distribution in the experiments.
At the receiver side, a Frame-Copy error concealment was used. 
Table~\ref{tab:MINT:parameters} shows the simulation parameters.

\begin{table}[!ht]
	{ \small
		\caption{MINT-FEC Simulation parameters}
		\vspace{-0.0in}
		\begin{center}
			\begin{tabular}{l|l}
				\hline \textbf{Parameters} & \textbf{Value} \\ 
				\hline
				\hline Display sizes & 1920x1080, 1280x720, and 800x600\\
				\hline Frame rate mode & Constant\\
				\hline Frame rate & 29.970 fps\\
				\hline GoP & 19:2 \\ 
				\hline Video format & H.264\\
				\hline Codec & x264 \\ 
				\hline Container & MP4 \\
				\hline Propagation model & FriisPropagationLossModel \\
				\hline Mobility model & Gauss-Markov \\
				\hline UAV velocity & 45-65 km/h (28-40 mph) \\
				
				\hline LTE Frequency band & 800MHz \\
				\hline LTE Mode & FDD \\
				\hline LTE Bandwidth & 5 MHz \\
				\hline eNodeB Operating Power  & 22 dBm \\
				\hline Antenna Gain & 16 dBi \\
				\hline
				
			\end{tabular}
			\label{tab:MINT:parameters}
		\end{center}
	}
	\vspace*{-0.0in}
\end{table}

Five different schemes were simulated as follows:~(1) without any FEC mechanism. This is only to serve as a baseline for comparison with the others;~(2) a non-adaptive video-aware FEC~(I- and P-Frames are equally protected) using a pre-set value of 75\% of redundancy~(Video-aware FEC). This value was chosen because it showed a good tradeoff between QoE and network overhead in several PLR;~(3) the adaptive FEC-based mechanism~(uavFEC), presented in Section~\ref{sec:uavFEC};~(4) a related work implementation of the Cross-Layer Mapping Unequal Error Protection~(CLM-UEP)~\cite{Lin2012}. At last,~(5) adopts the proposed MINT-FEC mechanism.

\subsubsection{QoE assessments}

\begin{figure}[!htb]
	\begin{center}
		\includegraphics[width=4.6in]{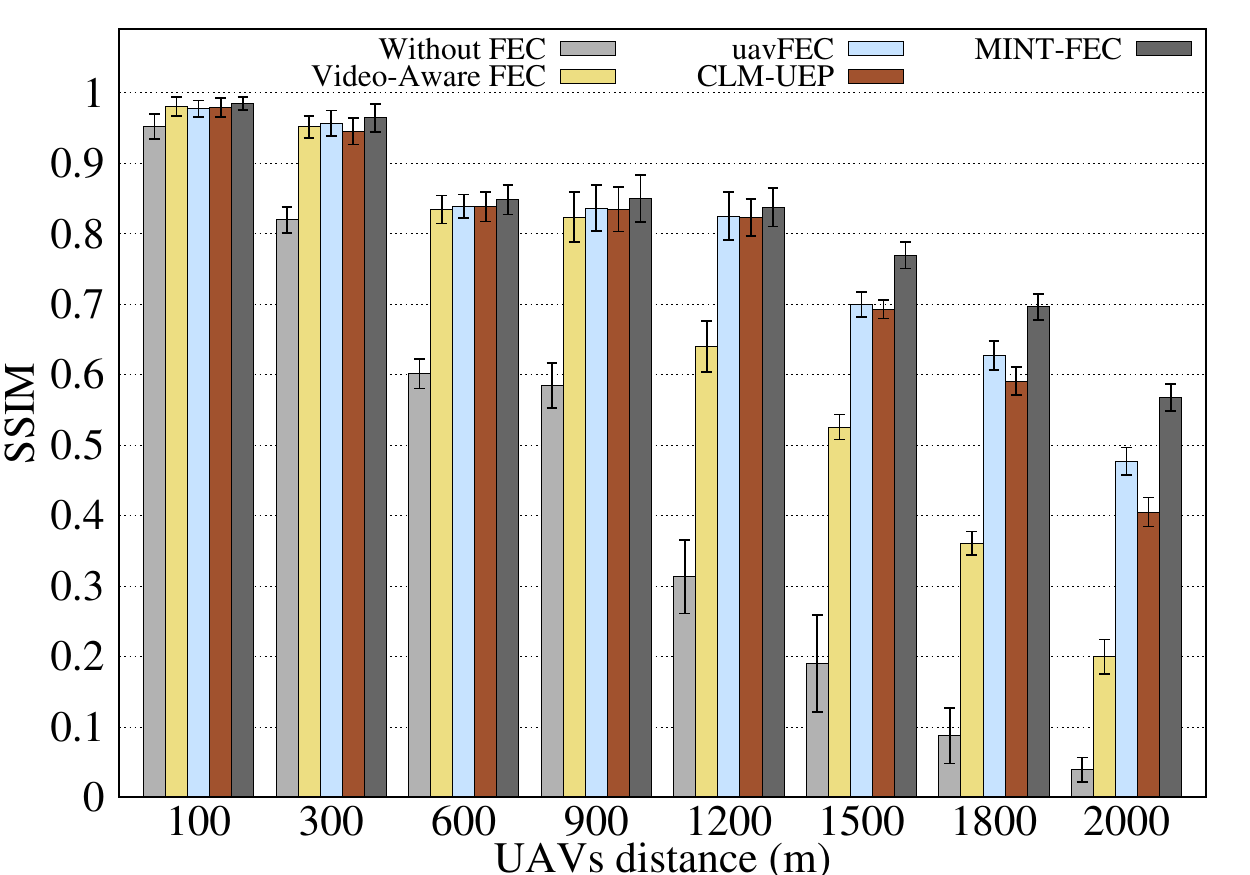}
	\end{center}
	\vspace{-0.0in}
	\caption{\small Average SSIM QoE for all scenarios}
	\label{fig:MINT:qoeSSIM}
\end{figure}

Figure~\ref{fig:MINT:qoeSSIM} shows the average of the SSIM results. 
A foreseen situation can be clearly noticed, the farther away the UAVs are from the ground control station the worst is the video quality. In the first case~(without FEC), a good video quality is noticed up to 600m. This is expected on the grounds that some video sequences tend to have a natural resiliency to packet loss. In particular, this situation is true in videos with low motion intensity, which usually scores higher results in QoE assessment. Between 600m and 900m the video quality is already affected and a sharp decline is perceived after that. At the same time, in the FEC-based schemes, such as Video-aware FEC, uavFEC, CLM-UEP and, MINT-FEC, the video quality was good for a long distance, until 1200m. Notwithstanding, the proposed mechanism outperforms all its competitors in terms of video quality, providing even better results over higher distances. A comprehensive comparison analysis is given further.

\subsubsection{Network footprint analysis}

\begin{figure}[!htb]
	\begin{center}
		\includegraphics[width=4.6in]{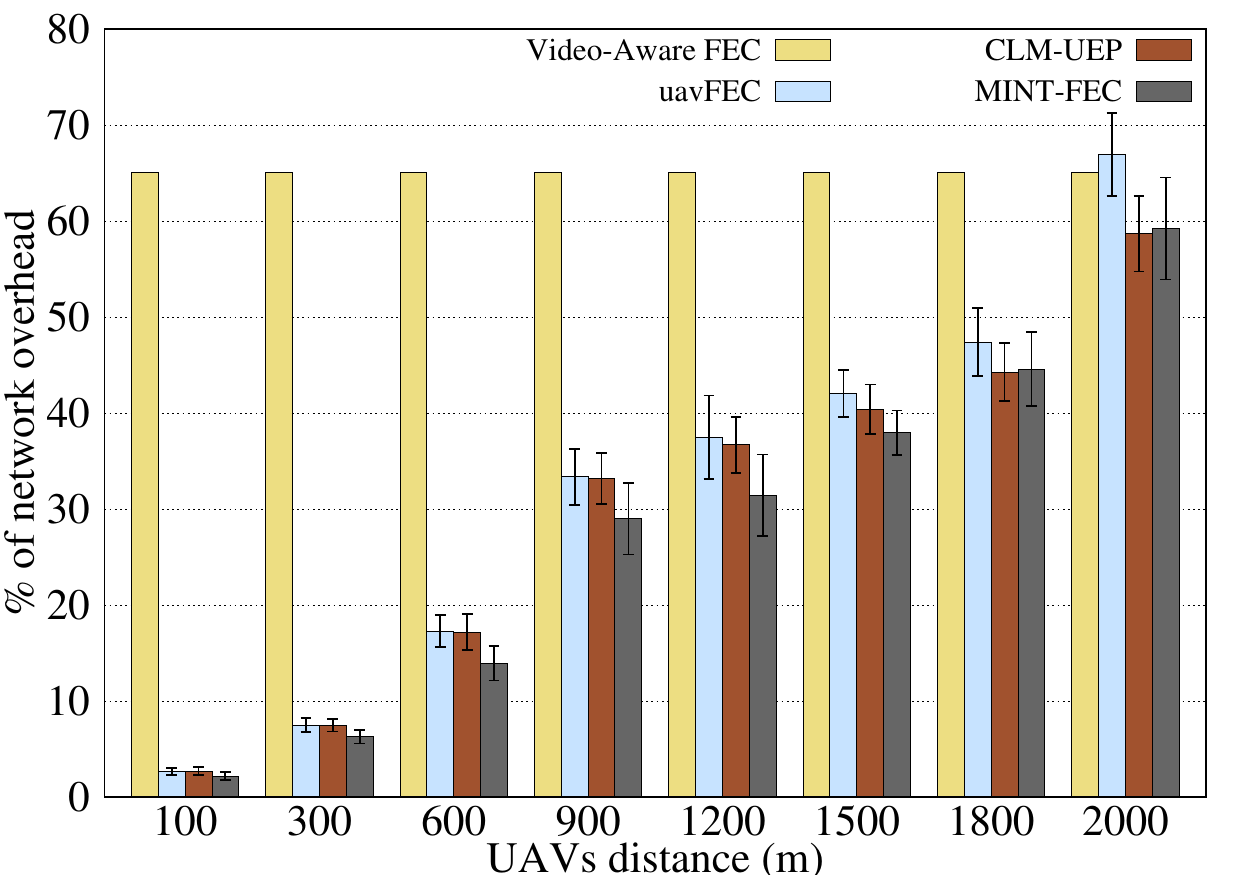}
	\end{center}
	\vspace{-0.0in}
	\caption{\small Average Network overhead}
	\label{fig:MINT:netOverhead}
\end{figure}

MINT-FEC can provide enhanced video quality, especially over higher distances, however, it is equally important to do so with lower network overhead. 
The Video-aware FEC scheme is non-adaptive and due to that, it has a constant network overhead, as showed in Figure~\ref{fig:MINT:netOverhead}. This is not suitable for UAVs because even when they are close to the base station, with a low PLR, a large amount of redundancy is added, wasting resources. On the other hand, the adaptive mechanisms~(uavFEC, CLM-UEP, and MINT-FEC) allow a better use of the network resources, as also shown in Figure~\ref{fig:MINT:netOverhead}. 

In all three mechanisms, the initial amount of redundancy is small and starts to become larger as the UAVs move away from the ground station. Both the uavFEC mechanism and CLM-UEP perform close to each other up to 1200m. After that, uavFEC starts to add more redundancy to provide better video quality. Here again, the MINT-FEC performs better than the others. Up to 1500m, it induces less network overhead, while providing higher video quality. Subsequently, it adds a slightly higher redundancy and still smaller than the other schemes, in favour of a considerably better video quality. This proves that it was possible to identify the most important video portions and protect them accordingly.

\subsubsection{Overall results}

\begin{figure}[!htb]
	\begin{center}
		\includegraphics[width=4.6in]{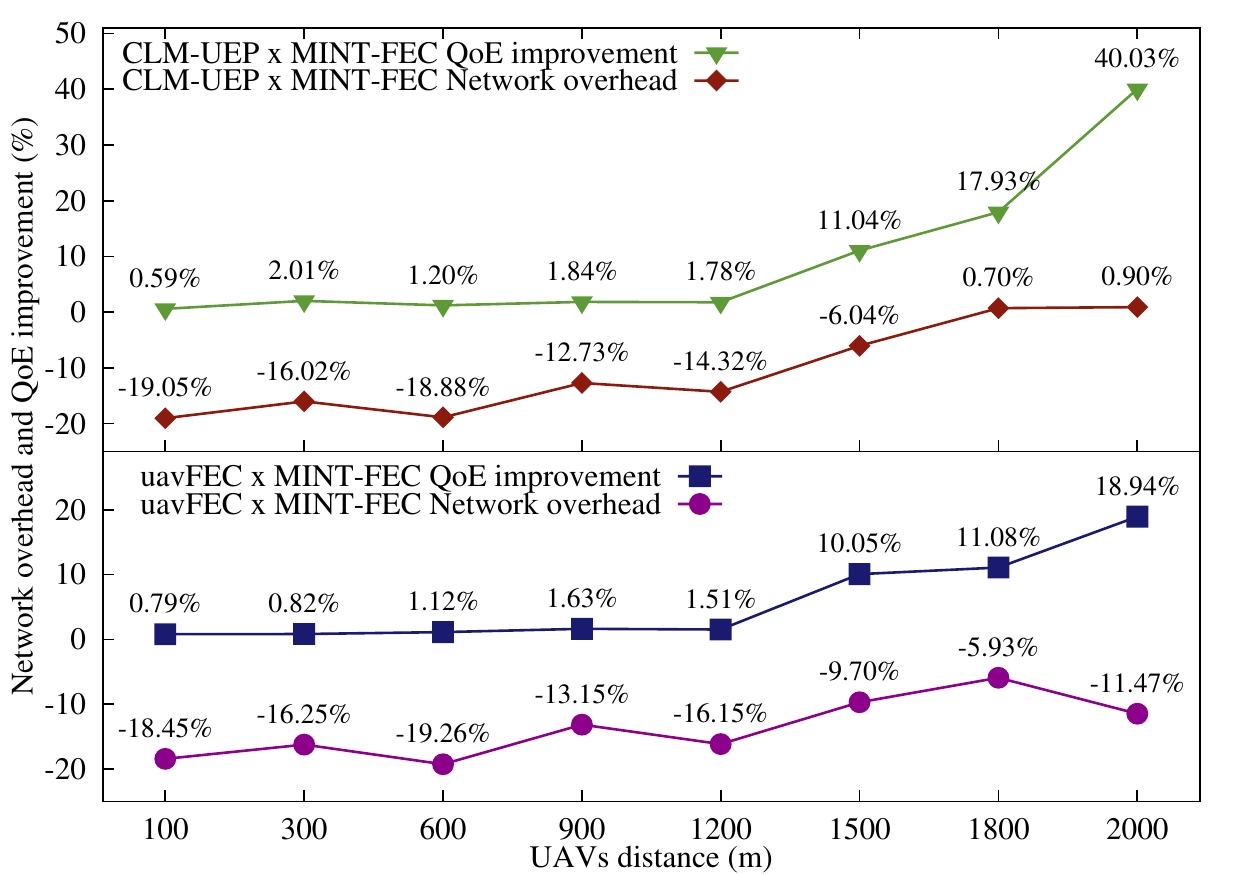}
	\end{center}
	\vspace{-0.0in}
	\caption{\small QoE and Redundancy comparison}
	\label{fig:MINT:qoeNetOverhead}
\end{figure}

To further understand the MINT-FEC achievements, a comparison against CLM-UEP and uavFEC is given in Figure~\ref{fig:MINT:qoeNetOverhead}. The first case is the comparison between CLM-UEP and MINT-FEC, and the second one, uavFEC against MINT-FEC. The graph shows the average percentage of QoE and network overhead improvement. In the QoE assessment, a positive percentage means that the proposed mechanism achieved higher video quality, which is desirable. On the other hand, in the network evaluation, a negative percentage means that the MINT-FEC generated less overhead, which is also advantageous.

In both cases, MINT-FEC presented a slightly better video quality until 1200m, which was between 0.59\% and 2.01\%, and between 0.69\% and 1.63\%, respectively. While outperforming the other mechanisms in terms of quality, the proposed mechanism was also able to considerably reduce the network overhead. It added on average 16.20\% less redundancy than CLM-UEP and 16.65\% less redundancy than uavFEC, up to 1200m. This proves that MINT-FEC is capable of better identifying the most QoE-sensitive data and adds a precise amount of QoE-aware redundancy to it, resulting in higher video quality and less network overhead. After this threshold, the MINT-FEC starts to increase the amount of redundancy to improve the video quality. This happens for the reason that the proposed mechanism was designed to sustain a higher video quality over long distances when the connection is more susceptible to errors. In doing that, the videos are received with up to 40\% better quality in comparison to the CLM-UEP mechanism and up to 18\% higher quality than the uavFEC mechanism. Another important advance of the MINT-FEC over the uavFEC was the network overhead reduction beyond 1200m. In this case, the MINT-FEC mechanism managed to reduce the overhead by up to 11.74\%. This is an additional proof that the proposed mechanism is doing a better work to infer the motion intensity using spatial complexity and temporal intensity, allowing the definition of a precise amount of redundant information to the most sensitive data. Thanks to that, it was possible to deliver higher video quality leading to a better user perception.

\section{Summary}

This chapter investigated the performance of the two proposed mechanisms to improve the resiliency to packet loss over FANETs using UAV-to-Ground model, namely uavFEC and MINT-FEC. 
The results shown in Section~\ref{sec:uavFEC:performance} demonstrated that the uavFEC mechanism improves video transmissions over highly dynamic networks, maximising the QoE without adding unnecessary network overhead. 
The uavFEC mechanisms was able to provide videos in the best-case scenario with, on average, between 11.59\% and 28.52\% better QoE than the main competitor~(CLM-UEP). 
Overall, the uavFEC mechanism attained good results. 
It had, nevertheless, some drawbacks especially due to its strict video resolution dependence and basic motion intensity classification.

To improve on these issues, the MINT-FEC mechanism was proposed.
The achieved results, shown in Section~\ref{sec:MINT-FEC:preformance}, attested that the proposed mechanism outperforms the other mechanisms in the experiments. 
In scenarios with up to 1200m, it was able to deliver videos with slightly higher quality and at the same time generating substantially less network overhead. 
Therefore, an enhanced video quality was perceived without wasting wireless resources. 
On the other hand, over 1200m, due to harsher conditions, the MINT-FEC mechanism starts to increase the redundancy providing a considerably higher QoE than the other mechanisms. 
This is a desired tradeoff between network overhead and video quality. 
The results support the claim that MINT-FEC was able to identify the motion intensity video sequences as well as to handle arbitrary video resolutions. 
At the end, MINT-FEC was able to better protect the most QoE-sensitive data, which is translated in higher video quality.

The studies and proposed mechanisms in this chapter resulted in the following publications:

\bigbreak
\textbf{Book chapter:}
\begin{itemize}

	\item {Immich}, R. and Cerqueira, E. and Curado, M., ``\textbf{Improving video QoE in Unmanned Aerial Vehicles using an adaptive FEC mechanism}'', in Wireless Networking for Moving Objects: Models, Approaches, Techniques, Protocols, Architectures, Tools, Applications and Services, Volume 8611, pp 198-216, Springer LNCS, 2014
	
\end{itemize}

\bigbreak

\textbf{Conference papers:}
\begin{itemize}

	\item {Immich}, R. and Cerqueira, E. and Curado, M., ``\textbf{Towards the Enhancement of UAV Video Transmission with Motion Intensity Awareness}'', in the IEEE IFIP Wireless Days, 2014

	\item Cerqueira, E. and Quadros, C. and Neto, A. and Riker, A. and {Immich}, R. and Curado, M. and Pescap\'{e}e, A., ``\textbf{A Quality of Experience Handover System for Heterogeneous Multimedia Wireless Networks}'', in the IEEE International Conference on Computing, Networking and Communications (ICNC), 2013
	
	\item Cerqueira, E. and Neto, A. and {Immich}, R. and Curado, M. and Riker, A. Barros, H., ``\textbf{A Parametric QoE Video Quality Estimator for Wireless Networks}'', in the GC'12 Workshop: IEEE Workshop on Quality of Experience for Multimedia Communications (QoEMC), 2012

\end{itemize}

\setcounter{mtc}{14}
\chapter{Mechanisms for Resilient Video Transmission over VANETs}
\chaptermark{Resilient Video Transmission over VANETs}
\label{ch:VANET}

\dictum[George Orwell, Animal Farm]{All animals are equal, but some animals are more equal than others.}

\minitoc

\lettrine[lines=3]{\color{gray}\bf{V}}{} ANETs are envisioned to offer support for a large variety of distributed applications such as road and traffic alerts, autonomous driving capabilities and video distribution. 
The use of video-equipped vehicles, with support for live transmission, unveils a set of challenges that can range from the scarce network resources and vehicles movement to the time-varying channel conditions and high error rates.
Here again, to overcome these challenges an adaptive FEC-based scheme can be tailored to shield the video transmission with QoE assurance.
This chapter details two proposed adaptive mechanisms to improve the video transmissions over VANETs.

\section{Introduction}

VANET is considered the core component of intelligent transportation systems, providing support to many applications. The endorsement of video-based services can be beneficial to a broad range of situations, such as road safety, driver awareness, traffic status, and infotainment applications. Besides the users' experience, the video quality is also important to allow a better assessment of each situation. For example, it can give police officers, paramedics, and firefighters an accurate representation of the scene they will attend, thereby reducing the response time. Beyond the traffic related services, a sports venue or a festival could broadcast a live feed to incoming fans caught in a traffic jam. These are only simple examples of a limitless number of alternatives to make available rich-format services. These services, however, have to deal with the unreliable wireless connection of the VANETs, which are highly dynamic in nature and strongly prone to packet loss~\cite{Zhou2011,Gerla2014}. Therefore, it is imperative to strengthen the video transmissions against losses~\cite{Immich2013,Immich2014c}. This calls for an adaptive mechanism to enhance the video delivery to provide higher QoE.

In VANETs, there is still a lack of adaptive QoE-driven mechanisms to better support live video transmissions~\cite{Soldo2011,Shen2011,Jiang2012,Bellalta2014}. This can be attributed to the challenging combination of the VANETs' dynamic topology and the stringent video requirements. In order to surpass these adversities, a good mechanism has to take into consideration several aspects of the intrinsic network characteristics and video details, being able to correctly identify and protect the most QoE-sensitive data. 

As mentioned before, several techniques have been proposed to tackle the VANETs challenges in the last few years. Some of them are trying to solve these issues throughout adaptive routing protocols~\cite{Marwaha2004,Zeng2013,Pham2014,Wu2014,Zhang2015}. The results show that a reliable routing protocol has a major influence on improving the video quality. This improvement, however, is restricted to a specific level. After this level, to increase or even sustain the video quality it is crucial to resort to some amount of redundant data, which allow reconstructing the original data set in case of packet losses. A known approach to supply this redundancy is using FEC techniques. However, due to the video requirement of a timely delivery of a considerable amount of data~\cite{Zhou2011a}, along with the shared wireless channel resources, a self-adaptive FEC-based mechanism is advisable. This mechanism needs to have the capability to operate under unforeseen conditions in order to increase the human perception while reducing the network overhead.

This chapter proposes and evaluates two QoE-aware mechanisms tailor-made to overcome any network mishaps, therefore providing the capability to deliver with high QoE for end-users. 
The first mechanism described is the adaptive QoE-driven COntent-awaRe VidEo Transmission opTimisation mEchanism~(CORVETTE) in Section~\ref{sec:corvette} and the latter is the self-adaptive FEC-based proactive error recovery mechanism to shield video transmissions over VANETs~(SHIELD), detailed in Section~\ref{sec:shield}.

\section{QoE-driven and Content-aware Video Delivery~(CORVETTE)}
\label{sec:corvette}

Owing to the open issues aforementioned, particularly the shortage of adaptive QoE-driven mechanisms that efficiently consider the motion activity of the videos together with VANETs features, this section outlines and evaluates the adaptive QoE-driven COntent-awaRe VidEo Transmission opTimisation mEchanism~(CORVETTE). 
The proposed mechanism enhances the work described in Sections~\ref{sec:uavFEC} and~\ref{sec:MINT-FEC}. The main improvements and new experiment setups are described next.

\subsection{CORVETTE Overview}

The main goal of the CORVETTE mechanism is to improve the resilience of video transmissions over VANETs.
One common issue found in the existing mechanisms is the lack of QoE-awareness. 
This means that important video characteristics, from the human perspective, are neglected, resulting in unnecessary redundancy. 
To tackle this issue, the CORVETTE uses a Hierarchical Fuzzy System~(HFS) together with intrinsic VANET characteristics, to add a precise amount of redundancy exclusively on QoE-sensitive data. 
This ensures high-quality video transmission, while downsizing the network overhead.

\begin{figure*}[!htb]
	\vspace{-0.0in}
	\begin{center}
		\ifBW \includegraphics[width=143mm]{./vanets_NEO_SHIELD_MOD_gray-eps-converted-to.pdf}
		\else \includegraphics[width=143mm]{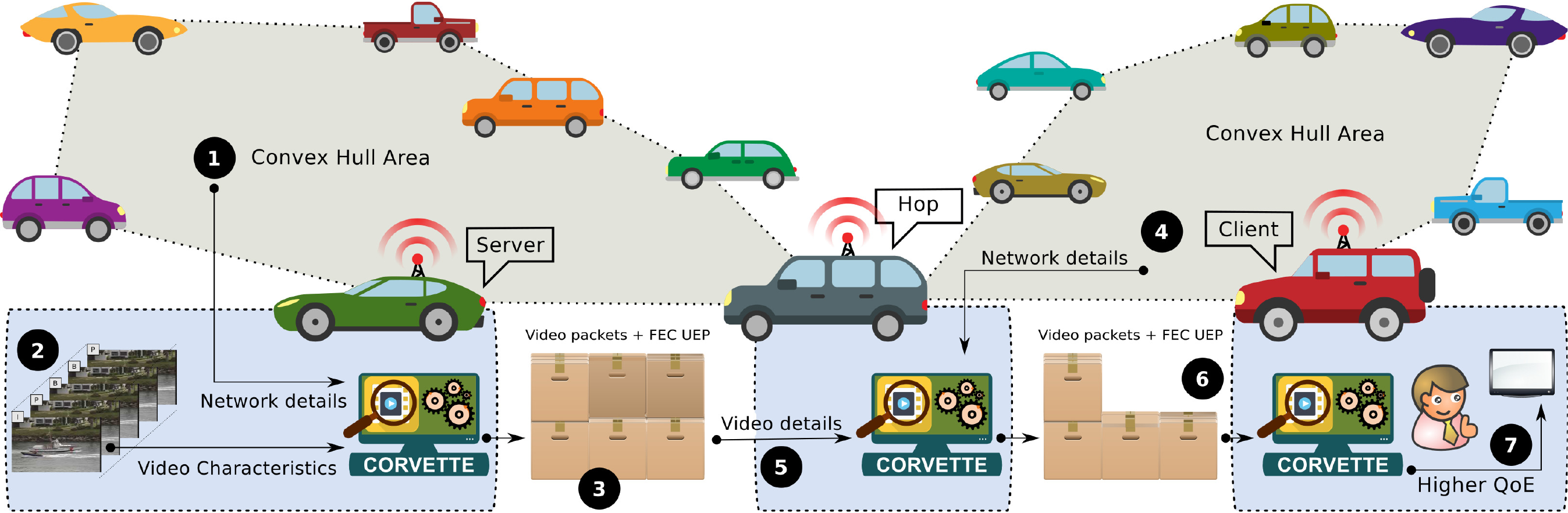}
		\fi
	\end{center}
	\vspace{-0.0in}
	\caption{\small General view of the CORVETTE mechanism}
	\label{fig:corv:mechanism}
	\vspace{-0.0in}
\end{figure*}

Figure~\ref{fig:corv:mechanism} depicts an overview of the proposed mechanism. The first step is to assess the network conditions~(1). In order to do that, different parameters are evaluated in a combined way, namely the network density, PLR, and the node's position. To calculate the density, first, the network area is found using a hull algorithm. After that, the total number of 1-hop nodes is divided by the area, which gives the network density. All these parameters are necessary because none of them by itself is accurate enough to characterise the quality of the network links~\cite{Vlavianos2008,Wan2015}. The combination of them, however, can provide a very good estimation of the network conditions. Thereafter, using cross-layer techniques, important details about the video characteristics are collected~(2). In the video-aware procedure of the mechanism, several details are analysed, such as the image resolution, frame type and size, motion vectors, and macroblock configuration. At the end, all the gathered data are fed to the fuzzy inference engine, which will compute a specific amount of redundancy~(3).

Provided that the network conditions are not the same at all intermediate nodes, this parameter has to be reassessed at each hop~(4). On the other hand, the video characteristics do not change during the transmission. In view of this, they are embedded in each packet header by the server node. This eliminates the need for processor intensive tasks~(e.g. deep packet inspection) on each and every packet. The IPv6 optional hop-by-hop header was chosen to store this information~\cite{Martini2007}. This means that it is always ready to use whenever needed~(5,6). Owing to this, the task of adjusting the redundancy amount on each hop is facilitated. The result is a higher video quality, and consequently, superior QoE is perceived by the end-users~(7).

\subsection{Towards the design of CORVETTE}

This section describes the manifold procedure and modules that the CORVETTE mechanism is composed of. Primarily, to enable the CORVETTE real-time capabilities a knowledge database is needed. The procedure to build this database uses the same core structure as in the mechanisms described before with the additional improvements to accommodate the new parameters. A comprehensive description of this process can be found in Chapters~\ref{ch:MESH} and~\ref{ch:UAV}.

An important feature to enable the CORVETTE real-time capabilities is the use of fuzzy logic. 
This allows building a dynamic and comprehensive scheme, which takes into consideration several network and video characteristics while still manages to perform in real-time. 
Nevertheless, in conventional fuzzy logic systems, the rules grow exponentially according to the number of variables. 
Therefore, it is common to have a rule-explosion situation when handling a lot of variables, making the fuzzy logic controller very hard to implement. 
To address this issue, the CORVETTE mechanism was designed to use a HFS, where low-dimensional fuzzy systems can be arranged in a hierarchical form, reducing the global number of rules because the system grows linearly.

\begin{figure}[!htb]
	\vspace{-0.0in}
	\begin{center}
		\includegraphics[width=102mm]{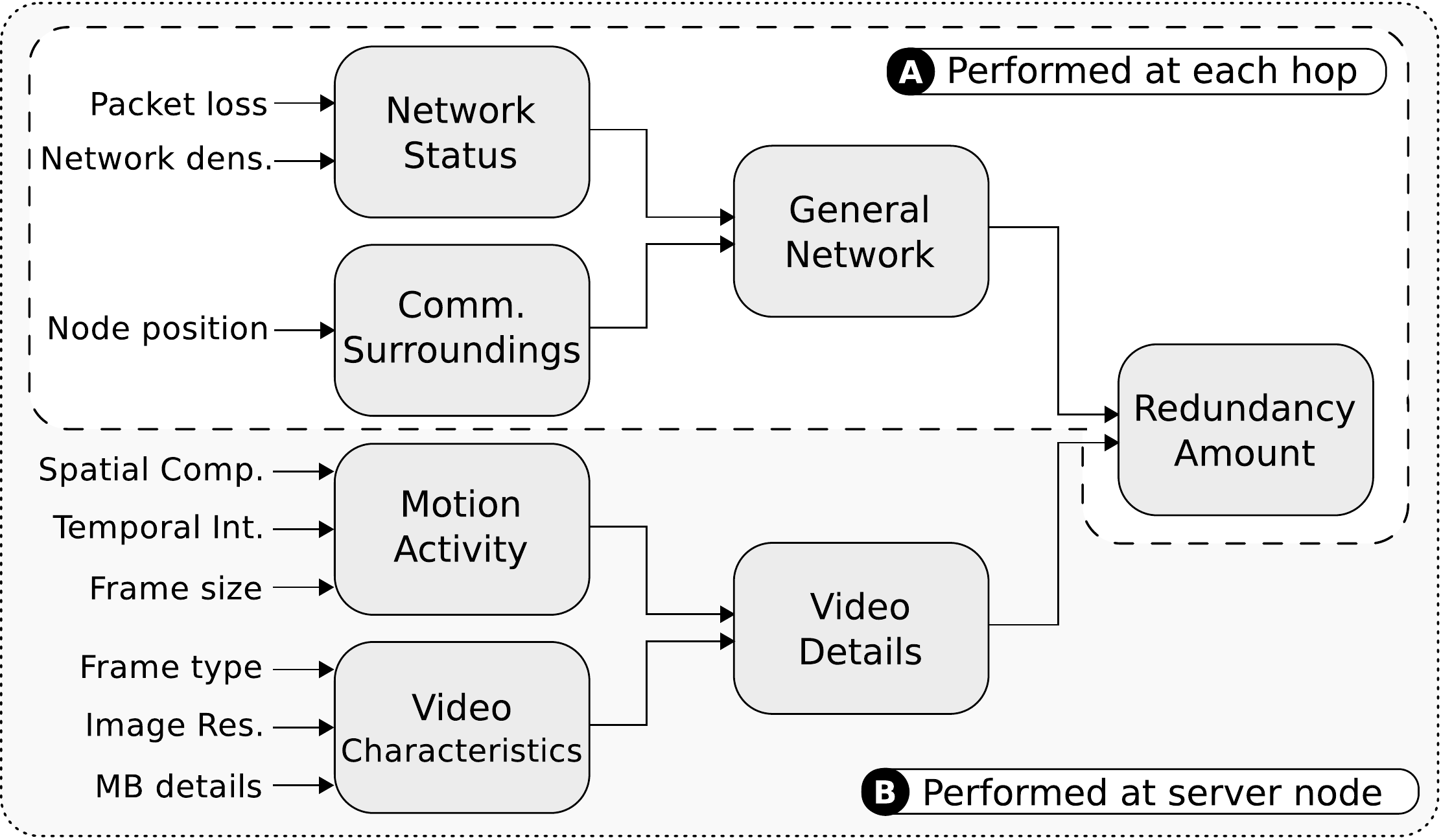}
	\end{center}
	\vspace{-0.0in}
	\caption{\small CORVETTE Hierarchical Fuzzy Logic structure}
	\label{fig:corv:hfuzzy}
	\vspace{-0.0in}
\end{figure}

The combined use of the knowledge database and human expertise enables setting up the fuzzy sets, rules, and hierarchical levels. Figure~\ref{fig:corv:hfuzzy} shows the several hierarchical levels used by CORVETTE. The output of each low-level component in the previous layer is used as input to the next layer components. As mentioned before, on each network hop the amount of redundancy is adjusted according to several factors, such as the network density, the PLR, and the distance to the next hop~(or final destination). This operation is represented by the~(A) portion of the HFS structure. On the other hand, at the server node, the full HFS structure is performed, which is composed of the~(A) and~(B) parts.

The design of HFS follows the same method as in standard fuzzy logic schemes. This means that several fuzzy components have to be defined, such as sets, rules, membership functions and the inference engine. 
Since this is a complex process, it has to be executed offline and only once. After loading all the generated information into the fuzzy engine, it can be used on-the-fly and with a high performance. A detailed explanation of this process is given below.

\subsubsection{The ``General Network'' layer conception}

The ``General Network'' component accounts for the definition of the network health. The characterization of a good or bad channel is not an easy task and it cannot rely upon a single metric, especially in wireless networks~\cite{Vlavianos2008}. With this in mind, the CORVETTE mechanism uses three metrics to better establish a network quality indicator. These metrics are divided into two specific components, namely ``Network status'' and ``Communication surroundings''. The former is defined by the combined assessment of the PLR and network density. The latter is given by the position of the vehicles. Each one of these metrics is described next.

To compute the network density, the number of nodes is divided by the network area. Because there is no fixed structure on VANETs and they can quickly change over time, it is challenging to estimate the network surface area. To solve this problem, the proposed mechanism uses a convex hull algorithm. In this type of algorithm, a convex polygon is drawn including all the nodes on the network. A polygon is convex when it is non-intersecting and a segment between any two points on the boundary lies entirely inside of it. Successively, a convex hull is the smallest polygon containing all the points.

There are several algorithms to find the convex hull, however, due to performance issues, the QuickHull method was chosen~\cite{Barber1996}. It uses a divide-and-conquer algorithm, and because of that, it is easier to find the convex hull of small sets, and at the end, the discovered hulls are merged. Once the convex polygon is found, it is possible to calculate the surface area and use it to find the network density.

Another component of the ``Network status'' is the PLR. The main goal in defining this parameter is to identify the impact of different PLRs in the QoE. As mentioned before, it is well-known that video sequences have a natural resiliency to packet loss~\cite{Immich2013a}.
To better define the PLR category, a broad number of network simulations with a large set of video sequences were carried out. The results show that the QoE is good from the PLR 0\% up to 11\%. Additionally, in most of the cases, from 5\% as far as 22\% a tolerable video quality for end-users was noticed. When exceeding 17\% however, a quick decrease in the video quality was observed, especially in high motion intensity videos. The majority of the cases, over 34\% of PLR, the QoE became unbearable. Using fuzzy logic it is possible to create classes of PLR exactly as found in the experiments, even with overlapping values. Algorithm~\ref{algo:corv:PLRset} shows the PLR set definition.

\iflatextortf
\else
\SetKwProg{Fn}{}{}{}
\vspace*{-0.0in}
\begin{algorithm}[!htb]
	{\small
		\Fn{InputLVar* \textbf{PLR} = new InputLVar("\textbf{PacketLossRate}");}{
			PLR $\rightarrow$ addTerm( TriangularTerm("\textit{LOW}", 0, 11))\;
			PLR $\rightarrow$ addTerm( TriangularTerm("\textit{MEDIUM}", 5, 22))\;
			PLR $\rightarrow$ addTerm( TriangularTerm("\textit{HIGH}", 17, 100))\; %
		}
		engine.addInputLVar(\textbf{PLR})\;
	}
	\caption{Packet loss rate input set}\label{algo:corv:PLRset}
\end{algorithm}
\vspace*{-0.0in}
\fi

The last component of the ``General network'' layer is the ``Communication surroundings''. In CORVETTE, this component is established by the node position. This is a straightforward, but very important information. Because of signal attenuation and radio-frequency interference, nodes further away from each other tend to require a higher amount of redundancy to preserve a good video quality. This information becomes even more valuable used in conjunction with the other input parameters. For example, a much higher amount of redundancy will be required if the network is very dense and the nodes are far apart than if the network was not so heavily populated.

\subsubsection{The ``Video Details'' layer conception}

Besides the network conditions, the video characteristics also play an important role when defining a precise amount of redundancy. In CORVETTE's hierarchical fuzzy system, the ``Video details'' criteria layer is divided into two specific components, namely ``Motion activity'' and ``Video characteristics''.

The motion activity parameter is defined by the combined use of temporal intensity and spatial complexity. The temporal intensity, in the CORVETTE mechanism, is given by the motion vector details. This structure is responsible for tracking the movement of macroblocks from one position, in any given frame, to another position, in the next one. Since the MPEG standard allows the use of distinct macroblock sizes, the CORVETTE mechanism adopts the same procedure of the uavFEC and MINT-FEC mechanisms by computing the area of each macroblock and using the number of pixels that are being moved. This enables a better representation of the intensity of the motion in arbitrary resolutions.

In addition, the Euclidean distance of each motion vector is also computed, resulting on how far each and every macroblock is being moved. This information gives more precise results than just counting the number of motion vectors. All the input parameters are normalised, allowing the protection of videos with arbitrary resolutions on the fly. Following the same idea as presented before, an exploratory analysis using hierarchical clustering is performed to find the best classes that represent the temporal intensity. After finding the classes, the fuzzy set and the membership function can be defined. Finding the best-fitted membership function is a complex and problem-dependent task~\cite{Wong2005}, being difficult to attain the optimal solution. Consequently, piecewise linear functions are desired.

As previously mentioned, the spatial complexity is also used to quantify the amount of the motion activity. This parameter represents the difference between the static information that the actual frame has when compared to the frame before. One way to compute this value is using the Sum of Absolute Differences~(SAD)~\cite{Vanne2006}. This process, however, compares each and every pixel of both frames resulting in a very complex and time-consuming operation. Taking this into consideration, the CORVETTE mechanism uses the normalised frame size to the same end. This enables a much faster operation and, on top of that, it also allows the use of arbitrary video resolutions.

\begin{figure}[!htb]
	\vspace{-0.0in}
	\begin{center}
		\ifBW \includegraphics[width=8.9cm]{./temporal_set_tt_CORV_gray-eps-converted-to.pdf}
		\else \includegraphics[width=8.9cm]{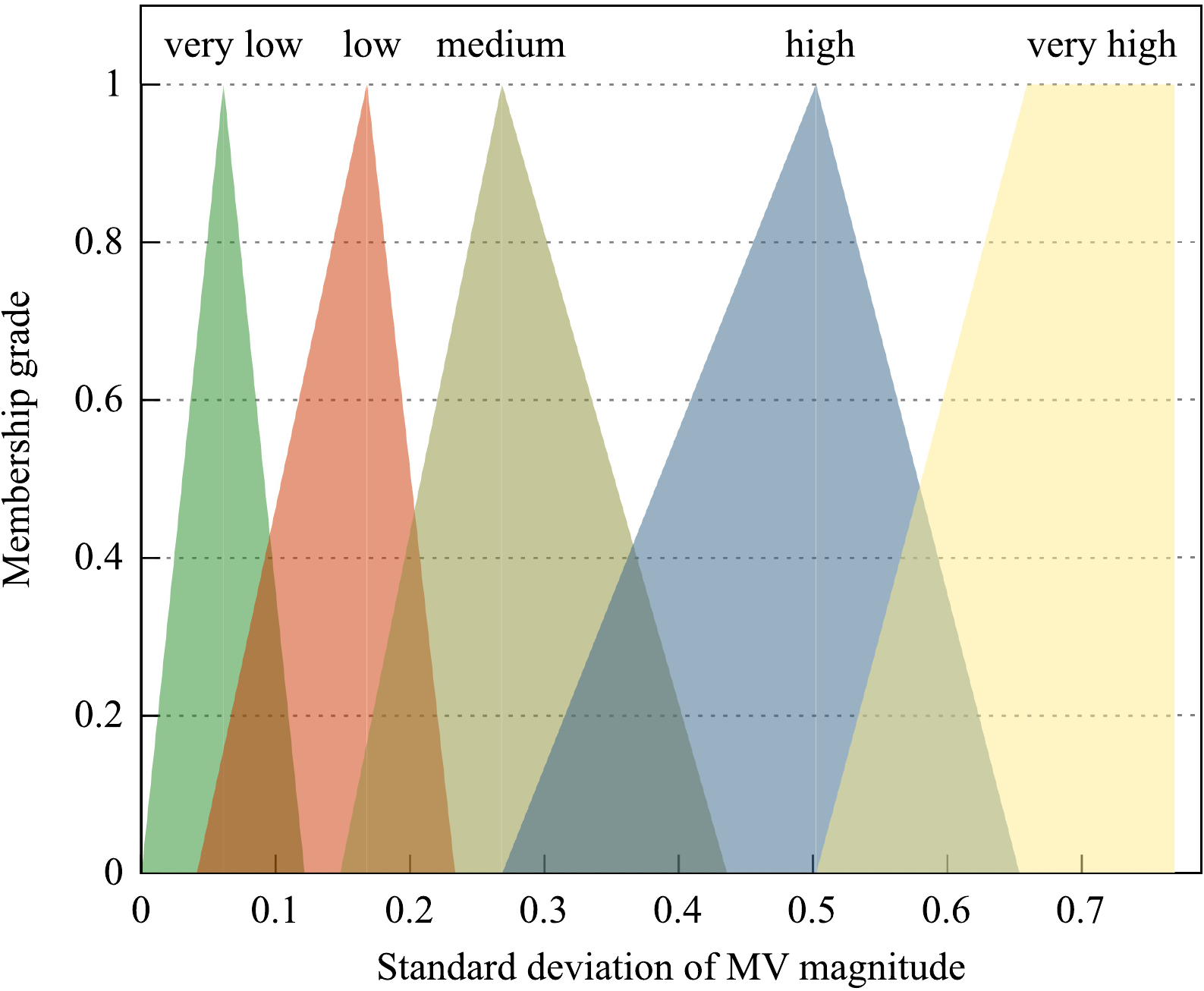}
		\fi
	\end{center}
	\vspace{-0.0in}
	\caption{Temporal intensity membership function}
	\label{fig:corv:membershipTemporal}
	\vspace{-0.0in}
\end{figure}

The same process used to find the different classes in the temporal intensity is also used to define the clusters here. This means that, once all the frame sizes are normalised, an exploratory analysis is performed to divide the data into the most homogeneous groups. After that, using the linkage distance between the clusters was possible to separate them into five distinct groups, namely ``very small'', ``small'', ``medium'', ``large'', and ``very large'', as showed in Figure~\ref{fig:corv:membershipTemporal}.

With the definition of the fuzzy sets completed, the fuzzy rules must be designed. As mentioned before, this is an intricate task, which requires a joint knowledge of the network details, VANETs properties, and video characteristics. Aiming to reduce this complexity during the design phase of the rules, as well as to have a better performance in real-time, the CORVETTE mechanism uses HFS. This layered system allows handling fewer input parameters at the same time. At the end, the result is a system with a small number of simple rules, which lead to better performance.
~\newline

\subsection{CORVETTE Performance Evaluation and Results}
\label{sec:corv:evaluation}
The main goal of the CORVETTE mechanism is to improve the QoE for end-users. At the same time, it avoids unnecessary network overhead, preserving wireless resources.

\subsubsection{Experiment settings}

The simulated scenario is composed of up to 360 vehicles, using IEEE 802.11p WAVE. The speed range varies from 5 to 17 m/s. 
Several real video sequences with three distinct resolutions, namely 1080p, 720p, and SVGA, were used in the experiments. The videos are examples of regular viewing material, covering different content and distortions. These sequences also include still and cut scenes, colour and luminance stress, as well as several motion intensities. They were all encoded with H.264 codec and GoP length of 19:2.

Using the OpenStreetMAP~\cite{Haklay2008}, a clipping of 2 by 2 km of the Manhattan borough~(New York City) was obtained. This clipping was used as input for the Simulation of Urban MObility~(SUMO)~\cite{Behrisch2011}, which considers the map structure, driving patterns, routes, crossings, roundabouts, traffic lights, to generate the mobility traces, which were used in the NS-3 simulations. 

Additionally, in this work, the V2V communication characteristics are explored to better adjust the proposed mechanism to the actual network conditions. Even though a VANET environment enables roadside infrastructure, the V2V communication was chosen because it is unlikely that such infrastructure will cover all the highways and cities in the near future. Consequently, if the infrastructure is available it can be used however, the optimisations will only be performed on the communication between the vehicles.

The adopted routing protocol is the Cross-Layer, Weighted, Position-based Routing~(CLWPR)~\cite{Katsaros2011}. As the name implies, this is a position-based routing protocol that uses mobility information from the nodes to tailor itself for VANETs environments.
Receiver nodes are using Frame-Copy error concealment; this means that if a frame is lost, the last good one will take its place. Table~\ref{tab:corv:parameters} shows the simulation parameters.

\begin{table}[!ht]
	\vspace{-0.0in}
	{ \small
		\caption{CORVETTE Simulation parameters}
		\vspace{-0.0in}
		\begin{center}
			\begin{tabular}{l|l}
				\hline 
				\textsc{\textbf{Parameters}} & \textsc{\textbf{Value}} \\ 
				\hline
				\hline Display sizes & 1920x1080, 1280x720, and 800x600\\
				\hline Frame rate mode & Constant\\
				\hline Frame rate & 29.970 fps\\
				\hline GoP & 19:2 \\ 
				\hline Video format & H.264\\
				\hline Codec & x264 \\ 
				\hline Container & MP4 \\
				
				\hline Propagation model & FriisPropagationLossModel \\
				\hline Mobility & SUMO mobility traces \\
				\hline Routing protocol & CLWPR \\
				\hline Wireless & IEEE 802.11p (WAVE)\\
				\hline Radio range & 250-300m\\
				\hline Internet layer & IPv6\\
				\hline Transport layer & UDP\\
				
				\hline Location & Manhattan borough~(New York City)\\
				\hline Map size & 2.000 m x 2.000 m\\
				\hline Vehicles speed & 18-61 km/h (11-38 mph) \\
				
				\hline
				
			\end{tabular}
			\label{tab:corv:parameters}
		\end{center}
	}
	\vspace{-0.0in}
\end{table}

Five distinct mechanisms were simulated as stated next: 
(1)~Without FEC, to be used as a baseline; 
(2)~Video-aware Equal Error Protection FEC~(VaEEP), where I- and P-frames are equally protected; 
(3)~Video-aware Unequal Error Protection FEC~(VaUEP). This mechanism protects I- and P-frames with a specific amount of redundancy according to the importance of each one; 
(4)~the adaptive FEC-based mechanism~(AdaptFEC), which takes into consideration several video characteristics and the network state~\cite{Immich2014}.
At last, (5)~the proposed CORVETTE mechanism.

\subsubsection{QoE assessments}

\begin{figure}[!htb]
	\vspace{-0.0in}
	\begin{center}
		\includegraphics[width=4.6in]{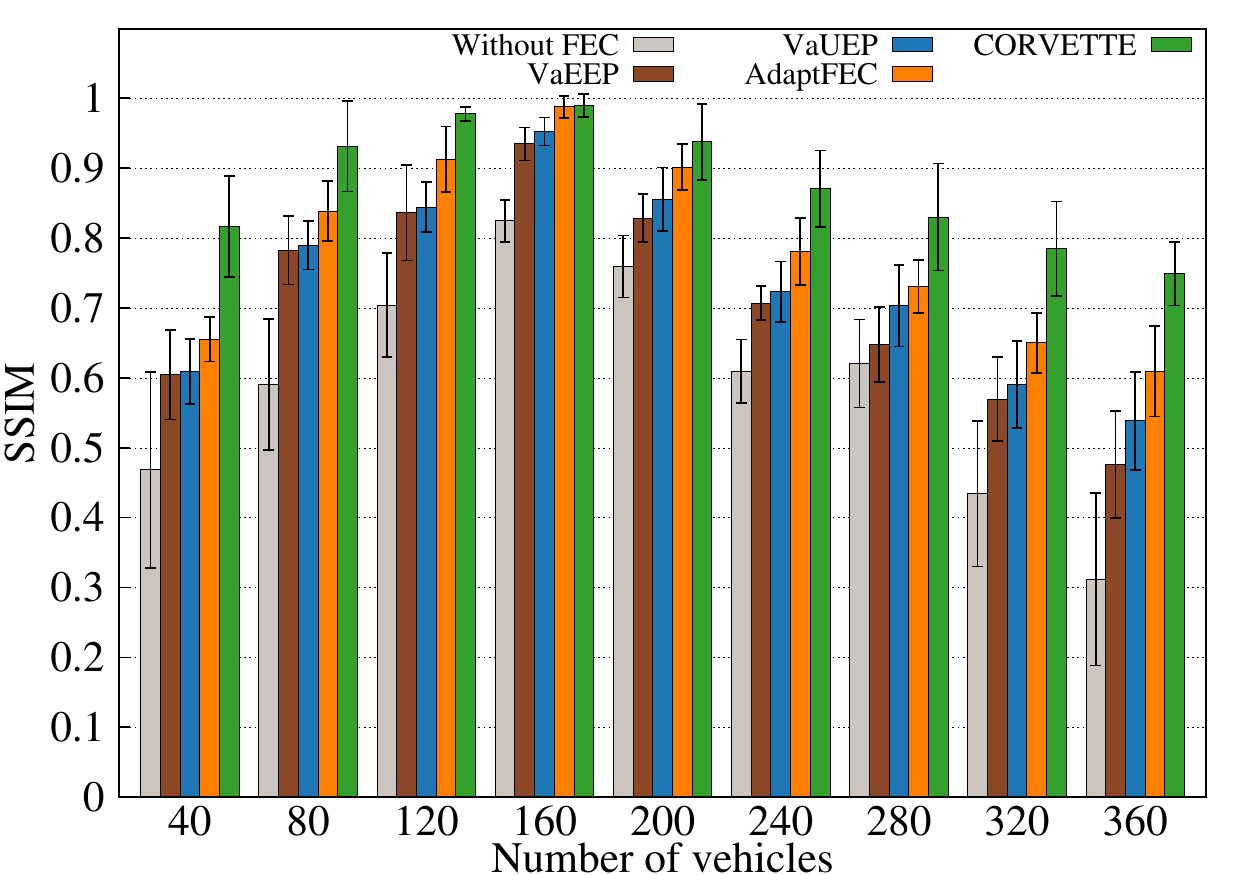}
	\end{center}
	\vspace{-0.0in}
	\caption{\small Average SSIM for all mechanisms}
	\label{fig:corv:qoeSSIM}
	\vspace{-0.0in}
\end{figure}

Figure~\ref{fig:corv:qoeSSIM} shows the SSIM results. The first thing to be noticed is that when the simulation is performed with a few vehicles, e.g., 40, the network is sparse and suffers from connectivity issues. The CORVETTE mechanism proves that it can handle those situations by adjusting the amount of redundancy correctly, thus, outperforming competitors. It is also important to perceive that the standard deviation is high in all mechanisms. This can be expected since some video sequences tend to be naturally more resilient to packet loss than others, resulting in distinctive quality scores. This is especially true in low motion intensity videos. On the other hand, when the network is very dense, e.g., above 300 cars, the mechanisms have to deal with interference and a degraded network connection. Here again, CORVETTE outperforms the other mechanisms, providing better video quality. 

Overall, the proposed mechanism exceeds by 48\% the without FEC scheme when it comes to video quality. 
Against VaEEP and VaUEP schemes, the CORVETTE mechanism granted over 23\% and 19\% better scores, respectively. 
In comparison to AdaptFEC, the proposed mechanism assured more than 11\% higher video quality. 
At first, this value does not seem to be very high however, it was able to achieve it while adding 41\% less network overhead. 

\begin{figure}[!htb]
	\vspace{-0.0in}
	\begin{center}
		\includegraphics[width=4.6in]{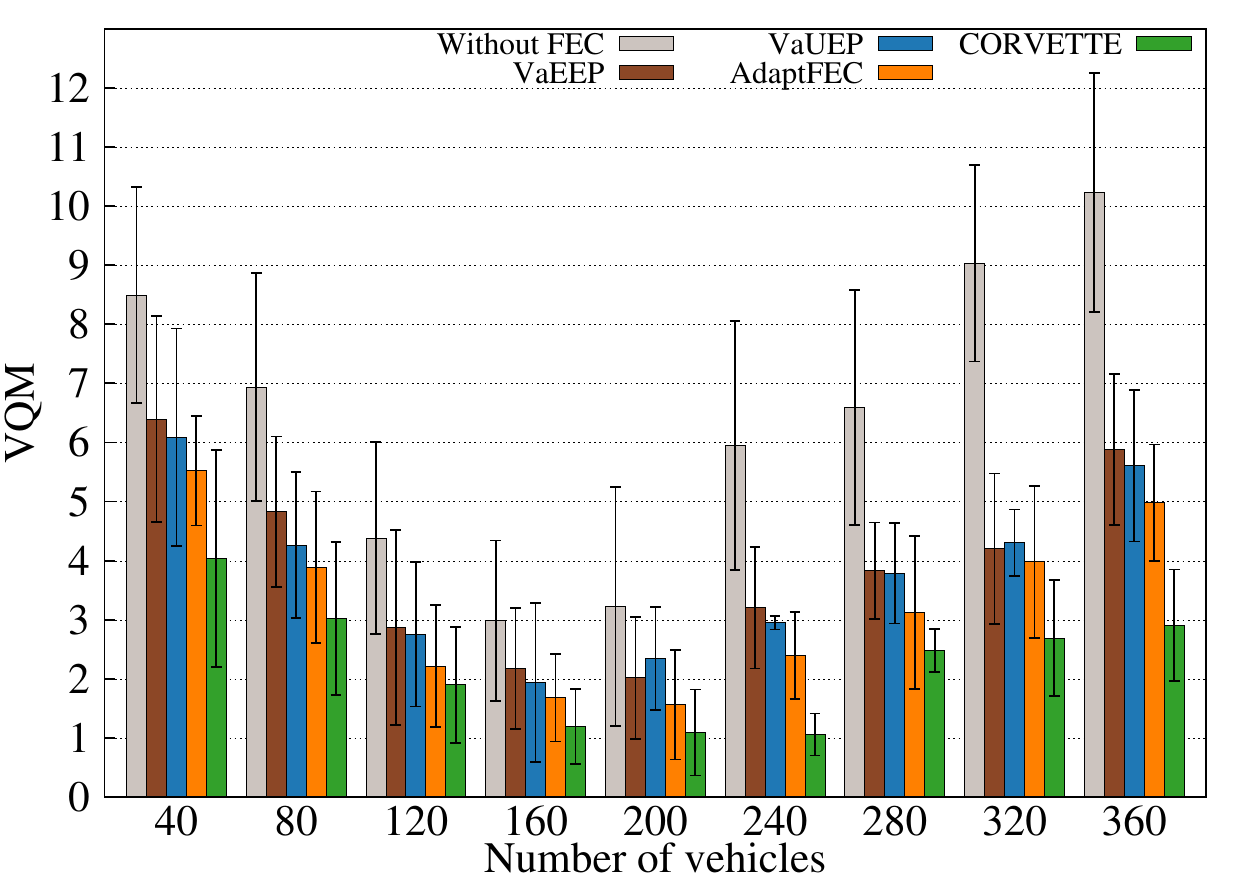}
	\end{center}
	\vspace{-0.0in}
	\caption{\small Average VQM for all mechanisms}
	\label{fig:corv:qoeVQM}
	\vspace{-0.0in}
\end{figure}

Figure~\ref{fig:corv:qoeVQM} presents the VQM scores. Almost the same pattern found in the SSIM metric is also present here. When the network is sparse, the videos tend to have lower quality, especially without any FEC-based mechanism. The best-case scenario in the experiments is between 160 and 240 vehicles. In this range, the videos were transmitted with better quality. Similarly to the SSIM assessment, using VQM the CORVETTE also outperformed all other mechanisms. Once again, this proves that the transmission enhancements performed by the proposed mechanism lead to a better video quality for the end-users. On average, the CORVETTE mechanism provided 65\% better video quality than the scheme without FEC, 42\% and 40\% higher scores than VaEEP and VaUEP, respectively. Against the AdaptFEC scheme, the proposed mechanism presented 30\% better results.

\subsubsection{Network footprint analysis}

\begin{figure}[!htb]
	\vspace{-0.0in}
	\begin{center}
		\includegraphics[width=4.5in]{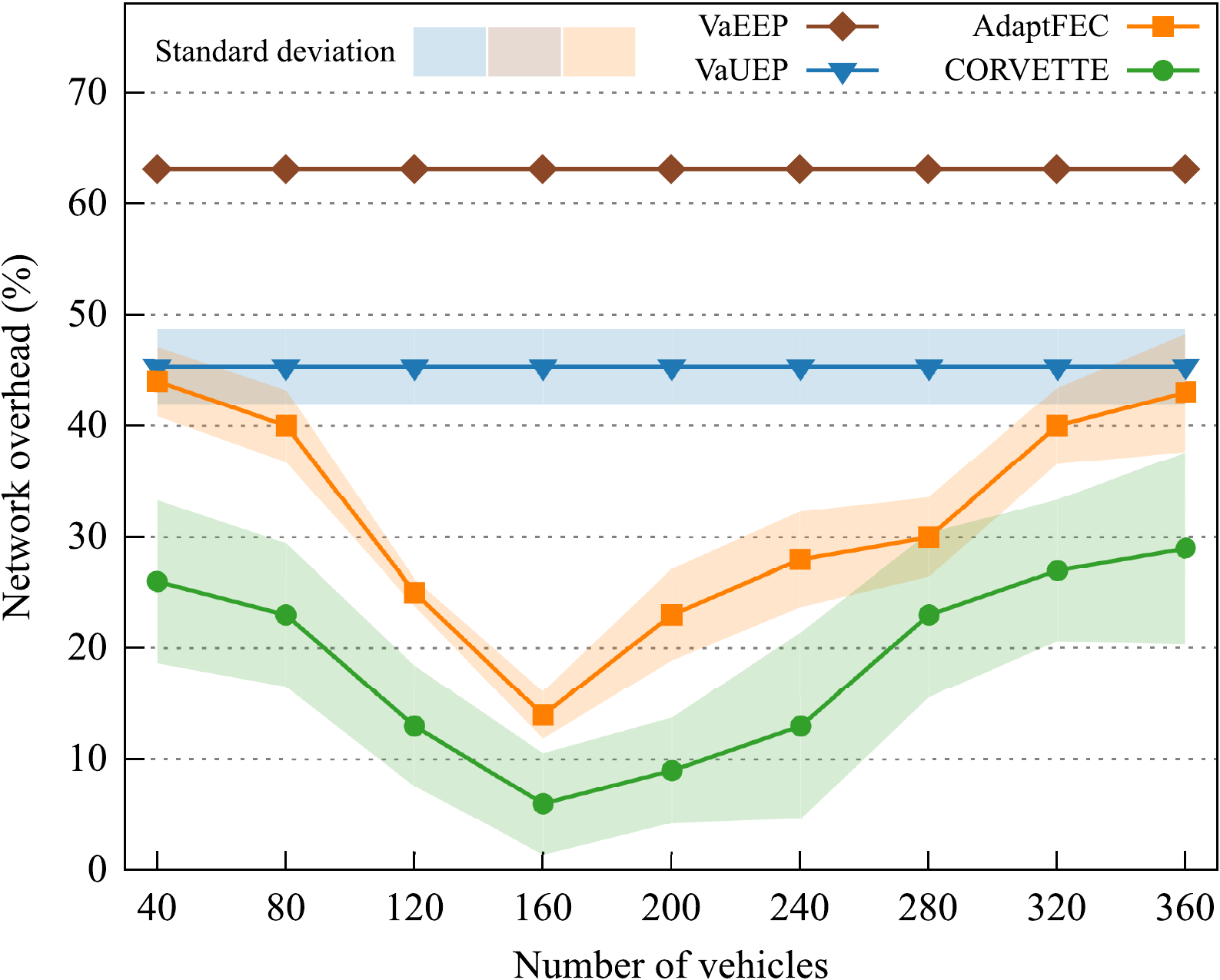}
	\end{center}
	\vspace{-0.0in}
	\caption{\small Average network overhead}
	\label{fig:corv:netOverhead}
	\vspace{-0.0in}
\end{figure}

Apart from higher video quality, a lower network overhead is equally desirable, which is shown in Figure~\ref{fig:corv:netOverhead}. The VaEEP and VaUEP schemes are non-adaptive, thus producing a constant network overhead through the experiments. As can be seen in the graph, the non-adaptive protection schemes produce a large amount of network overhead, especially in the VaEEP case. The VaEEP protection is also not very efficient because it shields certain parts that do not require protection. 
Therefore, the VaUEP scheme takes into consideration the video characteristics, namely the frame type, and adds a specific amount of redundancy to each one. This grants the reduction of the network footprint while improving the video quality. 
Unlike the VaEEP, the graph shows a standard deviation for the VaUEP values because different video sequences require distinctive amounts of redundancy. 

On the other hand, the adaptive mechanisms~(AdaptFEC and CORVETTE) yielded better results in terms of network overhead, allowing a proper usage of the wireless resources. Another important difference is the standard deviation, where the CORVETTE mechanism has higher values. In this case, this is very desirable, because it means that the proposed mechanism was able to better identify the most important portions of the video sequences and protect them accordingly. This demonstrated a tailored protection, providing both superior video quality and low network overhead. On average, the CORVETTE mechanism added 41\% less overhead than AdaptFEC. When compared to the VaEEP and VaUEP schemes, the proposed mechanism was able to generate 70\% and 58\% less overhead, respectively.

\subsubsection{Overall results}

Table~\ref{tab:corv:Sumary} summarises the average SSIM, VQM, and network overhead scores. It shows that the CORVETTE mechanism was able to provide a substantial reduction in network overhead. This was possible by adding a specific amount of redundancy to each video sequences, and thus, avoiding any unnecessary redundancy. Additionally, it did so while increasing the video quality, leading to higher QoE for end-users.

\begin{table}[!h]
	\vspace{-0.0in}
	{\small
		\caption{Average SSIM, VQM, and network overhead}
		\vspace{-0.0in}
		\begin{center}
				\begin{tabular}{l|c|c|c|c|c}
					\hline
					& \multicolumn{1}{|l}{\textbf{CORVETTE}} & \multicolumn{1}{|l}{\textbf{AdaptFEC}} & \multicolumn{1}{|l}{\textbf{VaUEP}} & \multicolumn{1}{|l}{\textbf{VaEEP}} & \multicolumn{1}{|l}{\textbf{Without FEC}} \\
					\hline
					\hline
					{SSIM}     & 0,876 & 0,785                                 & 0,734                                     & 0,709         & 0,591  \\
					\hline
					{VQM}      & 2,266 & 3,264                                  & 3,783                                     & 3,935         & 6,425   \\   
					\hline
					{Overhead} & 18,777\% & 31,888\%       & 45,342\%                                  & 63,102\%               & --                                  \\   
					\hline
				\end{tabular}
			\label{tab:corv:Sumary}
		\end{center}
	}
	\vspace{-0.0in}
\end{table}

Taking everything into consideration, the CORVETTE mechanism showed results with better video quality while reducing considerably the network overhead. 
This is only possible because it performs an accurate motion intensity classification, which is used along with other video and network characteristics. The combined use of all this information enables the mechanism to add an adequate protection to each video sequence improving the quality for end-users.

\section{Self-adaptive Mechanism to Shield Video Transmissions~(SHIELD)}
\label{sec:shield}

The CORVETTE mechanism was able to attain encouraging results, notwithstanding, there was still room for further improvement. 
For example, more parameters could be considered to enhance the reliability of the network conditions assessment, as well as, a reduced number of operations could be possible by changing the convex hull approach. 
To address these open issues, the self-adaptive FEC-based proactive error recovery mechanism to shield video transmissions over VANETs~(SHIELD) was proposed. 
The main enhancements are described next.

\subsection{SHIELD Overview}

The aim of the mechanism proposed in this section is to offer videos with higher QoE to video-equipped vehicles while, at the same time, downsizing the network overhead footprint. 
To this end, SHIELD extends the CORVETTE mechanism by using several video characteristics and specific VANETs details to safeguard real-time video streams against packet losses. 
This means that meaningful video aspects related to the human point-of-view, are not neglected, which leads to the addition of a very specific amount of redundancy. 
One of the main contributions of this work over the CORVETTE mechanism is the combined used of network density, SNR, PLR, and the vehicle's position.
This allows SHIELD to better protect the video sequences and enhance the QoE.

\subsection{Towards the design of SHIELD}

\begin{figure}[!htb]
	\begin{center}
		\ifBW \includegraphics[width=102mm]{./hfuzzy_v6_SHIELD_MOD_gray-eps-converted-to.pdf}
		\else \includegraphics[width=102mm]{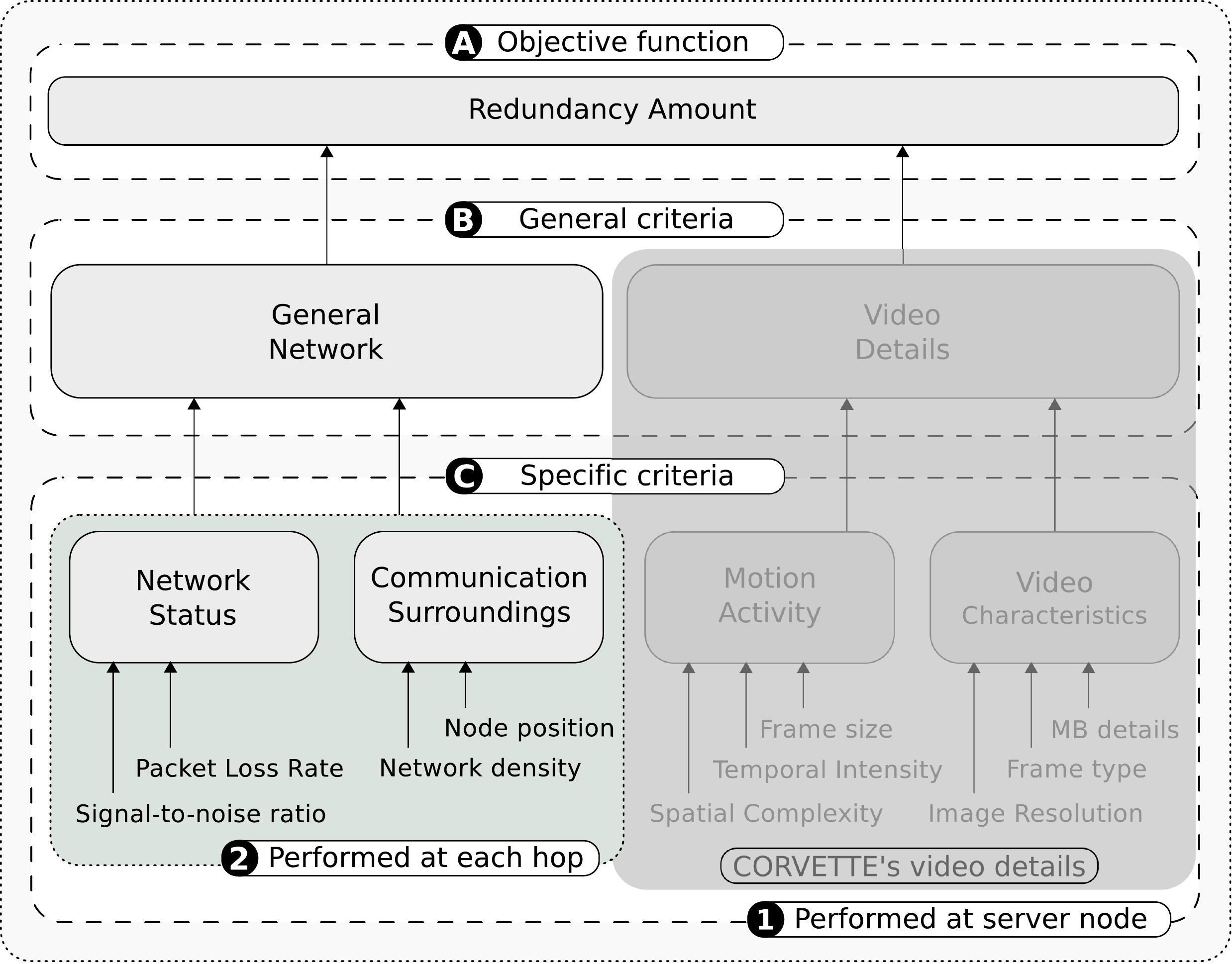}
		\fi
	\end{center}
	\caption{\small SHIELD HFS structure}
	\label{fig:shield:hfuzzy}
\end{figure}

The SHIELD mechanism is an improvement over the CORVETTE work presented in Section~\ref{sec:corvette}. In the same way as presented before, it needs a knowledge database which is populated before the mechanism execution.
Following the same basic core structure as before, SHIELD also employs a HFS. This allows it to reduce the global number of rules by arranging them in a hierarchical form, thus, making it possible to be performed in real-time. Despite the fact that the same HFS structure was used, all rules and sets regarding the network parameters were overhauled. On the other hand, the video related details were kept the same. Figure~\ref{fig:shield:hfuzzy} depicts the hierarchical levels adopted by SHIELD, as well as the components that were upgraded. The main improvements are detailed below.

There are three layers in SHIELD HFS, namely (A)~Objective function, (B)~General criteria, and (C)~Specific criteria. The output of each low-level layer is used as input to the next layer. The first layer~(A) represents the amount of redundancy that SHIELD will add to a specific portion of video data. The main goal is to find the amount of redundancy for a system that, given its constraints, results in less network overhead and better QoE. The second layer~(B) encompasses the overall details that the proposed mechanism uses to determine the redundancy amount, namely the network details and the video characteristics. The bottom layer~(C) is responsible for handling the input parameters of each feature used by the fuzzy logic inference system. This layer has a subdivision~(C)(2), which is performed at each network hop. All the input parameters~(C)(1) are only taken into consideration at the server node.

The ``General Network'' criteria account for the definition of the network conditions. This component was completely upgraded in the new mechanism. All the hierarchical rules were rewritten, the input parameters were modified, as well as a new one was added, namely the SNR.
This enables the mechanism to use more metrics to better establish a network quality indicator. In total, the SHIELD mechanism uses 4 metrics. These metrics are divided into two specific criteria, namely ``Network status'' and ``Communication surroundings''. The former is defined by the combined assessment of the SNR and the PLR. The latter is given by the network density and the position of the vehicles. Each one of these metrics is described next.

The SNR is the level of the desired signal against the level of background noise. This is a good indicator for the physical medium, especially for spectrum sensing. While this is true, it cannot be considered a reliable general network quality indicator by itself. This steams from the fact that a strong channel signal will not always produce a good network connection~\cite{Vlavianos2008}. On the other hand, a very weak signal will yield a low-quality network connection. As a result, to create a more holistic indicator more than one metric has to be used. Another obvious candidate to define the network quality is the PLR. In general, the SNR and PLR have a negative correlation, meaning that when one increases, the other decreases and vice versa. However, they complement each other because the SNR takes into consideration the physical spectrum part of the transmission and the PLR provides a point of view closer to the application layer.

The PLR input set was also redesigned. 
In the same way as before several experiments were carried out to find the relationship between the losses and QoE, using a more comprehensive number of video sequences.
The results made evident that it is possible to have a good QoE with packet loss between 0\% and 12\%, a small difference from CORVETTE's PLR input set~(between 0\% and 11\%), yet a step forward to better characterise the network conditions.
In most of the cases, an acceptable video quality for end-users was perceived with losses from 5\% up to 23\%. Here again, another small difference in the found values, which in CORVETTE case were between 5\% and 22\%.
Additionally, after a threshold of 19\%~(was 17\% in the CORVETTE mechanism) the video quality starts to decrease apace, particularly in videos with high resolution and motion intensity. 
In the experiments with more than 36\%~(34\% in CORVETTE implementation) of PLR, the QoE reached intolerable levels. 
Algorithm~\ref{algo:shield:PLRset} shows only one of the many fuzzy sets defined in the SHIELD mechanism. 

\iflatextortf
\else
\SetKwProg{Fn}{}{}{}
\begin{algorithm}[!htb]
	{\small
		\Fn{InputLVar* \textbf{PLR} = new InputLVar("\textbf{PacketLossRate}");}{
			PLR $\rightarrow$ addTerm( TriangularTerm("\textit{LOW}", 0, 12))\;
			PLR $\rightarrow$ addTerm( TriangularTerm("\textit{MEDIUM}", 5, 23))\;
			PLR $\rightarrow$ addTerm( TriangularTerm("\textit{HIGH}", 19, 100))\; %
		}
		engine.addInputLVar(\textbf{PLR})\;
	}
	\caption{SHIELD PLR input set}
	\label{algo:shield:PLRset}
\end{algorithm}
\fi

Another component of the ``General network'' criteria is the ``Communication surroundings''. 
This component also was enhanced during SHIELD design. It uses the network density and the position of the vehicles to provide more information about the network in which the video sequences are being transmitted. These parameters are updated at each beacon exchange in the routing protocol. The network density is given by the number of nodes, in this case vehicles, divided by the network area. It is important to notice that VANETs are very dynamic networks with a decentralised structure, proving to be a challenge for the estimation of the network surface area. To address this issue, the proposed mechanism uses an approximate convex hull algorithm.

\begin{figure*}[!htb]
	\begin{center}
		\begin{tabular}{cc}
			\includegraphics[width=69mm]{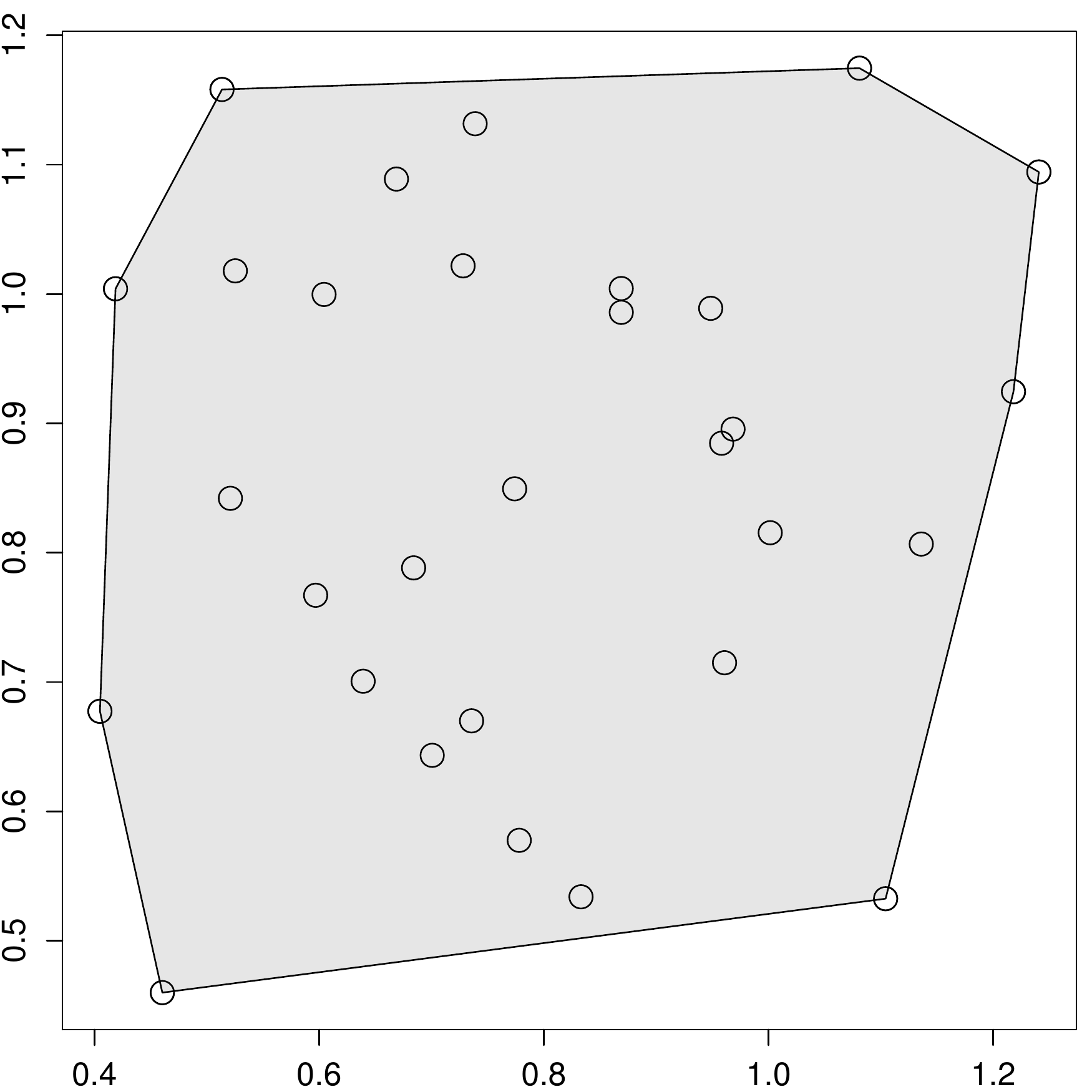} &
			\includegraphics[width=69mm]{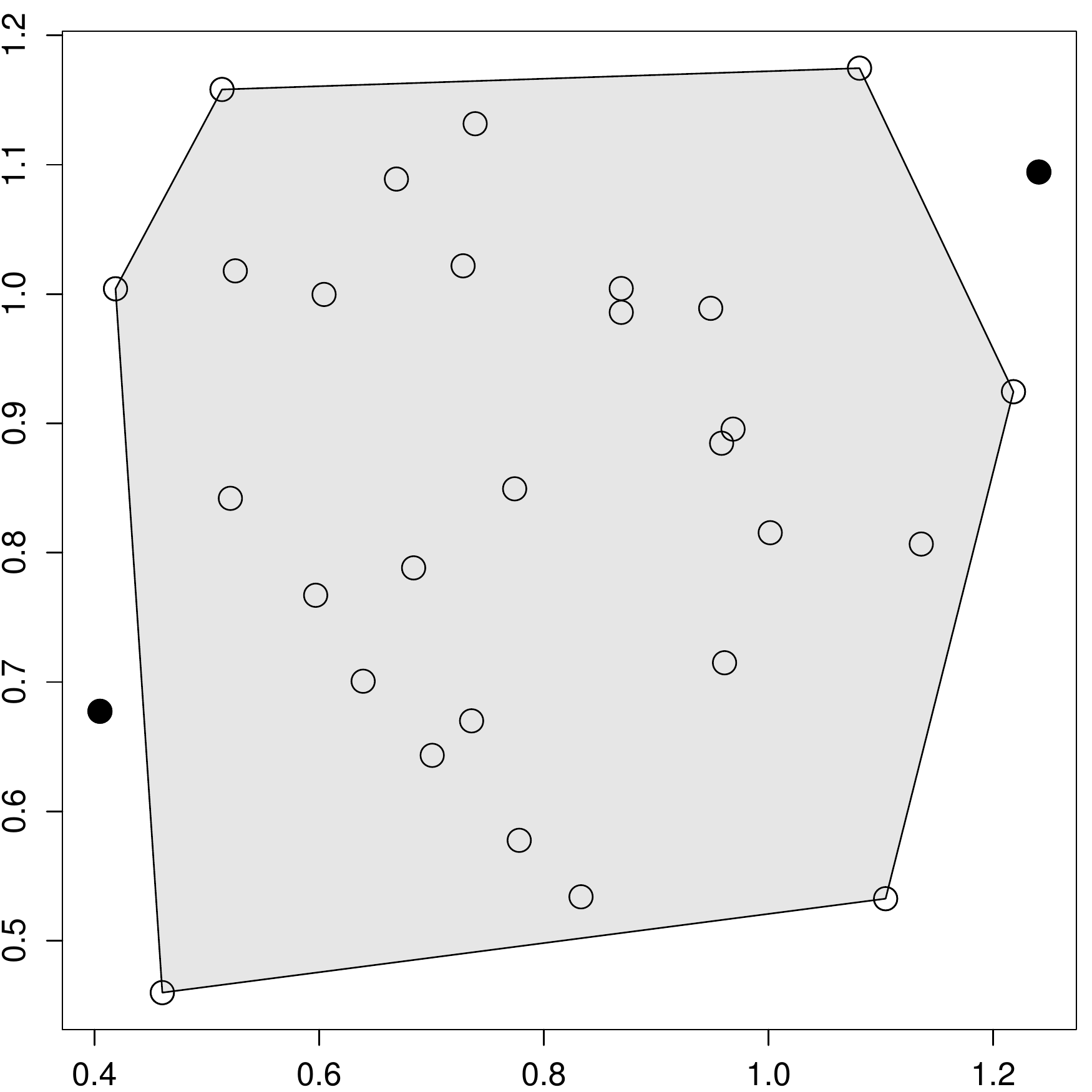} \\
			(a) Convex Hull & (b) Approximate Convex Hull
		\end{tabular}
		\caption{\small Convex Hull and Approximate Convex Hull }
		\label{fig:shield:convexhull}
	\end{center}
\end{figure*}

A convex hull algorithm is able to find the smallest boundary polygon containing all the points inside of it, using only non-intersecting segments, as showed in Figure~\ref{fig:shield:convexhull}(a). There are several algorithms to find the convex hull of a given set of points. 
In the CORVETTE mechanism, the QuickHull method was used. It uses a divide-and-conquer algorithm with average complexity of $\mathcal{O}(n \log{}n)$ and at the worst case, it could take $\mathcal{O}(n^2)$. 
However, it is not imperative for the mechanisms to use high-precision value for the area size, instead, a good approximation is sufficient to provide very good results. 
Owing to this, and to improve the general performance, SHIELD uses the Bentley Faust Preparata~(BFP)~\cite{Bentley1982} approximation convex hull algorithm as showed in Figure~\ref{fig:shield:convexhull}(b).

The BFP algorithm, which runs in $\mathcal{O}(n)$ time, replaces the sort operation by dividing the plane into vertical strips. In each strip, the minimum and maximum points are found and added to the boundary. This algorithm is an approximation because a non-extreme point, in a given strip, can be discarded even if it is on the convex hull boundary. Nevertheless, the point will not be far from the convex hull, resulting in a good approximation of the actual convex hull.

\begin{figure}[!htb]
	\begin{center}
		\ifBW \includegraphics[width=100mm]{./Fig5_shield_gray-eps-converted-to.pdf}
		\else \includegraphics[width=100mm]{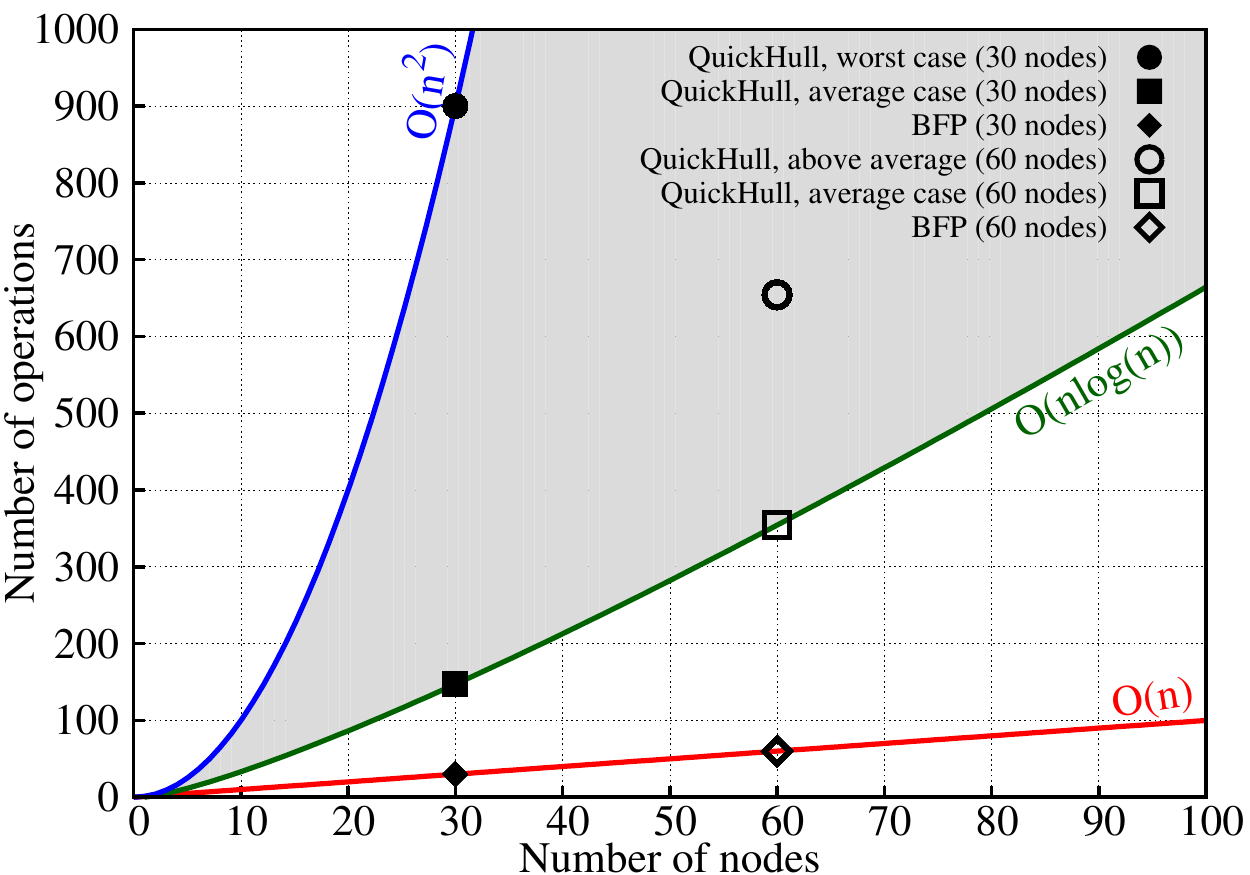}
		\fi
	\end{center}
	\caption{\small Complexity of QuickHull and BFP}
	\label{fig:shield:bigO}
\end{figure}

Figure~\ref{fig:shield:bigO} shows the comparison between the number of nodes and the resulting amount of operations in both QuickHull and BFP algorithm. On average, the QuickHull algorithm has fairly good performance, it can degrade, however, up to exponential in the worst case. On the contrary, the BFP algorithm has a steady linear performance, providing results more quickly even with a small number of nodes.

There is a multitude of advantages of performing fewer operations. First of all, due to the time-sensitive video data, it is important that the client node receives the information as soon as possible, thus performing fewer operations allows dispatching the video quickly. In addition, because of the fast time-varying network conditions, the faster this information is made available the more accurate it is. At last, performing a minimum number of operations means less energy consumption as well as more available processor power to perform other tasks.

Another noteworthy betterment in SHIELD is the approximation of the node position and the network density inputs. The latter parameter was moved from the ``Network status'' component to the ``Communication surroundings''. 
In doing that, a different set of rules can now be created to better correlate the movement of the nodes and the number of nodes with 1-hop distance in a given area.
These parameters provide more accurate results when analysed together than independently, enabling SHIELD to yield a precise network assessment.

In short, the adoption of another input~(SNR) and the adjustment of the ``Communication surroundings'' component led to a complete rearrangement of the ``General network'' module.  
Which in terms, led better-consolidated rules that express with higher precision the needs of the proposed mechanism. 
At the end, this new ``General network'' layer presented is responsible for the better integration of SHIELD in VANETs. In other words, it allows the proposed mechanism to ascertain the best-fitted QoE-aware amount of redundancy according to each video sequence.

\subsection{SHIELD Performance Evaluation and Results}
\label{sec:shield:evaluation}

The main goal of the SHIELD mechanism is to enhance the QoE while avoiding any unnecessary network overhead. In doing that, it improves end-users satisfaction and preserves the already scarce wireless resources at the same time.

\subsubsection{Experiment settings}

In order to better characterise the performance of the proposed mechanism two distinct environments were assessed: urban and highway. Each of these surroundings features a variety of unique challenges. 
In the urban environment, there are buildings and many other structures that will affect the signal propagation. 
On the other hand, in the highway environment, there is much more free space, which facilitates the signal propagation. Besides that, the mobility patterns are also very distinctive. 
The urban scenario presents a lot of driving options, such as avenues and streets close to each other. 
On the highway is quite different, as there are no crossroads and just a few exits and entrances. In addition, the speed of the vehicles has very particular properties in each one of these environments. 
In the urban case, the velocity usually is between 20~km/h and 60~km/h, and it changes frequently due to traffic lights, speed bumps, and crosswalks. Meanwhile, on the highway, the speed variance is very low, generally staying from 80~km/h to 120~km/h.

Several configurations are shared between the two environments, such as the wireless and network technology~(IEEE 802.11p WAVE with V2V and CLWPR routing protocol), as well as the video content and codification parameters. All videos were sent using Evalvid Tool and encoded with H.264, GoP length of 19:2. Additionally, three different resolutions were used, namely 1080p, 720p, and SVGA. For each resolution, 10 videos were chosen to be transmitted~\cite{Xiph.org}. A multi-flow scenario was adopted. This means that up to 10 videos are transmitted simultaneously\footnote{Samples of the transmitted videos are available in \url{http://www.youtube.com/channel/UCsB0SdKpCKD2GS6aXzB-FUQ/videos}}. All the receiver nodes are enabled with Frame-Copy error concealment.

The mobility traces for both environments were generated using SUMO. For the urban environment, a clipping of 2 by 2 km of the Manhattan borough~(New York City) was used. This environment was simulated with up to 360 vehicles at speeds ranging from 20~km/h and 60~km/h. To simulate this environment a clipping of 10 km of US Interstate Highway 78~(I-78) was used. The number of vehicles is the same, up to 360, with a velocity between 80~km/h and 120~km/h. 

Two different propagation models were used to better represent each environment. In the highway scenario, the logDistance propagation model was used~\cite{Mittag2011}. This is because of the open spaces and the reduced number of sources of interference existent in this environment. This leads to easier communication between the nodes. On the other hand, in the urban environment, there are plenty of sources generating interference. 
In view of this, the Nakagami-m propagation model was added on the top of the logDistance model.
This allows simulating the fast fading characteristics commonly found in this environment~\cite{Taliwal2004}. Table~\ref{tab:shield:parameters} summarises the simulation parameters.

\begin{table}[!ht]
	{ \small
		\caption{Simulation parameters}
		\begin{center}
				\begin{tabular}{l|l}
					\hline 
					\textsc{\textbf{Parameters}} & \textsc{\textbf{Value}} \\ 
					\hline
					\hline Display sizes & 1920x1080, 1280x720, and 800x600\\
					\hline Frame rate mode & Constant\\
					\hline Frame rate & 29.970 fps\\
					\hline GoP & 19:2 \\ 
					\hline Video format & H.264\\
					\hline Codec & x264 \\ 
					\hline Container & MP4 \\
					
					\hline Wireless technology & IEEE 802.11p (WAVE)\\
					\hline Communication &  Vehicle To Vehicle~(V2V) \\
					\hline Routing protocol & CLWPR \\
					\hline Mobility & SUMO mobility traces \\
					\hline Radio range & 250m\\
					\hline Internet layer & IPv6\\
					\hline Transport layer & UDP\\
					\hline
					
					\multicolumn{2}{c}{\textbf{Highway environment}} \\
					\hline
					\hline Propagation model & logDistance \\
					\hline Location & I-78\\
					\hline Map size & 10.000 m\\
					\hline Vehicles speed & 80-120 km/h (50-75 mph) \\
					\hline
					
					\multicolumn{2}{c}{\textbf{Urban environment}} \\
					\hline
					\hline Propagation model & logDistance + Nakagami-m \\
					\hline Location & Manhattan borough(New York City)\\
					\hline Map size & 2.000 m x 2.000 m\\
					\hline Vehicles speed & 20-60 km/h (12-37 mph) \\
					\hline
					
				\end{tabular}
			\label{tab:shield:parameters}
		\end{center}
	}
\end{table}

\begin{figure*}[!htb]
	\begin{center}
		\includegraphics[width=144mm]{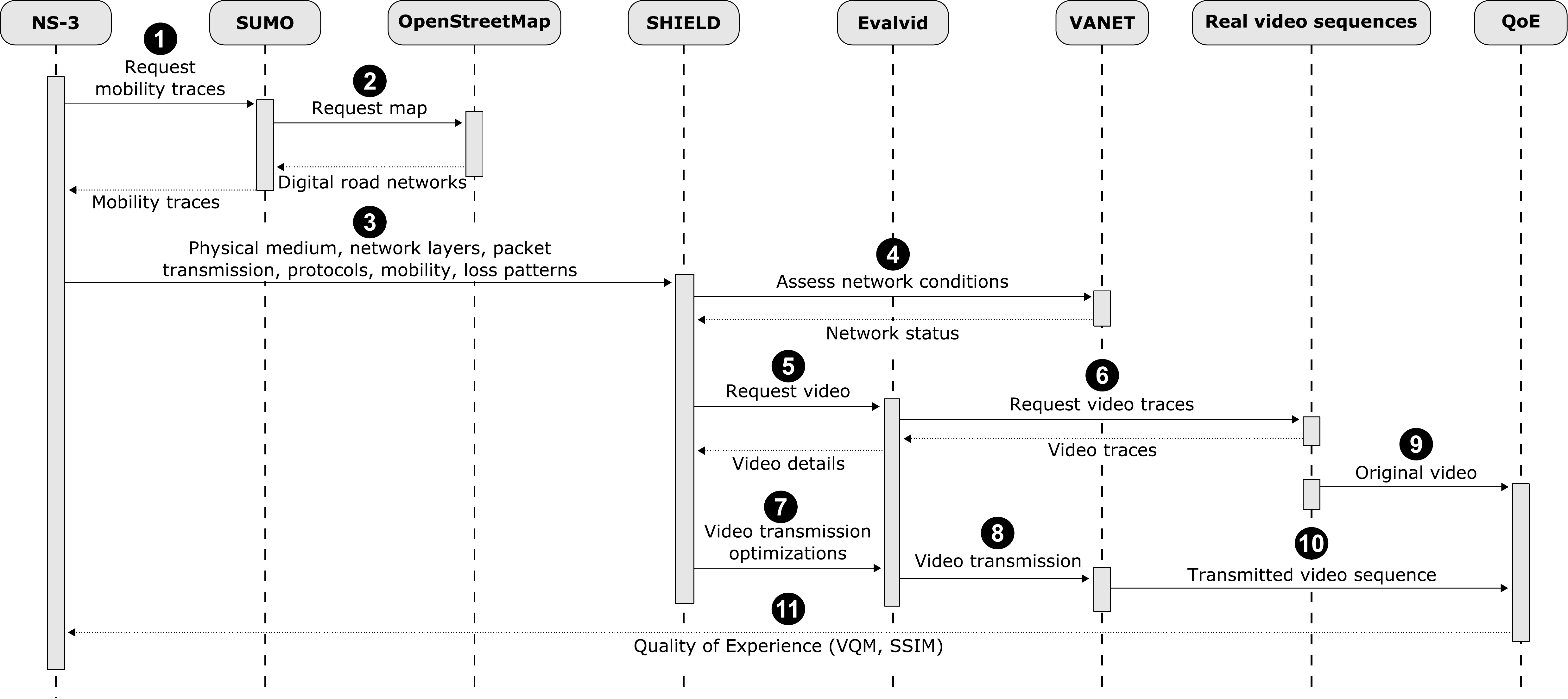}
	\end{center}
	\caption{Steps involved in the experiment}
	\label{fig:shield:sequenceDiagram}
\end{figure*}

Figure~\ref{fig:shield:sequenceDiagram} shows the steps involved in the experiment. First of all, the mobile traces are required from the SUMO application~(1). After that, SUMO will use real map clippings from the OpenStreetMap~(2) to generate the traces. The traces enable a realistic simulation, providing more accurate results. Following this, all the information is loaded in the SHIELD mechanism~(3). Next, the proposed mechanism will assess the network conditions~(4) and request the video to be transmitted~(5). Real video sequences are used in the experiment~(6). Afterwards, the SHIELD mechanism optimises and secures the video transmission against packet loss~(7). The next step is to deliver the video sequences to the receiver~(8). At the end, the original~(9) and the transmitted~(10) videos are assessed using objective QoE metrics~(11).

Five different scenarios were assessed in both urban and highway environments. The first one is without any type of FEC. The results of this experiment will be used as a baseline for the others. The second scenario is the Video-aware Equal Error Protection FEC~(VaEEP) mechanism. The Video-aware Unequal Error Protection FEC~(VaUEP) mechanism is the third scenario. The fourth scenario is using the adaptive QoE-driven COntent-awaRe VidEo Transmission opTimisation mEchanism~(CORVETTE), presented in Section~\ref{sec:corvette}. The fifth and last scenario is the proposed SHIELD mechanism.

\subsubsection{QoE assessments}

\begin{figure*}[!htb]
	\begin{center}
		\setlength\tabcolsep{0.5pt}
		\begin{tabular}{cc}
			\ifBW \includegraphics[width=72.5mm]{./Fig7_shield_gray-eps-converted-to.pdf} 
			\else \includegraphics[width=72.5mm]{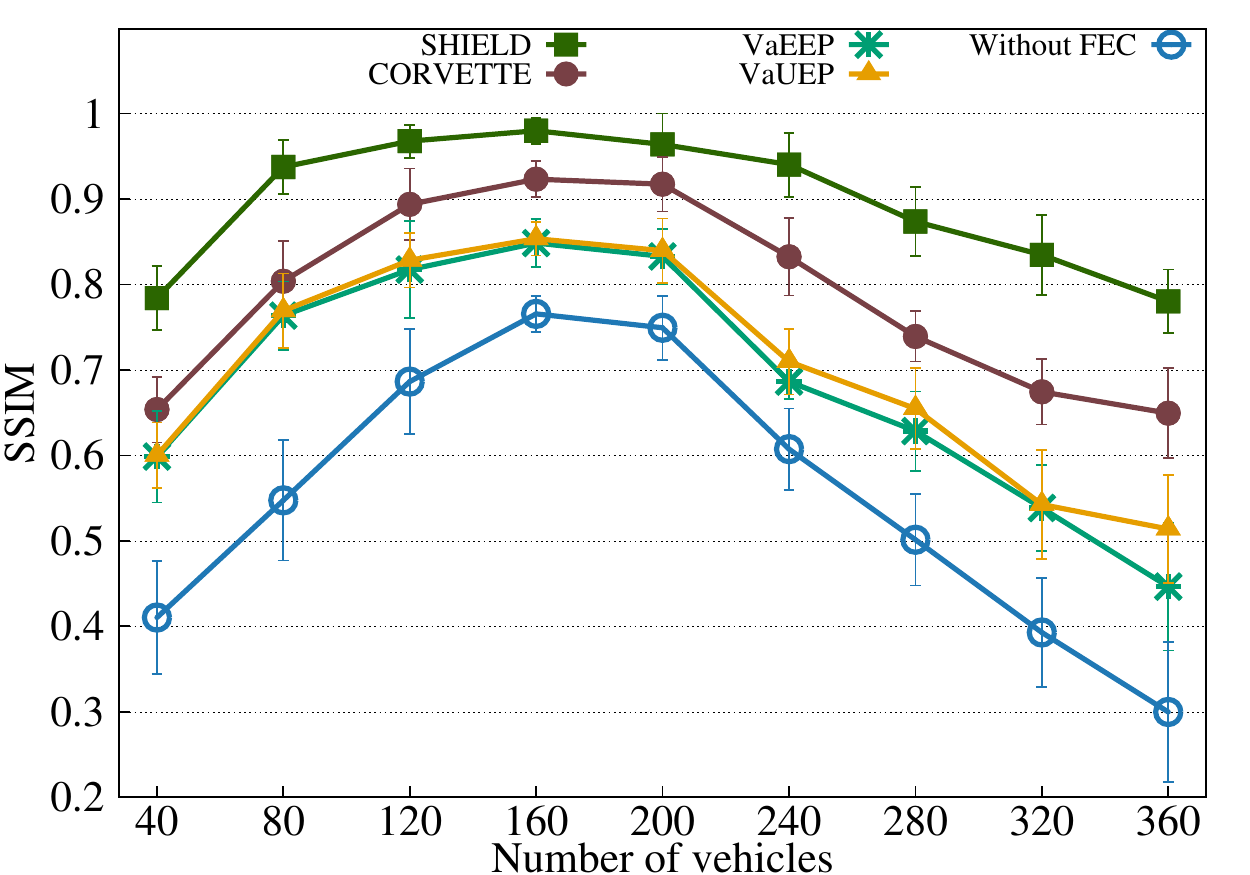}
			\fi &
			\ifBW \includegraphics[width=72.5mm]{./Fig8_shield_gray-eps-converted-to.pdf} 
			\else \includegraphics[width=72.5mm]{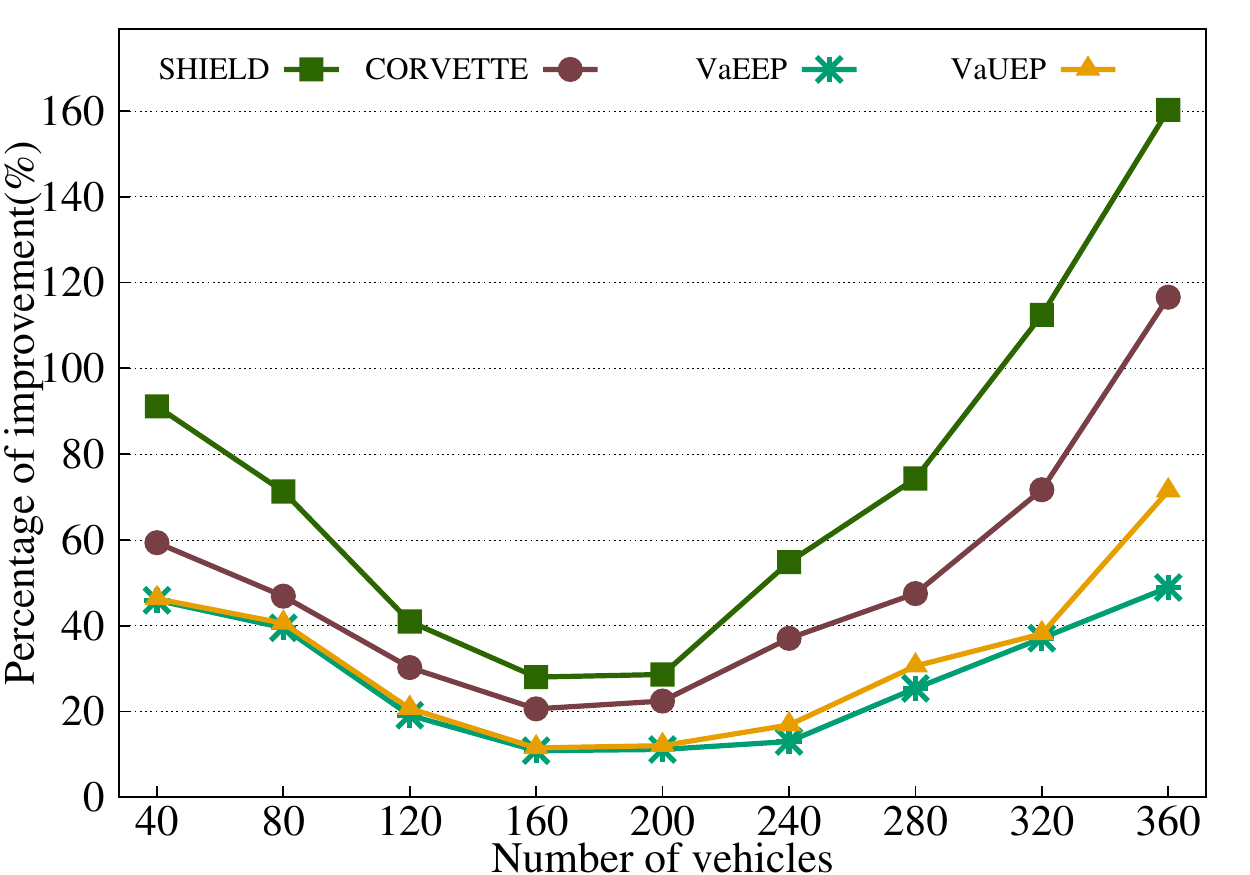}
			\fi \\
			(a) Average SSIM & (b) Percentage of SSIM improvement \\
		\end{tabular}
		\caption{\small SSIM assessment of the urban environment}
		\label{fig:shield:SSIMUrban}
	\end{center}
\end{figure*}

Figure~\ref{fig:shield:SSIMUrban} shows QoE results of the urban scenario.~(a) depicts the SSIM average and~(b) shows the QoE improvement in comparison to the base line. 
In~(a), it is possible to notice that the simulation starts with a few vehicles and the QoE results, for all mechanisms, can be considered low. This can be credited to the fact that the network is suffering from connectivity issues because it is relying on very few and scattered nodes to transmit all the video data. Even in this scenario, the SHIELD mechanism was able to protect the most important parts of the video sequences, producing better results. As showed in~(b), this led to an improvement of more than 90\% on the video quality when compared to the baseline~(without FEC). The second best result was the CORVETTE mechanism with 60\% of SSIM improvement.

Figure~\ref{fig:shield:SSIMUrban}(a) also shows that the best QoE results for all mechanisms are obtained when the network has 160 and 200 vehicles. This number of nodes provides the best coverage of the whole area, while it does not cause excessive interference. Because of the improved network conditions, the baseline also has better results, thus reducing the SSIM improvement perceived by the other mechanisms. This situation is clearly evidenced in Figure~\ref{fig:shield:SSIMUrban}(b) for 160 and 200 vehicles. On the other hand, when the network becomes very dense, e.g., above 280 vehicles, the mechanisms have to face increasingly degraded network connections. Once again, the SHIELD surpassed the other mechanisms, providing up to 160\% higher SSIM scores in comparison to the baseline.

\begin{figure*}[!htb]
	\begin{center}
		\setlength\tabcolsep{0.5pt}
		\begin{tabular}{cc}
			\ifBW \includegraphics[width=72.5mm]{./Fig9_shield_gray-eps-converted-to.pdf} 
			\else \includegraphics[width=72.5mm]{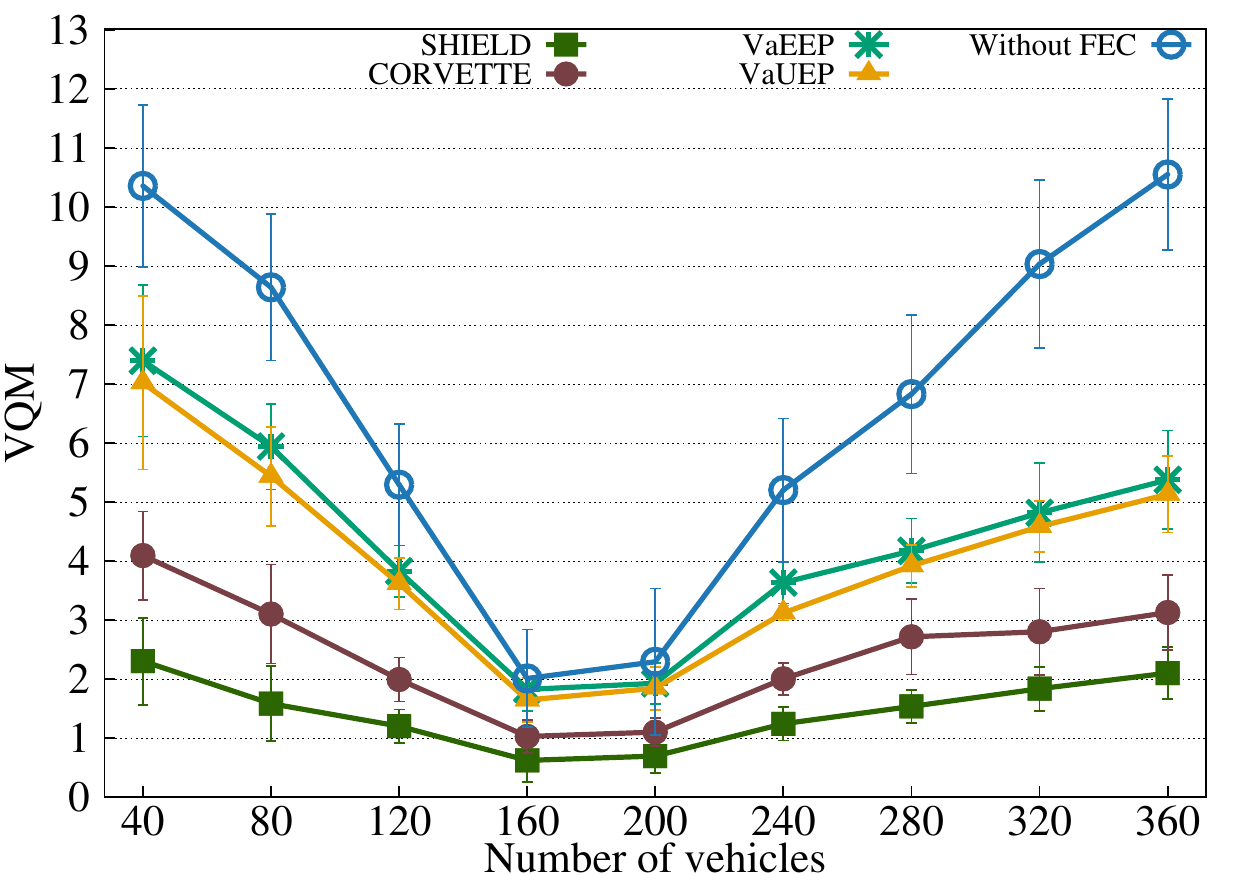}
			\fi &
			\ifBW \includegraphics[width=72.5mm]{./Fig10_shield_gray-eps-converted-to.pdf} 
			\else \includegraphics[width=72.5mm]{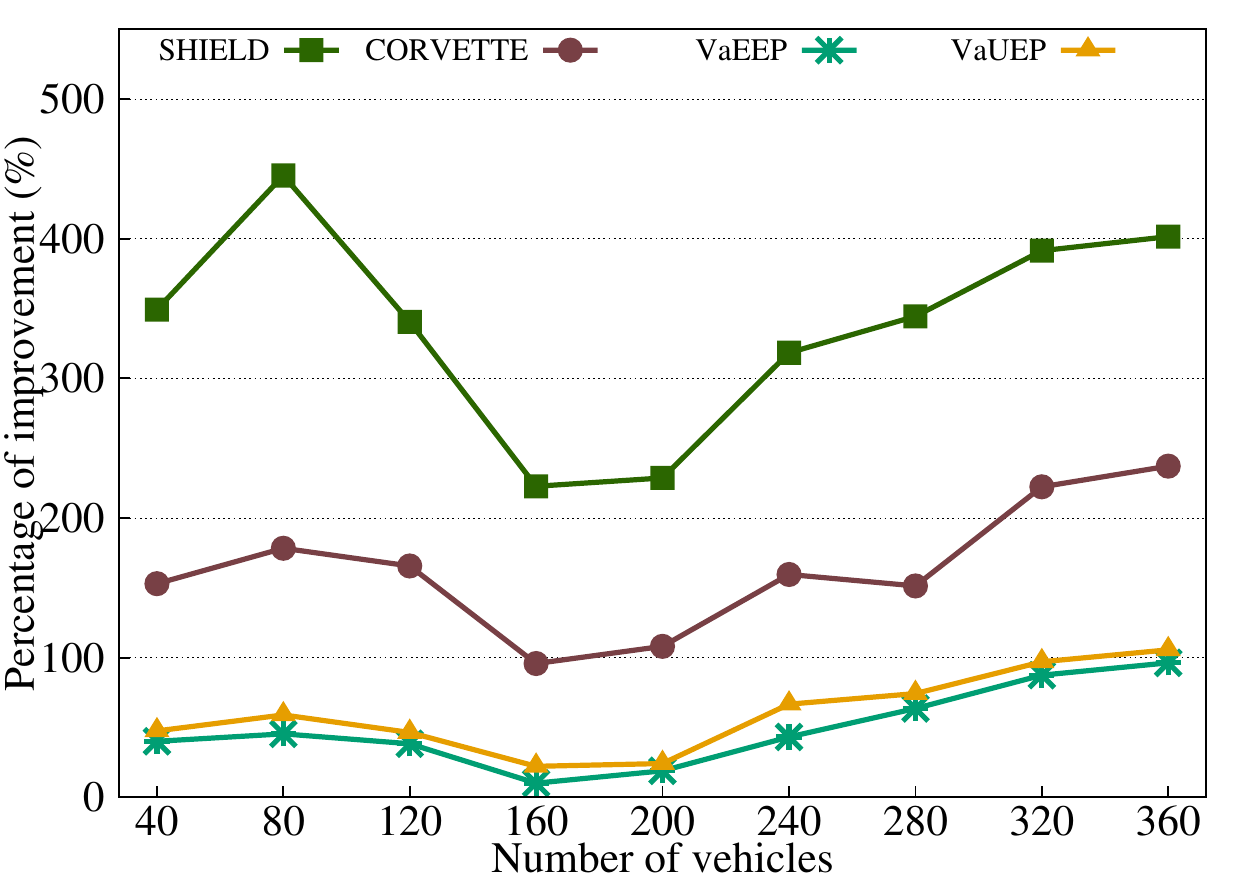}
			\fi \\
			(a) Average VQM & (b) Percentage of VQM improvement			
		\end{tabular}
		\caption{\small VQM assessment of the urban environment}
		\label{fig:shield:VQMUrban}
	\end{center}
\end{figure*}

Furthermore, Figure~\ref{fig:shield:VQMUrban}(a) presents the VQM average and~(b) depicts the percentage of QoE improvement of the mechanisms in comparison to the baseline. Although this metric differs from SSIM, almost the same pattern can be found in~(a). At the beginning of the experiment, the network is sparse and the videos have low quality. VQM gives them high scores, which in this case are not good. This is especially true for the baseline because it does not use any type of FEC-based mechanism to secure the transmissions. The best-case scenario in the VQM scores is the same as in the SSIM results, for 160 and 200 vehicles. This confirms the notion that the videos are transmitted with better quality with this configuration. In the same way as in the SSIM assessment, the VQM scores demonstrate that the SHIELD mechanism outperforms all other mechanisms. 

Additionally, Figure~\ref{fig:shield:VQMUrban}(b) shows a pattern similar to the SSIM results. The highest improvements are accomplished when the network is sparse, between 40 and 120 vehicles, or in a very dense network, above 240 vehicles. On average, the proposed mechanism provided scores 78\% better than the baseline. Additionally, it achieved 66\% and 63\% higher marks than VaEEP and VaUEP, respectively, and over 40\% better scores in comparison to the CORVETTE mechanism.

\begin{figure*}[!htb]
	\begin{center}
		\setlength\tabcolsep{0.5pt}
		\begin{tabular}{cc}
			\ifBW \includegraphics[width=72.5mm]{./Fig11_shield_gray-eps-converted-to.pdf} 
			\else \includegraphics[width=72.5mm]{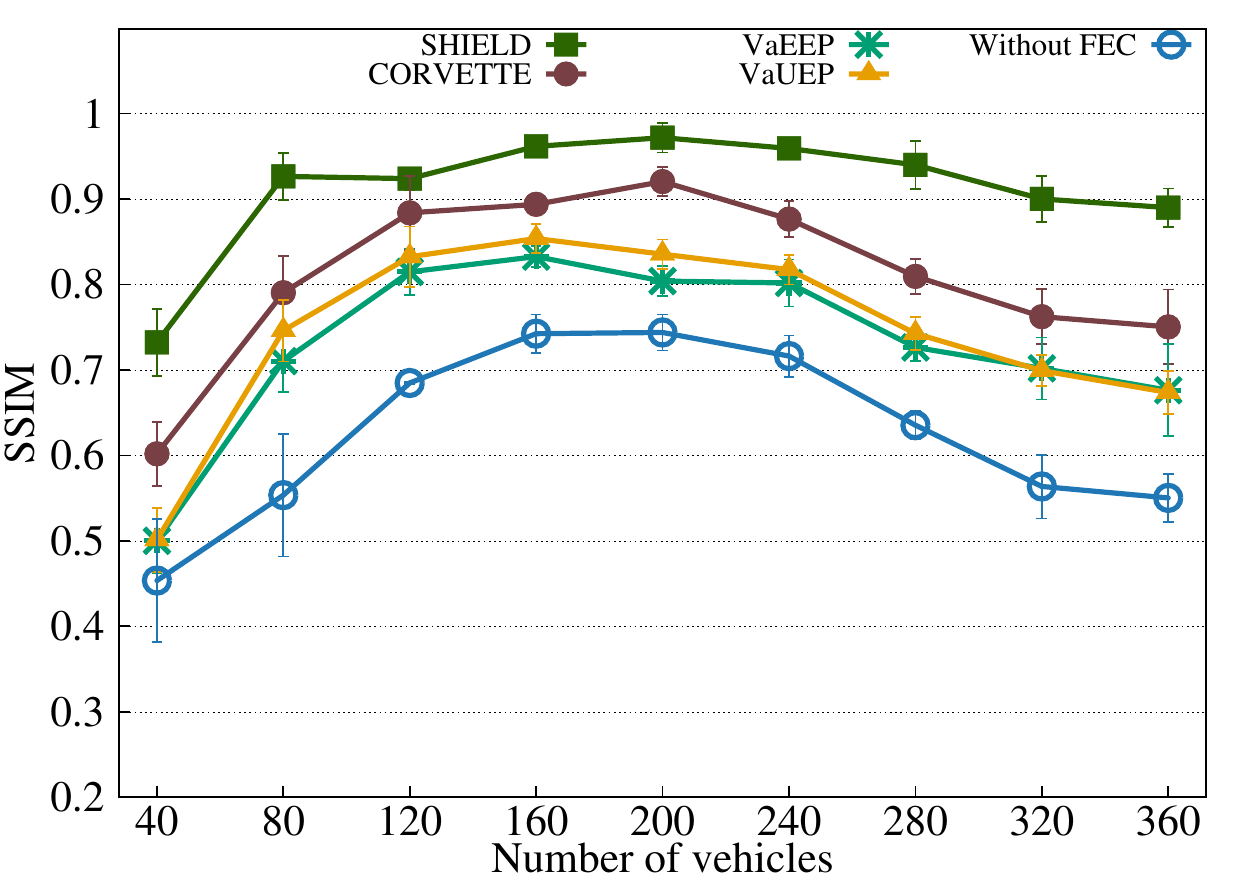}
			\fi &
			\ifBW \includegraphics[width=72.5mm]{./Fig12_shield_gray-eps-converted-to.pdf} 
			\else \includegraphics[width=72.5mm]{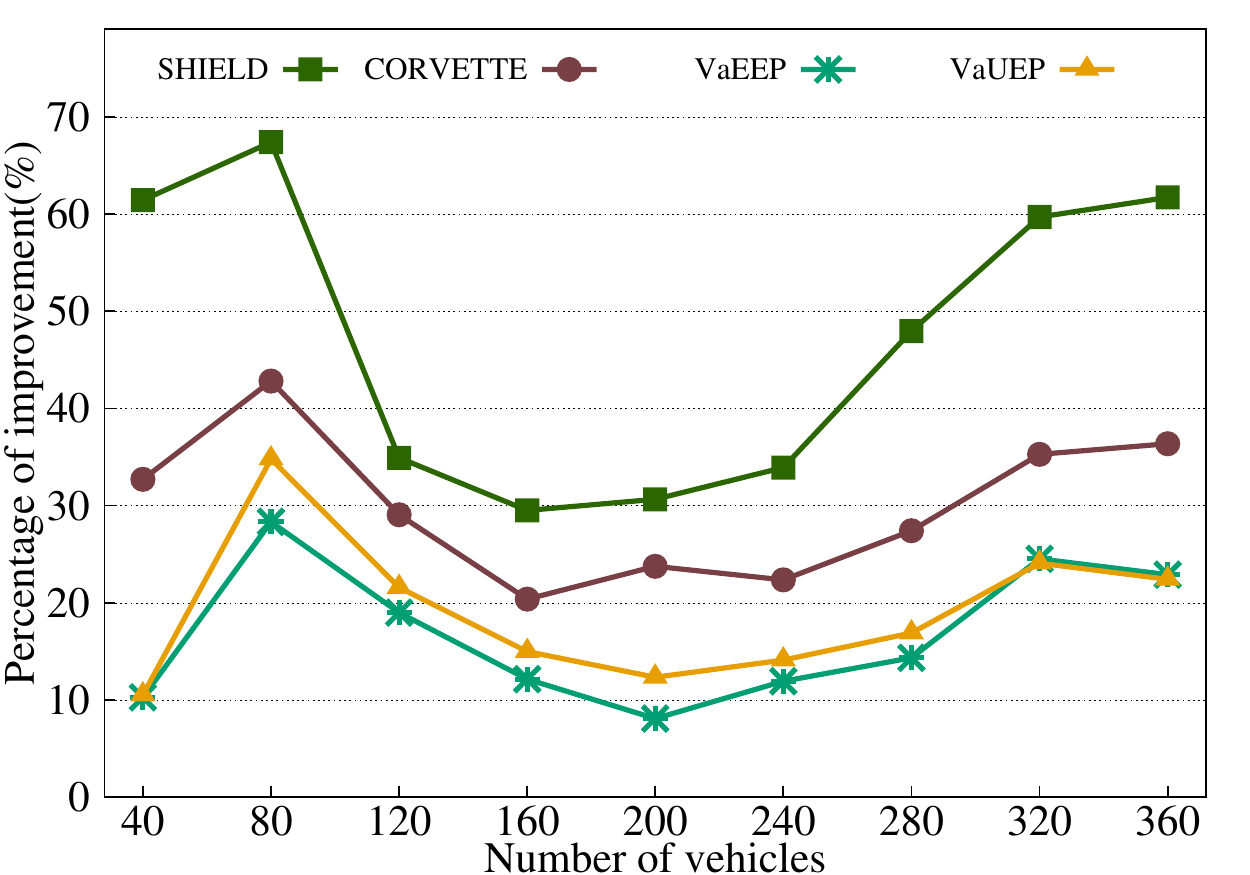}
			\fi \\
			(a) Average SSIM & (b) Percentage of SSIM improvement \\
		\end{tabular}
		\caption{\small SSIM assessment of the highway scenario}
		\label{fig:shield:SSIMHighway}
	\end{center}
\end{figure*}

In addition to the urban scenario, a highway environment was also assessed with both SSIM and VQM metrics. Figure~\ref{fig:shield:SSIMHighway} shows the QoE assessment, whereas~(a) depicts the average SSIM and~(b) shows the percentage of improvement achieved by the mechanism against the baseline. In~(a), the first thing to be noticed is that the QoE results are closer to one another in this environment. This happens because the network conditions are not as harsh as in the urban scenario. At first, there are some connectivity issues when the network is sparse, e.g., 40 vehicles. After this threshold, a better video quality is being provided. The best results are evidenced for 120 and 240 vehicles. 
In~(b), it is possible to notice that the highest improvements are reached when connectivity issues affect the network, for example, when the number of deployed vehicles is 40 and 80. 
In addition, major improvements are also perceived when there is a higher level of interference, such as above 280 vehicles. Here again, the SHIELD mechanism outperforms all its competitors.

\begin{figure*}[!htb]
	\begin{center}
		\setlength\tabcolsep{0.5pt}
		\begin{tabular}{cc}
			\ifBW \includegraphics[width=72.5mm]{./Fig13_shield_gray-eps-converted-to.pdf} 
			\else \includegraphics[width=72.5mm]{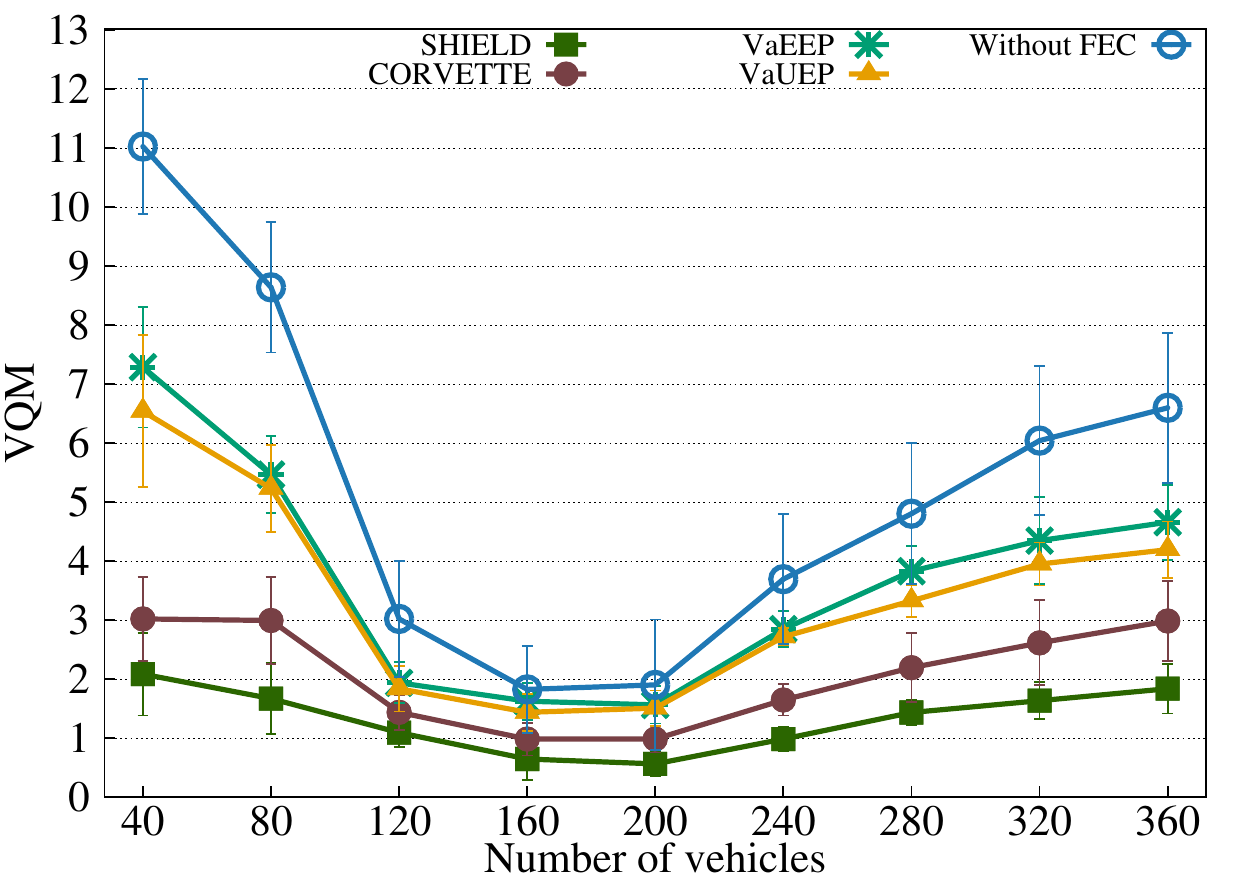}
			\fi &
			\ifBW \includegraphics[width=72.5mm]{./Fig14_shield_gray-eps-converted-to.pdf} 
			\else \includegraphics[width=72.5mm]{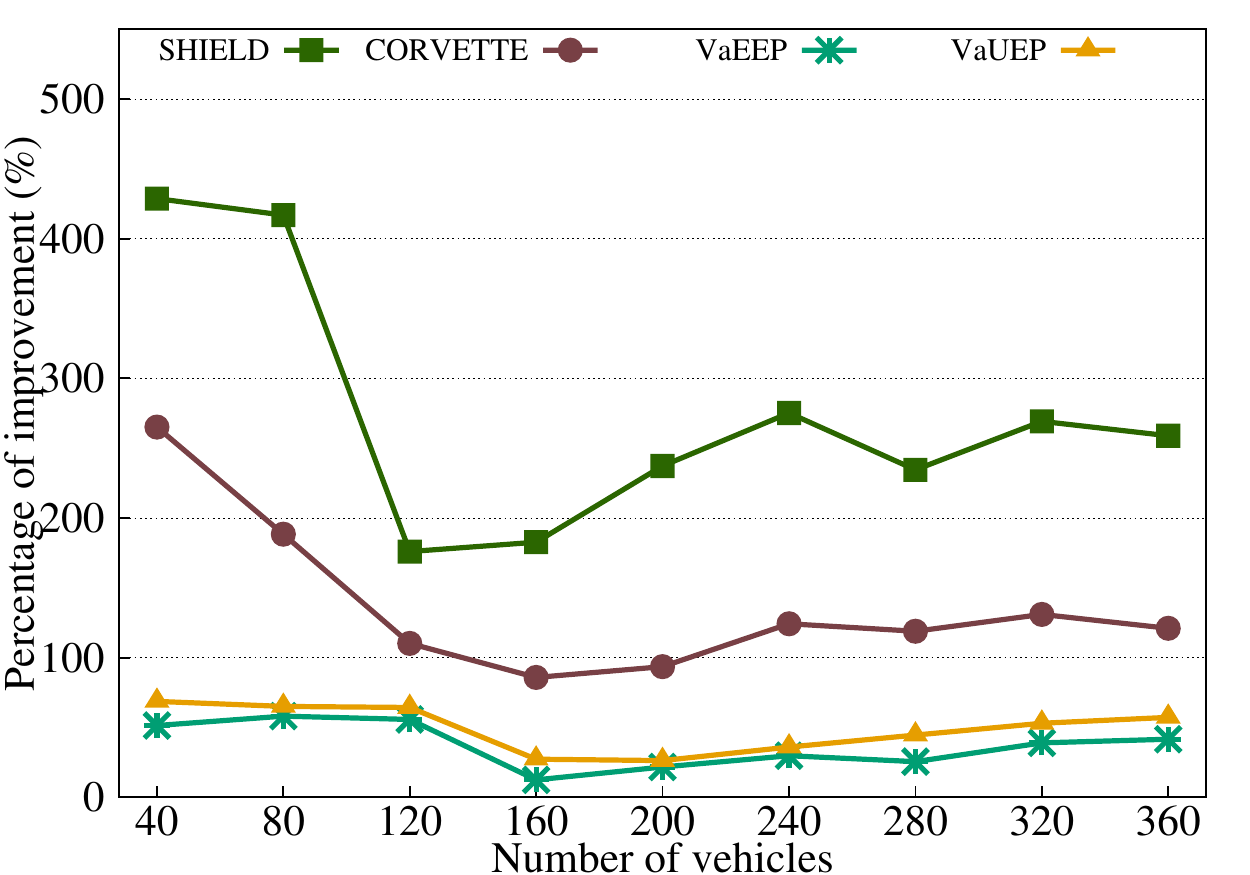} 
			\fi \\
			(a) Average VQM & (b) Percentage of VQM improvement			
		\end{tabular}
		\caption{\small VQM assessment of the highway scenario}
		\label{fig:shield:VQMHighway}
	\end{center}
\end{figure*}

The average VQM is shown in Figure~\ref{fig:shield:VQMHighway}(a) and the percentage of VQM improvement by each mechanism is shown in~(b). In~(a), the results follow the same tendency as the SSIM scores. This means that the VQM results are also closer to one another, especially above 120 vehicles. This is evidenced because the highway environment is not as rough as the urban setting. In~(b), it is clear that the highest percentage of improvement is achieved when the nodes are sparse. 
This means that there are connectivity issues in the network, e.g., for 40 and 80 vehicles. After this threshold, the network conditions improve and the enhancements provided by the mechanisms decrease. Nevertheless, the SHIELD mechanism is able to surpass the competitors. 

\subsubsection{Network footprint analysis}

\begin{figure*}[!htb]
	\begin{center}
		\setlength\tabcolsep{0.5pt}
		\begin{tabular}{cc}
			\ifBW \includegraphics[width=72.5mm]{./Fig15_shield_gray-eps-converted-to.pdf} 
			\else \includegraphics[width=72.5mm]{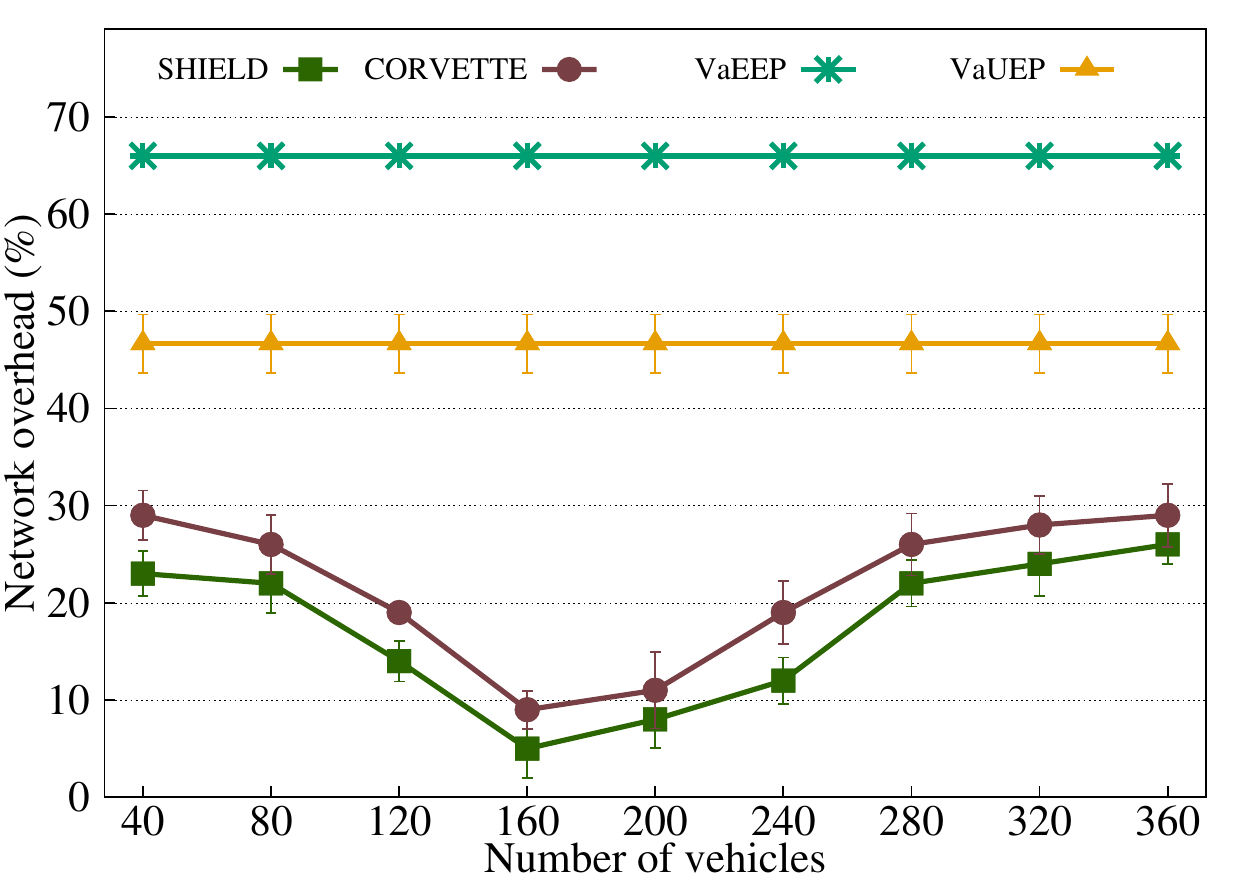}
			\fi &
			\ifBW \includegraphics[width=72.5mm]{./Fig16_shield_gray-eps-converted-to.pdf} 
			\else \includegraphics[width=72.5mm]{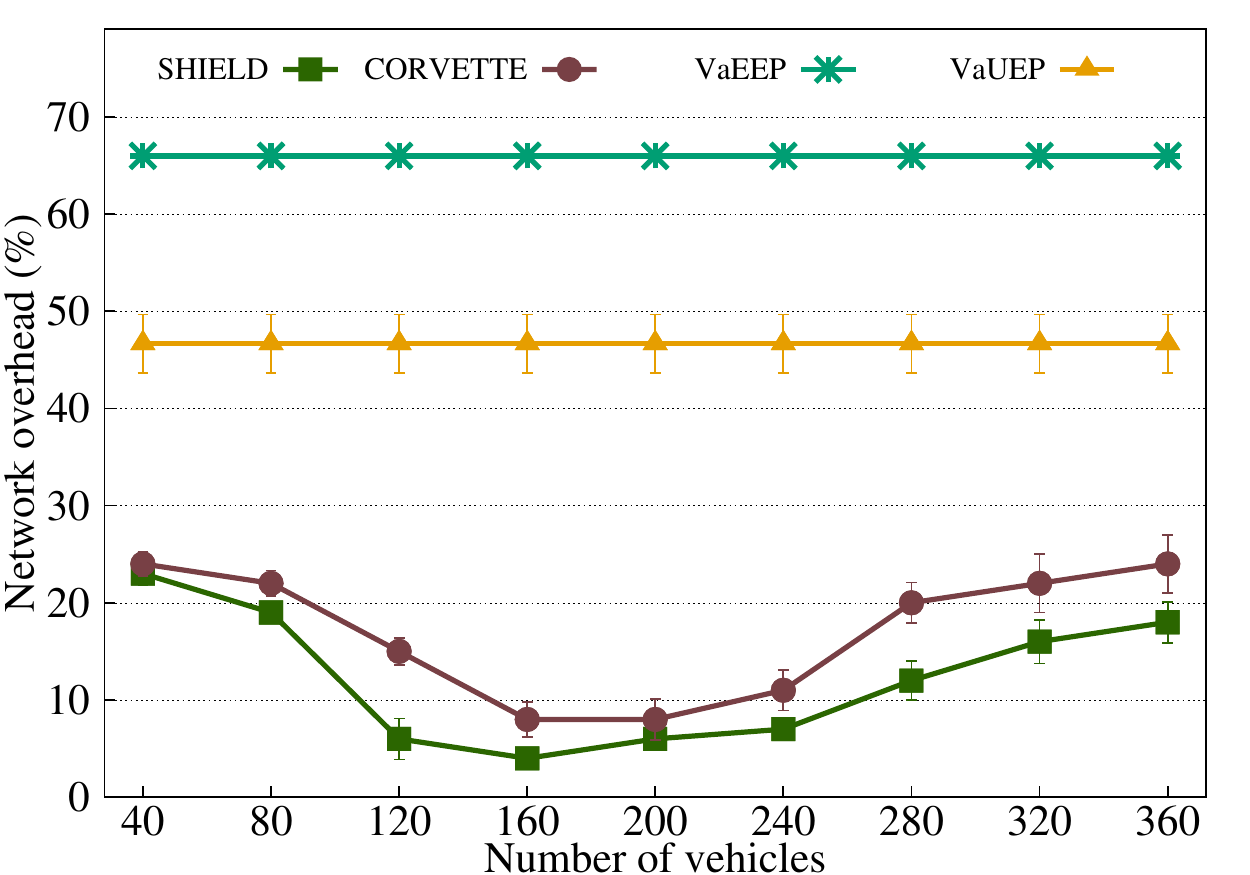}
			\fi \\
			(a) Urban scenario & (b) Highway scenario
		\end{tabular}
		\caption{\small SHIELD Network overhead}
		\label{fig:shield:netOverhead}
	\end{center}
\end{figure*}

Figure~\ref{fig:shield:netOverhead} shows the network overhead of all mechanisms in both~(a) urban and~(b) highway environments. The non-adaptive VaEEP and VaUEP schemes yield a constant network footprint in both scenarios because they do not adapt the amount of redundancy according to the network conditions. As depicted in the graph, these non-adaptive schemes add a considerably larger amount of redundancy. On top of that, the protection is not very efficient because, in the VaEEP case, the protection is added equally to all video data. As highlighted before, not all video packets need the same degree of protection. To tackle this issue VaUEP considers the frame type to add a specific amount of redundancy. This results in less network overhead and, at the same time, improves the video quality. 

The VaEEP mechanism does not have a standard deviation because it uses a unique and pre-defined amount of redundancy, which is applied equally to all videos. The VaUEP mechanism also has a pre-defined amount of redundancy, but it is not unique. This means that each type of video frame will have a specific amount of redundancy. Additionally, each video frame has a different size, leading to a variation in the amount of redundancy, and thus, a standard deviation is displayed.

The adaptive mechanisms, CORVETTE and SHIELD, were able to produce a lower network overhead, improving the wireless resources usage. In both mechanisms, when the network condition is better the footprint decreases. In the urban environment, this is evidenced when the simulation has 160 vehicles. The SHIELD mechanism produces a network overhead of only 5\%, while CORVETTE is producing 9\%. In the highway environment, the best conditions are experienced between 120 and 240 vehicles. The overhead produced by the SHIELD mechanism was between 4\% and 7\%, while CORVETTE is producing between 8\% and 15\%. 

On average, the SHIELD mechanism added 20\% less overhead in the urban environment and 28\% less in the highway scenario, in comparison to the CORVETTE mechanism. When compared to the VaEEP and VaUEP mechanisms, the SHIELD mechanism produced 73\% and 63\% less overhead, respectively in the urban scenario. In the highway scenario, the network overhead downsize was 81\% and 73\%, respectively. In the end, the proposed mechanism was able to produce a tailored protection, enabling a higher video quality and lower network overhead.

\subsubsection{Overall results}

The overall results demonstrate that the SHIELD mechanism outperforms all its competitors as showed in Table~\ref{tab:shield:Sumary}, which summarises the average SSIM, VQM, and the network footprint. 
The SHIELD mechanism enables downsizing the network footprint in both urban and highway environment. 
This stems from the fact that a tailored amount of redundancy, based upon video characteristics and the network conditions, is added to each video sequence, preventing any unnecessary redundancy. 
Furthermore, the proposed mechanism also enhanced the quality of the video delivered, thus providing higher QoE for the end-users.

\begin{table}[!h]
	{\small
		\caption{Average SSIM, VQM, and network footprint}
		\begin{center}
				\begin{tabular}{l|c|c|c|c|c}
					\hline
					& \multicolumn{1}{|l}{\textbf{SHIELD}} & \multicolumn{1}{|l}{\textbf{CORVETTE}} & \multicolumn{1}{|l}{\textbf{VaUEP}} & \multicolumn{1}{|l}{\textbf{VaEEP}} & \multicolumn{1}{|l}{\textbf{Without FEC}} \\
					\hline
					\hline
					\multicolumn{6}{c}{\textbf{Urban environment}} \\
					\hline
					{SSIM} & 0,895 & 0,787 & 0,701 & 0,684 & 0,551 \\
					\hline
					{VQM} & 1,459 & 2,441 & 4,034 & 4,323 & 6,688 \\
					\hline
					{Overhead} & 17,333\% & 21,778\% & 46,660\% & 65,984\% & -- \\   
					\hline
					\multicolumn{6}{c}{\textbf{Highway environment}} \\
					\hline
					{SSIM} & 0,911 & 0,809 & 0,744 & 0,729 & 0,627  \\
					\hline
					{VQM} & 1,328 & 2,095 & 3,414 & 3,728 & 5,281 \\
					\hline
					{Overhead} & 12,333\% & 17,112\% & 46,660\% & 65,984\% & -- \\   
					\hline
				\end{tabular}
			\label{tab:shield:Sumary}
		\end{center}
	}
\end{table}
\section{Summary}

This chapter described and assessed two proposed mechanisms to increase the video transmission resiliency over VANETs. 
The CORVETTE mechanism detailed in Section~\ref{sec:corvette} enabled safeguarding video delivery over highly dynamic networks, as well as improving the QoE perceived by end-users.
In the VQM assessments, it was able to surpass the AdaptFEC mechanism presenting results with 30\% higher QoE.
When it comes to the SSIM assessment, it generated videos with 11\% higher quality. 
This value was reached in conjunction with a 41\% decrease on the network overhead, which proves to be a good overall result.
Regardless of the CORVETTE mechanism satisfactory results, there was room for improvement in the way it handled the network-related parameters.

In order to improve on the aforementioned issues, the SHIELD mechanism, described in Section~\ref{sec:shield}, was proposed.
In the urban environment, SHIELD achieved between 27\% and 57\% better QoE in comparison to the CORVETTE mechanism. On the same scenario, it produced only 17\% of network overhead, at the same time, CORVETTE was producing 22\%.
In the highway environment, the SHIELD QoE improvement was between 20\% and 87\% better than CORVETTE, and also producing less network overhead, only 12\% in comparison to the 17\% of CORVETTE.
The assessment results clearly show that the latter mechanism was able to outperform the other competitors, including the CORVETTE mechanism.
This corroborates that the improvements made to the network module result in both, better QoE and lower network footprint.

The following publications resulted from the work carried out in this chapter:

\bigbreak
\textbf{Journal paper:}
\begin{itemize}

	\item {Immich}, R. and Cerqueira, E. and Curado, M., ``\textbf{Shielding video streaming against packet losses over VANETs}'', The Journal of Mobile Communication, Computation and Information, Wireless Networks, Volume 22, Issue 8, pp 2563-2577, Springer, 2015
	
\end{itemize}

\bigbreak

\textbf{Conference papers:}
\begin{itemize}

	\item {Immich}, R. and Cerqueira, E. and Curado, M., ``\textbf{Towards a QoE-driven Mechanism for Improved H.265 Video Delivery}'', in the 15th IFIP Annual Mediterranean Ad Hoc Networking Workshop (MED-HOC-NET), 2016

	\item {Immich}, R. and Cerqueira, E. and Curado, M., ``\textbf{Adaptive QoE-driven video transmission over Vehicular Ad-hoc Networks}'', in the IEEE Conference on Computer Communications Workshops (INFOCOM), 2015

	\item Cerqueira, E. and Curado, M. and Neto, A. and Riker, A. and {Immich}, R. and Quadros, C., ``\textbf{A Mobile QoE Architecture for Heterogeneous Multimedia Wireless Networks}'', in the GC'12 Workshop: The 4th IEEE International Workshop on Mobility Management in the Networks of the Future World (MobiWorld), 2012
	
\end{itemize}

\setcounter{mtc}{15}
\chapter{Conclusions and Future Work}
\label{ch:Conclusions}

\dictum[Aldous Huxley, Brave New World]{As if one believed anything by instinct! One believes things because one has been conditioned to believe them.}

\minitoc

\lettrine[lines=3]{\color{gray}\bf{T}}{} he video delivery over wireless networks is growing apace, every day hundreds of thousands of users and companies produce, share, and consume a vast amount of this type of content.
This is an important way to convey educational information and news reports, as well as the most varied forms of entertainment.
Despite numerous efforts on this field, several open issues still remain to be solved.
This thesis proposed several mechanisms to improve the quality of the video transmissions in diverse wireless networks contributing to the advancement of the state-of-the-art in this topic.
This chapter revises the addressed problems emphasising the major contributions accomplished in this thesis.
At the end, future research opportunities and directions are outlined.

\section{Synthesis of the Thesis}
\label{sec:con:synthesis}

An effective method for making video transmission resilient to losses is of critical importance for the success of video transmission over wireless networks, this is even more evident in multi-hops networks.
Due to the recent technological advances, the end-users of today desire high-quality video stream, services that do not follow this tendency will have poor acceptance. 
However, to guarantee a video transmission with high perceived quality is a complex task that depends on a multitude of elements. 

One of the most evident parameter that has an impact on the video quality are the network-related details.
Taking this into consideration, Chapter~\ref{ch:networkStuff} presented an overview of the network technologies involved in this thesis. 
It started with a review of cross-layer techniques, which are necessary to provide the flexibility needed to implement any network transmission optimisation.
After that, particular details of each multi-hop wireless networks are explored, such as the routing protocols and channel characteristics, as well as the mobility and traffic patterns.

In addition to the network-related features, the video characteristics also play an important role on how the perceived quality is impaired by losses.
In the light of this, Chapter~\ref{ch:videoStuff} discussed several video-related peculiarities that are known to have an impact on the quality, such as the video format, the type of the frames and the hierarchical structure of the video format, as well as the motion intensity classification.
It is also reviewed how the human vision system perceives the impairments and the assessment approaches to quantify the video quality.

Under the given conditions, several mechanisms have been proposed to tackle the open issues related to the video transmission.
Chapter~\ref{ch:rw} presented the most relevant ones in the literature. 
The main characteristics, advantages, and weakness of each one were discussed. 
In addition, this chapter also described the off-the-shelf techniques to improve the video transmission quality and why they are not enough to ensure best possible perceived quality to end-users.
Owing to the open issues in the literature review the main contributions of this thesis were proposed.

Chapter~\ref{ch:MESH} introduced three proposed mechanisms to shield video transmissions.
These mechanisms adopted different approaches, such as how to classify the motion intensity and how to compute the most suited amount of redundancy, to enable high-quality video delivery over wireless networks.
Several techniques are used to this end, e.g., a heuristic method, random neural networks, and ant colony optimisation.
The results were assessed using quality of experience metrics, as well as the network footprint impact.

Chapter~\ref{ch:UAV} corresponds to the advancements made on improving video transmissions using unmanned aerial vehicles.
The proposed mechanisms take into consideration several video characteristics, such as the frame type, the motion vectors, the spatial video complexity, and the temporal video intensity, as well as been capable of handling videos with arbitrary size.
In addition, fuzzy logic methods are used to both classify the video details and to set the redundancy amount.
In the same way as before, the experiments were assessed using quality of experience metrics.

Chapter~\ref{ch:VANET} specified the achievements on enhancing video delivery over vehicular ad-hoc networks.
The mechanisms proposed in this chapter contemplated both the video characteristics and the intrinsic details of this type of network, including even the prediction of future packet loss rates.
This was only possible by the combined use of several methods, such as fuzzy logic and hierarchical fuzzy logic, as well as convex hull and approximated convex hull algorithms.

\section{Contributions}
\label{sec:con:contrib}

The main goals of this thesis were delineated in Chapter~\ref{ch:Introduction} and are briefly revisited here. 
One objective was to define a method to provide video content characterization.
This should be done according to the motion intensity and details of each video sequence.
Another objective was to propose a series of adaptive FEC-based content- and video-aware mechanisms. 
The objective of these mechanisms is to shield video transmission over wireless networks providing both high QoE and low network overhead.
The third general goal was to study the performance of the proposed mechanisms in different network environments, such as WMNs, FANETs, and VANETs.

These goals have tailored the work presented in this thesis and have led to the following contributions:

\begin{description}

	\item[\protect{%
		\parbox[t][2\baselineskip][t]{\textwidth}{
			Contribution 1, Assessing the impact of the video characteristics on the video quality level}
	}
	]
	This is a transversal contribution, which is present in all mechanisms described in Chapters~\ref{ch:MESH}, \ref{ch:UAV}, and~\ref{ch:VANET}.
	The proposed method uses several video characteristics to assess the impact of them in the perceived video quality.
	The video details used were the codec type, the frame type and size, the length and format of the group of pictures as well as the motion vectors.
	According to this information, the proposed mechanisms were able to correctly identify the most important parts of the video sequences.

	\item[Contribution 2, A method to characterise the motion intensity of videos]~\\
	This is the second transversal contribution in this thesis that is also used by all mechanisms presented in Chapters~\ref{ch:MESH}, \ref{ch:UAV}, and~\ref{ch:VANET}.
	In the proposed method, the video content is put through a series of procedures to characterise the intensity pace.
	This is important because different intensities have distinct impacts on the perceived quality if some information is lost.
	The proposed mechanisms use this knowledge to better protect the most QoE-sensitive data, which allows providing high perceived quality without unnecessary network overhead.

	\item[Contribution 3, Heuristic mechanism]~\\
	This contribution was achieved in Chapter~\ref{ch:MESH}, Section~\ref{sec:viewfec}.
	The main goal of the proposed mechanism was to provide a practical solution with satisfactory results.
	In order to do that, the mechanism used human experience and knowledge to delineate several strategies to shield the video transmission against losses. 
	The results show that it was possible to maintain or even increase the video quality without adding unnecessary network redundancy. 

	\item[Contribution 4, Random neural networks mechanism]~\\
	Chapter~\ref{ch:MESH}, more specifically Section~\ref{sec:neuralFEC}, describe how this contribution was achieved.
	In the proposed mechanism, random neural networks were used to both categorise the video content according to its motion intensity and to set, in real-time, an adaptive amount of redundancy.
	The mechanism has a training and validation phase, which is offline and after that, it is able to perform at run-time.
	The QoE assessment shows that it outperform others mechanisms, generating also, less network overhead.

	\item[Contribution 5, Ant colony optimisation mechanism]~\\
	Section~\ref{sec:PredictiveAnts} describes the mechanism built to support this contribution.
	It uses an ant colony metaheuristic to assess and classify several video characteristics as well as network conditions, which allows choosing a precise amount of redundancy.
	In doing this, the perceived video quality is kept at the maximum level possible while the network footprint stays low.

	\item[Contribution 6, Fuzzy logic mechanism]~\\
	This contribution is spread in Chapters~\ref{ch:UAV} and~\ref{ch:VANET}.
	Several of the proposed mechanisms use Fuzzy logic as well as Hierarchical fuzzy logic.
	These techniques were broadly adopted because of their flexibility, allowing the mechanisms to better adapt to a variety of situations.
	They are also easy to reconfigure and appropriate to manage abstract concepts.
	Additionally, it allows building complex systems, which are able to handle a large number of inputs at low computation costs.
	The proposed mechanism's results show that they outperform other competitors, producing less network overhead while increasing the QoE for end-users.
	
\end{description}

The following subsection outlines further research directions in the fields addressed in this thesis.

\section{Future Work}
\label{sec:con:future}

The relevance and the applicability of mechanisms to improve the quality of video transmissions have been discussed in this thesis, highlighting the need for reliable solutions. 
The methods and procedures contemplated in this work have provided motivating results, as well as showing the ability to outperform existing solutions in a varied of scenarios. 
However, there are still several upgrades that could be addressed in the future including, but not limited to, concurrent multipath transmission, opportunistic routing, network coding, and path interleaving.

A distinct aspect that could be addressed is the impact of the proposed mechanisms in an environment where they need to compete for resources against other applications. 

The adoption of a smart retransmission scheme could also be considered. 
Such technique could use a number of information in the decision-making processes, assessing whether it is worth to retransmit a specific packet. 
The information can range from video characteristics, such as frame type, motion activity and play-out time, and network information, such as delay and link quality, to FEC-related details, for example, what is the minimum number of packets needed to reconstruct the original data. 
In doing that, it would enable the retransmission of only what is really needed from the user-point-of-view, respecting the strict constraints commonly found in these services. 

Additionally, owing to the ever growing number of wireless-capable devices and the broader amount of communication technologies that they can use, future mechanisms will have to accommodate these heterogeneous networks and devices allowing them to coexist while providing satisfactory levels of user satisfaction. 
In multihoming environments, where a single device with several interfaces can be connected to multiple communication networks, the necessity of mechanisms to assess the multiple link qualities and adjust the transmission parameters is even more evident. 
In this context, wireless mobile telecommunications technologies such as Long-Term Evolution Advanced~(LTE-Advanced) and 5th Generation Networks~(5G) can be used together with Wireless Local Area Network~(WLAN). 
The mechanisms should consider each communication channel's characteristic to adapt the video transmissions. 
Furthermore, load-balancing schemes can also be considered to further improve the reliability of the transmissions. 

Along with the wireless communication, the underlying network infrastructure is also changing. 
The adoption of Software-defined network~(SDN) and Cloud based systems provides a way to build dynamically adaptable and cost-effective networks by decoupling the network control from the forwarding functions. 
However, one of the main challenges is how to autonomic configure these networks and protect the transmissions in order to provide the best possible service while not using unnecessary network resources. 
In order to achieve that, several link quality indicators, such as the bit-error rate and signal-to-noise and distortion ratio, as well as mobility patterns, signal strength, and the importance of the data that is being transmitted have to be taken into consideration to programmatically configure the network control layer. 
This allows maximising the resources usage in a quick and dynamic fashion leading to a better transmission quality. 

Finally, sustainable computing with energy-aware devices has gained importance in recent years. 
The improvement of the overall energy-efficient video transmission over wireless networks can be accomplished, for example, by minimising the amount of redundant information, as proposed in this thesis and thus reducing the network activity. 
However, there are several other ways to improve on this issue. 
In a multihoming environment, for instance, it is possible to take into consideration the specific wireless technology's energy consumption when choosing the more appropriated communication method. 
Furthermore, it is also possible to incorporate the end-user preferences and behaviours to adapt the video transmission mechanisms in order to find the most desirable energy/quality ratio.


%
%
\bibliographystyle{apalike}
\bibliography{./myReferences}

\end{document}